\documentclass[a4paper,fleqn,usenatbib]{mnras}

\usepackage{newtxtext,newtxmath}

\UseRawInputEncoding

\usepackage[T1]{fontenc}
\usepackage{ae,aecompl}
\usepackage{pdflscape}
\usepackage{rotating}
\usepackage{tablefootnote}
\usepackage{longtable}
\usepackage{threeparttable}
\usepackage{caption}
\usepackage{comment}

\usepackage{graphicx}	
\usepackage{amsmath}	
\usepackage{amssymb}	
\usepackage{xcolor}
\usepackage[LGRgreek]{mathastext}

\title[Methanol in proto-BDs]{First observations of warm and cold methanol in Class 0/I proto-brown dwarfs}

\author[Riaz, Thi, Machida]{
Riaz, B.$^{1}$\thanks{E-mail: briaz@usm.lmu.de}
Thi, W.-F.$^{2}$
Machida, M. N.$^{3}$
\\
$^{1}$ Universit\"{a}ts-Sternwarte M\"{u}nchen, Ludwig Maximilians Universit\"{a}t, Scheinerstra$\beta$e 1, 81679 M\"{u}nchen, Germany
\\
$^{2}$ Max-Planck-Institut f\"{u}r Extraterrestrische Physik, Giessenbachstrasse 1, 85748 Garching, Germany
\\
$^{3}$ Department of Earth and Planetary Sciences, Faculty of Sciences, Kyushu University, Fukuoka, Japan
\\
}

\date{Accepted XXX. Received YYY; in original form ZZZ}

\pubyear{2022}

\begin{document}
\label{firstpage}
\pagerange{\pageref{firstpage}--\pageref{lastpage}}
\maketitle

\begin{abstract}

We present results from the first molecular line survey to search for the fundamental complex organic molecule, methanol (CH$_{3}$OH), in 14 Class 0/I proto-brown dwarfs (proto-BDs). IRAM 30-m observations over the frequency range of 92-116 GHz and 213-280 GHz have revealed emission in 14 CH$_{3}$OH transition lines, at upper state energy level, E$_{upper}\sim$7-49 K, and critical densities, $n_{crit}$ of 10$^{5}$ to 10$^{9}$ cm$^{-3}$. The most commonly detected lines are at E$_{upper} <$ 20 K, while 11 proto-BDs also show emission in the higher excitation lines at E$_{upper}\sim$21-49 K and $n_{crit}\sim$10$^{5}$ to 10$^{8}$ cm$^{-3}$. In comparison with the brown dwarf formation models, the high excitation lines likely probe the warm ($\sim$25-50 K) corino region at $\sim$10-50 au in the proto-BDs, while the low-excitation lines trace the cold ($<$ 20 K) gas at $\sim$50-150 au. The column density for the cold component is an order of magnitude higher than the warm component. The CH$_{3}$OH ortho-to-para ratios range between $\sim$0.3-2.3. The volume-averaged CH$_{3}$OH column densities show a rise with decreasing bolometric luminosity among the proto-BDs, with the median column density higher by a factor of $\sim$3 compared to low-mass protostars. Emission in high-excitation (E$_{upper}>$ 25 K) CH$_{3}$OH lines together with the model predictions suggest that a warm corino is present in $\sim$78\% of the proto-BDs in our sample. The remaining show evidence of only the cold component, possibly due to the absence of a strong, high-velocity jet that can stir up the warm gas around it.

\end{abstract}


\section{Introduction}
\label{intro}

Complex organic molecules (COMs) have been widely observed across different stages of star formation towards high-mass and low-mass star-forming regions (e.g., Ceccarelli et 2022). Their chemistry is of great interest because organic molecules are the proposed starting point of an even more complex prebiotic chemistry during star and planet formation, thus linking interstellar chemistry with the origins of life (e.g., Ceccarelli et 2022, Mathew et al. 2022, \"{O}berg et al. 2009, M\"{u}noz Caro et al. 2014). Methanol (CH$_{3}$OH) is a fundamental COM and a precursor of more complex COMs. 

Purely gas-phase models cannot reproduce the abundance of methanol in dark clouds (e.g., Garrod et al. 2006). Gas-grain models have shown that under the physical conditions prevalent in dense and cold molecular clouds and pre-stellar cores, with n(H$_{2}$)$\geq$10$^{5}$ cm$^{-3}$ and T $\sim$10 K, methanol can efficiently form on ice grains via surface chemistry and is then released into the gas phase via thermal and/or non-thermal desorption mechanisms (e.g., Tielens et al. 1991; Charnley et al. 1992; Watanabe \& Kouchi 2002; Hidaka et al. 2004). In chemical models that consider only grain surface chemistry, the icy mantles covering the interstellar grains are processed by UV/cosmic ray irradiation forming radicals that react and form methanol and other COMs at warm temperatures of $\sim$20-30 K (e.g., Garrod et al. 2008). At cooler temperatures of $\sim$10-20 K, non-diffusive processes have been proposed for radicals to react and form methanol and other COMs (e.g., Garrod \& Hebst 2006; Garrod et al. 2008; Jin \& Garrod 2020). 

The main challenge in the chemical models is to understand the thermal and non-thermal desorption mechanisms at low temperatures that are able to remove COMs from the grain mantles and inject them into the gas-phase where they are detected (e.g., Herbst \& Garrod 2022). Thermal evaporation is extraordinarily slow and obeys a standard Boltzmann law using desorption energy barriers for species heavier than hydrogen and helium typically on the order of 1000 K, or 0.1 eV. In case of methanol, the desorption or the binding energy is $\sim$3770--8618 K (e.g., Minissale et al. 2022). Dark cloud temperatures ($\sim$10 K) are far too cold for this mechanism to be very significant, and thus non-thermal desorption processes must be more effective.

In the inner regions of highly accreting objects with accretion to bolometric luminosity ratios of $\sim$50\%-70\%, dust grains can be heated to $>$20-40 K. Only weakly-bound species such as CO ice can desorb. Non-thermal desorption mechanisms such as photo-desorption triggered by absorption of UV photons and cosmic-rays induced sputtering are required. However for methanol, the UV absorption does not result in intact methanol molecules but rather in smaller species (Cruz-Diaz et al. 2016). Species such as ions and radicals resulting from UV absorption that remain in the grain mantle can react further to form more complex species. Sputtering by strong jets and shock emission knots can spot raise the grain temperature and inject frozen intact methanol into the gas-phase (e.g., Draine 1995, Wakelam et al. 2021, Paulive et al. 2022). 

An additional desorption mechanism is the excess formation energy of newly formed species that allows them to desorb immediately into the gas phase, the so-called chemidesorption (Minissale et al. 2016, Cazaux et al. 2016). The exothermicity of the hydrogenation reaction that ultimately forms methanol on grains, CH$_{2}$OH + H $\rightarrow$ CH$_{3}$OH, is $\sim$4.1 eV, which is about an order of magnitude larger than the desorption energies of the products ($\sim$0.18 eV). Laboratory experiments performed for a standard dark cloud model conditions have shown that only about 1\% of the chemically desorbed methanol can reproduce the observed gas phase methanol levels (e.g., Garrod et al. 2006).

We present here results from the first molecular line survey to search for methanol in early-stage Class 0/I proto-brown dwarfs (proto-BDs), with an aim to understand the chemical complexity during the early phases of brown dwarf formation. Proto-BDs are high-density (n(H$_{2}$)$\geq$10$^{5}$ cm$^{-3}$) and cold ($\leq$10 K) objects and are also predicted to posses a warm corino ($\sim$20-70 K) in the inner dense regions (Machida et al. 2009; 2019) that can be observed using high-excitation molecular lines. Our recent survey of deuterium chemistry in proto-BDs has revealed emission in various deuterated species, including the doubly-deuterated D$_{2}$CO molecule (Riaz \& Thi 2022abc). These surveys have shown that proto-BDs show enhanced deuterium fractionation and high molecular D/H ratios in the range of $\sim$0.02-4.5. Low temperature and high densities can significantly enhance the molecular D/H ratio compared to the ISM elemental value ($\sim$2.8$\times$10$^{-5}$). The high ratios for the proto-BDs thus confirm a cold and dense environment in these objects. Since methanol is proposed to form on the surfaces of interstellar dust grains and in icy grain mantles, the physical conditions in proto-BDs should be ripe enough for the efficient formation of this fundamental COM, and the predicted presence of a warm corino should result in thermal/non-thermal desorption of the methanol thus formed to the gas phase.

Section~\ref{obs} describes the sample and the IRAM 30m observations of multiple methanol transition lines. Data analysis and results are presented in Sections~\ref{analysis} and~\ref{results}. Section~\ref{discussion} presents a comparison of the observations with the predictions from the brown dwarf physical model, a discussion on the CH$_{3}$OH ortho-to-para ratio, and the dependence of the methanol column densities on the extent of CO depletion and the bolometric luminosity of the system.

\section{Sample, Observations, Data Reduction}
\label{obs}

\begin{table*}
\centering
\caption{Sample}
\label{sample}
\begin{threeparttable}
\begin{tabular}{lcccccc} 
\hline
Object & L$_{bol}$ (L$_{\sun}$)\tnote{a} & Classification\tnote{b} & Region \\
\hline
SSTc2d J182854.9+001833 (J182854) &  0.05 &  Stage 0+I, Stage I, Class 0/I & Serpens \\ 
SSTc2d J182844.8+005126 (J182844) &  0.04 &  Stage 0+I, Stage 0, Class 0/I & Serpens \\ 
SSTc2d J182959.4+011041 (J182959) &  0.008 &  Stage 0+I, Stage I, Class 0/I & Serpens \\ 
SSTc2d J182856.6+003008 (J182856) & 0.004 &  Stage 0+I, Stage 0, Class 0/I & Serpens \\ 
SSTc2d J182952.2+011559 (J182952) & 0.024 &  Stage 0+I, Stage 0, Class 0/I & Serpens \\ 
SSTc2d J163143.8-245525 (J163143) &  0.09 &   Stage 0+I, Stage I, Class Flat & Ophiuchus \\ 
SSTc2d J163136.8-240420 (J163136) &  0.09 &   Stage 0+I, Stage I, Class Flat & Ophiuchus \\ 
SSTc2d J163152.06-245726.0 (J163152) & 0.009 &  Stage 0+I, Stage 0, Class 0/I & Ophiuchus \\ 
SSTc2d J162625.62-242428.9 (J162625) & 0.04 &  Stage 0+I, Stage 0, Class 0/I & Ophiuchus \\ 
SSTc2d J032838.78+311806.6 (J032838) & 0.017 &  Stage 0+I, Stage I, Class 0/I & Perseus \\   
SSTc2d J032848.77+311608.8 (J032848) & 0.013 &  Stage 0+I, Stage I, Class 0/I & Perseus \\   
SSTc2d J032851.26+311739.3 (J032851) & 0.011 &  Stage 0+I, Stage 0, Class 0/I & Perseus \\  
SSTc2d J032859.23+312032.5 (J032859) & 0.006 &  Stage 0+I, Stage I, Class 0/I & Perseus \\   
SSTc2d J032911.89+312127.0 (J032911) & 0.06 &  Stage 0+I, Stage I, Class 0/I & Perseus \\   
\hline
\end{tabular}
\begin{tablenotes}
\item[a] Errors on L$_{bol}$ is estimated to be $\sim$20\%. 
\item[b] The first, second, and third values are using the classification criteria based on the integrated intensity in the HCO$^{+}$ (3-2) line, the physical characteristics, and the SED slope, respectively.
\end{tablenotes}
\end{threeparttable}
\end{table*}

Our sample consists of 14 Class 0/I proto-BDs (Table~\ref{sample}). A detailed discussion on the target selection, classification and measurements of mass, luminosity, H$_{2}$ number and column density for these objects can be found in Riaz et al. (2018) and Riaz \& Thi (2022a). The distance to the targets is 436$\pm$9 pc in Serpens, 144$\pm$6 pc in Ophiuchus, and 294$\pm$17 pc in Perseus (Ortiz-Leon et al. 2017a, b; Dzib et al. 2010; Mamajek 2008; Schlafly et al. 2014; Zucker et al. 2019). The observations were obtained at the IRAM 30-m telescope between 2017 and 2022. We used the EMIR heterodyne receiver (E090 and E230 bands) and the FTS backend in the wide mode, with a spectral resolution of 200 kHz. The observations were taken in the frequency switching mode with a frequency throw of approximately 7 MHz. The source integration times ranged from 3 to 4 hours per source per tuning reaching a typical RMS (on {\it T}$_{A}^{*}$ scale) of $\sim$0.01-0.03 K. The telescope absolute RMS pointing accuracy is better than 3$\arcsec$ (Greve et al. 1996). The absolute calibration accuracy for the EMIR receiver is around 10\% (Carter et al. 2012). The telescope intensity scale was converted into the main beam brightness temperature ({\it T}$_{mb}$) using standard beam efficiency of $\sim$76\% at 86 GHz and $\sim$57\% at 230 GHz. The spectral reduction was conducted using the CLASS software (Hily-Blant et al. 2005) of the GILDAS facility. The standard data reduction process consisted of averaging multiple observations, extracting a subset around the line rest frequency, and fitting a low-order polynomial baseline which was then subtracted from the average spectrum.

\begin{table*}
\centering
\caption{CH$_{3}$OH molecular line observations}
\label{lines}
\begin{threeparttable}
\begin{tabular}{lllll} 
\hline
Frequency  & Transition	& $E_{upper}$ & $A_{jk}$   & $n_{crit}^{thin}$	\\
(GHz)	  &			& (K) 		&  (s$^{-1}$) & (10, 20, 30, 40, 50 K) (cm$^{-3}$)   	\\
\hline

95.9143 & 2$_{12}$-1$_{11}$ A & 21.4 & 0.25$\times$10$^{-5}$ & (0.45, 0.47, 0.5, 0.53, 0.55)$\times$10$^{5}$	\\
96.7393 & 2$_{12}$-1$_{11}$ E & 12.5 & 0.26$\times$10$^{-5}$ & (0.8, 1.0, 1.2, 1.4, 1.6)$\times$10$^{7}$	\\
96.7414 & 2$_{02}$-1$_{01}$ A & 6.9 & 0.34$\times$10$^{-5}$ & (0.45, 0.48, 0.51, 0.54, 0.56)$\times$10$^{5}$\\
96.7445 & 2$_{02}$-1$_{01}$ E & 20.1 & 0.34$\times$10$^{-5}$ & (3.4, 3.1, 3.1, 3.4, 3.5)$\times$10$^{5}$	\\
97.5828 & 2$_{11}$-1$_{10}$ A & 21.6 & 0.26$\times$10$^{-5}$ & (1.3, 1.8, 2.6, 3.2, 3.8)$\times$10$^{6}$	\\
108.8939 & 0$_{00}$-1$_{11}$ E & 13.1 & 1.5$\times$10$^{-5}$ & (0.8, 1.1, 1.5, 1.8, 2.1)$\times$10$^{8}$	\\
218.4401 & 4$_{23}$-3$_{12}$ E & 45.5 & 4.7$\times$10$^{-5}$ & (4.7, 6.3, 7.7, 9.0, 10.2)$\times$10$^{6}$	\\
230.0270 & 3$_{21}$-4$_{14}$ E  & 39.8 & 1.5$\times$10$^{-5}$ & (1.15, 1.0, 1.1, 1.15, 1.25)$\times$10$^{8}$\\
239.7462 & 5$_{15}$-4$_{14}$ A & 49.1 & 5.6$\times$10$^{-5}$ & (8.6, 10.8, 12.4, 13.3, 14.0)$\times$10$^{6}$\\
241.7002 & 5$_{05}$-4$_{04}$ E & 47.9 & 6.0$\times$10$^{-5}$ & (1.4, 1.6, 1.7, 1.8, 1.9)$\times$10$^{6}$	\\
241.7672 & 5$_{15}$-4$_{14}$ E & 40.4 & 5.8$\times$10$^{-5}$ & (2.9, 3.9, 4.8, 5.8, 6.5)$\times$10$^{7}$	\\
241.7913 & 5$_{05}$-4$_{04}$ A & 34.8 & 6.0$\times$10$^{-5}$ & (0.4, 0.43, 0.46, 0.46, 0.5)$\times$10$^{6}$\\
254.0154 & 2$_{02}$-1$_{11}$ E & 20.1 & 2.0$\times$10$^{-5}$ & (3.0, 3.0, 3.3, 3.6, 4.0)$\times$10$^{7}$	\\
261.8057 & 2$_{11}$-1$_{01}$ E & 28.0 & 5.6$\times$10$^{-5}$ & (1.6, 1.8, 2.0, 2.2, 2.4)$\times$10$^{9}$	\\
\hline
\end{tabular}
\end{threeparttable}
\end{table*}

\section{Data Analysis}
\label{analysis}

Our observations cover the frequency range of 92--116 GHz and 213--280 GHz. While this frequency range covers several low- and high-excitation methanol lines, none of the CH$_{3}$OH transition lines with upper state energy level E$_{upper} >$ 50 K were detected for our sample. Table~\ref{lines} lists the CH$_{3}$OH transition lines that were detected towards at least one of the targets. The CH$_{3}$OH spectra are shown in Section~\ref{all-spectra}. We have measured the parameters of the line center, line width, the peak and integrated intensities using a single- or double-peaked Gaussian fit. A double-peaked profile was required for the case of J182952 and J162625 that show an extended wing in some of the spectra (Fig.~\ref{spectra}). The wing may be the signature of wind/outflow. The line parameters are listed in Section~\ref{all-spectra}. The uncertainty is estimated to be $\sim$10\%-20\% for the peak and integrated intensities and $\Delta${\it v}, and $\sim$0.02-0.04 km s$^{-1}$ for {\it V}$_{lsr}$. The errors on the line parameters are due to uncertainties in fitting the line profile, and mainly arise from the end points chosen for the Gaussian fit. The half power beam width of the telescope beam is $\sim$10$\arcsec$ at 230 GHz and $\sim$28$\arcsec$ at 100 GHz. The fluxes have been corrected for the beam filling factor, which is the ratio of the source size to the beam size. The source sizes are in the range of $\sim$3$\arcsec$-7$\arcsec$. The beam dilution factors range between $\sim$0.3$\arcsec$-0.7$\arcsec$ at 230 GHz and $\sim$0.1$\arcsec$-0.25$\arcsec$ at 100 GHz.

The CH$_{3}$OH column density and rotational temperature, $T_{rot}$, were determined using the method of constructing a rotational diagram, as described in Goldsmith \& Langer (1999). Both the upper level degeneracies and the partition function values to construct the rotational diagrams are taken from the CDMS database for consistency. The degeneracies in the CDMS database accounts for the spin degeneracy ($g_{\rm{I}}$=4). We have assumed optically thin lines at a single temperature. The assumption of CH$_{3}$OH lines to be optically thin has been tested in low-mass protostars using $^{13}$CH$_{3}$OH observations and is found to be valid (e.g., \"{O}berg et al. 2014). There are no $^{13}$CH$_{3}$OH line detections in our sample so this assumption cannot be tested for the proto-BD targets. Figure~\ref{rot-diag} shows the rotational diagrams for the proto-BDs with detection in at least two CH$_{3}$OH transition lines. The beam-averaged column density and $T_{rot}$ (or $T_{ex}$) determined from these diagrams are listed in Tables~\ref{Trot};~\ref{Ncol}. 

The uncertainty on $T_{rot}$ and N$_{CH_{3}OH}$ is estimated to be approximately 10\%-20\%, and is propagated from the errors on the line intensities, the slope and intercept of the linear fit to the data. The errors are larger for the weaker lines with a $\sim$3$\sigma$-5$\sigma$ detection. We have included the uncertainties to obtain the linear fits. The CH$_{3}$OH molecular abundances relative to H$_{2}$, [N(X)/N(H$_{2}$)] are listed in Table~\ref{abund}, and have a measurement uncertainty of approximately 20\%-25\%. The measurements on N(H$_{2}$) are provided in Riaz \& Thi (2022a). Note that these are beam-averaged abundances derived from single-dish observations. We have assumed that the methanol lines trace the same volume for the proto-BDs, considering their small source sizes compared to the large beam sizes. Interferometric observations can provide a better insight into the physical scale of the peak emission in these lines.

\section{Results}
\label{results}

Most objects show emission in only the low-excitation (E$_{upper}\leq$ 20 K) CH$_{3}$OH lines, and these can be fitted at a single $T_{ex}$ in the range of $\sim$5-15 K (Table~\ref{Trot}). This indicates that a large mass fraction of the gas is cold in these proto-BDs. For the cases of J182959, J182952, J182856, J032838, and J032859, all of the points cannot be fit at a single $T_{ex}$, and we see a break in the relation around E$_{upper}\sim$ 20 K (Fig.~\ref{rot-diag}). The fit to the points at E$_{upper}\leq$ 20 K is steeper than the flat fit for E$_{upper}>$ 20 K points. The $T_{ex}$ derived for the E$_{upper}\leq$ 20 K points is $\sim$5-7 K, while that for the high-excitation (E$_{upper}\sim$ 21-50 K) lines is $\sim$17-28 K (Table~\ref{Trot}). This indicates the presence of both a cold and warm gas component in these proto-BD systems. The $\chi^{2}$ value for the fits are notably different; for e.g., the fit to the cold (E$_{upper}\leq$ 20 K) and warm (E$_{upper}>$ 20 K) points for J182952 have a $\chi^{2}$ value of 1.7 and 0.95, respectively, while the fit to all points has $\chi^{2}$ = 6.6. The quality of the fit is thus better using two different fits. A detailed discussion on the sinlge versus two temperature components is provided in Section~\ref{radexgrid}.

The column density for the cold component is about an order of magnitude higher than the warm component, indicating a large mass fraction of the gas is at a cold temperature (Table~\ref{Ncol}). An interesting feature is a rise in the column density at $\sim$30-40 K seen for J182856 and J032859 (Fig.~\ref{rot-diag}), which is in contrast to the gradual decline with increasing E$_{upper}$ seen for the other objects. A sudden jump in the column density is also seen for J182959 at E$_{upper}$$\sim$45 K. Such anomalies could be explained by the presence of warm clumpy material or possible shock emission knots in the line of sight.

We note that for the sources with few line detections (e.g., J182854, J163136), fits to just two or three data points cannot provide a good constraint to the derived values for $T_{rot}$ and column density. However, the derived values for these sources are consistent within the uncertainties with the average values of $T_{rot}$ = 9.8$\pm$2.9 K and N(CH$_{3}$OH) = (0.4$\pm$0.3)$\times$10$^{14}$ cm$^{-2}$ for the whole sample.


The strongest emission is seen in the 2$_{02}$-1$_{01}$ A line at the lowest E$_{upper}\sim$7 K, and this line is detected in all objects. Among the high-excitation CH$_{3}$OH lines with E$_{upper}\geq$40 K detected in this work, half of the proto-BDs show emission in the 4$_{23}$-3$_{12}$ E (E$_{upper}\sim$45 K) and 5$_{15}$-4$_{14}$ E (E$_{upper}\sim$40 K) lines, 4 objects show emission in the 5$_{15}$-4$_{14}$ A (E$_{upper}\sim$49 K) line, while the 5$_{05}$-4$_{04}$ E (E$_{upper}\sim$48 K) line is detected in only two objects. 

Among the high-density tracers at n$_{crit}$ of the order of 10$^{8}$-10$^{9}$ cm$^{-3}$, 12 out of 14 proto-BDs show emission in the 0$_{00}$-1$_{11}$ E line (E$_{upper}\sim$13 K; n$_{crit}\sim$1$\times$10$^{8}$ cm$^{-3}$). In addition, 4 proto-BDs show emission in the 2$_{11}$-1$_{01}$ E (E$_{upper}\sim$28 K; n$_{crit}\sim$2$\times$10$^{9}$ cm$^{-3}$) line, while the 3$_{21}$-4$_{14}$ E (E$_{upper}\sim$40 K; n$_{crit}\sim$1.1$\times$10$^{8}$ cm$^{-3}$) line is detected in just one object. 

The detection in a particular transition line and the peak line strength are dependent on E$_{upper}$, A$_{jk}$, and the critical density, n$_{crit}$ (Table~\ref{lines}). Another factor is the location of the origin of emission. For instance, consider the case of J182856 that shows emission in 11/14 CH$_{3}$OH lines probed in this work. This object shows strong emission in the 2$_{02}$-1$_{01}$ A line with a peak intensity of $\sim$0.8 K, while the peak intensity in the 0$_{00}$-1$_{11}$ E line is $\sim$0.15 K (Table~\ref{line-pars}). These lines are at a similar E$_{upper}\sim$13 K but the 0$_{00}$-1$_{11}$ E line has an order of magnitude higher n$_{crit}$ and a $\sim$6 times higher A$_{jk}$ than the 2$_{02}$-1$_{01}$ A line, making the detection more difficult in the former case. J182856 shows the weakest emission in the 2$_{11}$-1$_{10}$ A, 4$_{23}$-3$_{12}$ E and 5$_{05}$-4$_{04}$ E lines (Table~\ref{line-pars}). While the 4$_{23}$-3$_{12}$ E and 5$_{05}$-4$_{04}$ E transitions have E$_{upper} >$ 40 K and so are more difficult to detect, weak emission in 2$_{11}$-1$_{10}$ A could be due to ortho vs. para emission (Sect.~\ref{discussion}).

 \begin{figure*}
  \centering 
     \includegraphics[width=2.5in]{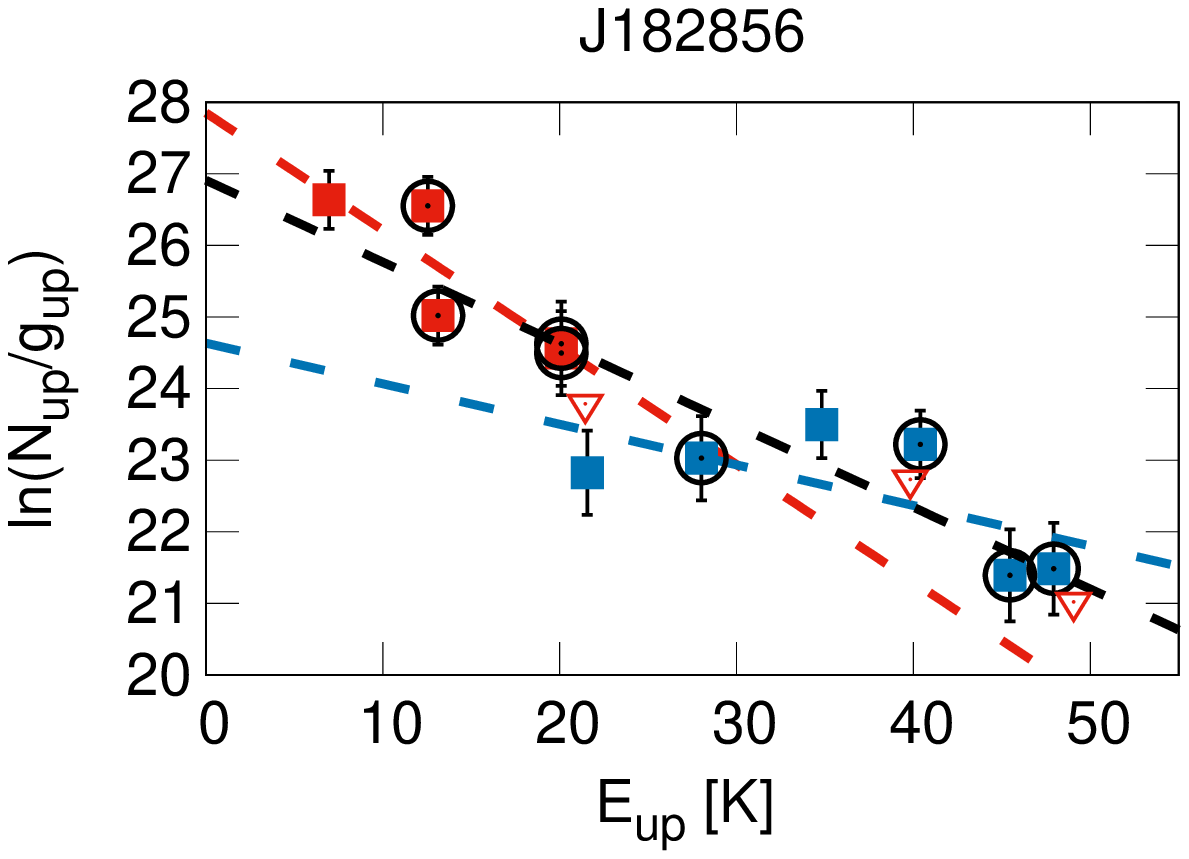}	\vspace{0.1in}  
     \includegraphics[width=2.5in]{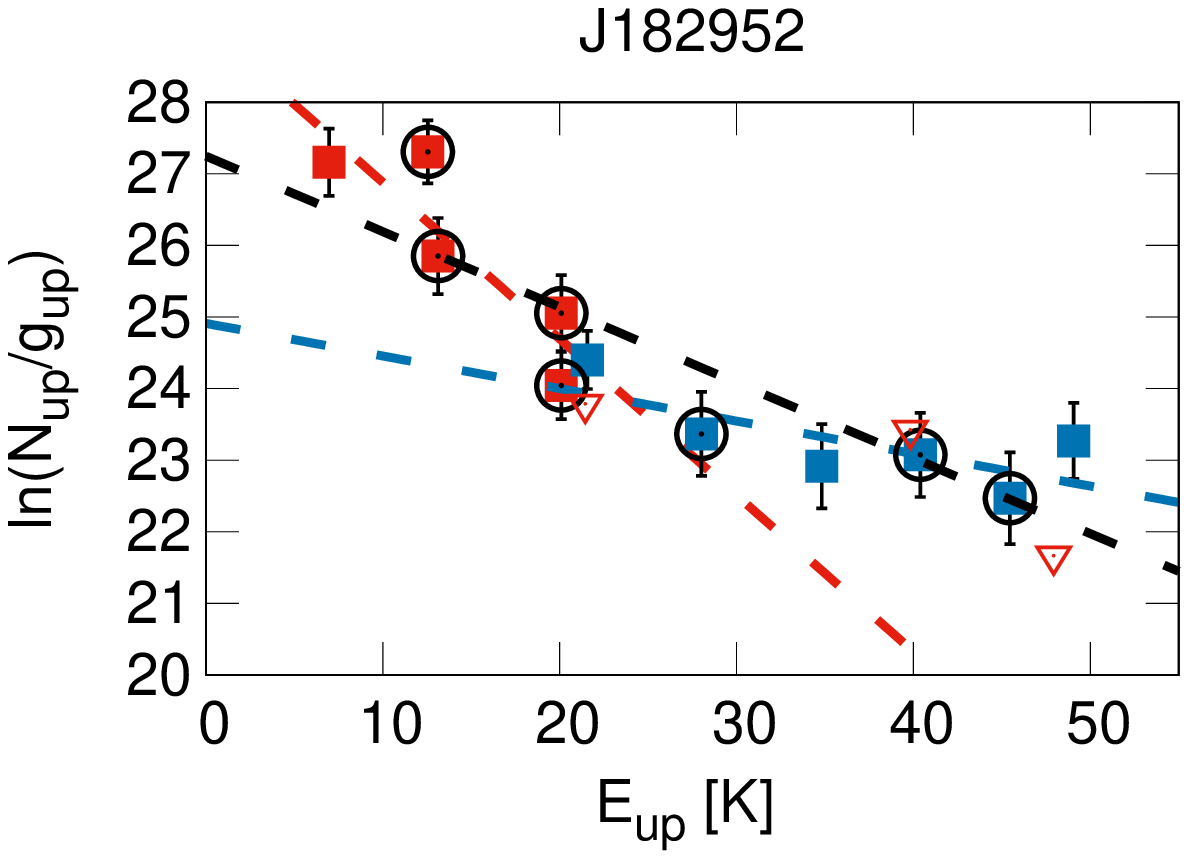}	\vspace{0.1in} 
     \includegraphics[width=2.5in]{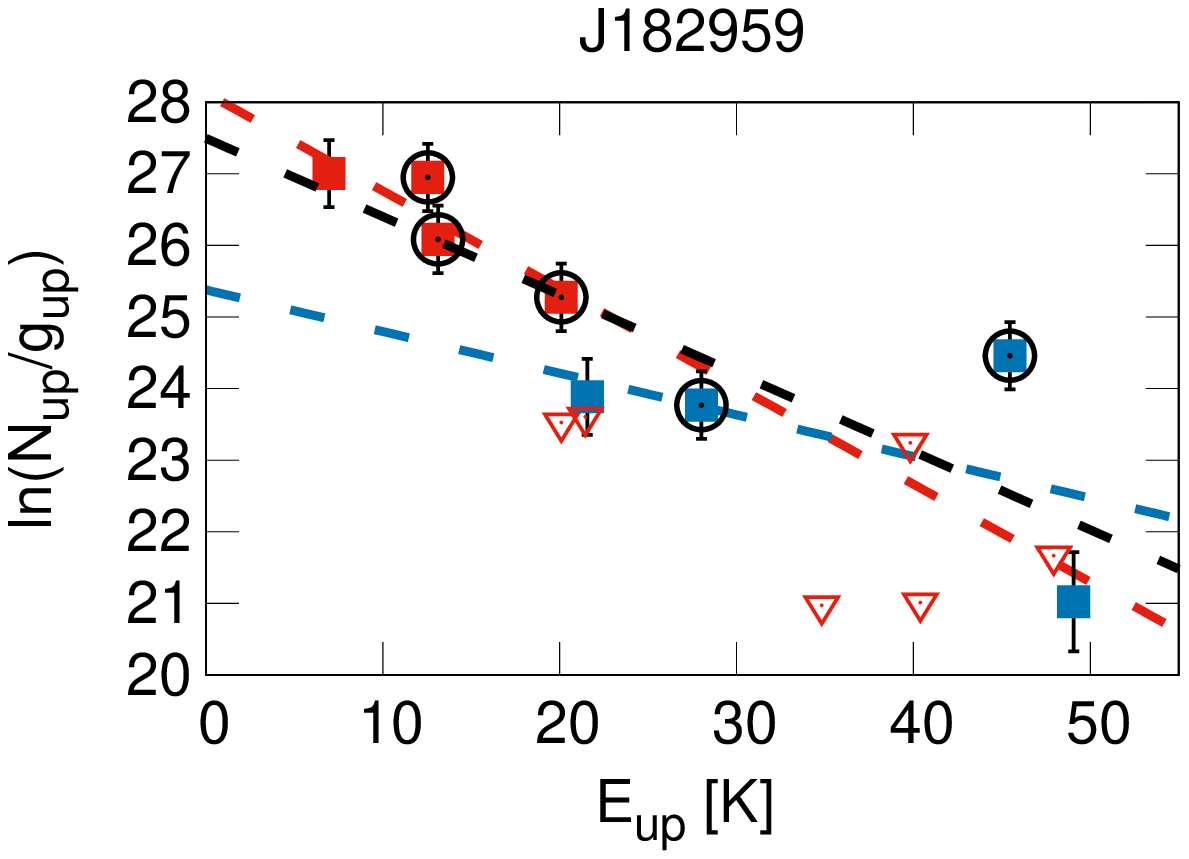}		 \vspace{0.1in} 
     \includegraphics[width=2.5in]{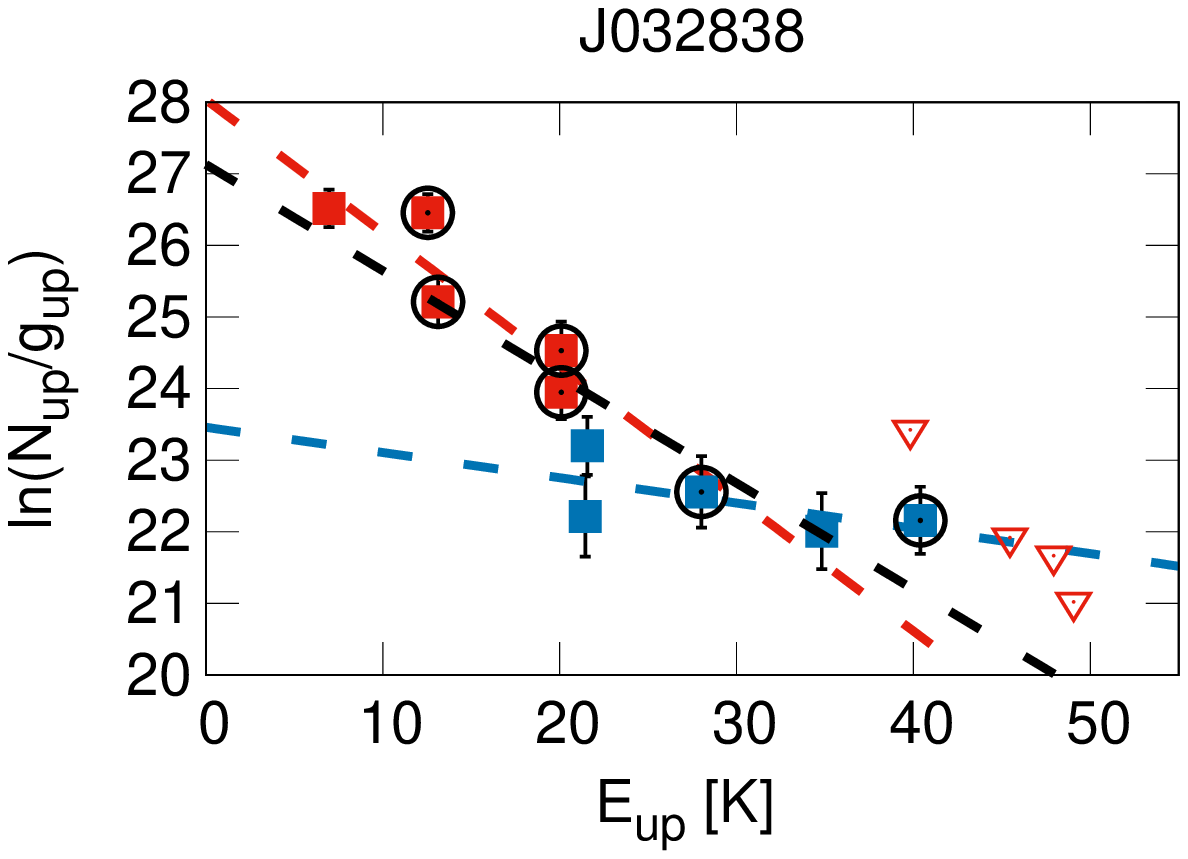}	 \vspace{0.1in}
     \includegraphics[width=2.5in]{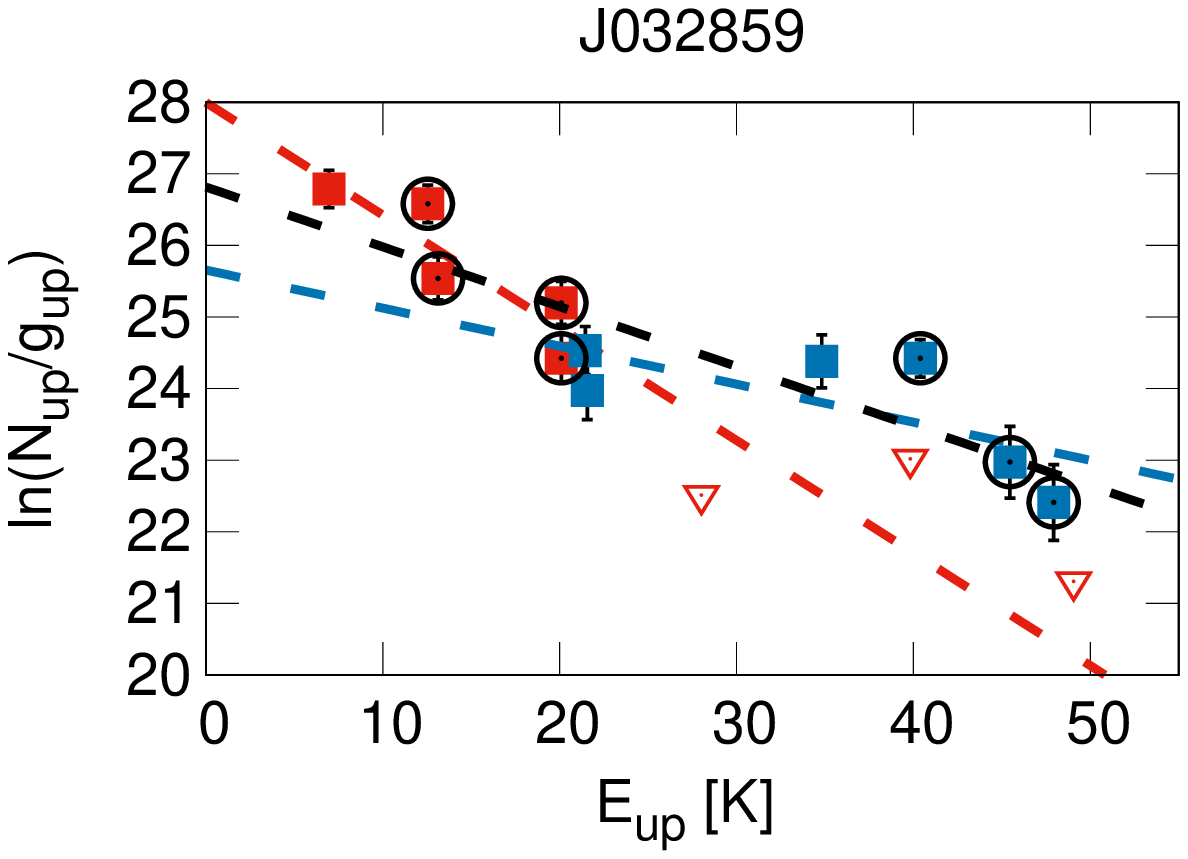}     \vspace{0.1in} 
     \includegraphics[width=2.5in]{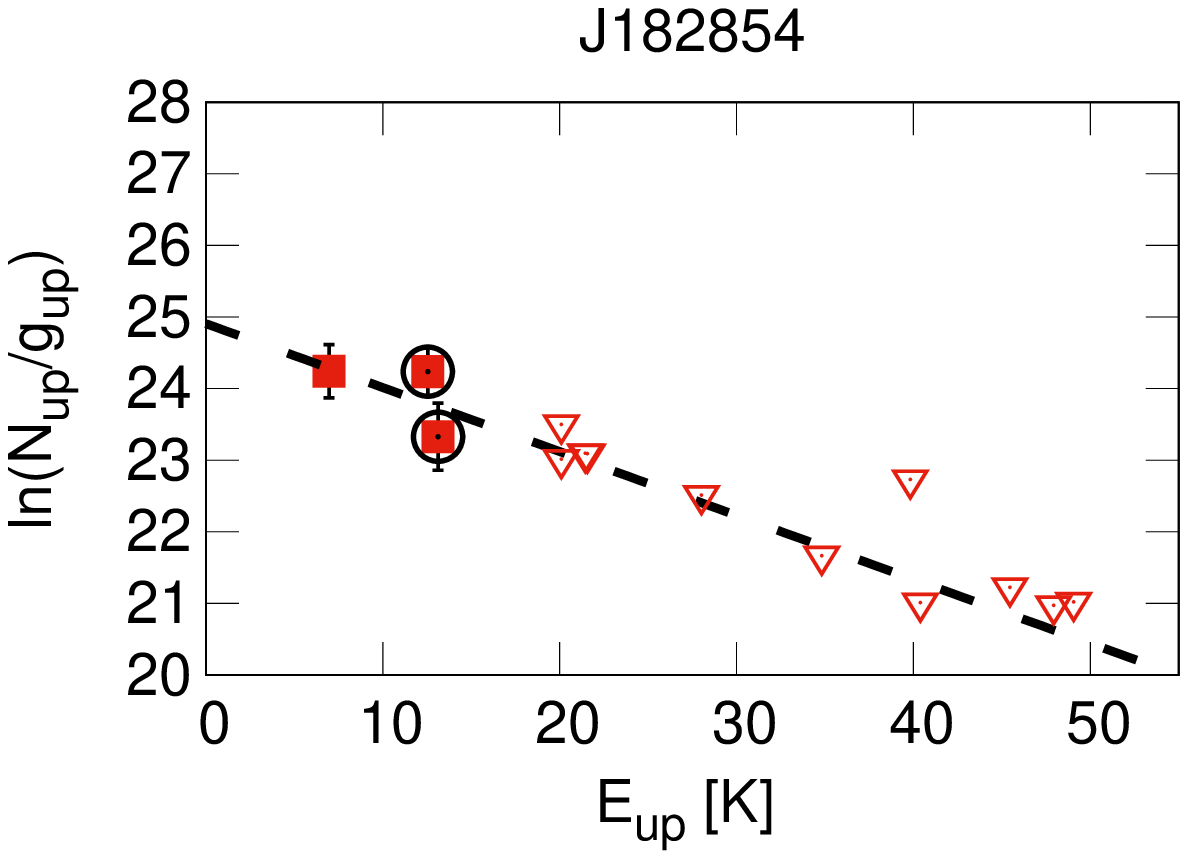}		\vspace{0.1in} 
     \includegraphics[width=2.5in]{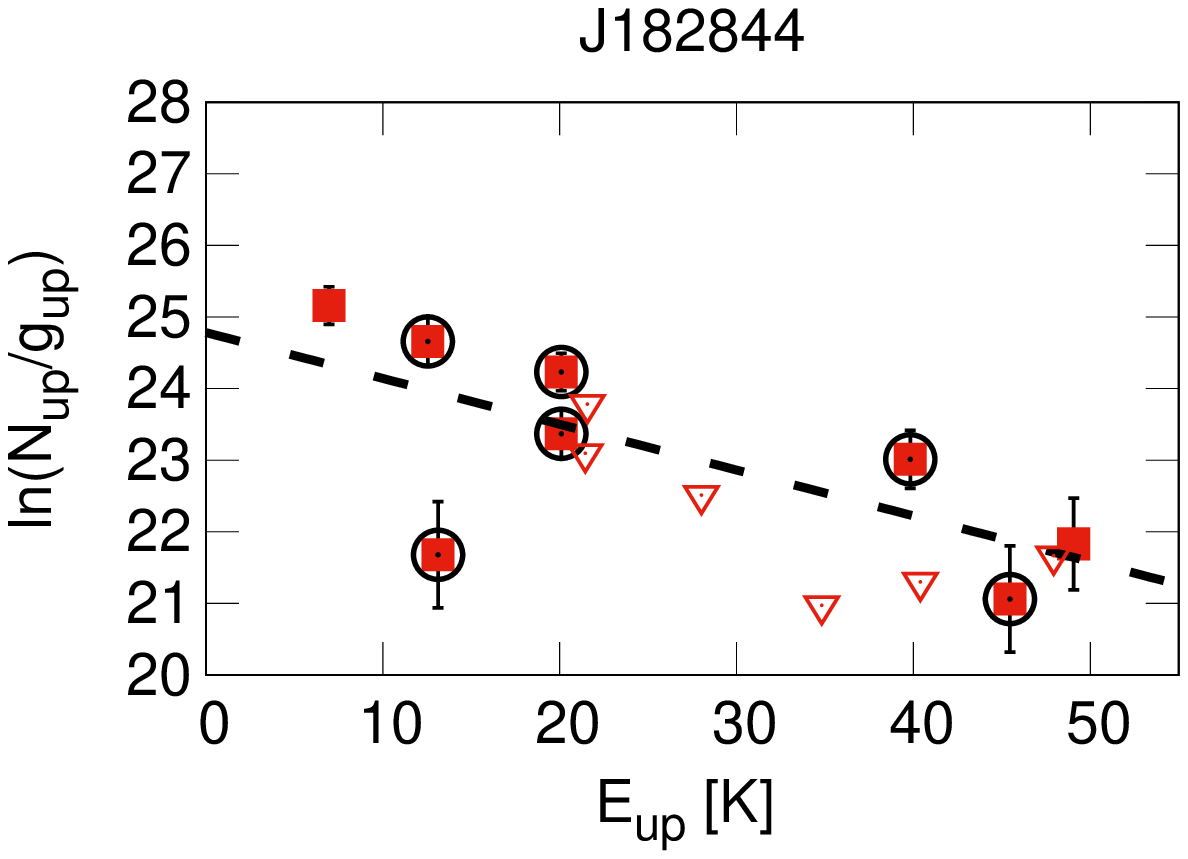}  	 \vspace{0.1in} 
     \includegraphics[width=2.5in]{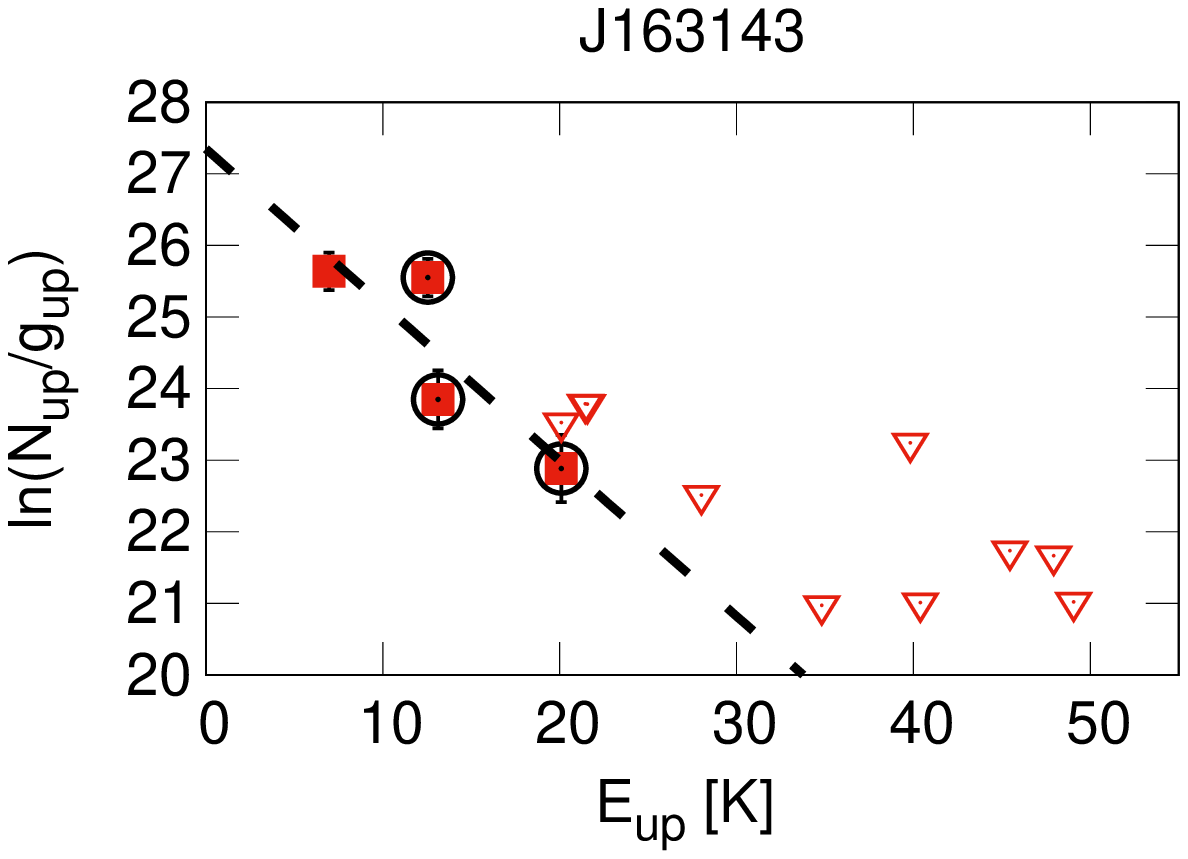}	     
   \caption{CH$_{3}$OH rotational diagrams. Red and blue dashed lines in the top panels are the fits to the cold (E$_{upper}\leq$ 20 K; red box) and warm (E$_{upper}>$ 20 K; blue box) points, respectively. Black dashed line in all diagrams is the fit to all data points. The data points for the ortho lines are marked by a black circle. The 3-$\sigma$ upper limits are denoted by red open diamonds and have not been included in the fits.} 
     \label{rot-diag}     
  \end{figure*}

\setcounter{figure}{0}    
 \begin{figure*}
  \centering       
      \includegraphics[width=2.5in]{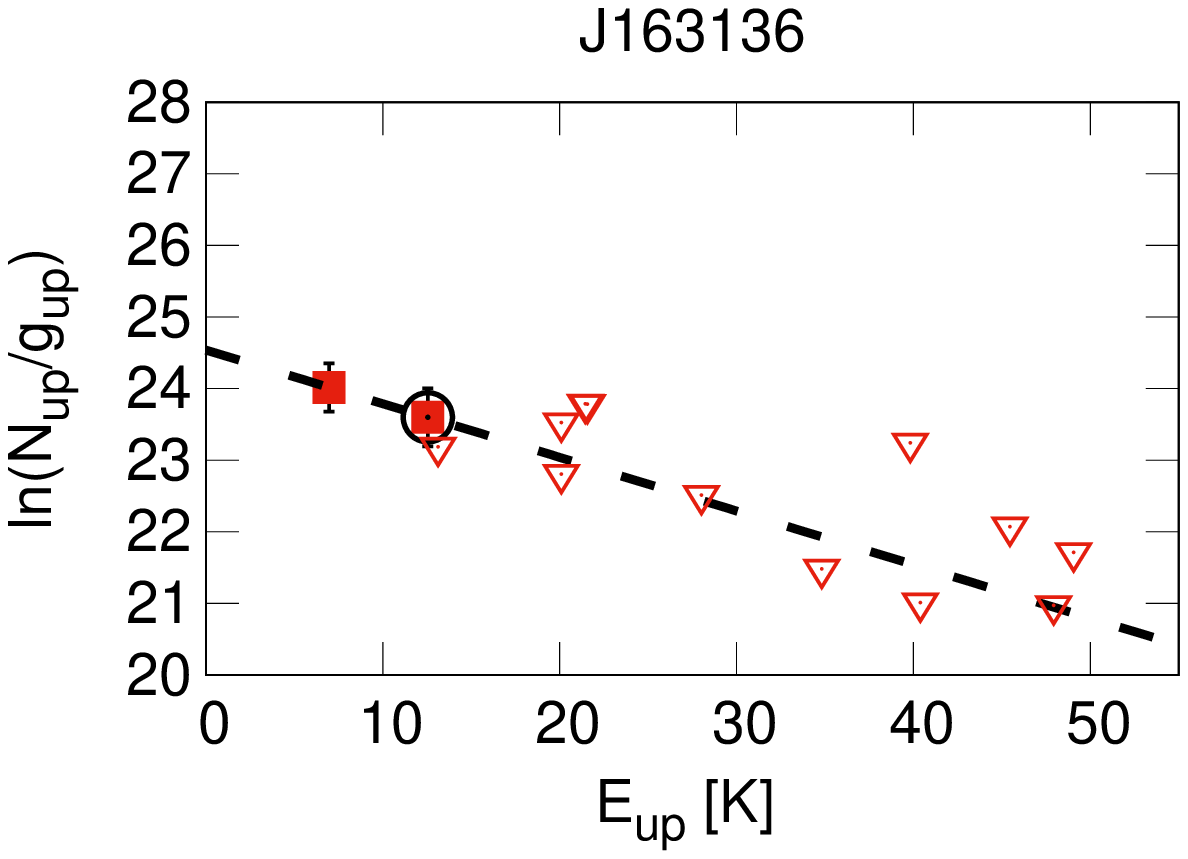}     \vspace{0.1in} 
     \includegraphics[width=2.5in]{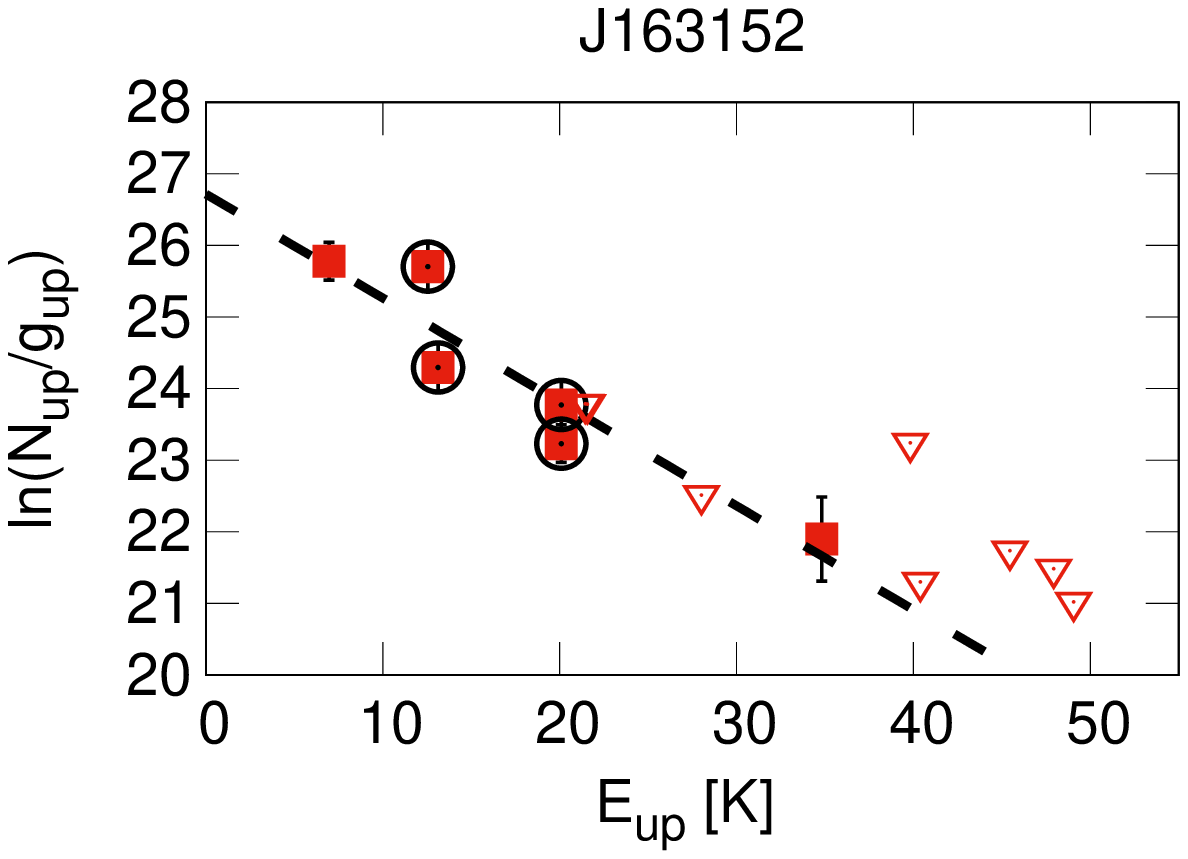}	  	\vspace{0.1in} 
     \includegraphics[width=2.5in]{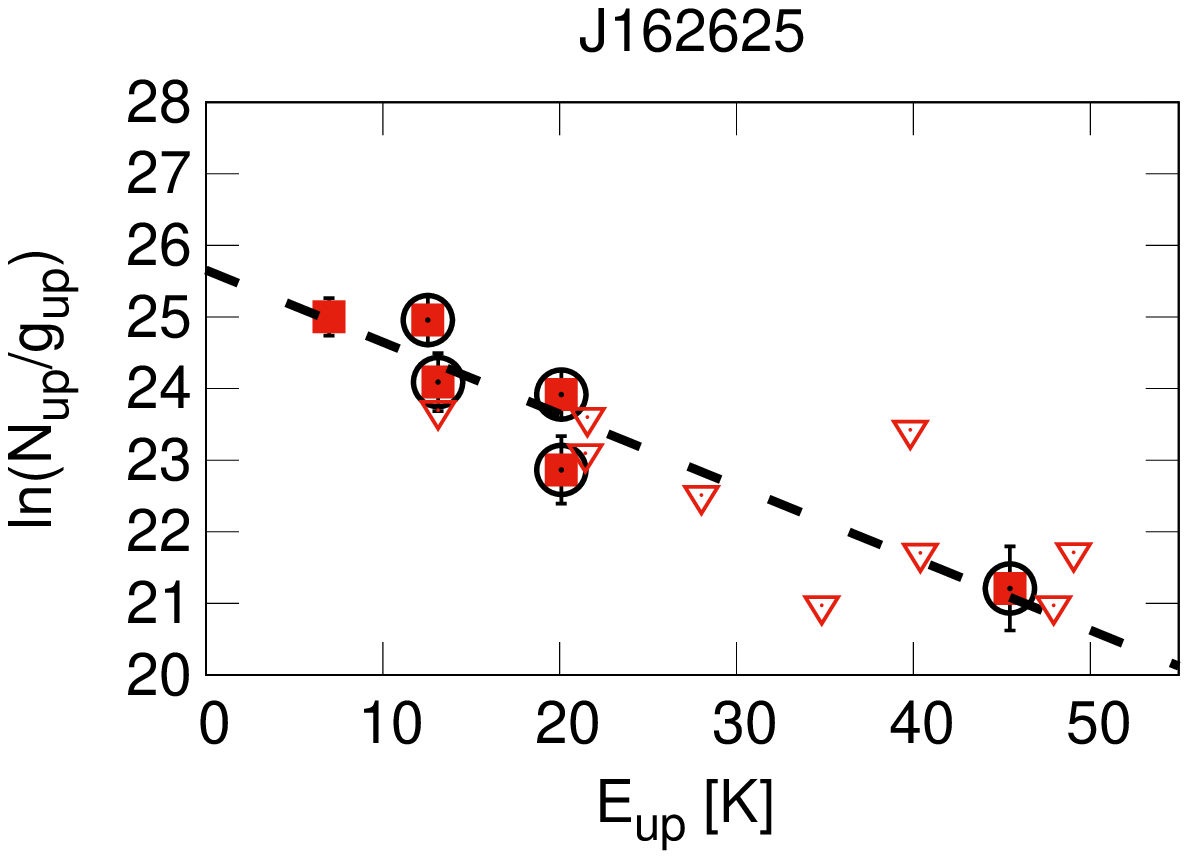}    	\vspace{0.1in}  
     \includegraphics[width=2.5in]{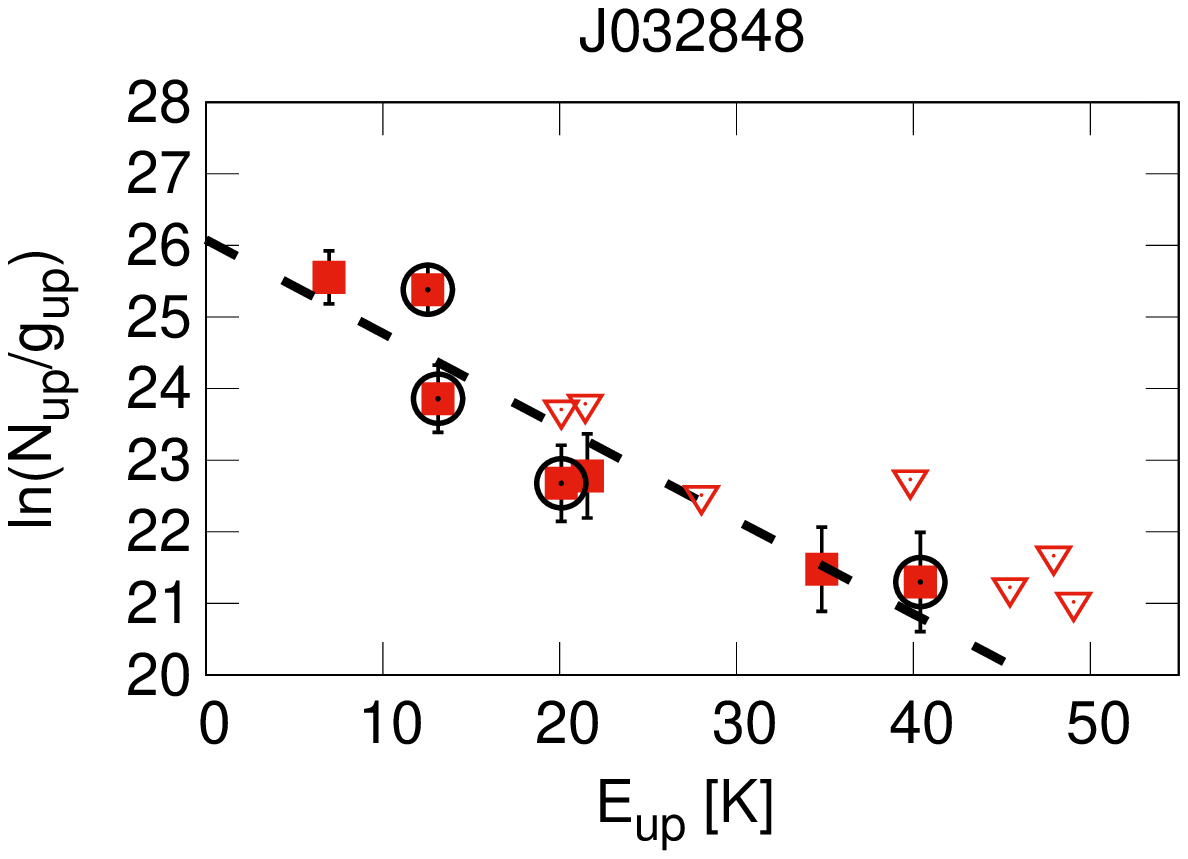}	\vspace{0.1in} 	
     \includegraphics[width=2.5in]{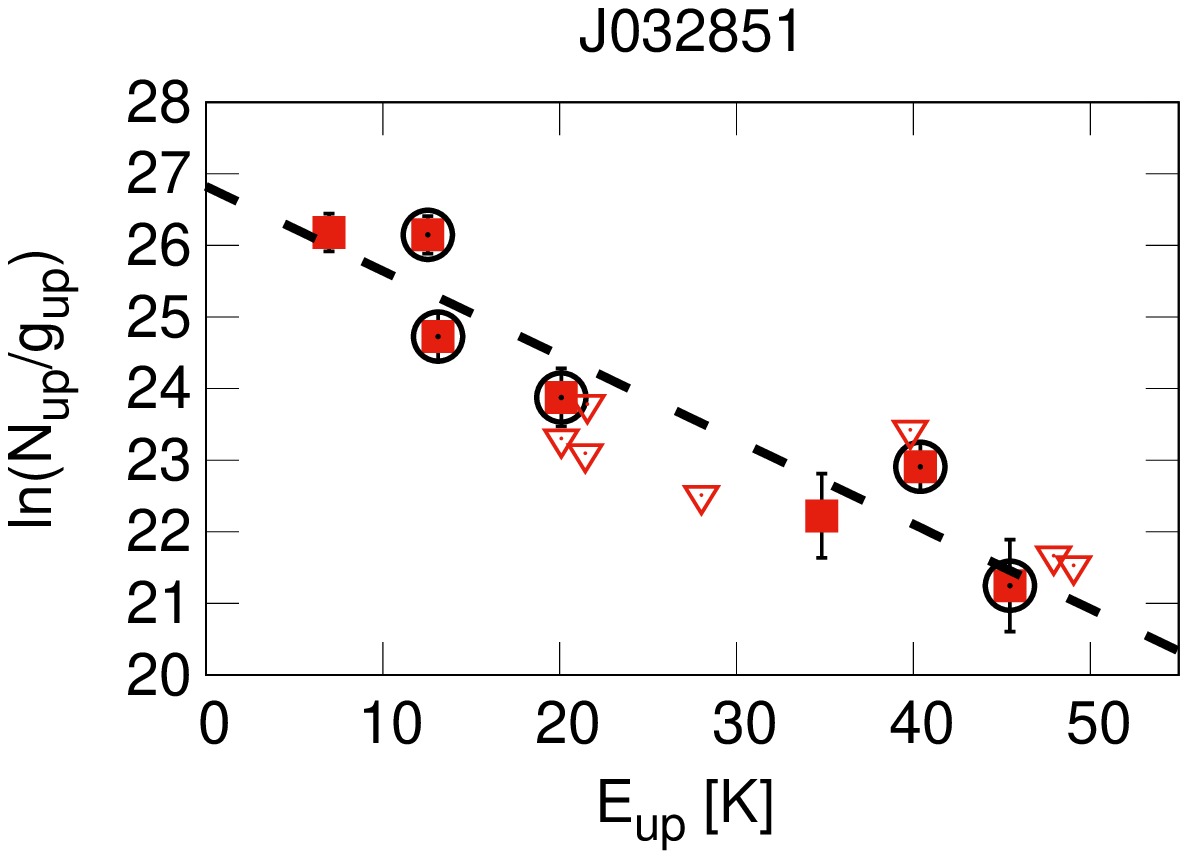}	\vspace{0.1in} 
     \includegraphics[width=2.5in]{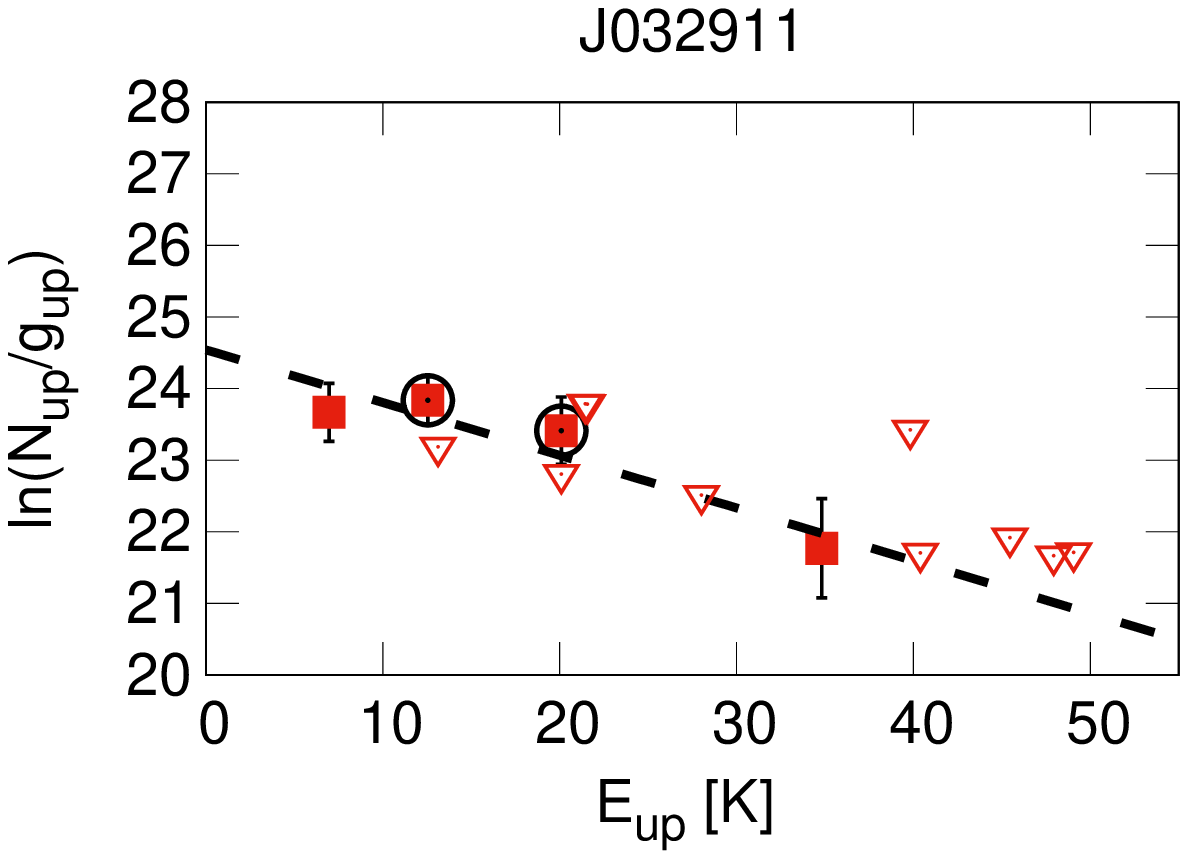}   
   \caption{Continued.}
  \end{figure*}

\begin{table*}
\centering
\caption{CH$_{3}$OH rotational temperatures}
\label{Trot}
\begin{threeparttable}
\begin{tabular}{lccccc} 
\hline
Object	& \multicolumn{5}{c}{$T_{rot}$ [K] }	 \\	\hline
		&  All points	&  Cold  & Warm  &  Ortho  &  Para    \\
\hline
J182854 	& 11.20$\pm$2.2 	& -- 		& -- 		& -- 		& -- 		\\
J182844 	& 15.63$\pm$2.8 	& -- 		& -- 		& -- 		& --		 \\
J182959 	& 9.14$\pm$1.6 	& 7.32$\pm$1.5 	& 17.20$\pm$3.0 & 14.42$\pm$2.3 & 7.30$\pm$1.3	\\	
J182952 	& 9.48$\pm$1.5 	& 4.58$\pm$0.8 	& 22.00$\pm$3.5 & 8.52$\pm$1.4  & 10.63$\pm$1.6 \\	
J182856 	& 8.70$\pm$1.6 	& 6.11$\pm$1.1 	& 24.98$\pm$4.6 & 8.50$\pm$1.5 & 8.73$\pm$1.6		\\	
J163143 	& 4.60$\pm$0.9 	& -- 		& -- 		& -- 		& -- 		\\
J163136 	& 13.35$\pm$2.5 	& -- 		& -- 		& -- 		& --		 \\
J163152 	& 6.89$\pm$1.3 	& -- 		& -- 		& -- 		& -- 		\\
J162625 	& 9.94$\pm$1.8 	& -- 		& -- 		& -- 		& -- 		 \\   
J032838 	& 6.74$\pm$1.0 	& 5.39$\pm$0.8 & 28.35$\pm$4.5 & 7.12$\pm$1.1 	& 6.14$\pm$0.9		\\	
J032848 	& 7.65$\pm$1.4 	& -- 		& -- 		&   8.44$\pm$1.5 	& 6.82$\pm$1.2 	\\	
J032851 	& 8.48$\pm$1.5 	& -- 		& -- 		& -- 		& -- 		\\	
J032859 	& 11.98$\pm$1.9 & 6.36$\pm$1.0 & 18.82$\pm$3.2  & 11.83$\pm$1.8 	& 11.33$\pm$1.8   \\	
J032911 	& 13.58$\pm$2.7 	& -- 		& -- 		& -- 		& -- 		\\	
\hline
\end{tabular}
\end{threeparttable}
\end{table*}

\begin{table*}
\centering
\caption{CH$_{3}$OH column densities}
\label{Ncol}
\begin{threeparttable}
\begin{tabular}{lccccc} 
\hline
Object	& \multicolumn{5}{c}{N(CH$_{3}$OH) ($\times$10$^{14}$ cm$^{-2}$)}	 \\	\hline
		&  All points	&  Cold  	& Warm  	&  Ortho  	&  Para    \\
\hline
J182854 	& 0.10$\pm$0.02 	& --		&	--	& --		& --		\\	
J182844 	& 0.13$\pm$0.02 	& --		&	--	& --		& --	 	\\	
J182959 	& 0.99$\pm$0.16 & 1.42$\pm$0.2  &	0.27$\pm$0.05	& 1.04$\pm$0.15 & 0.82$\pm$0.12 \\	
J182952 	& 0.80$\pm$0.12 & 2.14$\pm$0.3 &	0.25$\pm$0.03	& 0.93$\pm$0.13 & 0.83$\pm$0.12 \\	
J182856 	& 0.52$\pm$0.09 & 0.87$\pm$0.12 &  	0.14$\pm$0.015	& 0.63$\pm$0.09 & 0.42$\pm$0.06 \\	
J163143 	& 0.39$\pm$0.08 	& --		&	--	& --		& --		\\	
J163136 	& 0.08$\pm$0.02 	& --		&	--	& --		& --		\\	
J163152 	& 0.32$\pm$0.05 	& --		&	--	& --		& --		\\	
J162625 	& 0.17$\pm$0.04 	& --		&	--	& --		& --		\\   	
J032838 	& 0.47$\pm$0.07 & 0.91$\pm$0.13 &	0.08$\pm$0.015	& 0.62$\pm$0.09 & 0.40$\pm$0.06 \\	
J032848 	& 0.19$\pm$0.03 	& --		&	--	& 0.21$\pm$0.03	& 0.22$\pm$ 0.03	\\	
J032851 	& 0.46$\pm$0.09	& --		&	--	& --		& --		\\	
J032859 	& 0.70$\pm$0.1 	& 1.06$\pm$0.15 & 0.41$\pm$0.06 & 0.72$\pm$0.1 & 0.61$\pm$0.09	\\	
J032911 	& 0.09$\pm$0.02 	& --		&	--	& --		& --		\\	
\hline
\end{tabular}
\end{threeparttable}
\end{table*}

\begin{table*}
\centering
\caption{CH$_{3}$OH abundances}
\label{abund}
\begin{threeparttable}
\begin{tabular}{lccccc} 
\hline
Object	& \multicolumn{5}{c}{[CH$_{3}$OH] ($\times$10$^{-10}$) }	 \\	\hline
		&  All points	&  Cold  	& Warm  	&  Ortho  	&  Para    \\
\hline
J182854 	& 2.7$\pm$0.6 	& --		&	--	& --		& --		\\	
J182844 	& 2.7$\pm$0.6 	& --		&	--	& --		& --	 	\\	
J182959 	& 22.0$\pm$4.5 & 31.0$\pm$5.6  &	6.6$\pm$1.5 & 23.0$\pm$5.2	& 17.0$\pm$3.6 \\	
J182952 	& 9.2$\pm$2.0 	& 24.0$\pm$5.0 & 2.3$\pm$0.5	& 10.0$\pm$2.0 & 9.1$\pm$2.0		\\	
J182856 	& 15.0$\pm$3.5 & 23.0$\pm$5.5 & 2.9$\pm$0.7 & 17.0$\pm$4.0 & 13.0$\pm$3.0	\\	
J163143 	& 22.0$\pm$5.5 	& --		&	--	& --		& --		\\	
J163136 	& 3.4$\pm$0.8 	& --		&	--	& --		& --		\\	
J163152 	& 30.0$\pm$7.5 	& --		&	--	& --		& --		\\	
J162625 	& 5.0$\pm$1.2 	& --		&	--	& --		& --		\\   	
J032838 	& 2.3$\pm$0.5 	& 4.2$\pm$1.0		& 0.5$\pm$0.1	& 2.8$\pm$0.6	& 1.7	$\pm$0.4	\\	
J032848 	& 1.9$\pm$0.5 	& --		&	--	& 1.6$\pm$0.4		& 2.2$\pm$0.5 	\\	
J032851 	& 6.8	$\pm$1.7	& --		&	--	& --		& --		\\	
J032859 	& 2.0$\pm$0.4 	& 3.2	$\pm$0.7	&	1.2$\pm$0.3	& 2.2	$\pm$0.5	& 1.8$\pm$0.4	\\	
J032911 	& 0.8$\pm$0.2 	& --		&	--	& --		& --		\\	
\hline
\end{tabular}
\end{threeparttable}
\end{table*}

\section{Discussion}
\label{discussion}

\subsection{Cold and warm gas component: predictions from brown dwarf formation model}
\label{model}

 \begin{figure*}
  \centering 
     \includegraphics[width=6in]{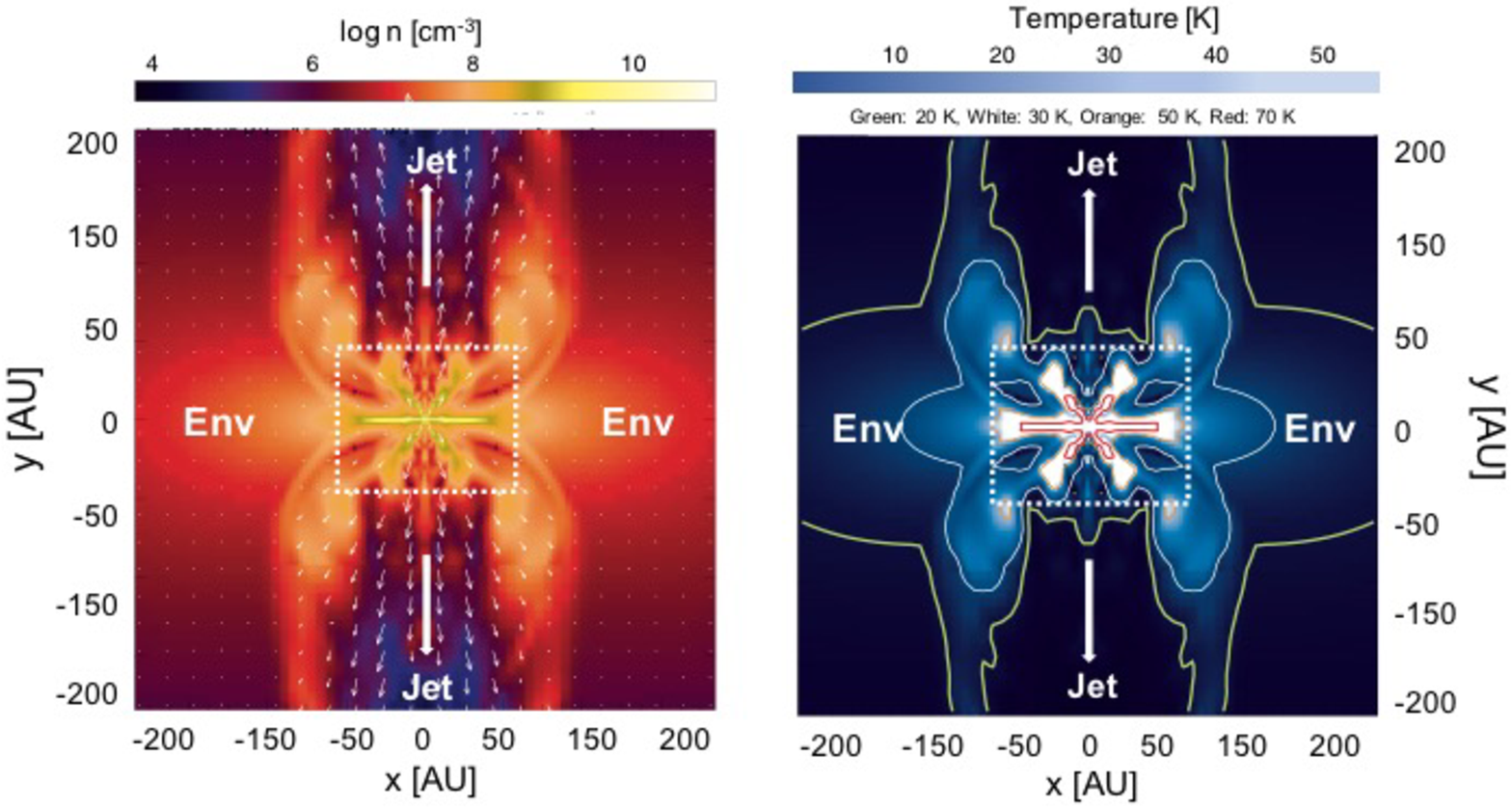}
   \caption{An edge-on view of the internal density (left) and temperature (right) structure of a proto-BD predicted by the brown dwarf formation model (Machida \& Basu 2019). White dotted lines mark the expected warm corino region close to the jet launching zone. }
   \label{model}     
  \end{figure*}

Simulations on the formation of brown dwarfs via gravitational collapsing core predict a ``warm corino'' region in close proximity to the central proto-BD (Machida et al. 2009; Machida \& Basu 2019). Figure~\ref{model} shows the internal density and temperature structure of a proto-BD predicted by the brown dwarf formation model by Machida \& Basu (2019).\footnote{The protostellar mass is 0.04\,M$_\odot$ at the end of the simulation, which is in the brown dwarf mass range.} The figure shows that, in addition to a low-velocity outflow, a strong high-velocity jet is driven by the inner Keplerian disk that is enclosed by the pseudo-disk.

The density and temperature in the envelope and the outer pseudo-disk regions is 10$^{5}$-10$^{7}$ cm$^{-3}$ and $\sim$10 K, respectively. In the region near the proto-BD ($<$50 au), there is a sudden rise in the temperature to $>$20 K and the density is as high as 10$^{8}$-10$^{10}$ cm$^{-3}$. This is the predicted warm corino in the innermost, high-density and warmest regions in proto-BDs. The temperature in the warm corino may not be as high as hot corinos ($>$100 K) in protostars, but it is predicted to reach temperatures as high as $\sim$70-80 K.

The temperature rise is mainly due to the adiabatic contraction of the gas and subsequent ejection of the high-density and high-temperature gas by the jet. Without the appearance of the jet, the hot gas is distributed only in a tiny domain near the protostar ($r \lesssim 10$\,au, for details, see Masunaga \& Inutsuka 2000), which should be difficult to be observed. The jet can blow the high-temperature and high-density gas from the region in close proximity to the central proto-BD. Thus, the high-temperature and high-density gas is entrained and pumped up by the jet from the region near the proto-BD, as seen in the white dotted box in the right panel of Figure~\ref{model}. In addition, part of such gas falls again and again in the region far from the proto-BD (Tsukamoto, Machida \& Inutsuka 2021). 

Although the thermal heating or radiative feedback from the central object is not considered in the simulation, these may not alone heat up the central region to temperatures $>$20 K due to the very low luminosity of the proto-BD. This implies that the warm corino region at temperatures $>$20 K wouldn't exist if there is no jet. However, we cannot definitely deny that when the radiative feedback from the central object is included, the temperature could rise to warmer temperatures ($>$30 K) at large radii of $>$50 au. 

We explain our scenario by comparing the observations and the theoretical studies. The barotropic equation of state in the brown dwarf simulations (Machida et al. 2009, Machida \& Basu 2019) was modelled according to the one-dimensional calculation perfomed by Masunaga \& Inutsuka (2000). The gas temperature in the models is a function of density, due to which the maximum temperature is determined by the maximum density that can be reached in the innermost regions. The highest temperature in the proto-BD model is $\sim$70-80 K and is reached in a narrow region at close-in radii of within $\lesssim$10 au of the central proto-BD without considering the jet.

If we consider the pumping-up of the high-temperature gas by the jet, the temperature of $\sim$25-50 K appears at a radius of $\sim$10-50 au, and further to $\sim$7-20 K at $\sim$50-150 au, as seen in the right panel of Fig.~\ref{model}. At larger radii ($>$150 au) or in the outer envelope, the temperature remains $<$6 K. The high-temperature ($\sim$70-80 K) gas component probes high densities of $\geq$10$^{10}$ cm$^{-3}$. The warm-temperature ($\sim$25-50 K) gas component traces moderate densities of the order 10$^{8}$--10$^{10}$ cm$^{-3}$, and the low-temperature or the cold ($\sim$7-20 K) gas component probes the low-density material at $\leq$10$^{8}$ cm$^{-3}$. However, the mass of the high-temperature component is expected to be quite low ($\sim$0.0005-0.015 M$_{\sun}$) compared to the warm-temperature ($\sim$0.006-0.02 M$_{\sun}$) and the cold gas component ($\sim$0.015-0.04 M$_{\sun}$), indicating the pumping-up of the high- and warm-temperature components by the jet is not very effective.

Comparing these models with our observations, the low-excitation (E$_{upper}\leq$20 K) CH$_{3}$OH lines with n$_{crit}$ of the order of 10$^{5}$--10$^{8}$ cm$^{-3}$ (Table~\ref{lines}) are probing the low-density cold gas component at $\sim$50-150 au in the proto-BD systems (Fig.~\ref{model}). Most notably, all proto-BDs show emission in the 2$_{02}$-1$_{01}$ A line (E$_{upper}\sim$7 K; n$_{crit}\sim$0.5$\times$10$^{5}$ cm$^{-3}$) and $\sim$86\% show emission in the 0$_{00}$-1$_{11}$ E line (E$_{upper}\sim$13 K; n$_{crit}\sim$1$\times$10$^{8}$ cm$^{-3}$). This indicates that a considerable gas mass is cold and at a low-density in these objects.

The remaining CH$_{3}$OH lines detected in this work with E$_{upper}\sim$28-49 K are probing the warm-temperature gas component at $\sim$10-50 au in the proto-BDs. This is the warm corino region surrounding the jet launching zone. In particular, detection in the 2$_{11}$-1$_{01}$ E line (n$_{crit}\sim$2$\times$10$^{9}$ cm$^{-3}$; E$_{upper}\sim$28 K) indicates that despite being a small fraction by mass, a high-density warm-temperature gas component originating from the warm corino region is present in the proto-BDs, as predicted by the models. The critical densities for some of these warm lines are, however, at 10$^{6}$--10$^{7}$ cm$^{-3}$ (Table~\ref{lines}), and thus they do not all trace the moderately high density material. As noted, if the radiative feedback is included, the central proto-BD can heat the low-density gas distributed at larger radii of $>$50 au and also above and below the pseudo-disk midplane to $>$30 K (e.g., Tomida et al. 2015). This could make the low-density gas component warmer than expected. For the objects with evidence of both a cold and warm gas component as seen in the rotational diagrams, the column density in the cold component is about an order of magnitude higher than the warm component. This is consistent with the model predictions of a higher mass fraction of cold compared to warm gas component.

We have shown here through observations of multiple transitions of CH$_{3}$OH that a warm corino exists in $\sim$78\% of the proto-BDs (Fig.~\ref{rot-diag}). The remaining $\sim$22\% proto-BDs show evidence of only the cold gas component. If all proto-BDs show a similar temperature structure then the relatively higher excitation CH$_{3}$OH lines with E$_{upper}>$ 20 K should be detected in all of them. The non-detection does not directly imply the absence of a warm corino but rather that the detection of the warm corino in a few proto-BDs could be due to favourable inclination that allows a more direct view of the jet launching zone. Episodic accretion as well as a high-velocity jet can also raise the temperature around a proto-BD. If we only see the cold gas component then this suggests that either the system is very young and embedded so that thermal and non-thermal heating is not yet efficient or the inclination is close to edge-on that gives the least view of the warm corino zone while there is more colder material in envelope/pseudo-disk regions in the line of sight.

None of the CH$_{3}$OH transitions at E$_{upper}>$50 K are detected in any of the proto-BDs. This suggests an upper limit on the gas kinetic temperature of $\sim$50 K in these objects. A sub-thermal population is less likely since we have detected lines from very high critical densities. The high-excitation methanol lines can be observed if a high-velocity powerful jet is associated with the proto-BD, because the jet stirs up the high-temperature gas around it. The strength (or momentum flux) of the jet is also related to the existence of the hot gas. Therefore, the non-detection in the high-excitation lines may be related to whether a powerful jet is associated with a proto-BD or not. It may also be the case that the gas with a high temperature of $>$50 K is embedded in high-density gas and thus optically thick. As predicted by the models, the high-temperature ($\sim$70-80 K) gas component has the lowest mass fraction and thus very low in abundance than the cooler gas to be detected. It also traces the innermost and narrower regions at $\leq$10 au in the proto-BD than the cold component, and thus is more likely to be beam-diluted in single-dish observations.

\subsection{CH$_{3}$OH ortho-to-para ratios}
\label{ratios}

In general, we see more ortho ($E$) lines detected than para ($A$) CH$_{3}$OH (Table~\ref{lines}) at a similar range in E$_{upper}\sim$7-49 K and A$_{jk}\sim$ (0.3-6)$\times$10$^{-5}$. The para lines have n$_{crit}\sim$ 10$^{5}$--10$^{6}$ while the ortho have 10$^{5}$--2$\times$10$^{9}$. The ortho CH$_{3}$OH lines are thus tracing the warmer and denser gas mass, and likely a larger mass reservoir or column density than the para lines. Basically, detection in 5 para but 9 ortho transitions indicates that both spin symmetries may not trace common physical conditions. We can roughly estimate the fraction of the total CH$_{3}$OH emission from the integrated fluxes of the $A$ and $E$ isomers. The CH$_{3}$OH ortho-to-para (o/p) ratios for the proto-BDs range between $\sim$0.3--2.3 (Table~\ref{op-fd}). We have used the ortho and para column densities (Table~\ref{Ncol}) to calculate the o/p ratios for the objects that show emission in at least three ortho and para lines. For the rest of the objects, the o/p ratios are calculated from the integrated fluxes (Appendix A). The o/p ratios are consistent within 10\% when calculated from the column densities or integrated fluxes, likely due to the lines being optically thin (Riaz et al. 2019).

The CH$_{3}$OH o/p ratios of less than unity suggest the presence of spin conversion processes that may be induced by molecular interactions on molecular ices or collisions in the gas phase, favouring an overabundance in the para spin symmetry (e.g., Sun et al. 2015). As shown in this laboratory study, in the gas phase, molecular collisions could induce the interconversion of spin in methanol molecules exhibiting a rate that decreases as the pressure increases with the number of collisions. Perhaps strong shock emission could result in the nuclear spin conversion of A and E symmetries, which is considered to be a rare event and could only be affected by non-reactive collisions and grain surface mechanisms (e.g., Willacy et al. 1993). This effect is expected to be more pronounced at earlier star formation stages (e.g., Minh et al. 1993; Wirstrom et al. 2011). Although there are some claims that different ortho-to-para ratio may reflect conditions where the molecule is formed, it is not clear whether nuclear spin temperature is kept as it is during the formation and desorption of this molecule. Such hypothesis are denied at least for H$_{2}$O molecule recently (e.g. Hama et al. 2018).


\begin{table}
\centering
\caption{Ortho-to-para ratios and CO depletion factors}
\label{op-fd}
\begin{threeparttable}
\begin{tabular}{lccc} 
\hline
Object	& 	CH$_{3}$OH o/p \tnote{a}	&	H$_{2}$CO o/p	\tnote{a}	&	f$_{D}$	\\
\hline
J182854 	&	0.9 (F)			&	 $<$1.5 (N)		&	5.6		\\	
J182844 	&	0.9 (F)			&	 3.8 (N)			&	4.6 		\\
J182959 	&	1.3 (N)			&	 3.1 (N)			&	0.2 		\\	
J182952 	&	1.1 (N)			&	 0.7 (N)			&	1.9		\\	
J182856 	&	1.3 (N)			&	1.1 (N)			&	2.1		\\	
J163143 	&	0.9 (F)			&	4.0 (N)			&	2.1		\\
J163136 	&	0.3 (F)			&	$<$2.0 (N)		&	1.3 		\\
J163152 	&	0.9 (F)			&	1.0 (N)			&	0.9 		\\	
J162625	&	1.5 (F)			&	--				&	2.4 		\\   	
J032838 	&	1.6 (N)			&	0.8 (N)			&	34.3 		\\	
J032848 	&	0.7 (N)			&	1.7 (N)			&	12.0		\\	
J032851 	&	1.1 (F)			&	1.5 (N)			&	5.0		\\	
J032859 	&	1.2 (N)			&	1.6 (N)			&	18.5 		\\	
J032911	&	2.3 (F)			&	1.7 (N)		 	&	2.0		\\	
\hline
\end{tabular}
\begin{tablenotes}
\item[a] The uncertainty on the o/p ratios is approximately 20\%-25\%. The CH$_{3}$OH o/p ratios have been calculated using either the column densities (N) or integrated fluxes (F). The H$_{2}$CO o/p ratios are from Riaz et al. (2019). The uncertainty on the $f_{D}$ measurements is estimated to be $\sim$20\%-30\% (Riaz \& Thi 2022a).
\end{tablenotes}
\end{threeparttable}
\end{table}

\subsection{Correlations}

 \begin{figure}
  \centering 
     \includegraphics[width=3in]{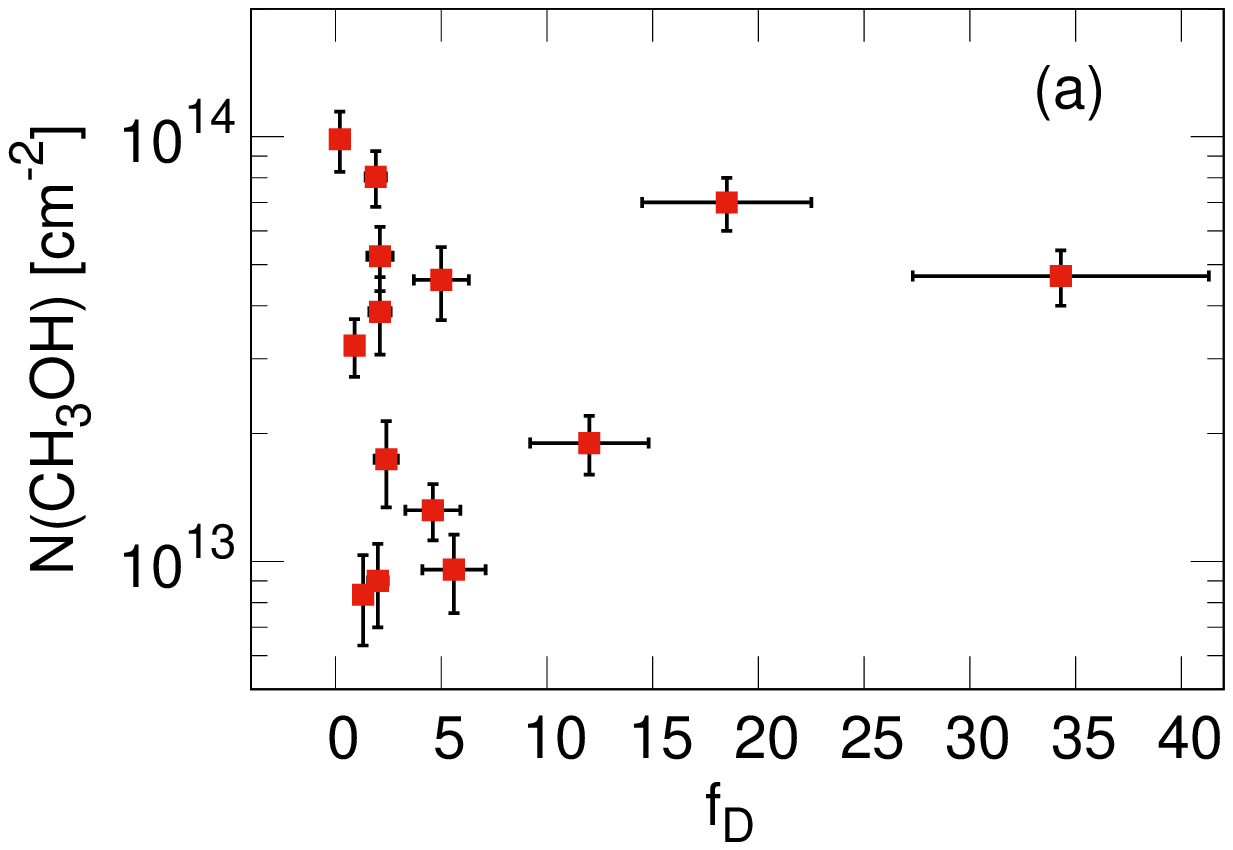}
     \includegraphics[width=3in]{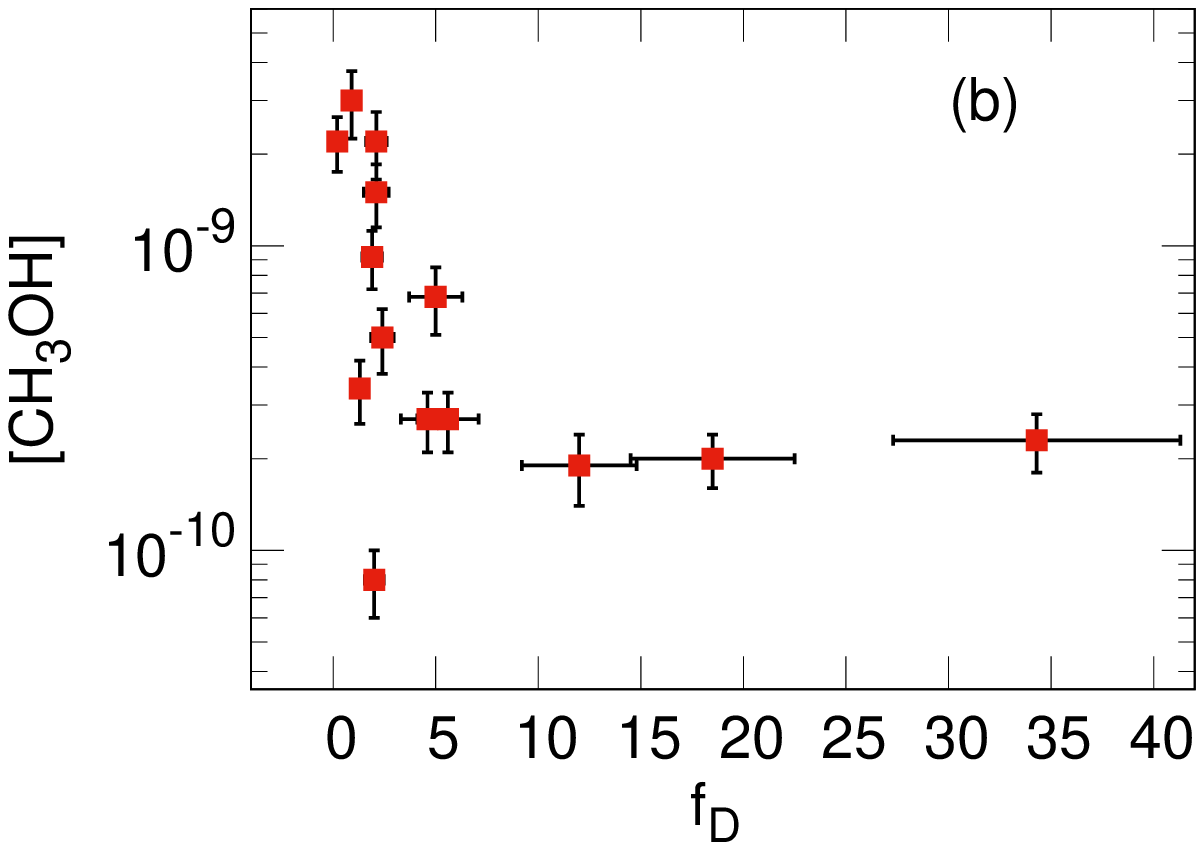}
     \includegraphics[width=3in]{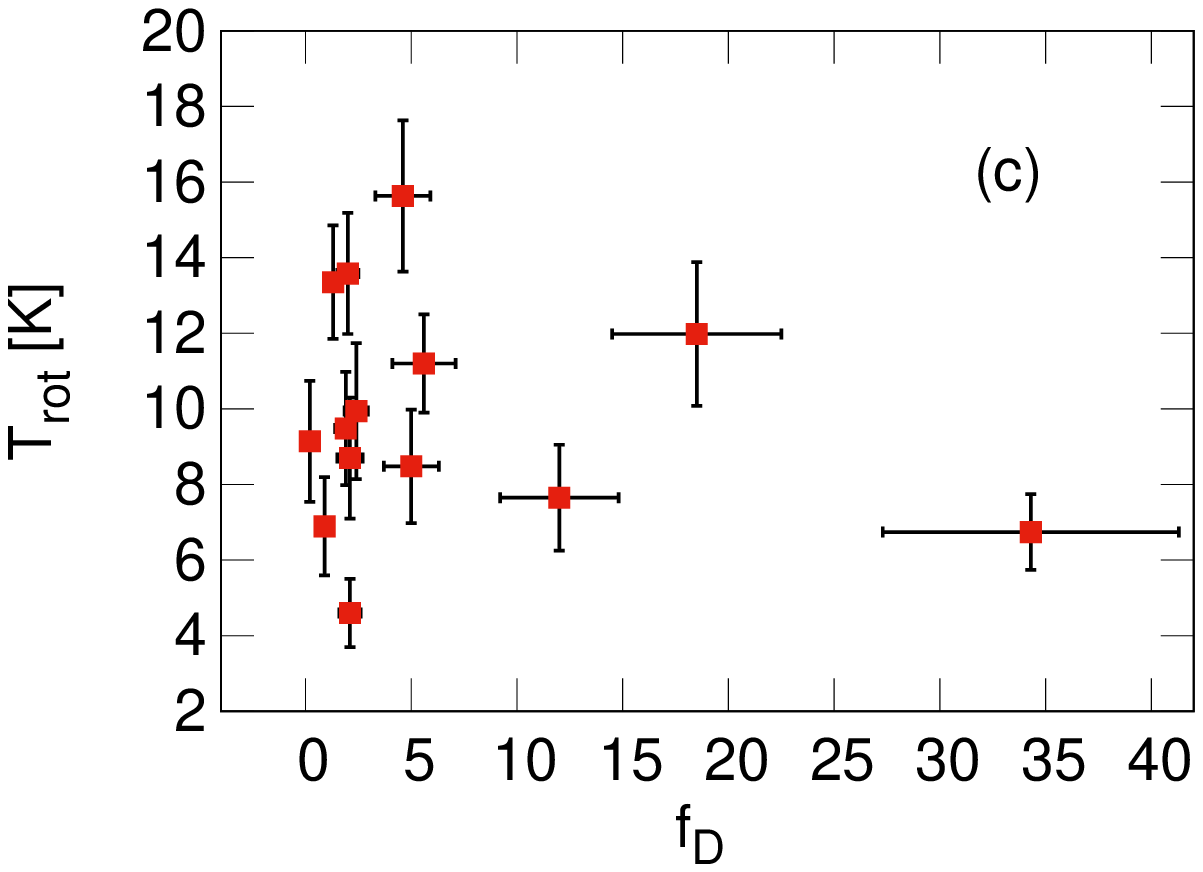}
   \caption{The CH$_{3}$OH column densities (a), abundances (b), and $T_{ex}$ (b) vs. the CO depletion factor f$_{D}$ for the proto-BDs. These are the values derived from fitting all data points in the rotational diagrams.}   
   \label{co-depl}     
  \end{figure}


Figure~\ref{co-depl} probes the dependence of the CH$_{3}$OH column densities, abundances, and $T_{ex}$ for the proto-BDs on the CO depletion factor, $f_{D}$, listed in Table~\ref{op-fd}. The details on the $f_{D}$ measurements are provided in Riaz \& Thi (2022a). The uncertainty on the $f_{D}$ measurements is estimated to be $\sim$20\%-30\% (Riaz \& Thi 2022a). For $f_{D} <$1, CO is expected to be in the gas-phase, while higher values of $f_{D} \sim$10 indicate a large fraction ($>$80\%) of CO is frozen onto dust grains. Most proto-BDs have f$_{D}\leq$5, except for the cases of J032838, J032848, and J032859, which have f$_{D}$ between $\sim$10-35, indicating high CO depletion. No clear correlation is seen between $f_{D}$ and the CH$_{3}$OH column densities, abundances, or $T_{ex}$ (Fig.~\ref{co-depl}ab), and the full range in values measured for these parameters are observed for f$_{D}\leq$5. The lack of any definitive trend suggests that the gas-phase abundances of CO and CH$_{3}$OH may not reflect their ice-phase abundances, possibly due to the different desorption timescales of these species.

 \begin{figure}
  \centering 
     \includegraphics[width=3in]{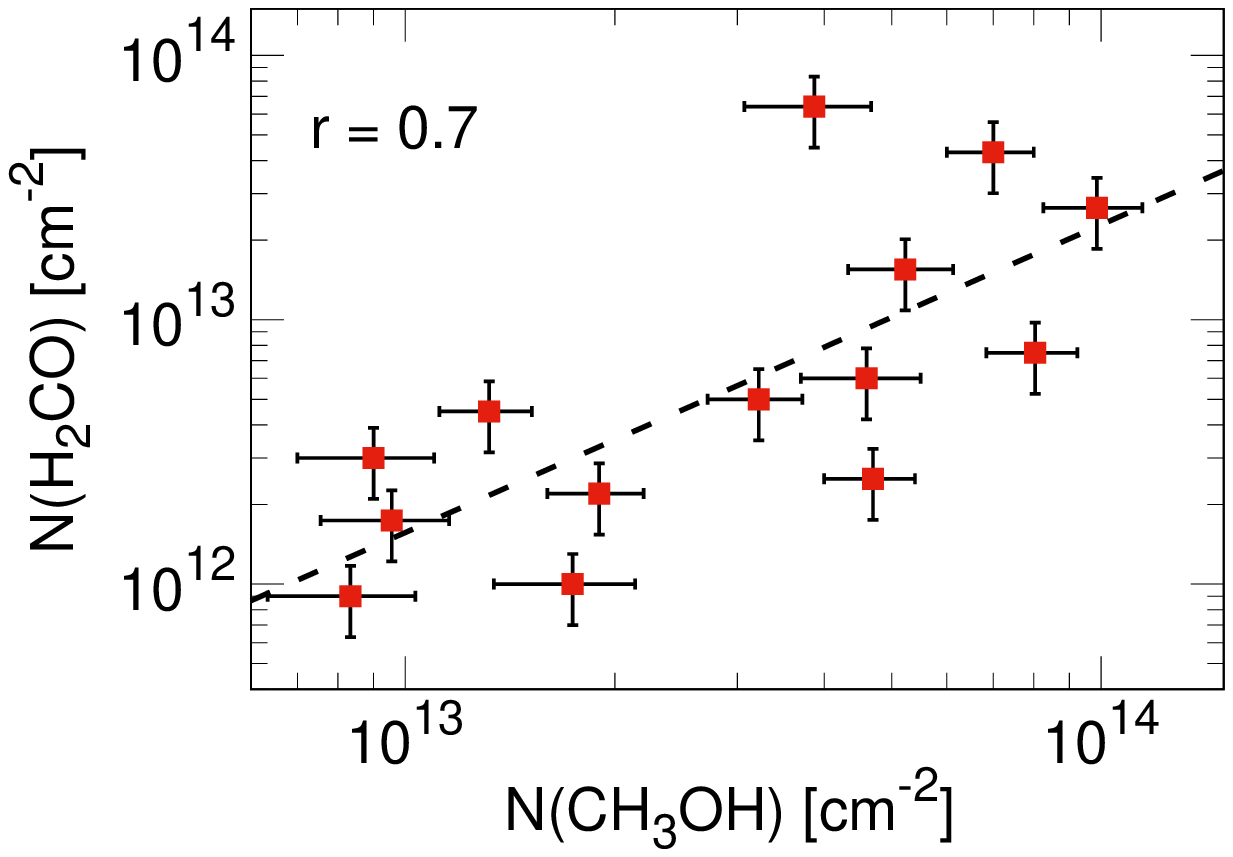}
\caption{The CH$_{3}$OH vs. H$_{2}$CO column densities for the proto-BDs. Dashed line is the fit to the data points. The correlation coefficient from the fit is noted in the plot. These are the values derived from fitting all data points in the rotational diagrams.}
\label{h2co}     
\end{figure}

Figure~\ref{h2co} shows a tight correlation between the CH$_{3}$OH and H$_{2}$CO column densities for the proto-BDs. The H$_{2}$CO data is from Riaz et al. (2019). The H$_{2}$CO column densities are lower than CH$_{3}$OH by a factor of $\sim$3-17. A tight correlation is expected since H$_{2}$CO is an intermediary in the formation of CH$_{3}$OH. The H$_{2}$CO o/p ratios for the proto-BDs range between $\sim$0.8-4 (Table~\ref{op-fd}). Low H$_{2}$CO o/p ratios of $<$3 suggest formation on dust grains while H$_{2}$CO o/p $\geq$3 indicate gas-phase formation of this molecule (e.g., Riaz et al. 2019). No clear correlation is seen between the CH$_{3}$OH and H$_{2}$CO o/p ratios; for e.g., J163143 has a CH$_{3}$OH and H$_{2}$CO o/p ratio of $\sim$0.9 and $\sim$4, respectively, while J163152 has similar CH$_{3}$OH and H$_{2}$CO o/p ratios (Table~\ref{op-fd}). A possible explanation for the lack of correlation between these ratios could be due to the presence of spin conversion processes, as discussed in Section~\ref{ratios}.

 \begin{figure}
  \centering 
     \includegraphics[width=3in]{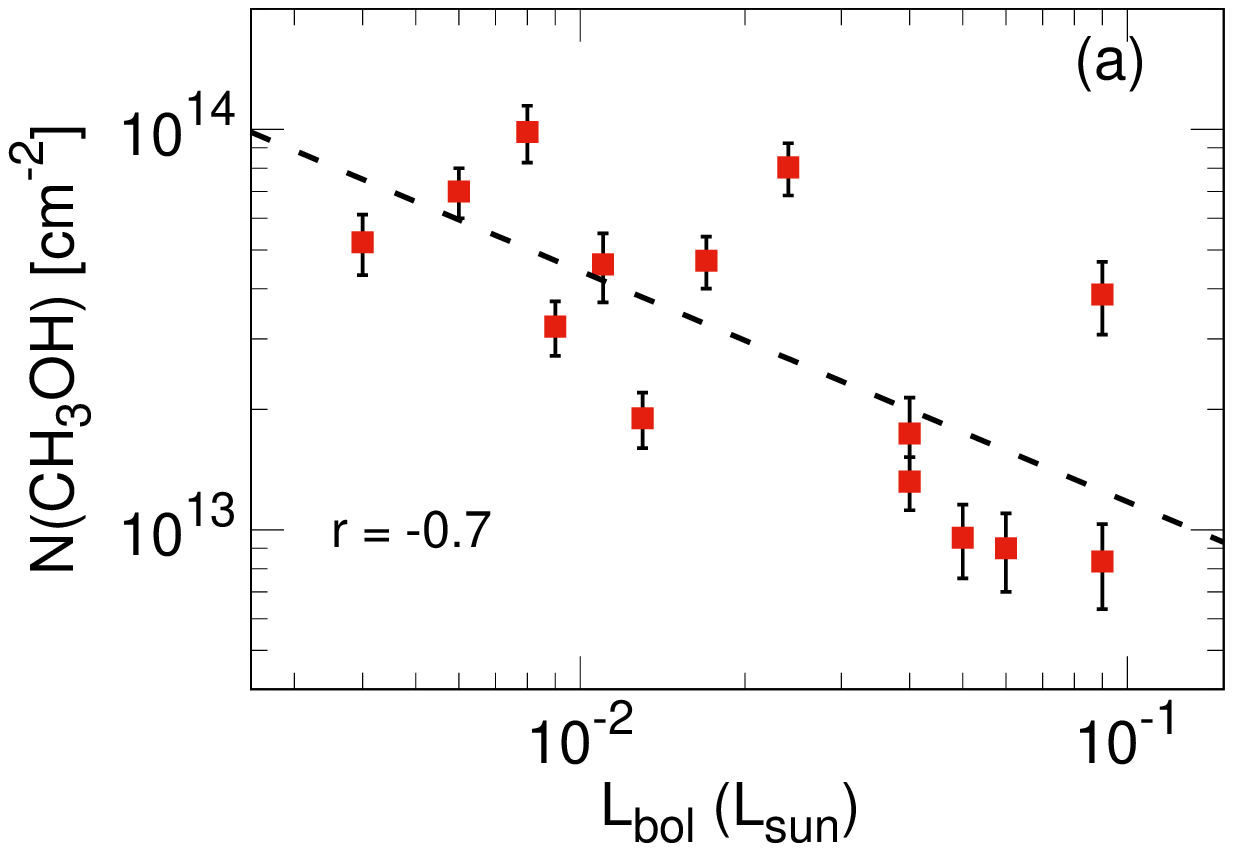}
     \includegraphics[width=3in]{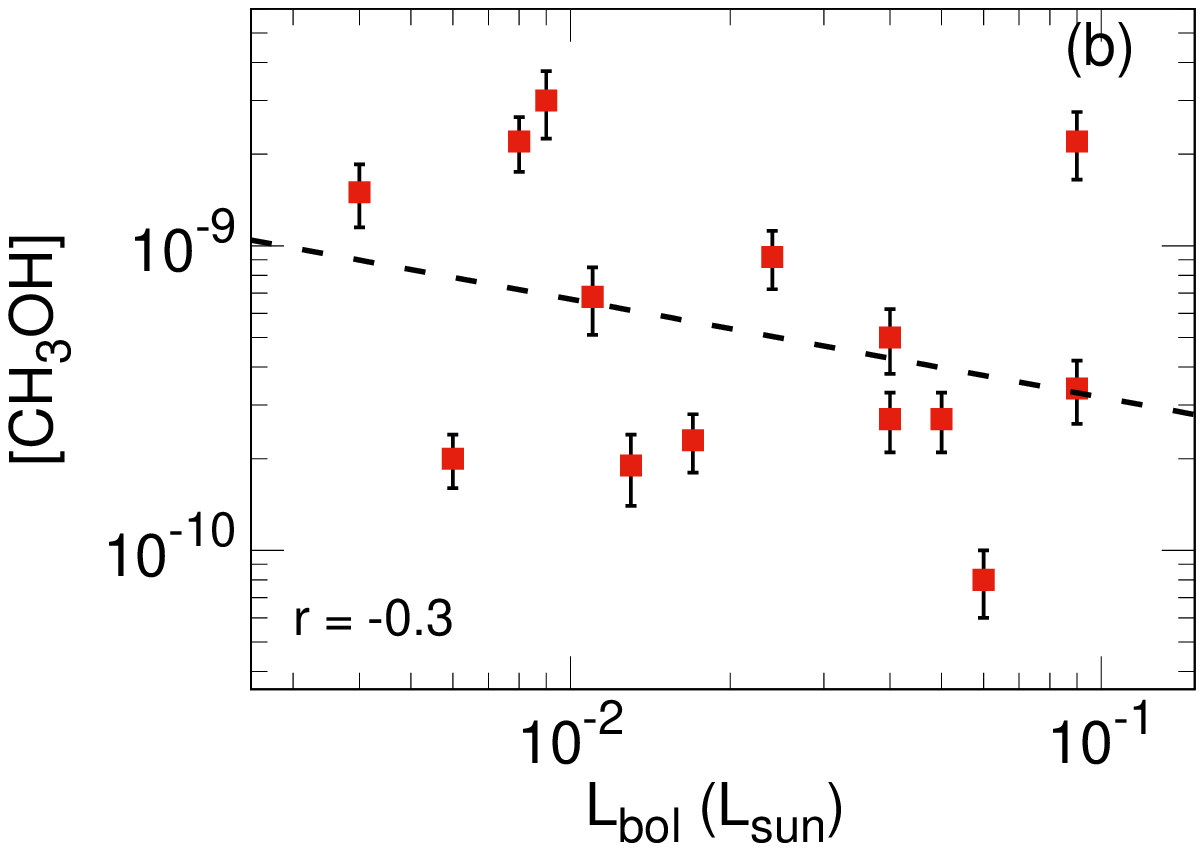}
     \includegraphics[width=3in]{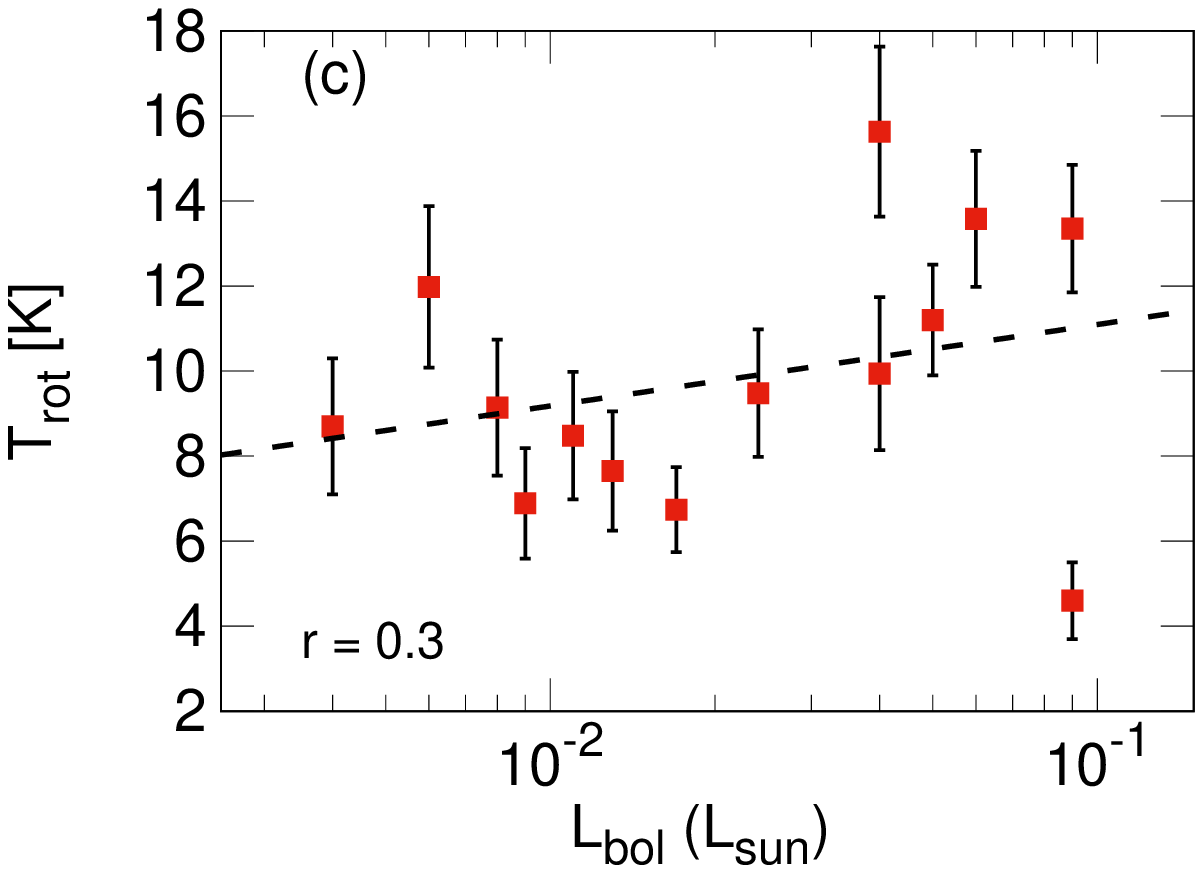}
   \caption{The CH$_{3}$OH column densities (a), abundances (b), and $T_{ex}$ (b) vs. the bolometric luminosity for the proto-BDs. These are the values derived from fitting all data points in the rotational diagrams.}
   \label{protoBDs}     
  \end{figure}

 \begin{figure}
  \centering 
     \includegraphics[width=3in]{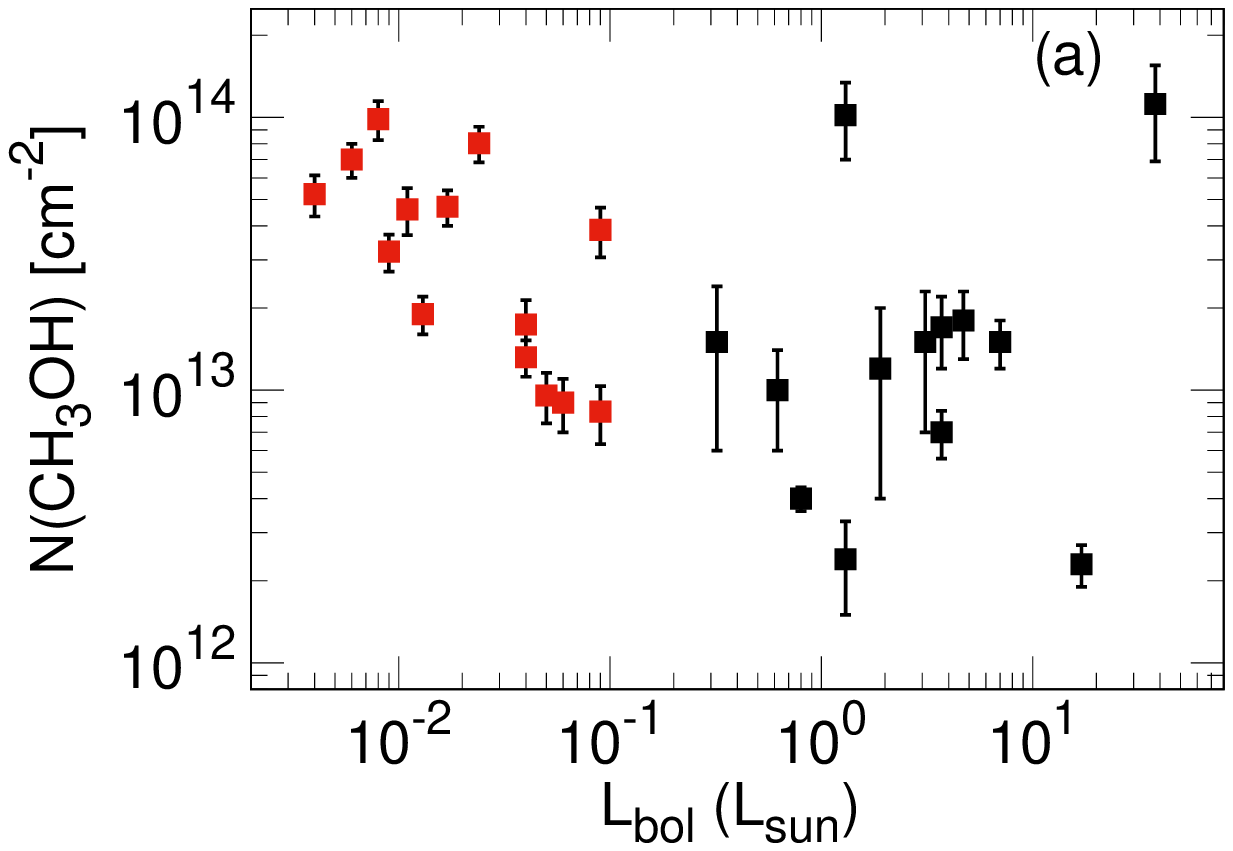}     
     \includegraphics[width=3in]{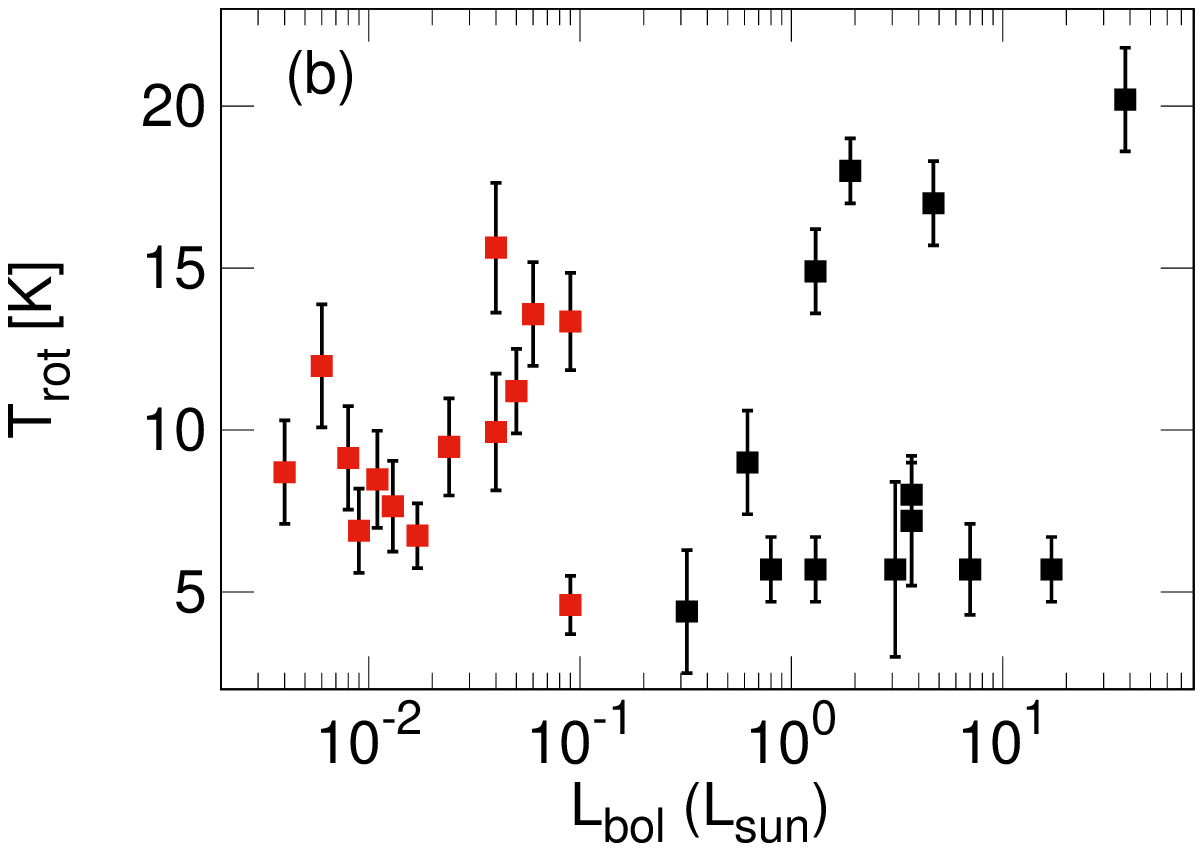}
   \caption{The CH$_{3}$OH column densities (a) and $T_{ex}$ (b) vs. the bolometric luminosity for the proto-BDs (red points) and low-mass protostars (black points). These are the values derived from fitting all data points in the rotational diagrams.  }
   \label{proto}     
  \end{figure}

Figure~\ref{protoBDs} plots the N(CH$_{3}$OH) and $T_{ex}$ for the proto-BDs. For a quantitative comparison of the correlations between the various parameters, we have calculated the Pearson correlation coefficient, {\it r}, from the best-fit to the observed data points, and the value for {\it r} is noted in each plot. A strong anti-correlation ({\it r} = -0.7) is seen between N(CH$_{3}$OH) and L$_{bol}$, with a clear rise in N(CH$_{3}$OH) with decreasing bolometric luminosity, whereas the CH$_{3}$OH abundances show a weaker anti-correlation ({\it r} = -0.3) with L$_{bol}$. A weak correlation ({\it r} = 0.3) is also seen between $T_{ex}$ and L$_{bol}$, with a slight decline in $T_{ex}$ with decreasing bolometric luminosity. Together, these trends suggest that the cool and dense physical conditions are ripe for efficient formation of methanol in proto-BDs, and the presence of a warm corino allows a large fraction of the methanol formed to be released in the gas-phase. 


Figure~\ref{proto} shows a comparison of N(CH$_{3}$OH) and T$_{rot}$ for the proto-BDs with low-mass protostars. Data for the low-mass protostars is also from IRAM 30m observations presented in \"{O}berg et al. (2014) and Graninger et al. (2016). A comparison of the methodology used in our work for deriving the column densities and T$_{rot}$ with Graninger et al. (2016) is presented in Section~\ref{proto-2}. Despite the large spread in the values for these parameters, the median N(CH$_{3}$OH) and $T_{ex}$ for the proto-BDs are 4.0$\times$10$^{13}$ cm$^{-2}$ and 9.5 K, respectively. These values are comparatively higher than the median N(CH$_{3}$OH) and $T_{ex}$ of 1.5$\times$10$^{13}$ cm$^{-2}$ and 7.2 K, respectively, for the protostars. There may be a tentative trend of a rise in N(CH$_{3}$OH) and a decline in $T_{ex}$ with decreasing bolometric luminosity.


\section{Summary}

We have conducted a molecular line survey with IRAM 30-m to search for methanol in 14 Class 0/I proto-BDs, with an aim to understand the chemical complexity during the early phases of brown dwarf formation. Our observations cover the frequency range of 92-116 GHz and 213-280 GHz. We have detected 14 CH$_{3}$OH transition lines with E$_{upper}\sim$7-49 K, and $n_{crit}\sim$10$^{5}$-10$^{9}$ cm$^{-3}$. Detection in a particular transition line is more a function of E$_{upper}$ than $n_{crit}$. 

\begin{itemize}

\item All proto-BDs show emission in the transition lines at E$_{upper}<$ 20 K, indicating that a considerable gas mass is cold in these objects. None of the CH$_{3}$OH transitions at E$_{upper}>$50 K are detected in any of the proto-BDs. 

\item Comparing the observations with the brown dwarf formation models indicates that the low-excitation (E$_{upper}\leq$20 K) CH$_{3}$OH lines likely trace the cold gas component at $\sim$50-150 au, while the higher excitation lines at E$_{upper}\sim$25-49 K probe the warm-temperature gas component at $\sim$10-50 au in the proto-BDs. This is the warm corino region surrounding the jet launching zone. 

\item The CH$_{3}$OH column density for the cold component is at least an order of magnitude higher than the warm gas component. 

\item The CH3OH ortho-to-para ratios for the proto-BDs are in the range of $\sim$0.3-2.3. Ratios of less than unity suggest the presence of spin conversion processes.

\item We find a clear rise in the CH$_{3}$OH column densities with decreasing bolometric luminosity among the proto-BDs, indicating that the cool and dense physical conditions are ripe for efficient formation of methanol in these objects. 

\item The median CH$_{3}$OH column density for the proto-BDs is a factor of $\sim$3 higher than the median for low-mass protostars. 

\item Observations of CH$_{3}$OH lines from levels with upper energy $>$ 25 K together with the model predictions suggest that a warm corino is likely present in $\sim$78\% of the proto-BDs in our sample. The remaining targets show evidence of only the cold ($<$ 20 K) gas component, possibly due to the absence of a high-velocity jet that can stir up the warm-temperature gas around it. Further non-LTE studies with a radiative-transfer code as well as interferometric observations are needed to confirm these suggestions.

\end{itemize}

\section*{Acknowledgements}

B.R. acknowledges funding from the Deutsche Forschungsgemeinschaft (DFG) - Projekt number RI-2919/2-3. This work is based on observations carried out with the IRAM 30m telescope. IRAM is supported by INSU/CNRS (France), MPG (Germany) and IGN (Spain).

\section{Data Availability}

The data underlying this article are available in the IRAM archives through the VizieR online database.


\newpage

\onecolumn

\title{Supplementary Material}

\appendix

\section{Spectra and Line Parameters}
\label{all-spectra}

The observed CH$_{3}$OH spectra are shown in Fig.~A1. The line parameters derived from these spectra are listed in Table~\ref{line-pars}. The uncertainty is estimated to be $\sim$10\%-20\% for the peak and integrated intensities and $\Delta${\it v}, and $\sim$0.02-0.04 km s$^{-1}$ for {\it V}$_{lsr}$. 


\begin{figure*}
  \centering 
     \includegraphics[width=5.6in]{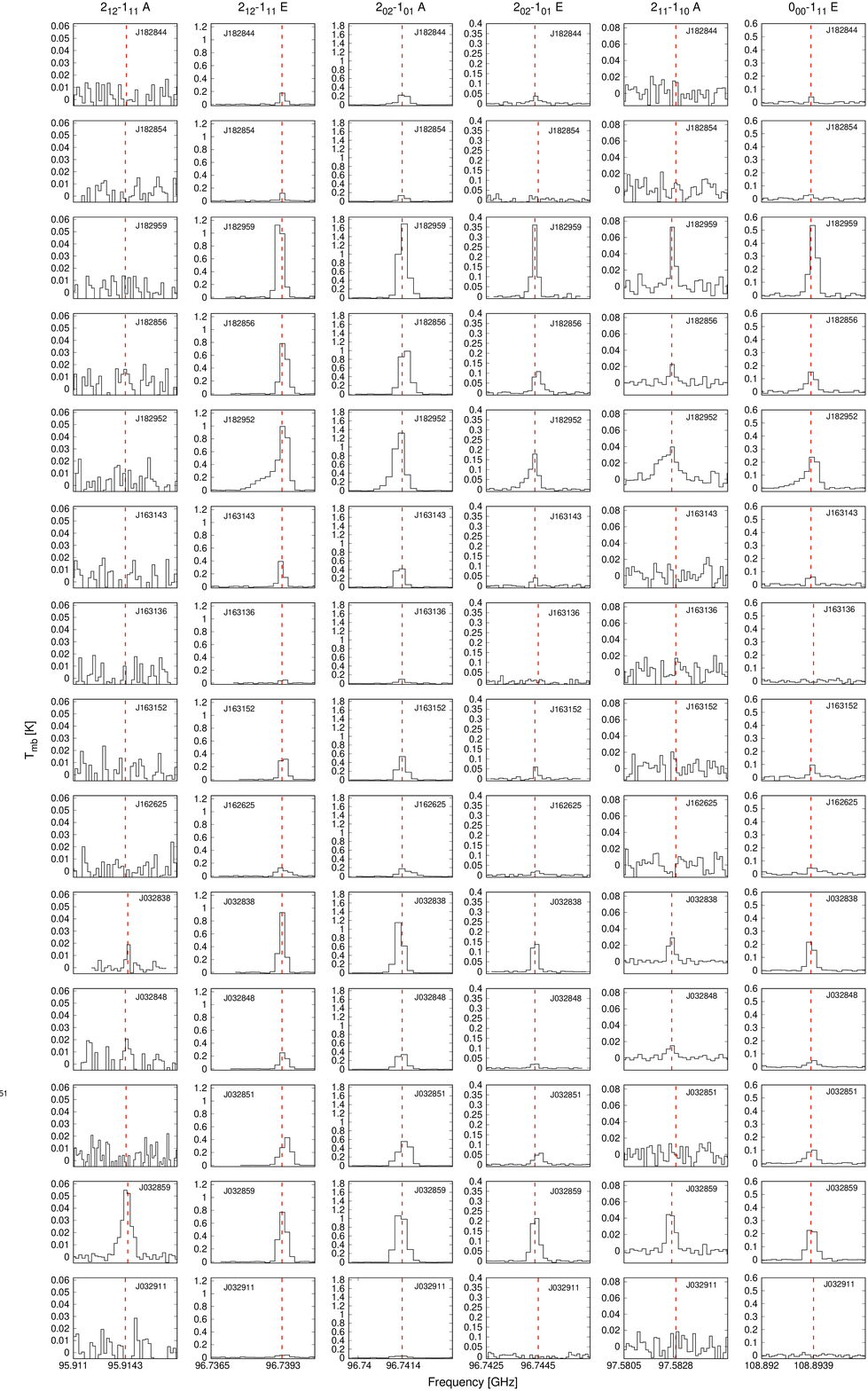}
     \caption{The CH$_{3}$OH spectra for the targets in the E090 band. Red dashed line marks the rest frequency of the transition line. }
   \label{spectraE090}     
  \end{figure*}

\begin{figure*}
  \centering 
     \includegraphics[width=7.2in]{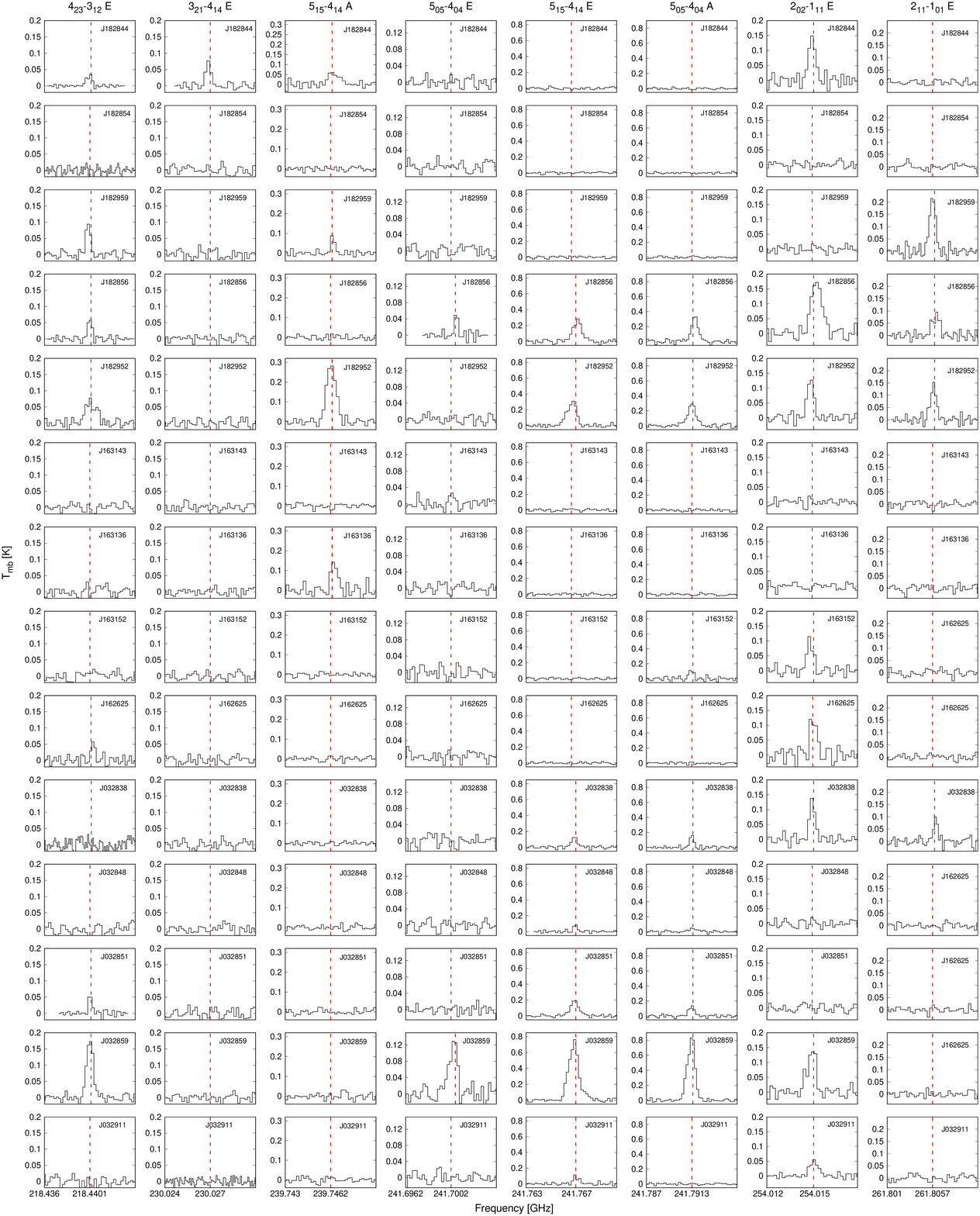}
     \caption{The CH$_{3}$OH spectra for the targets in the E230 band. Red dashed line marks the rest frequency of the transition line. }
   \label{spectraE230}     
  \end{figure*}

 \begin{figure*}
  \centering 
     \includegraphics[width=1.5in]{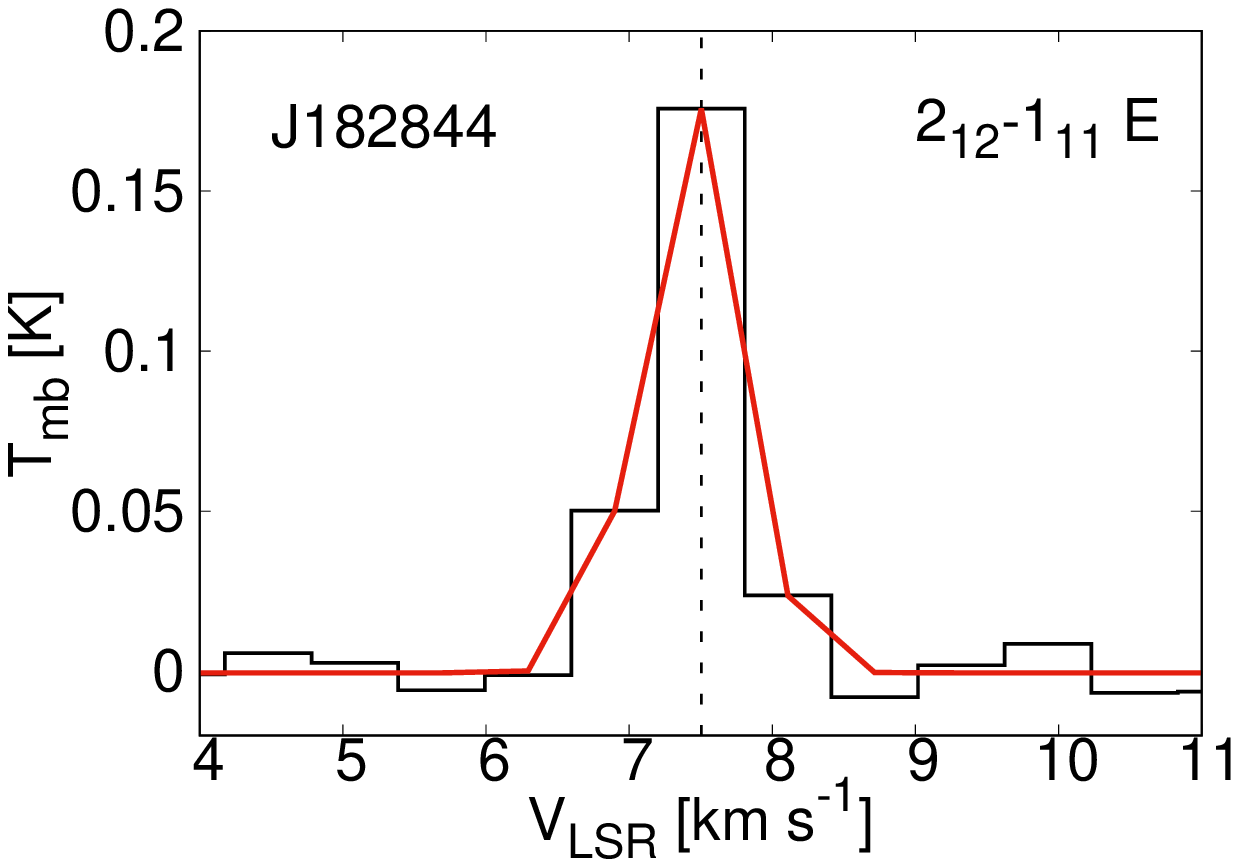}
     \includegraphics[width=1.5in]{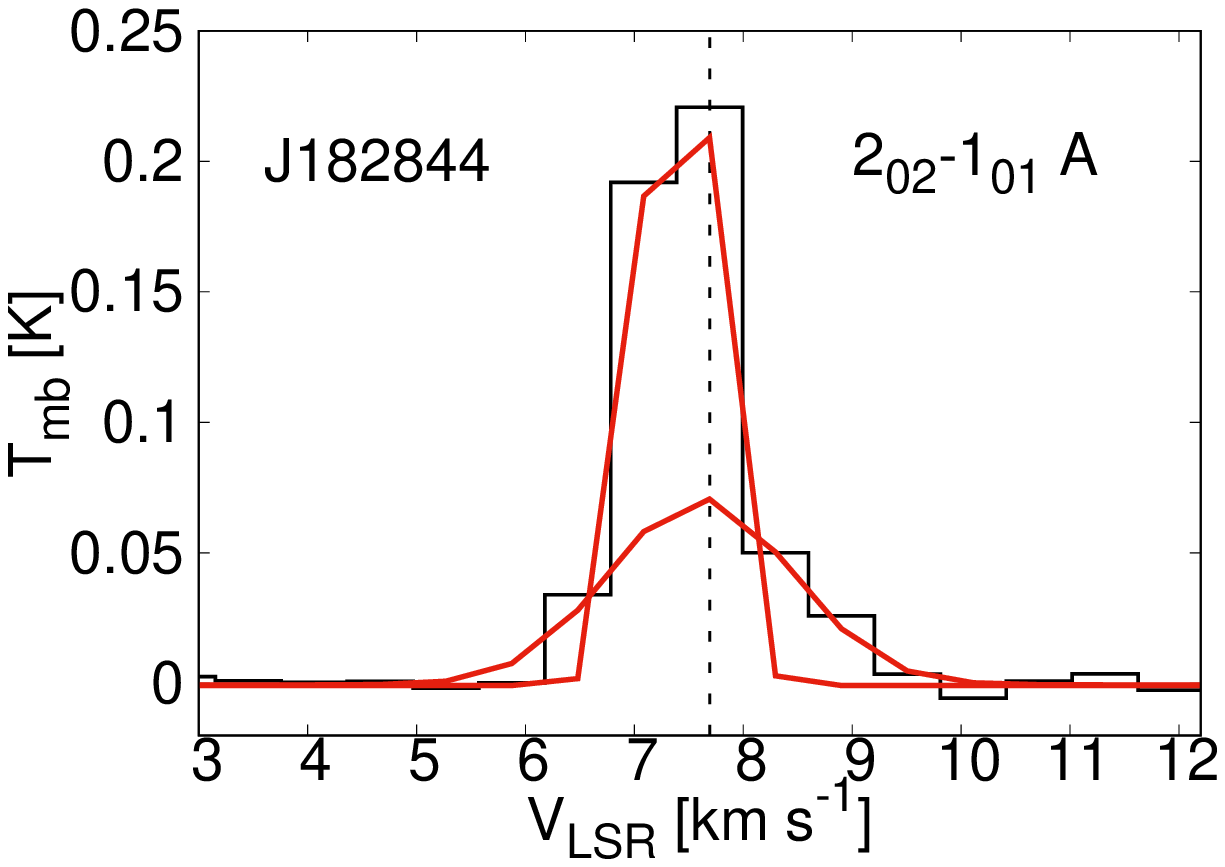}
     \includegraphics[width=1.5in]{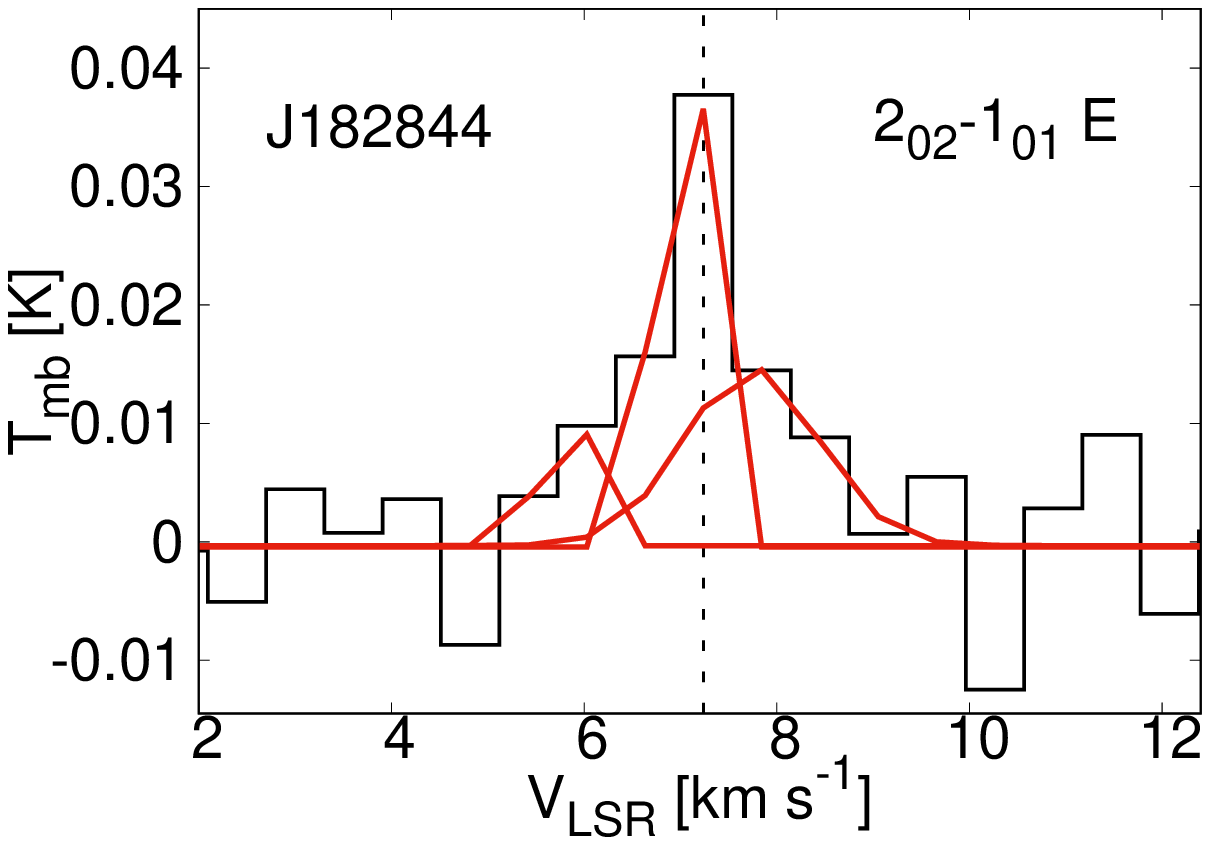}
     \includegraphics[width=1.5in]{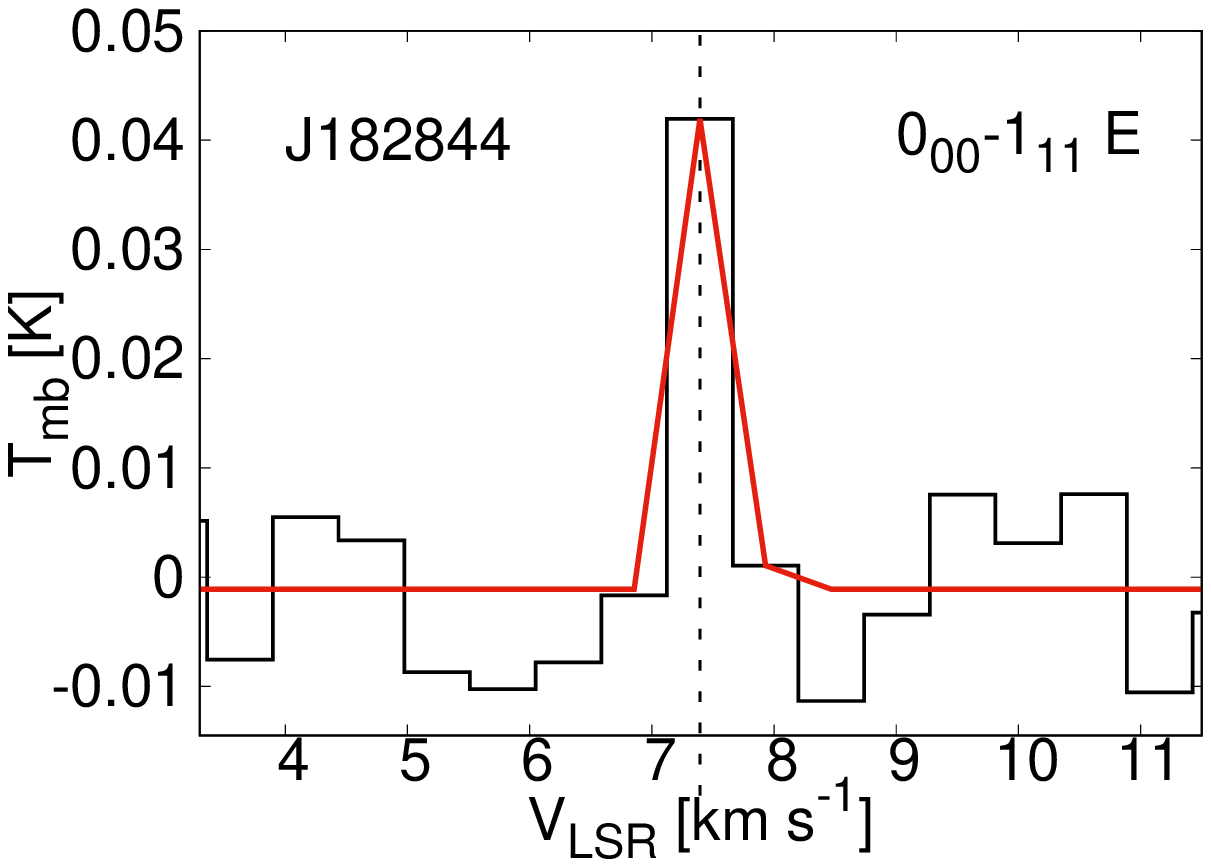}
     \includegraphics[width=1.5in]{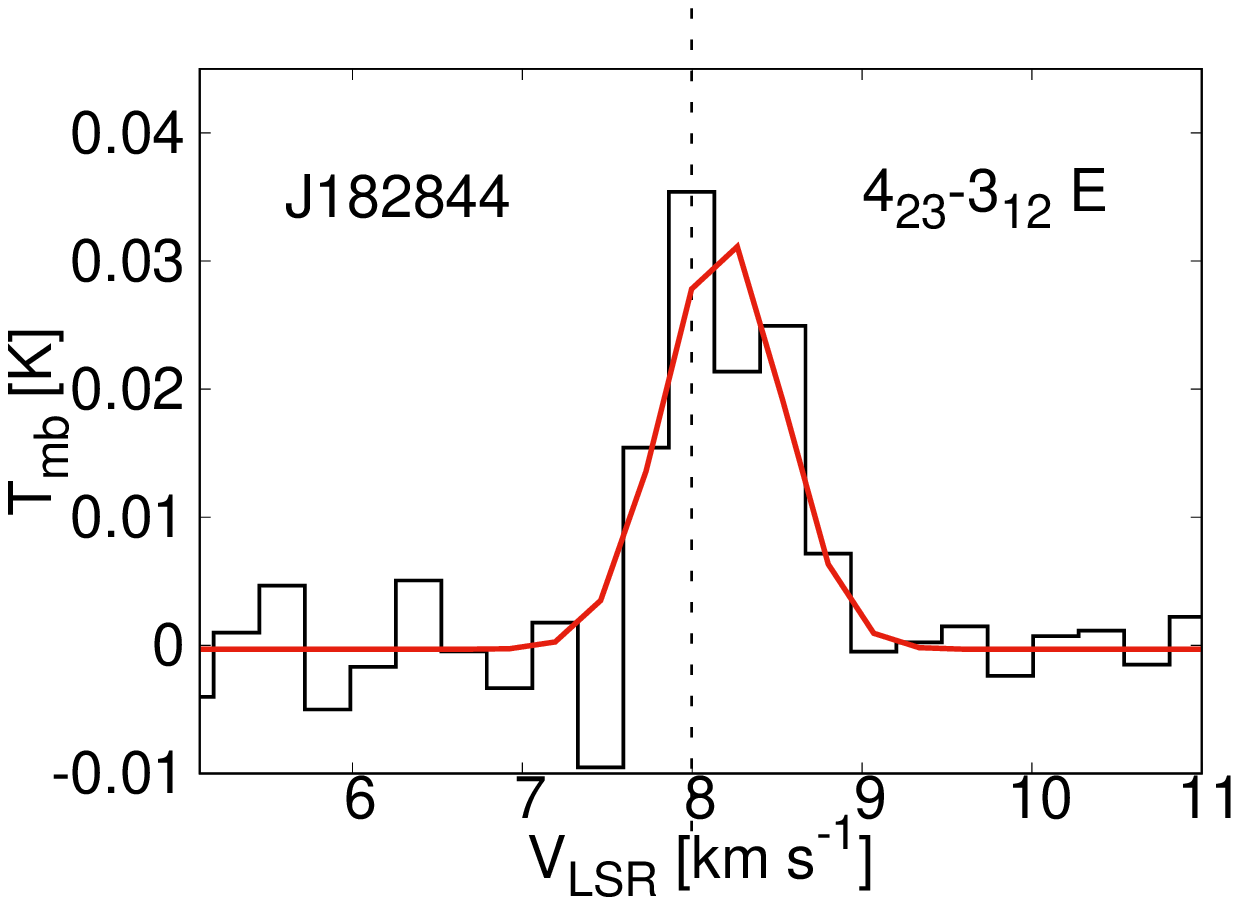}
     \includegraphics[width=1.5in]{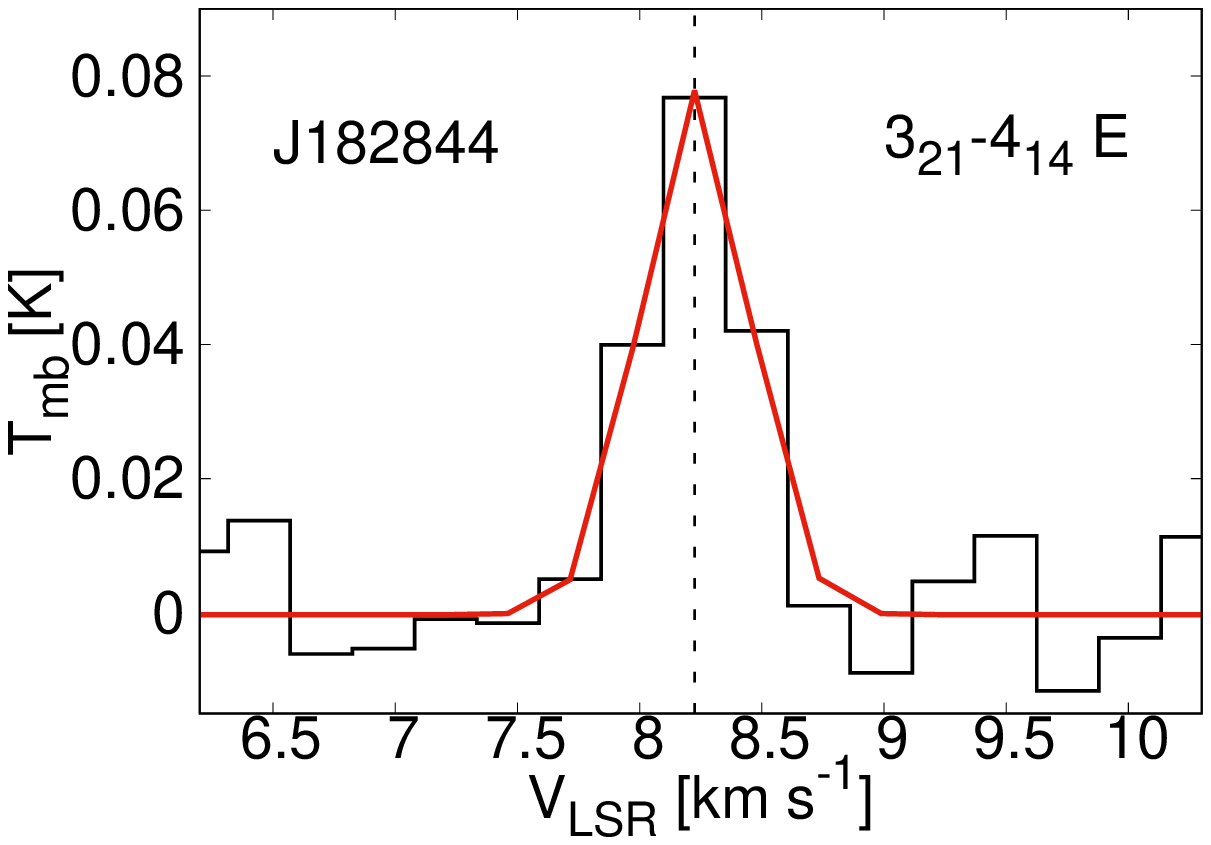}	
     \includegraphics[width=1.5in]{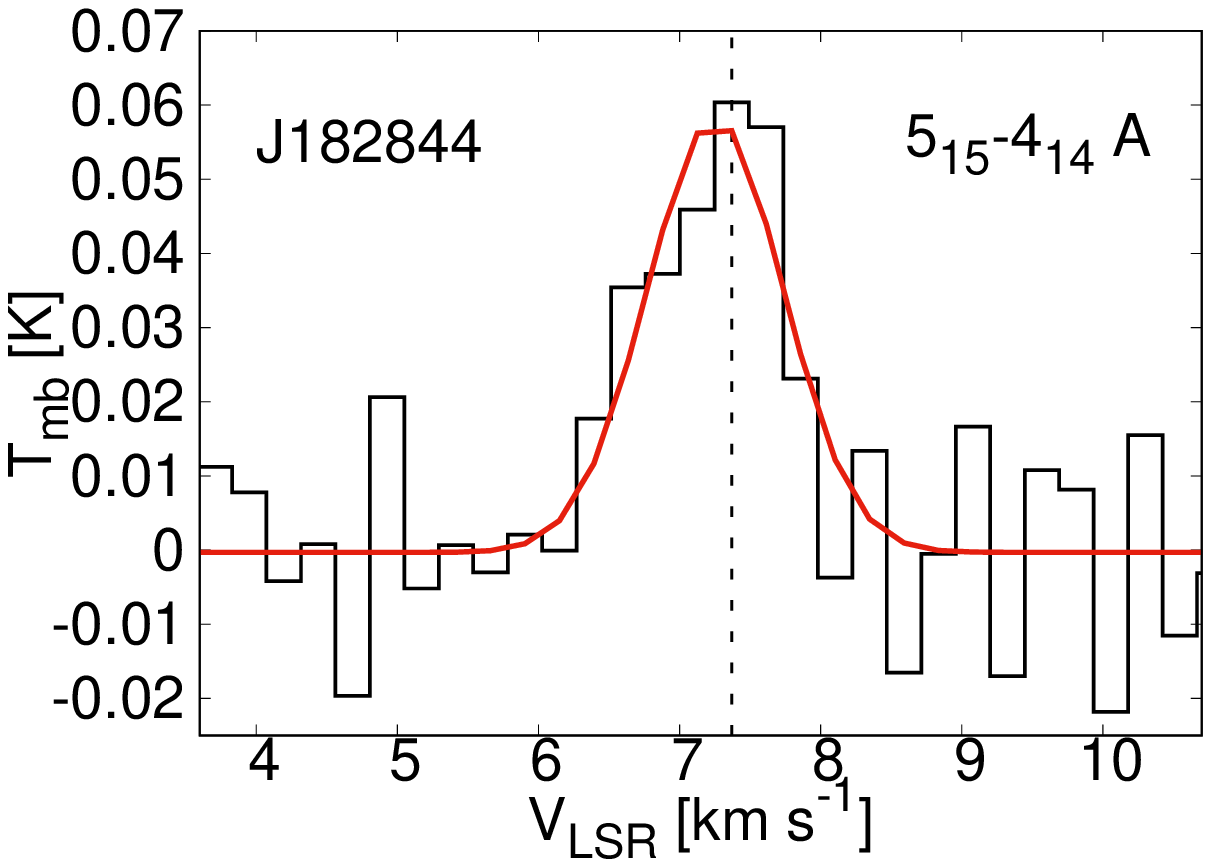}	
     \includegraphics[width=1.5in]{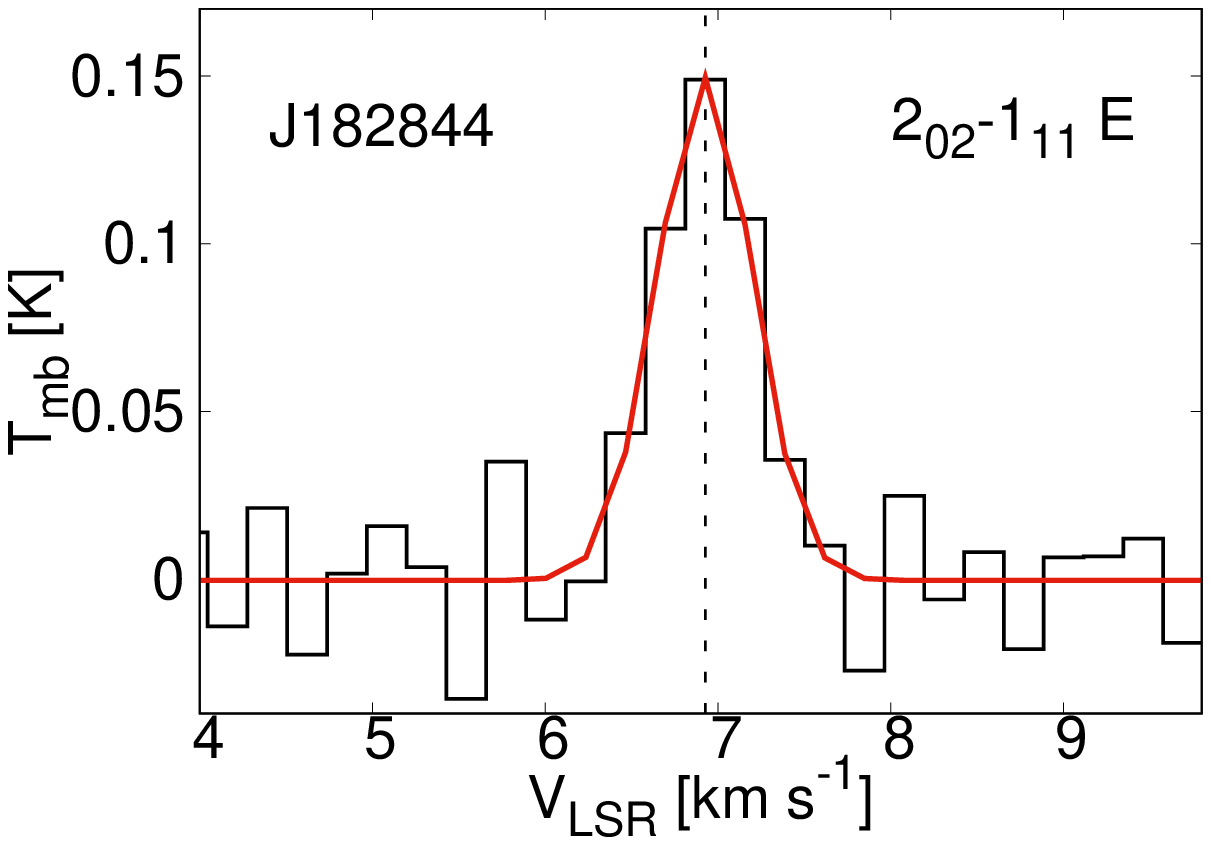}	\\	\vspace{0.2in}

     \includegraphics[width=1.5in]{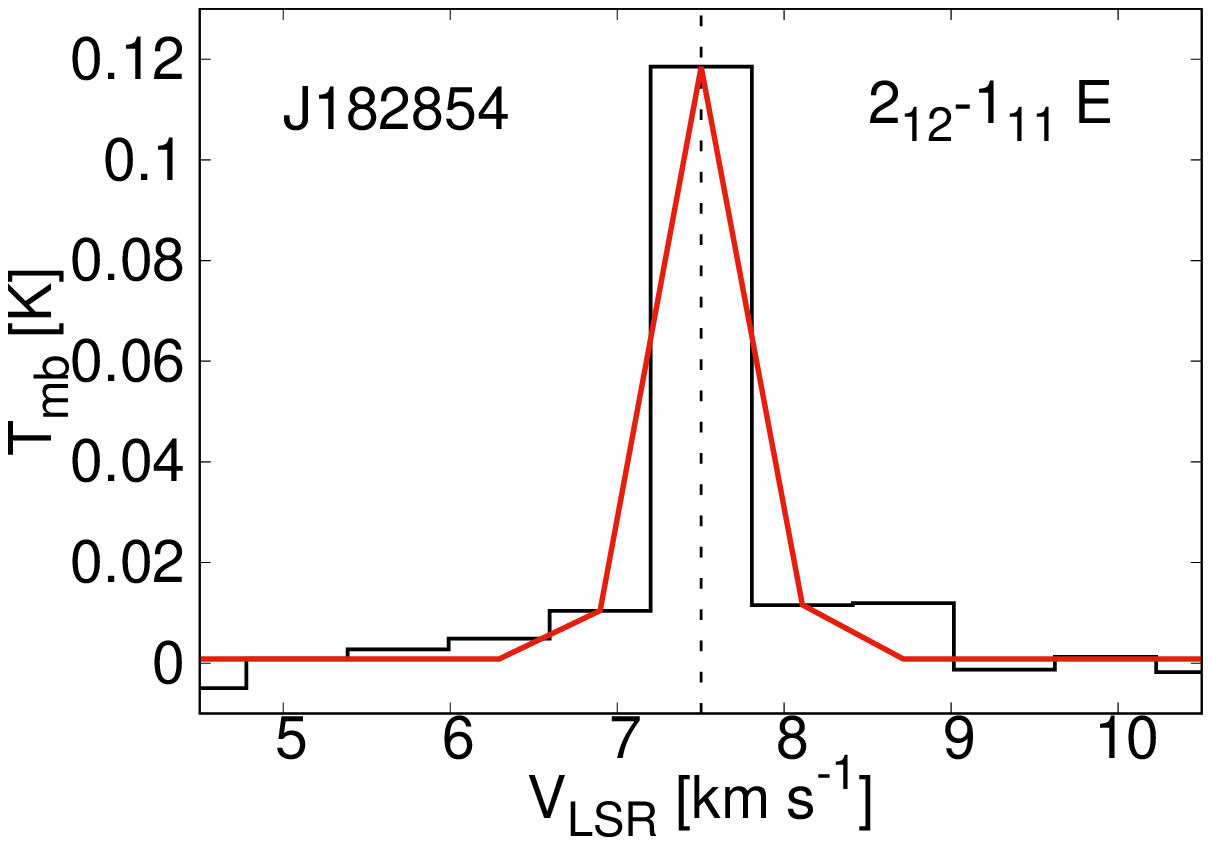}	
     \includegraphics[width=1.5in]{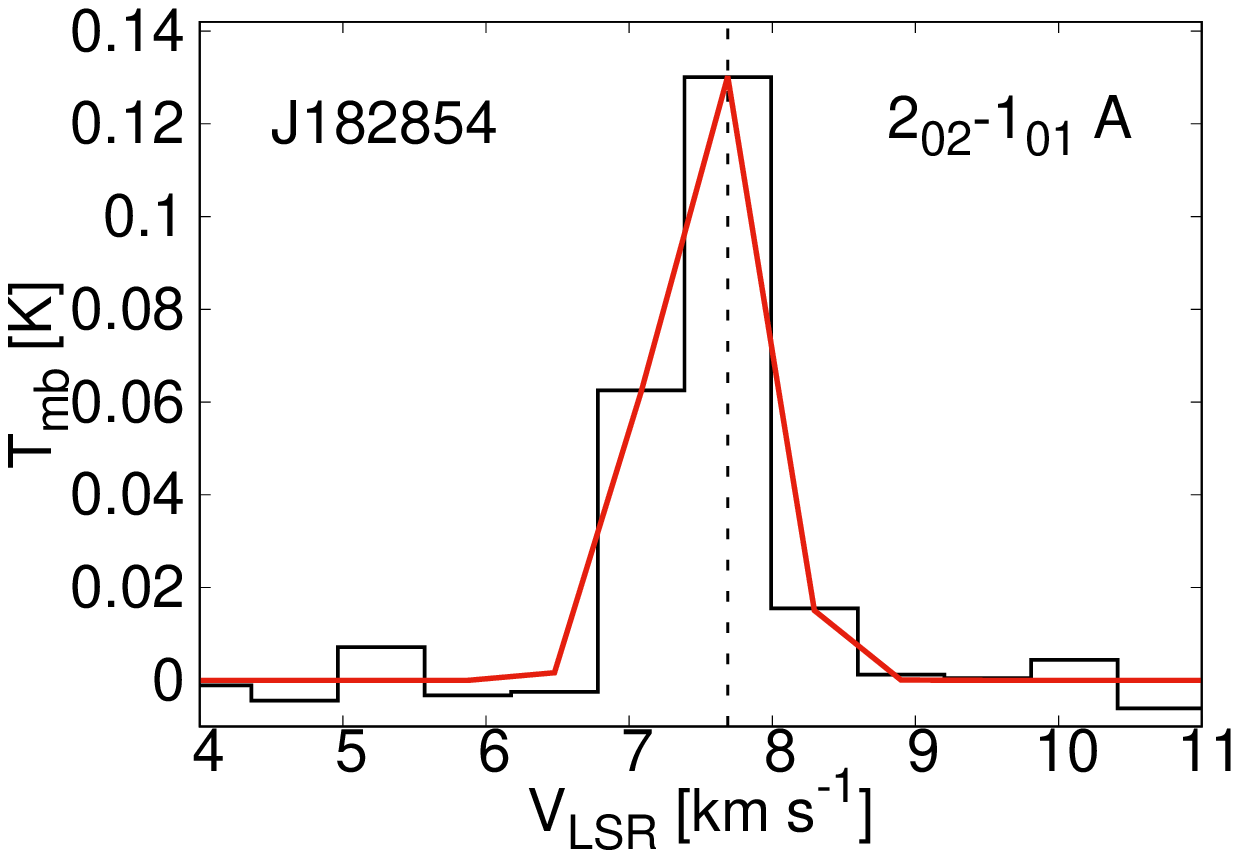}
     \includegraphics[width=1.5in]{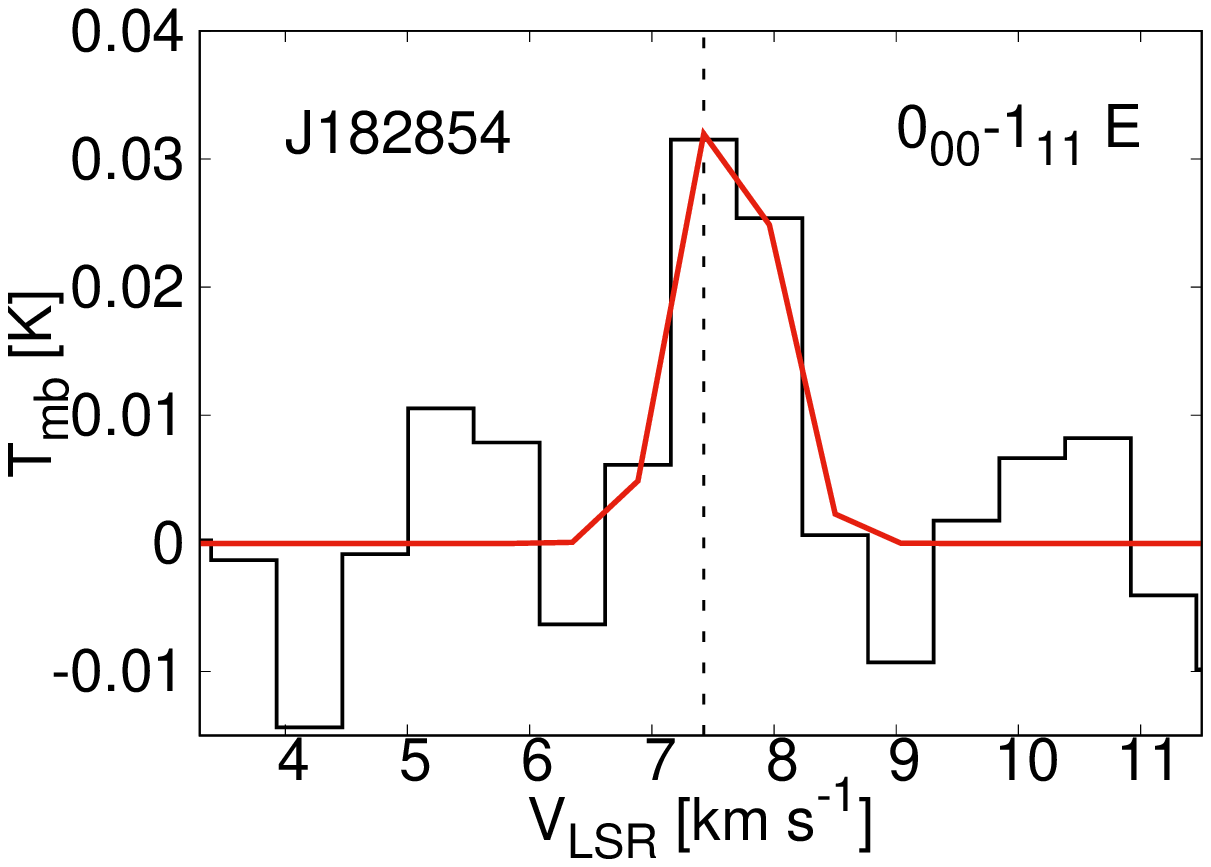}	\\	\vspace{0.2in}

    \includegraphics[width=1.5in]{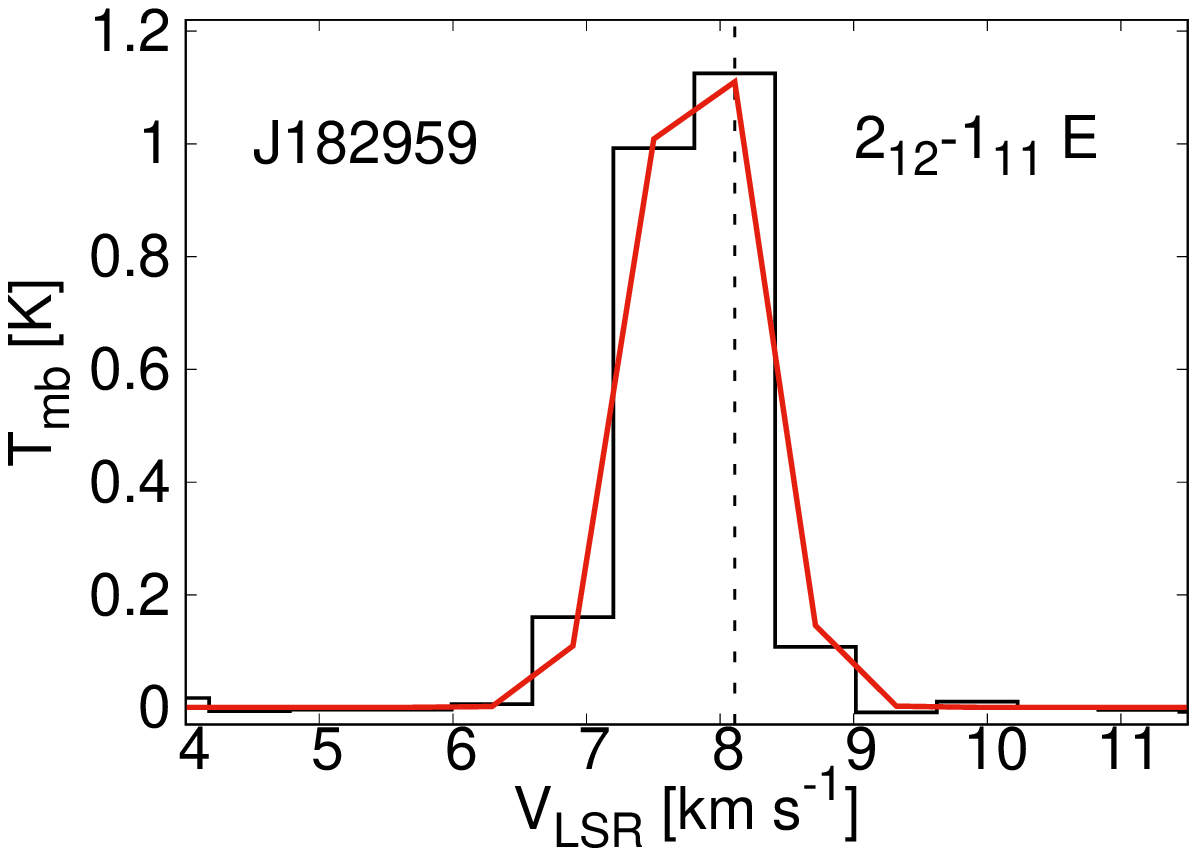}	
     \includegraphics[width=1.5in]{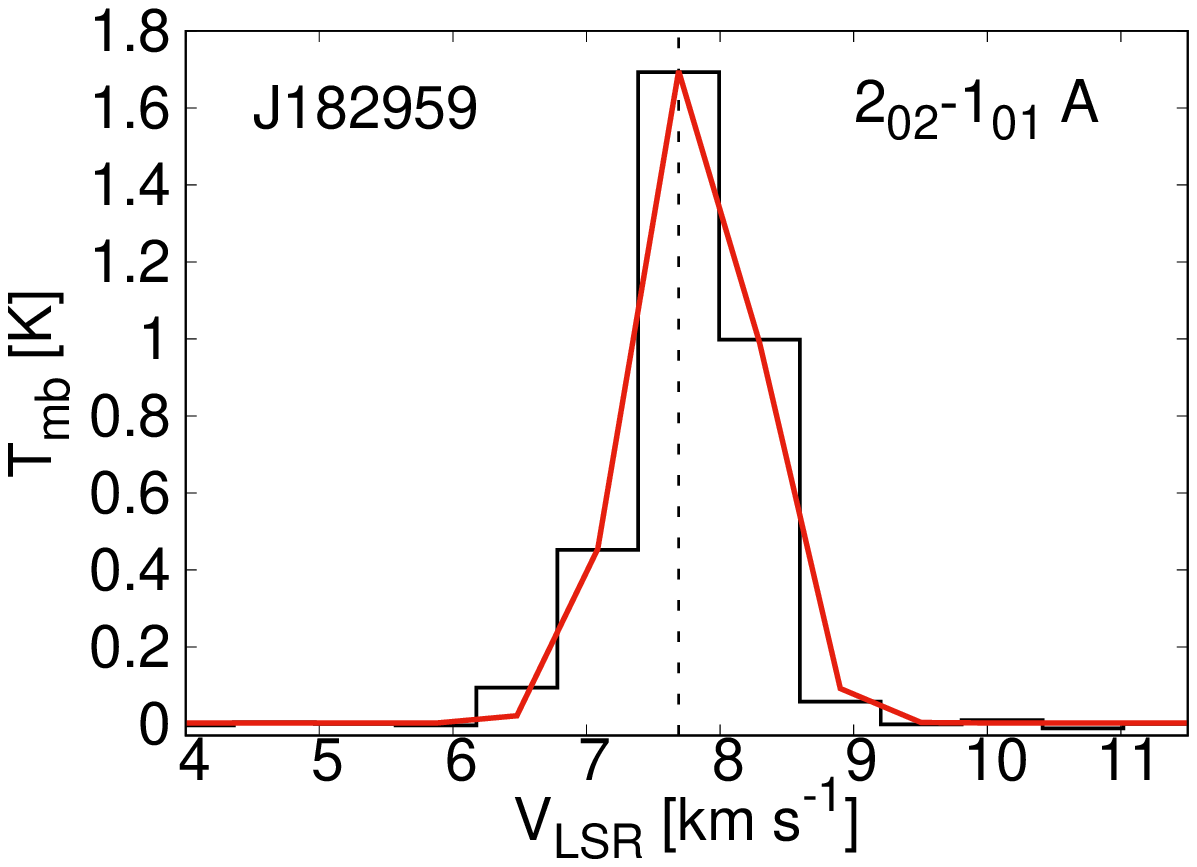}	
     \includegraphics[width=1.5in]{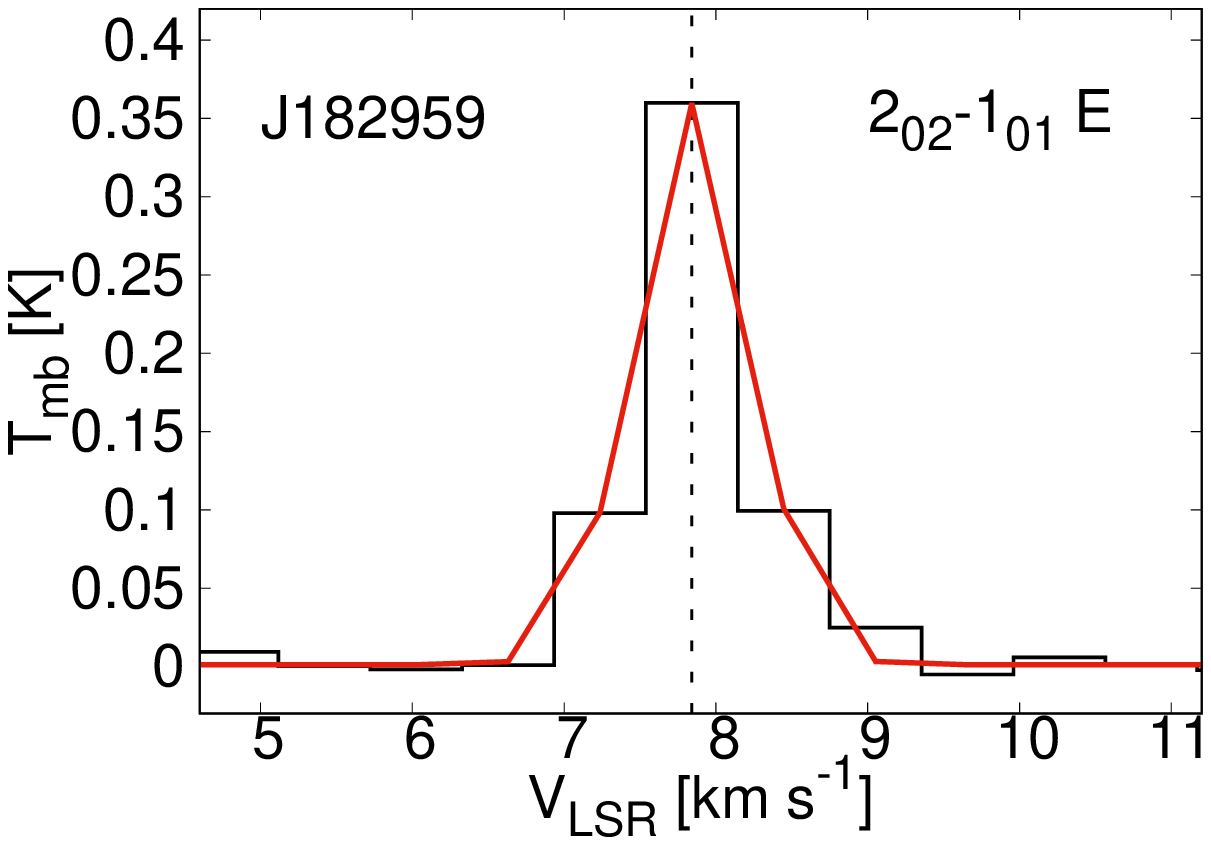}
     \includegraphics[width=1.5in]{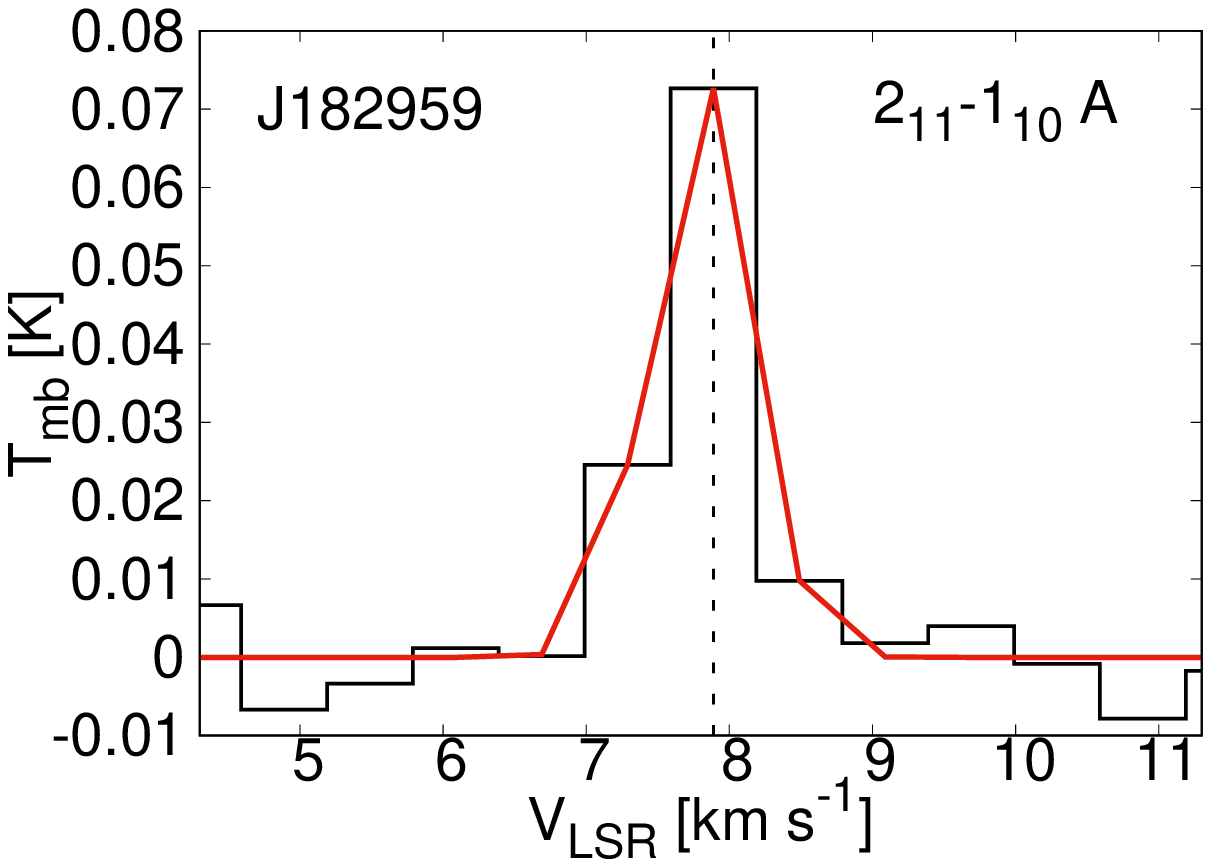}
     \includegraphics[width=1.5in]{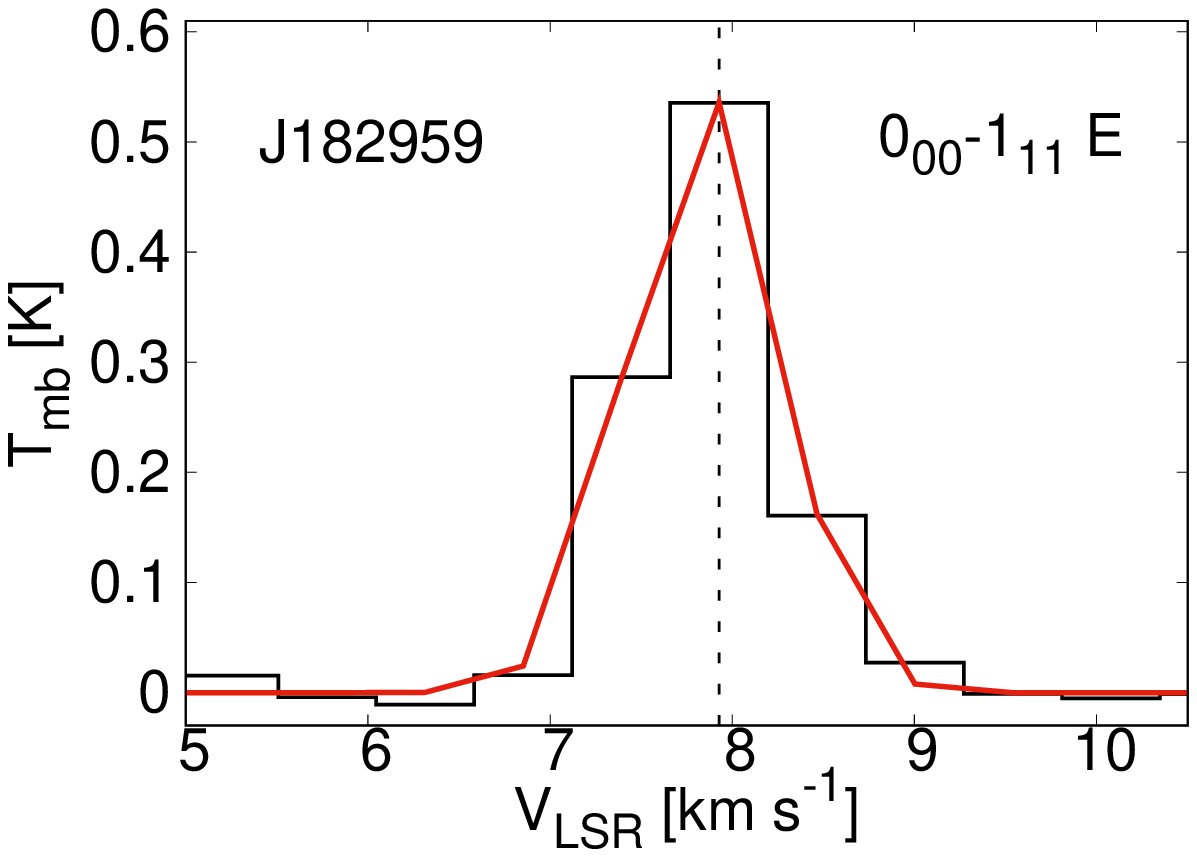}
     \includegraphics[width=1.5in]{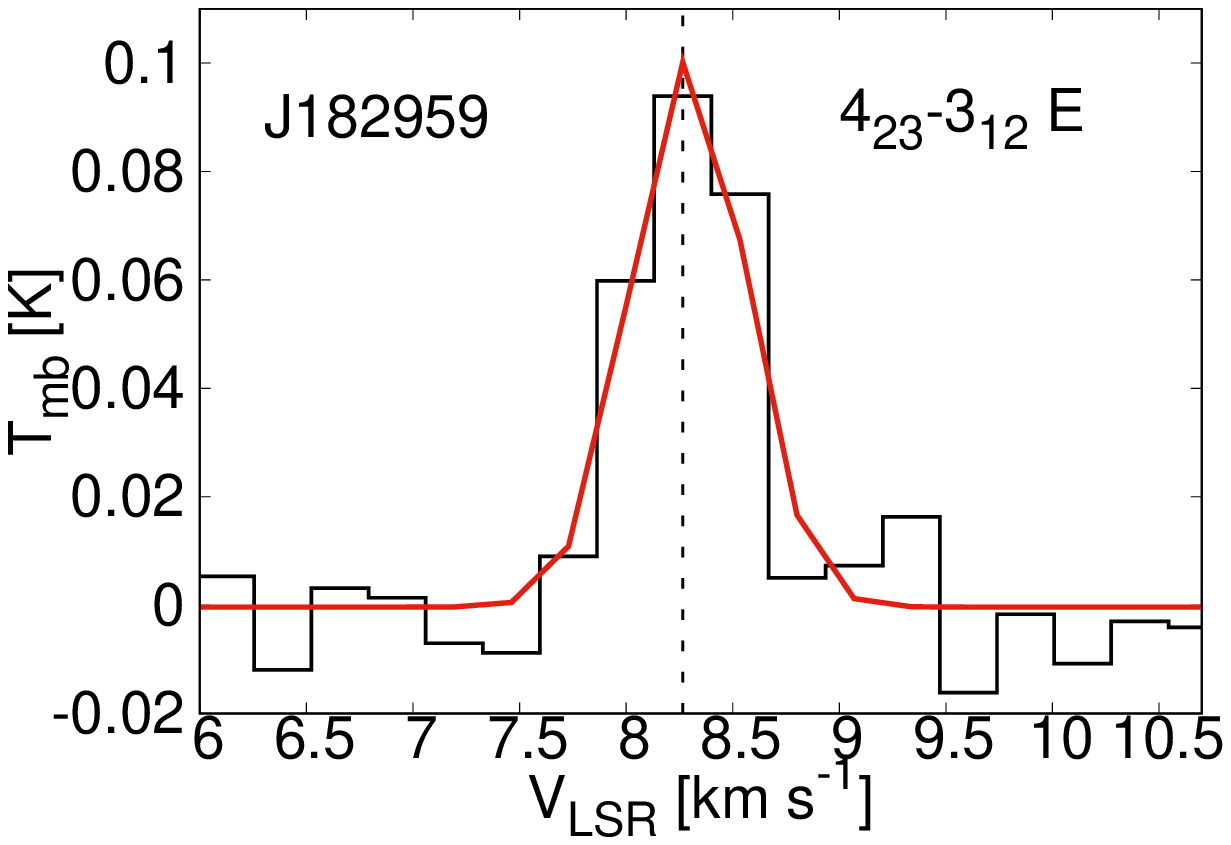}
     \includegraphics[width=1.5in]{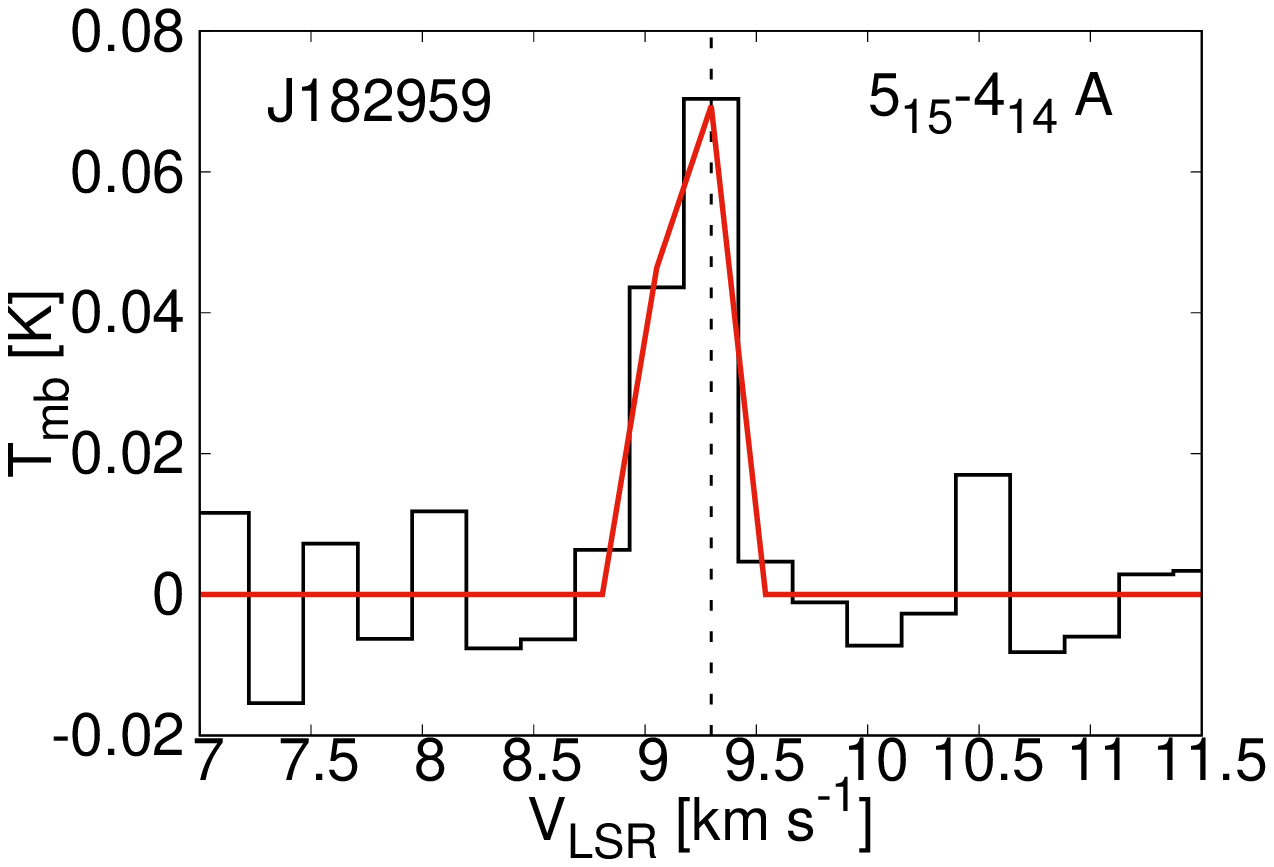}
     \includegraphics[width=1.5in]{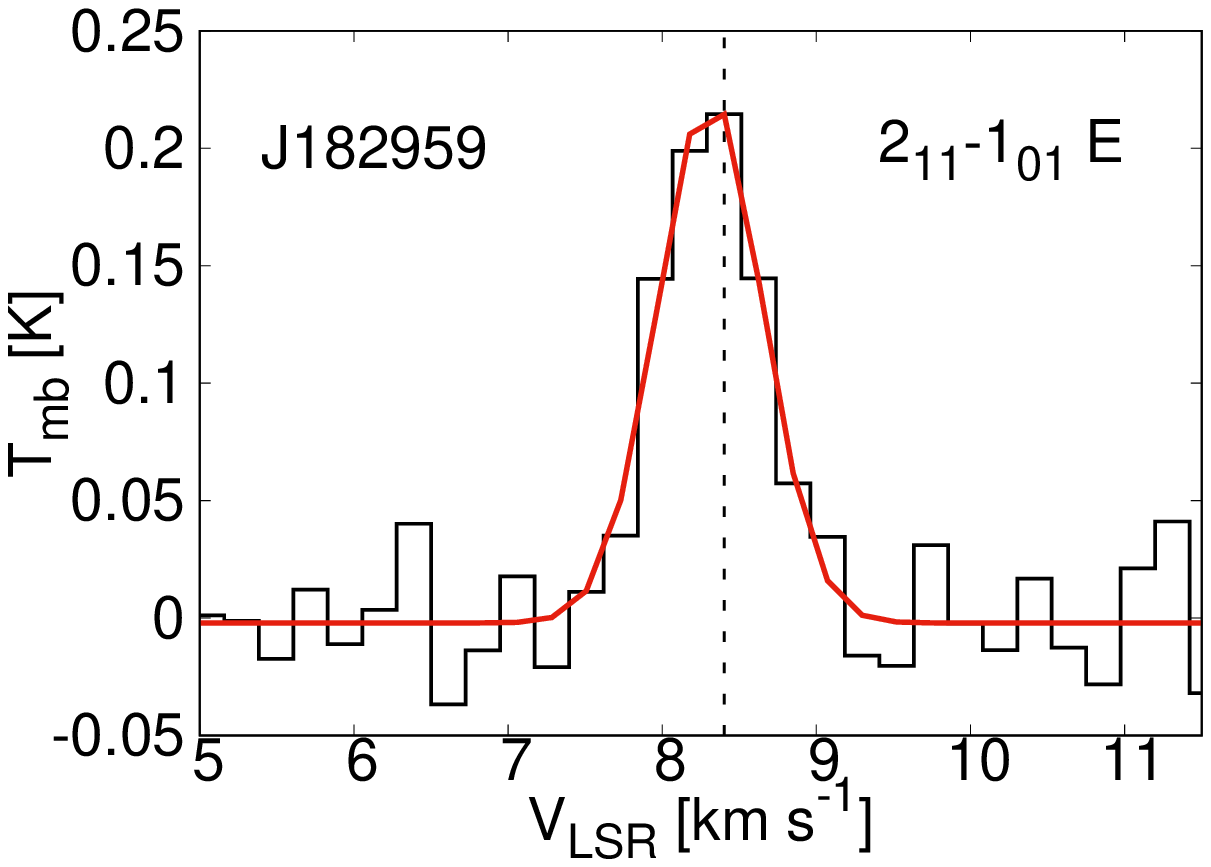}		\\	\vspace{0.2in}
     
    \includegraphics[width=1.5in]{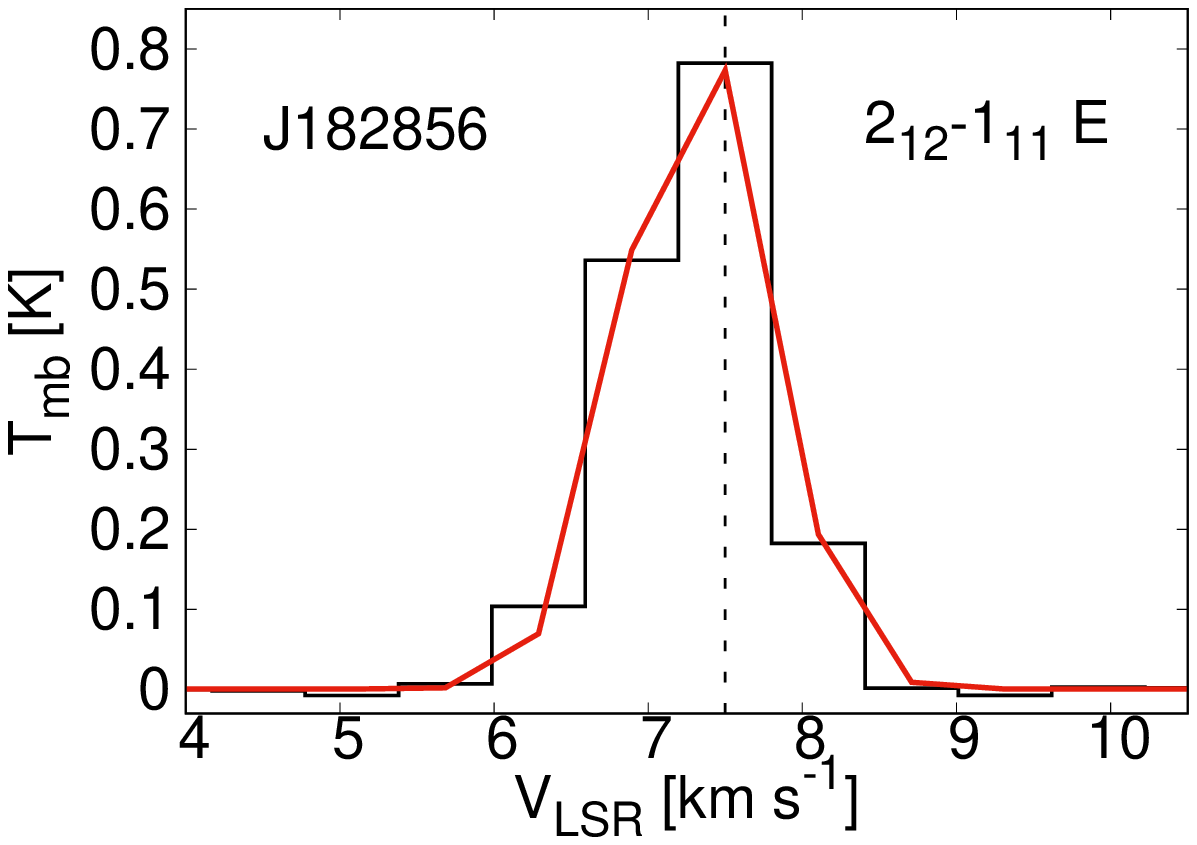}	
    \includegraphics[width=1.5in]{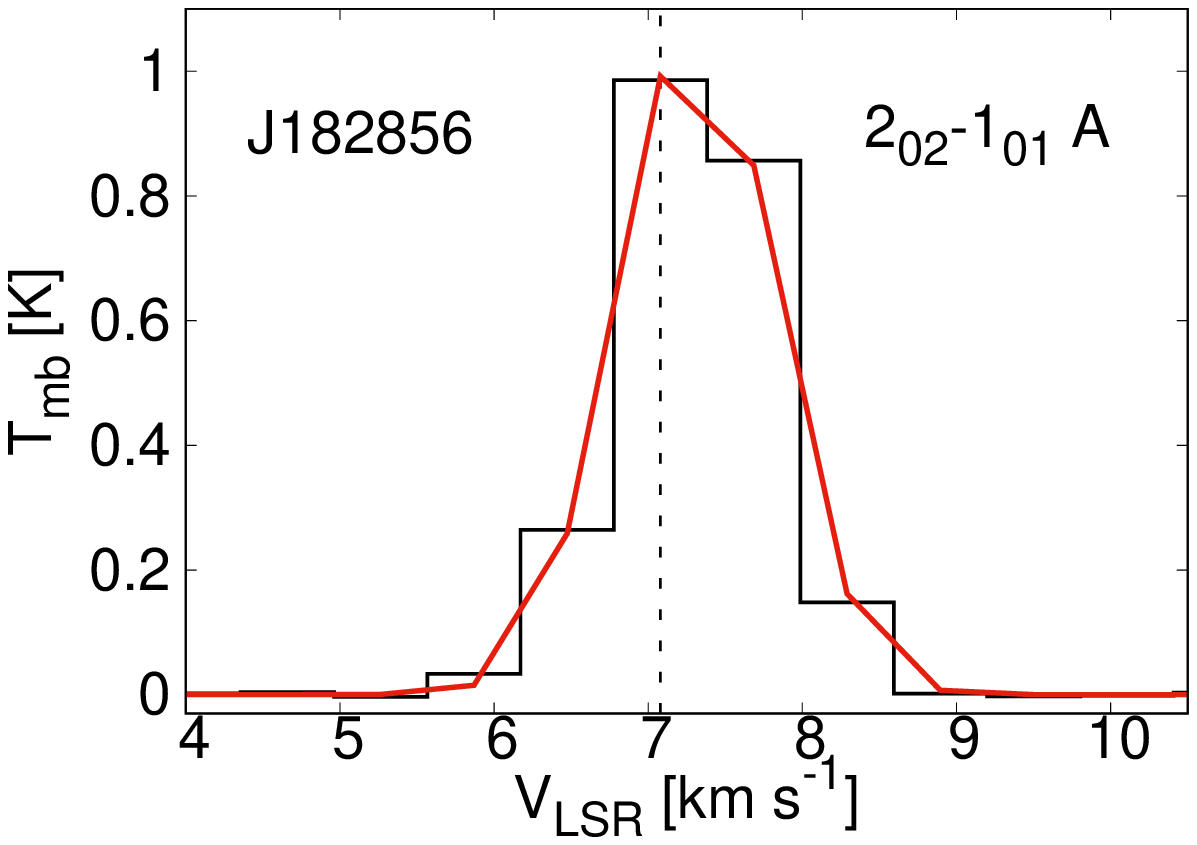}	
    \includegraphics[width=1.5in]{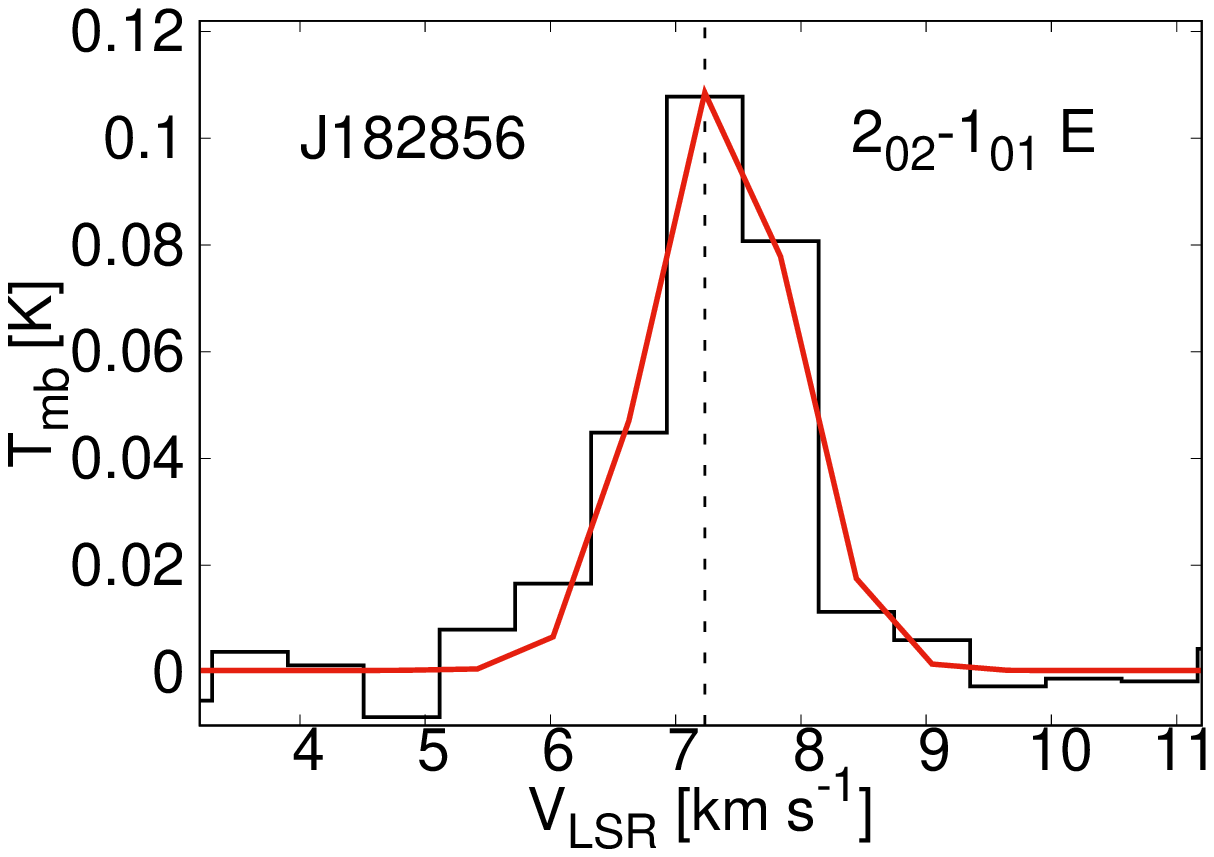} 
    \includegraphics[width=1.5in]{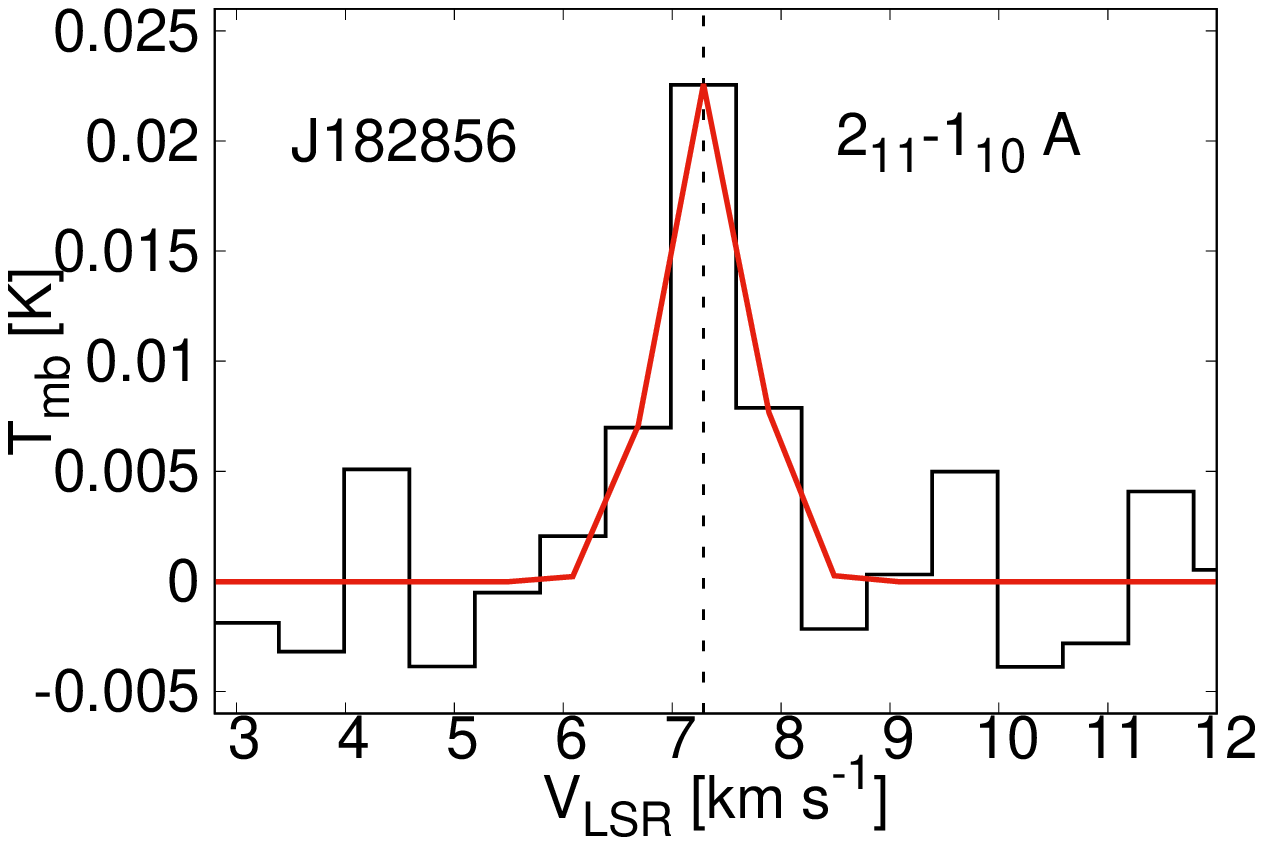}
    \includegraphics[width=1.5in]{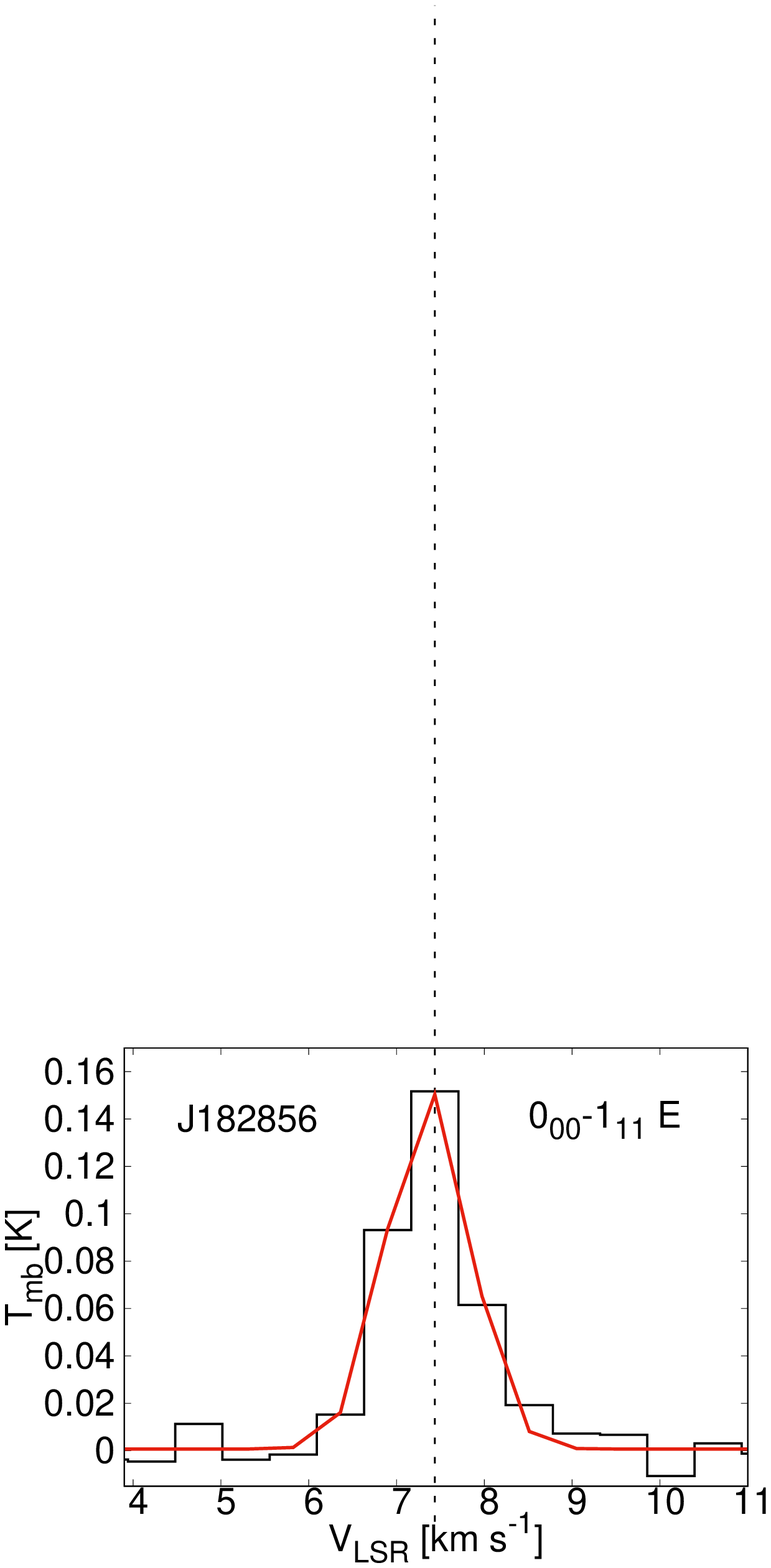}  
    \includegraphics[width=1.5in]{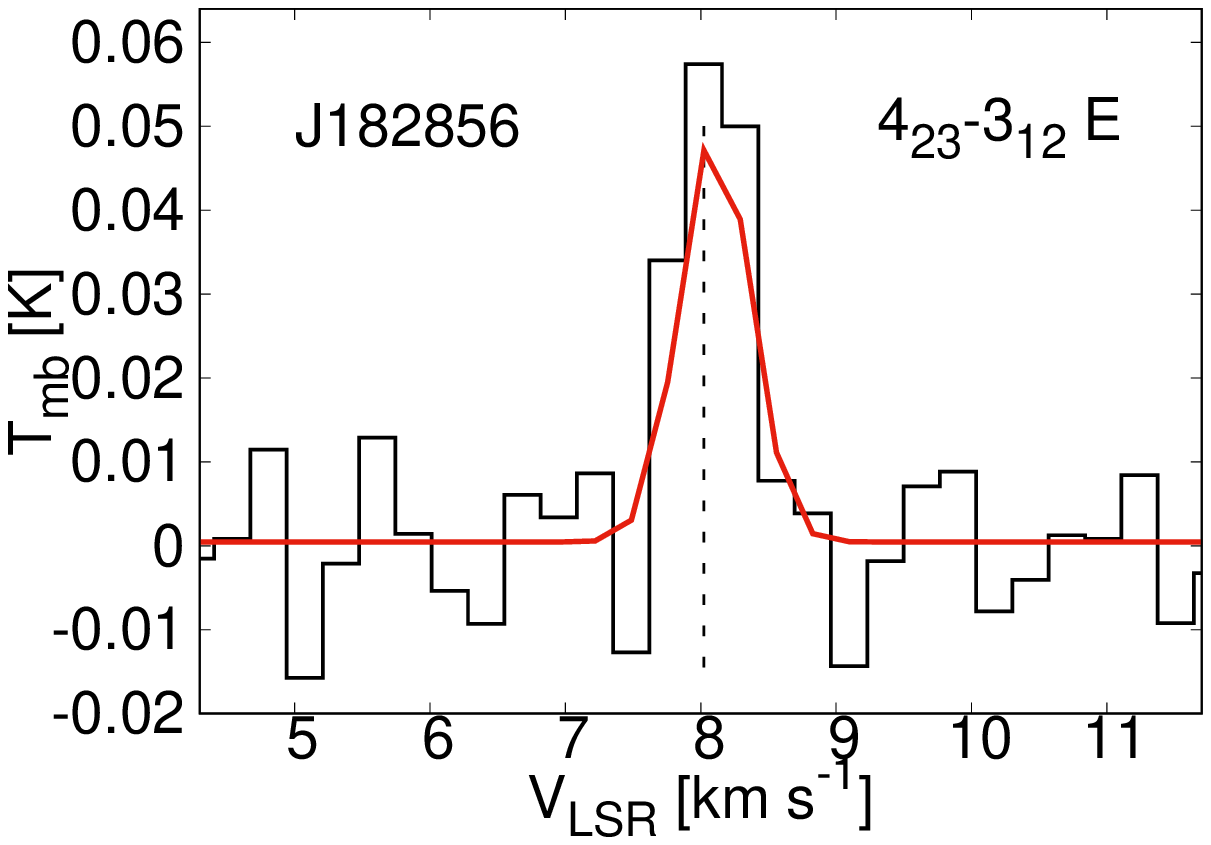}
    \includegraphics[width=1.5in]{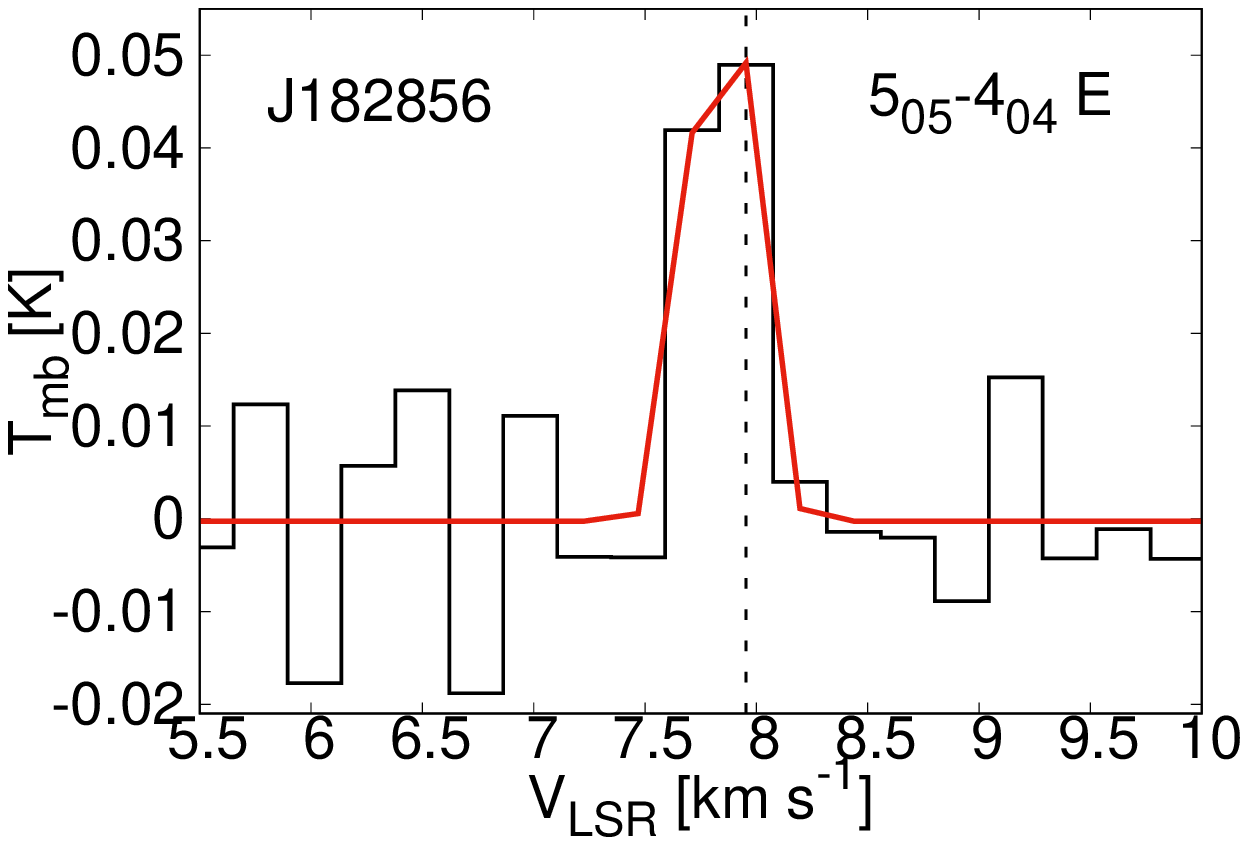}
    \includegraphics[width=1.5in]{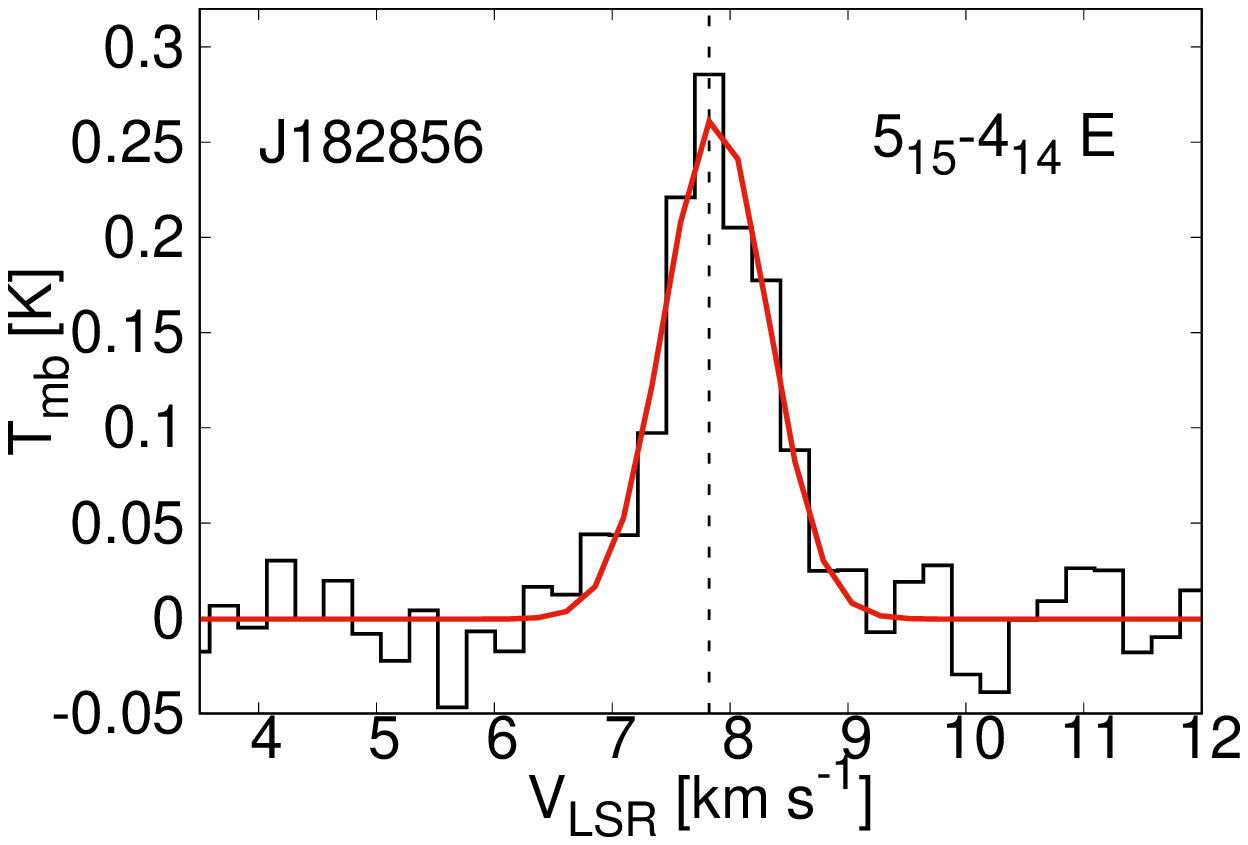}
    \includegraphics[width=1.5in]{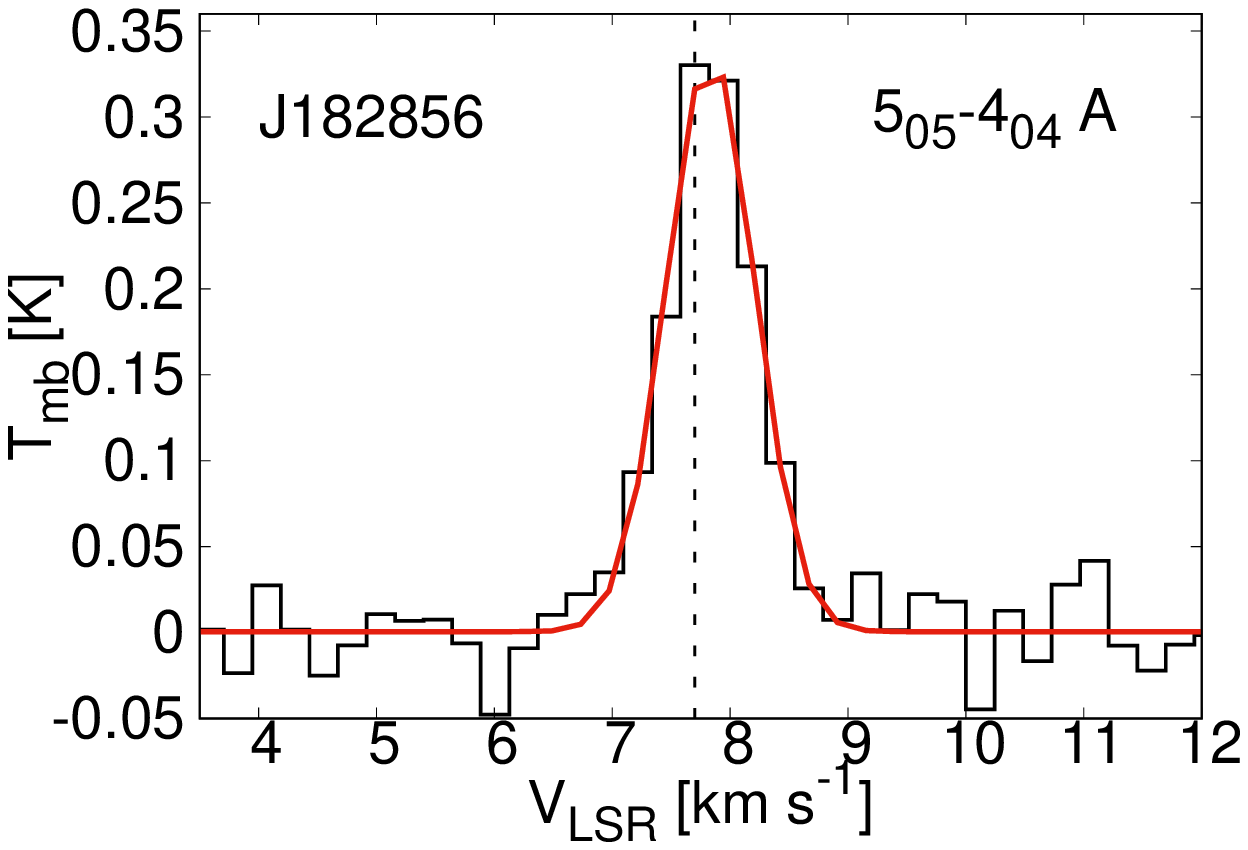}	
    \includegraphics[width=1.5in]{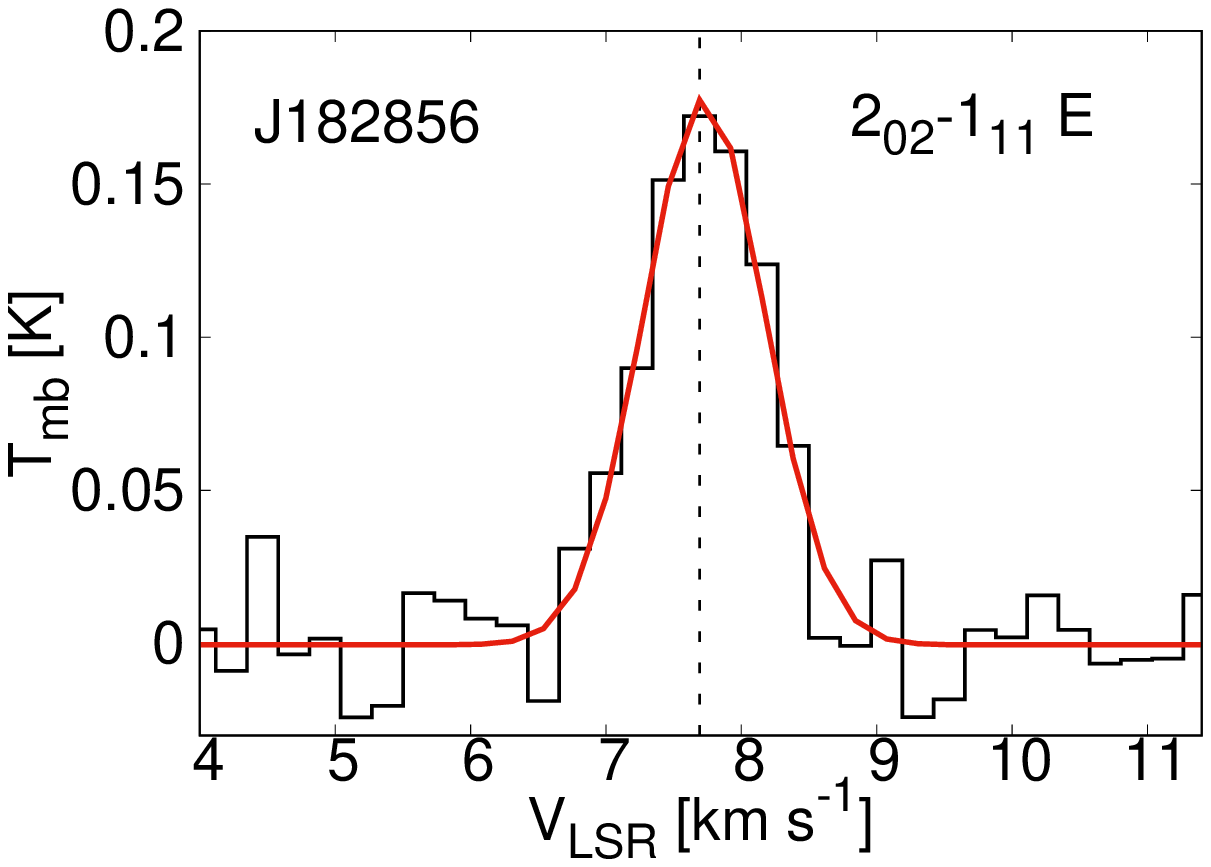}    
    \includegraphics[width=1.5in]{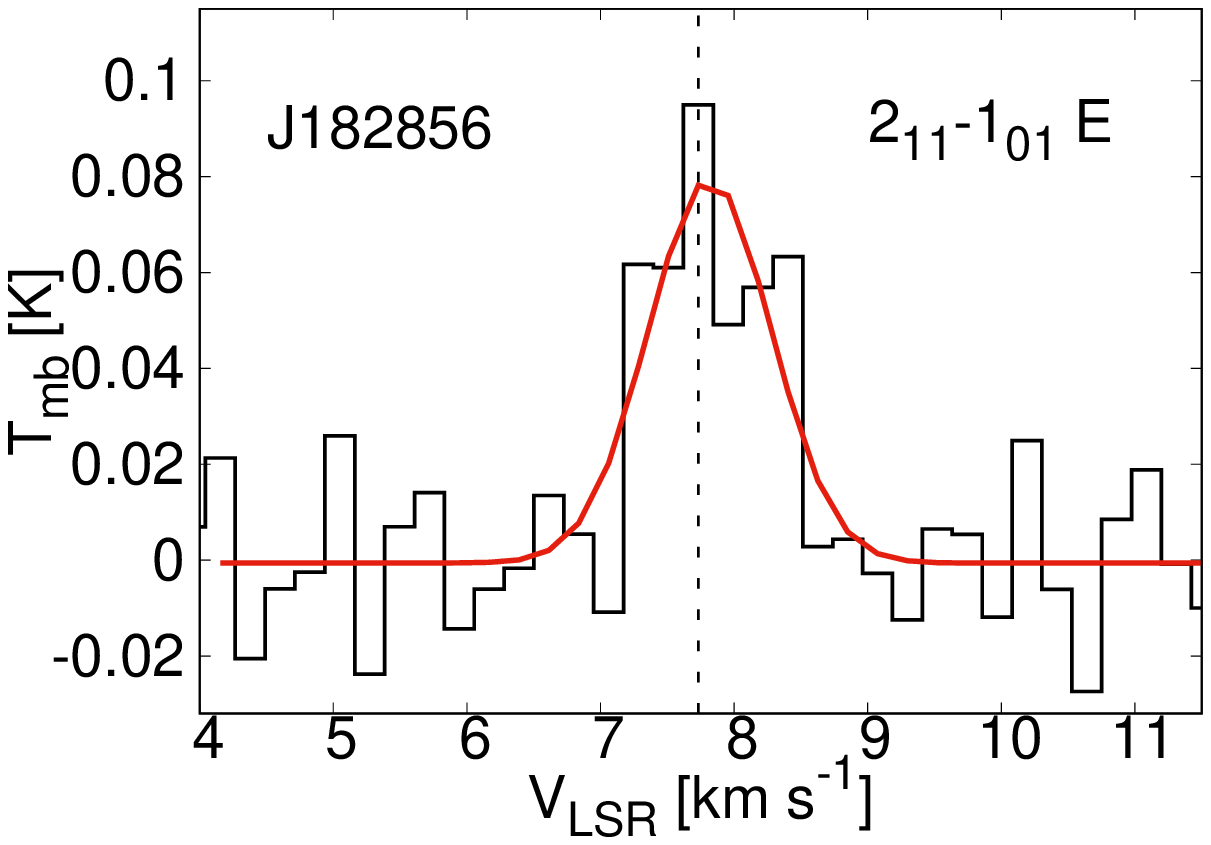}	
   \caption{The observed CH$_{3}$OH spectra (black) with the single or double Gaussian fits (red). Black dashed line marks the V$_{lsr}$ listed in Table~\ref{line-pars}}.  
   \label{spectra}     
  \end{figure*}

\setcounter{figure}{2}    
 \begin{figure*}
  \centering         
    \includegraphics[width=1.5in]{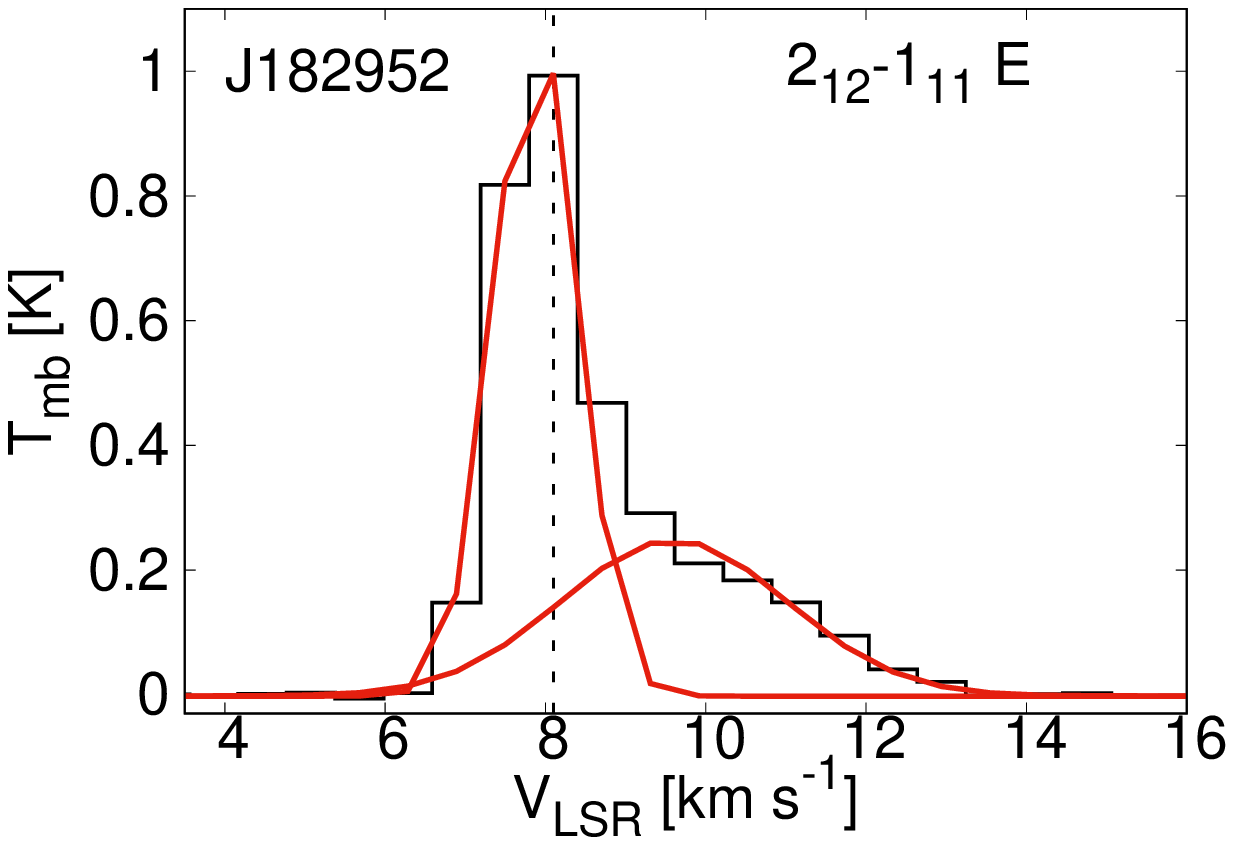}	
    \includegraphics[width=1.5in]{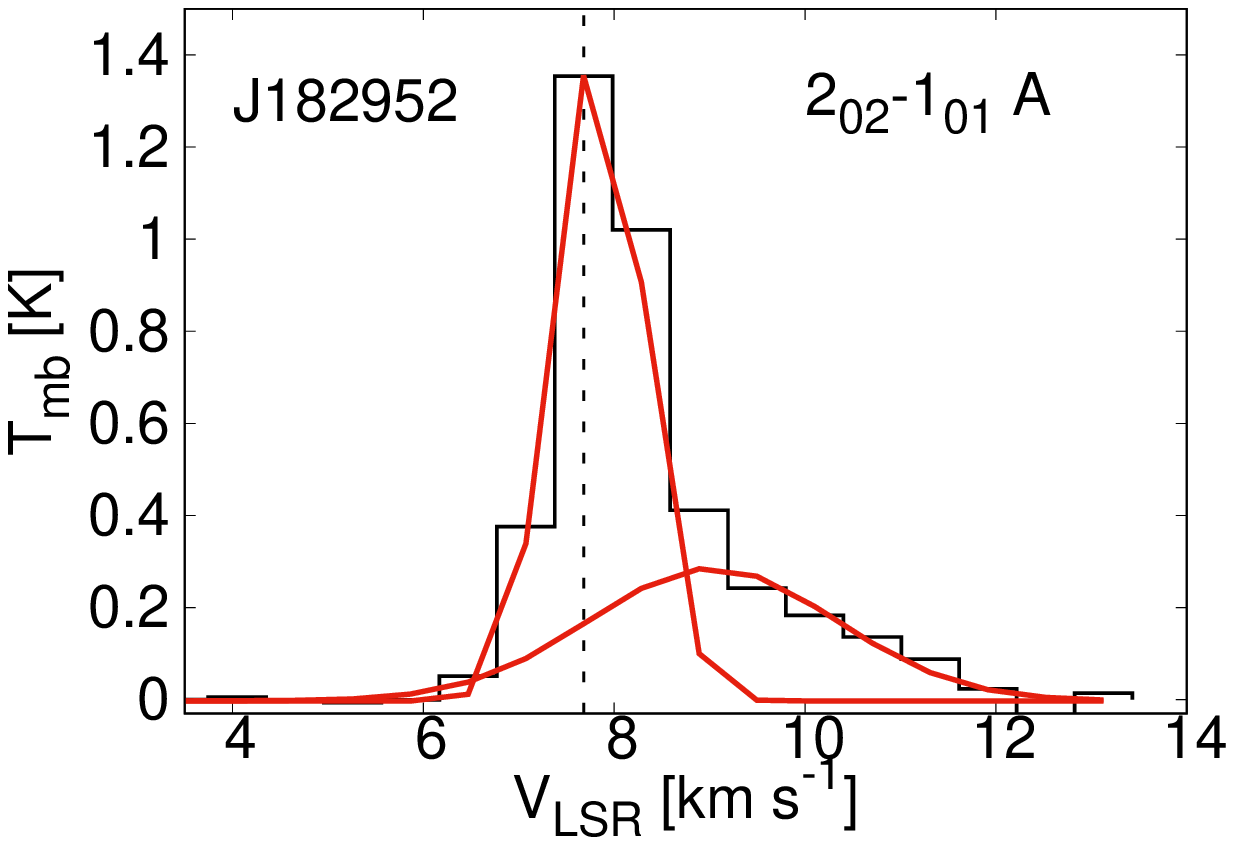}	
    \includegraphics[width=1.5in]{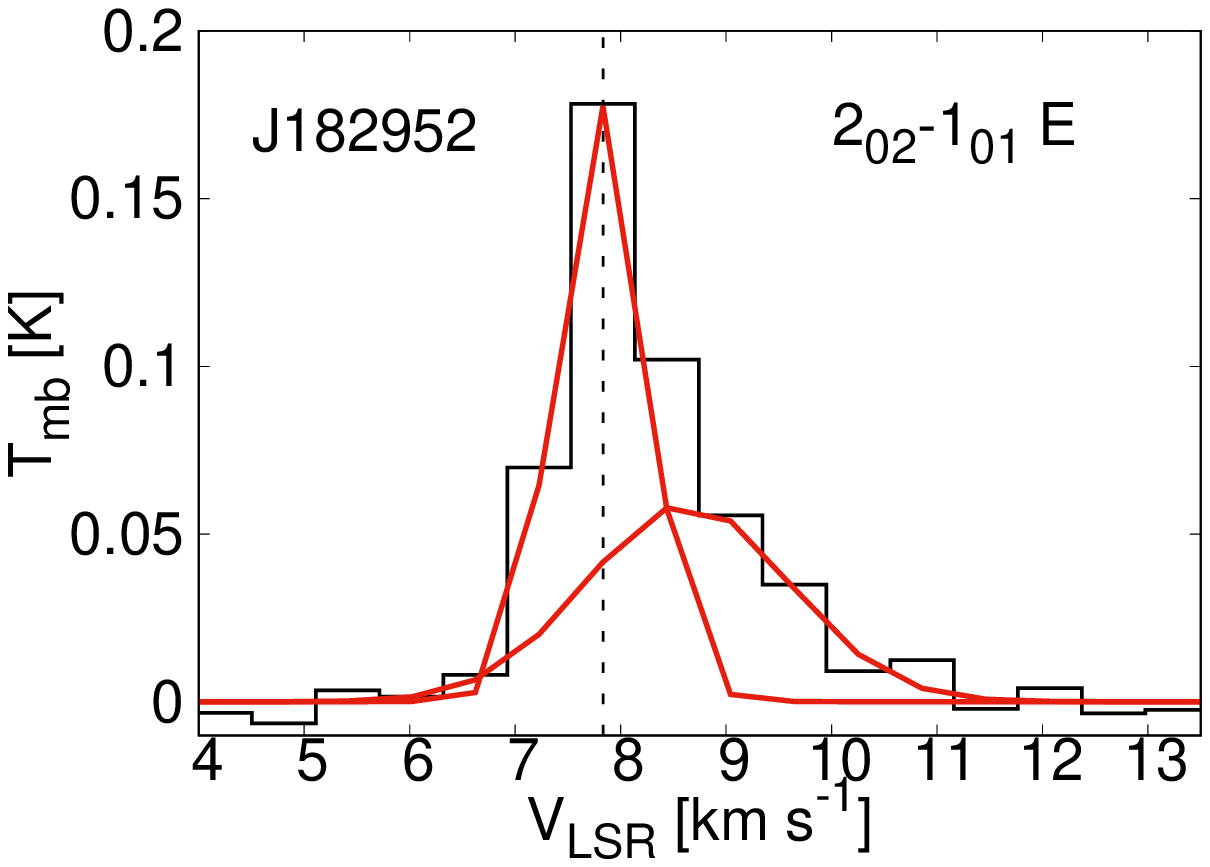}
    \includegraphics[width=1.5in]{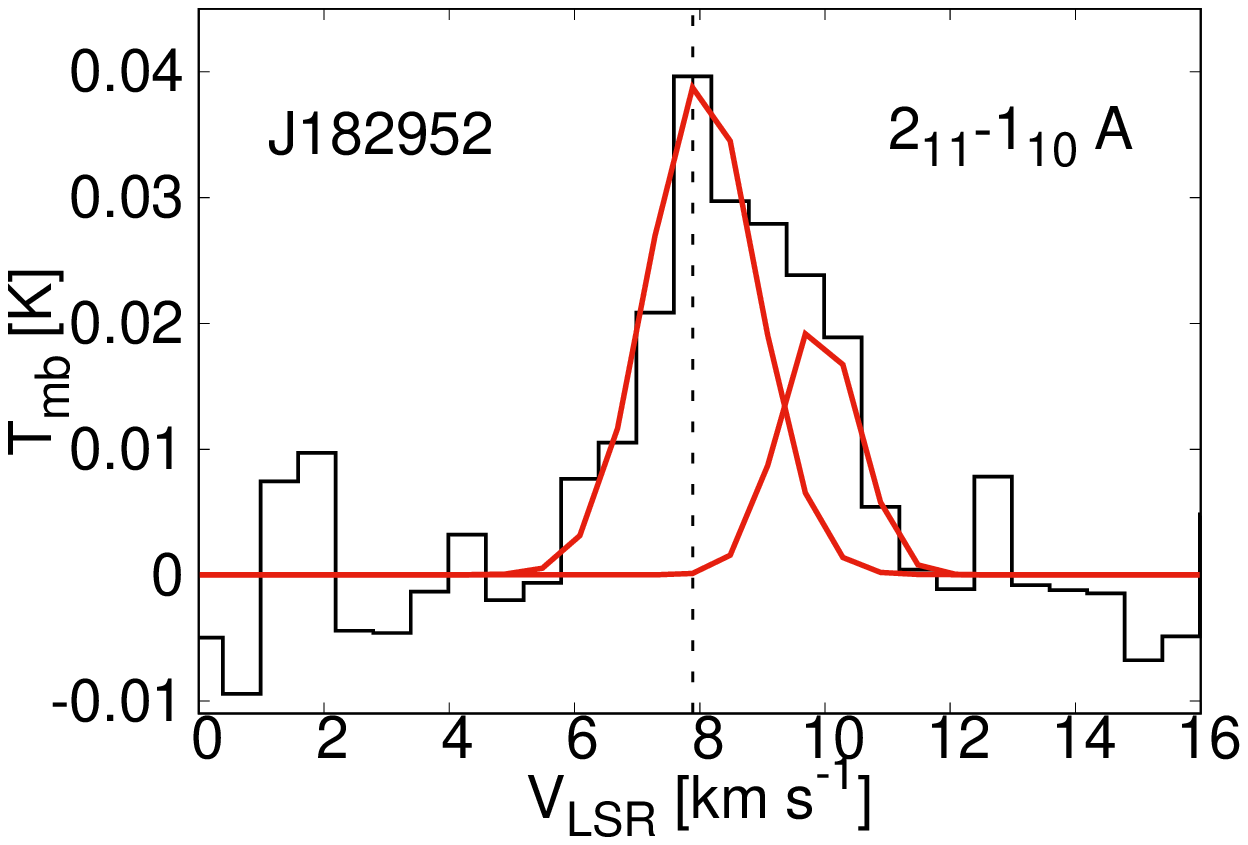}
    \includegraphics[width=1.5in]{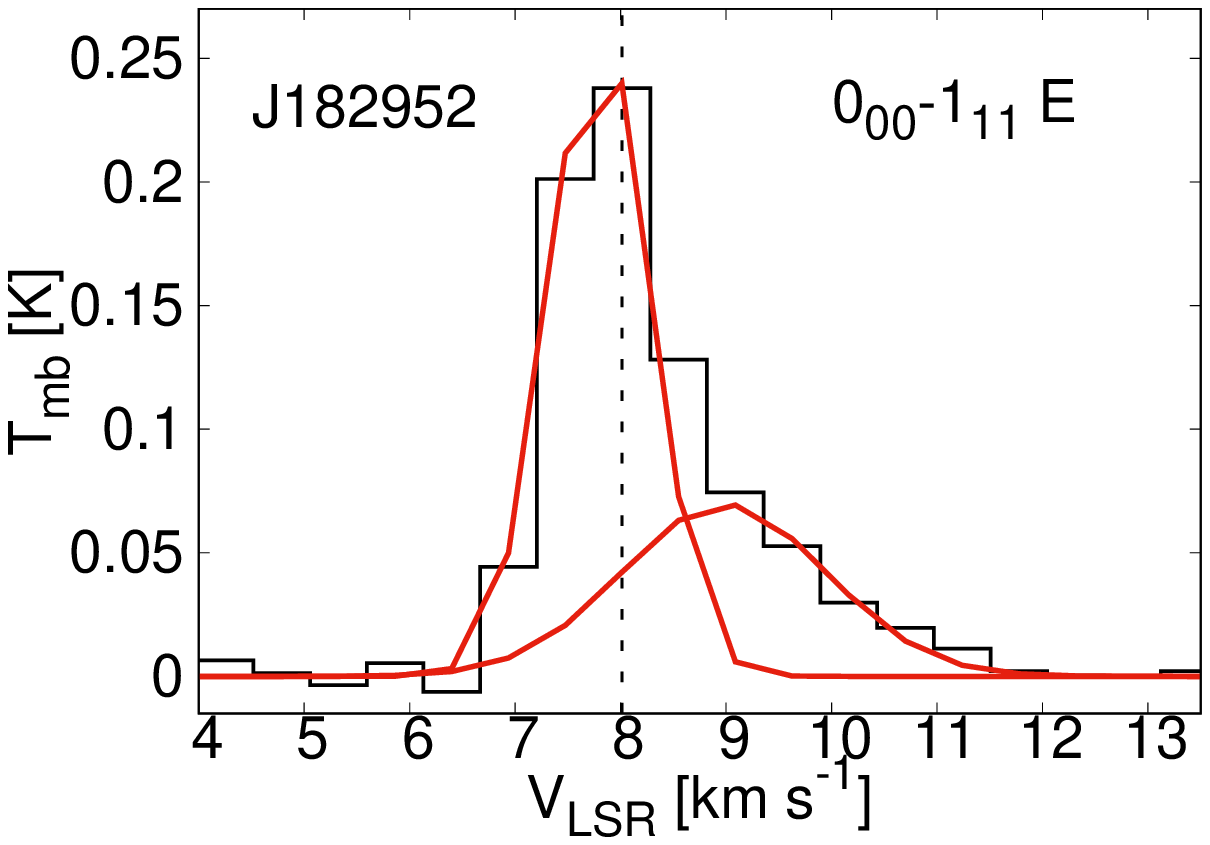}
    \includegraphics[width=1.5in]{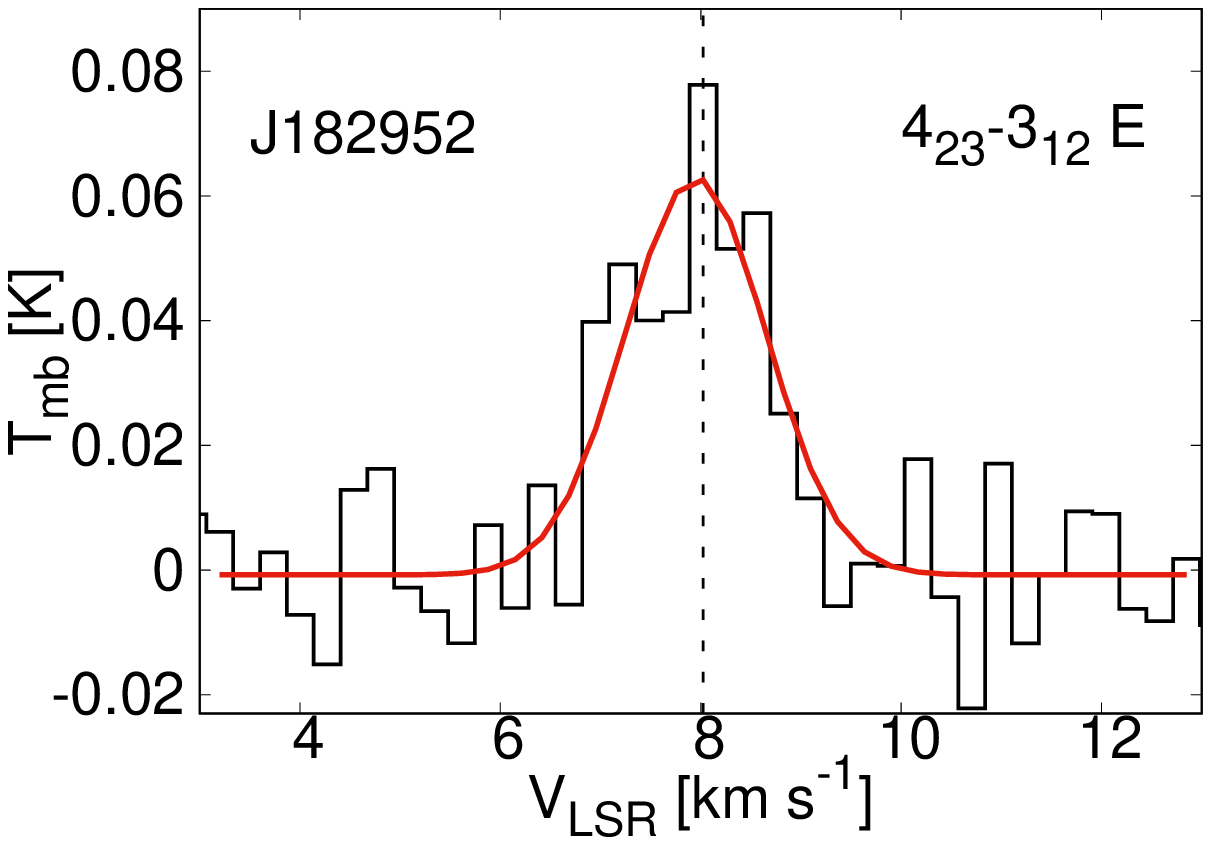}
    \includegraphics[width=1.5in]{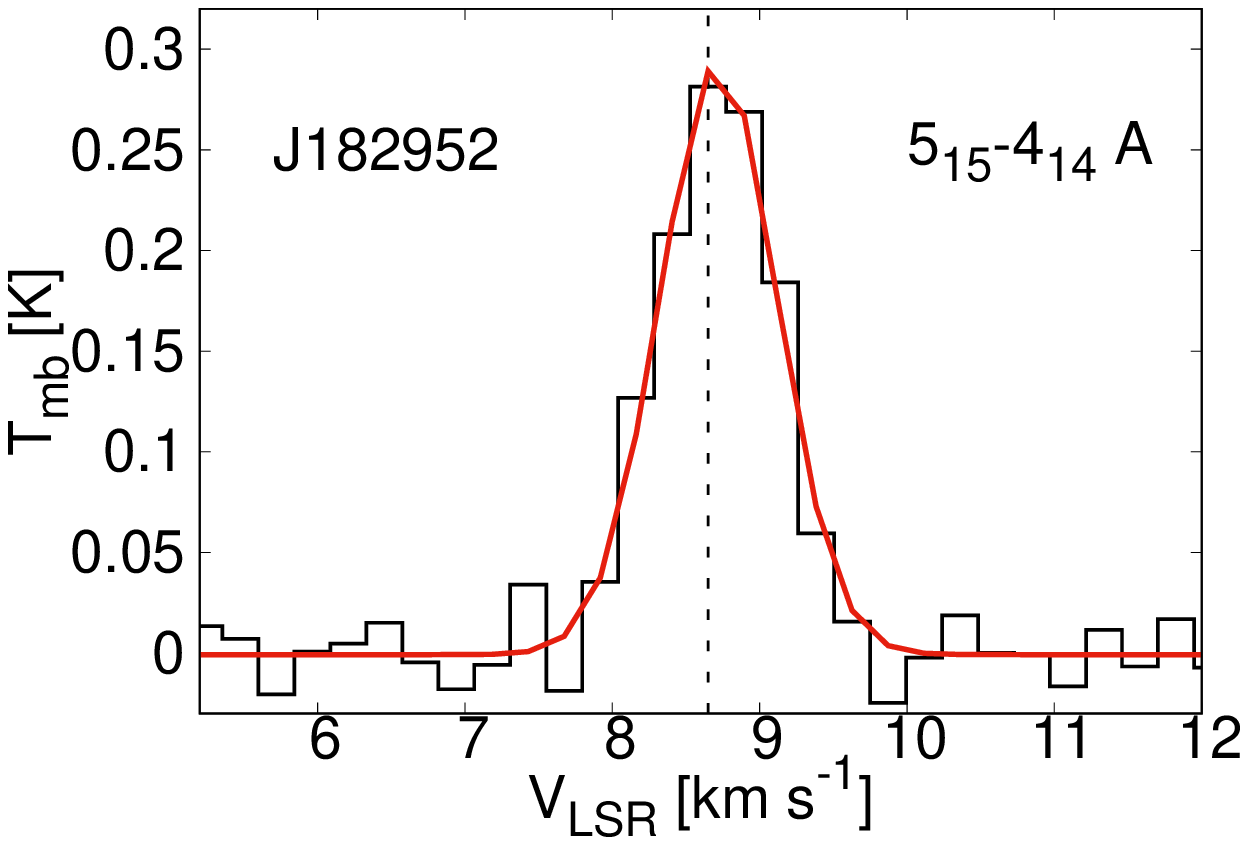}	
    \includegraphics[width=1.5in]{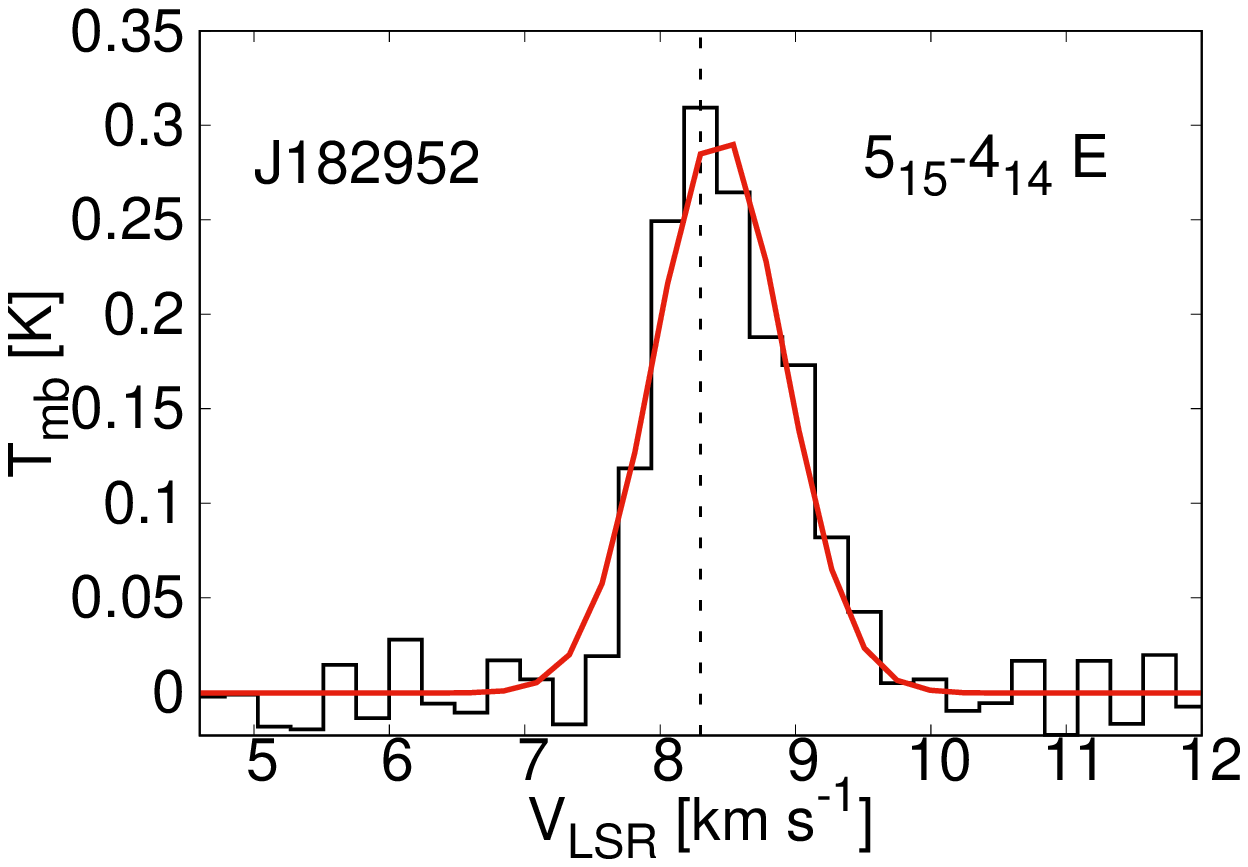}
    \includegraphics[width=1.5in]{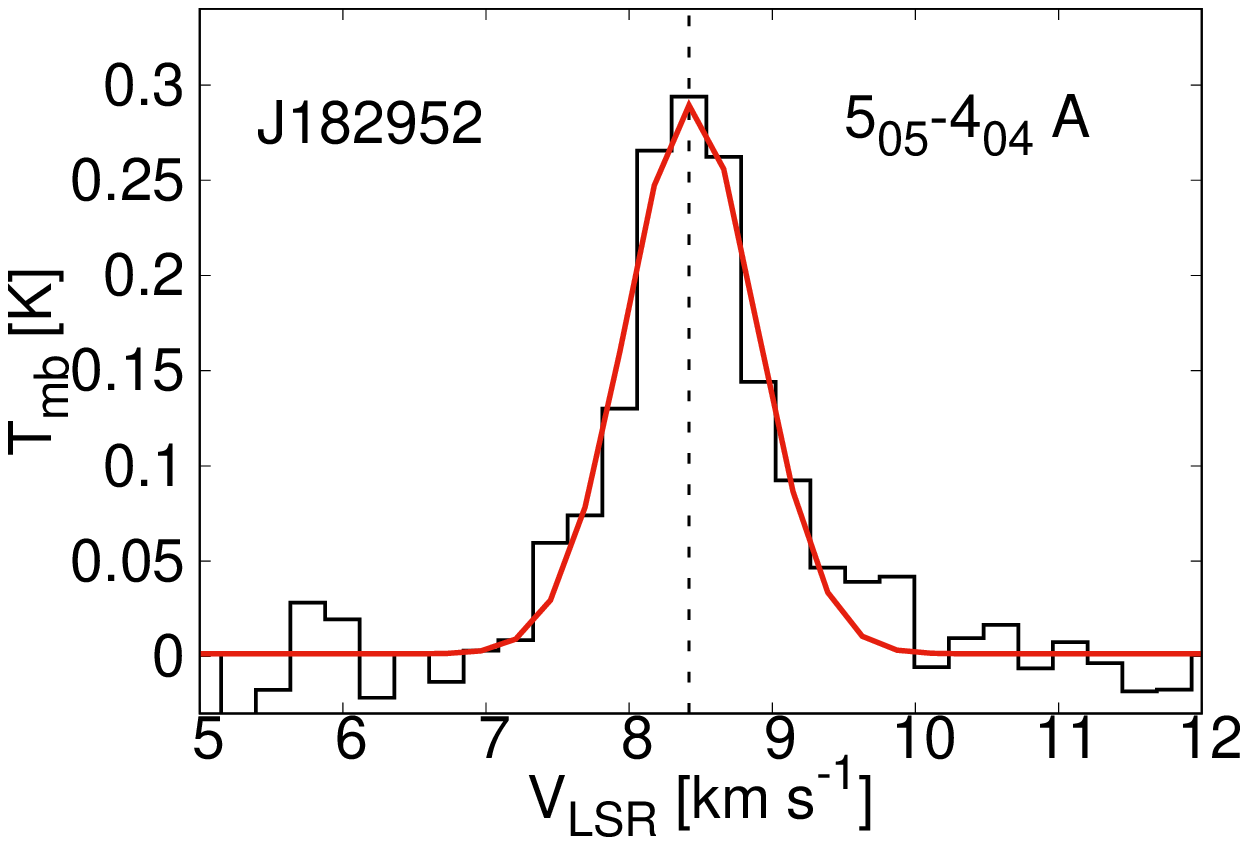}    
    \includegraphics[width=1.5in]{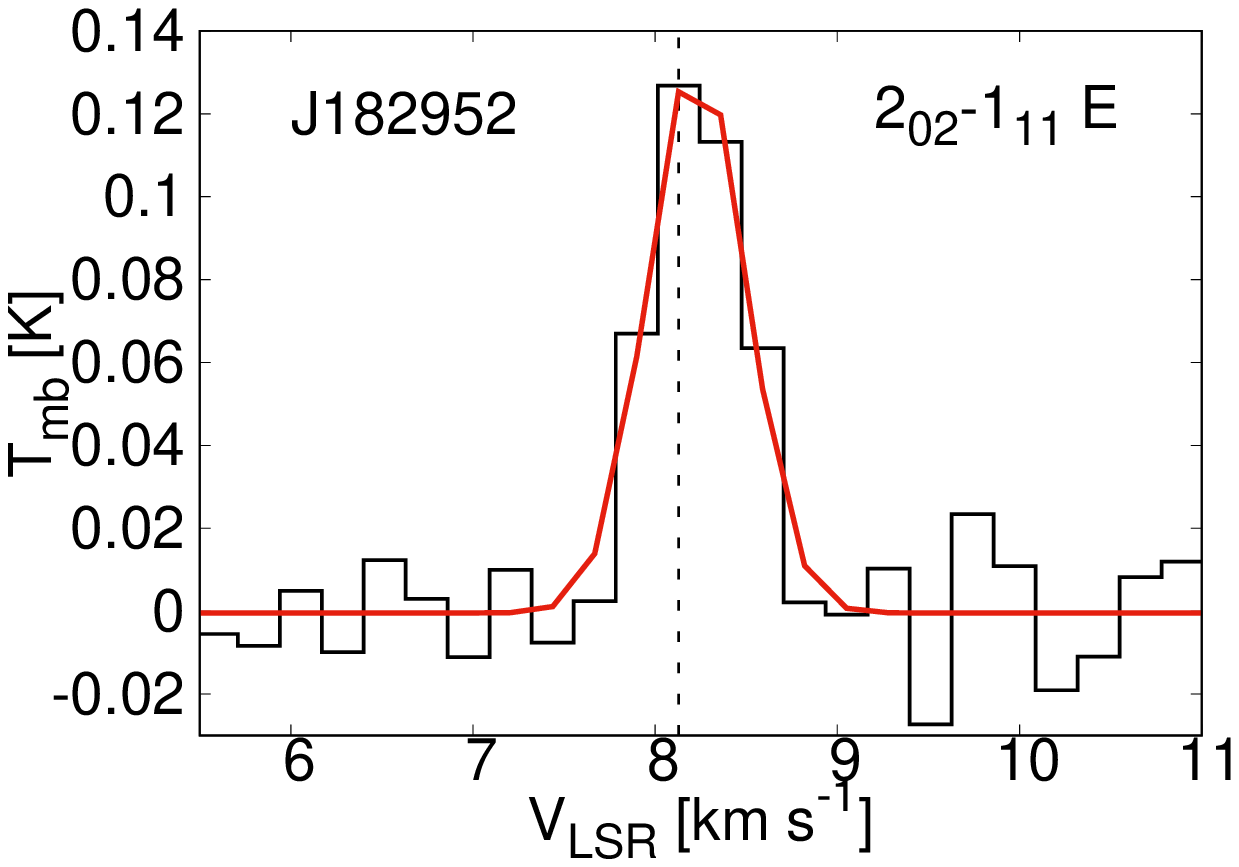}
    \includegraphics[width=1.5in]{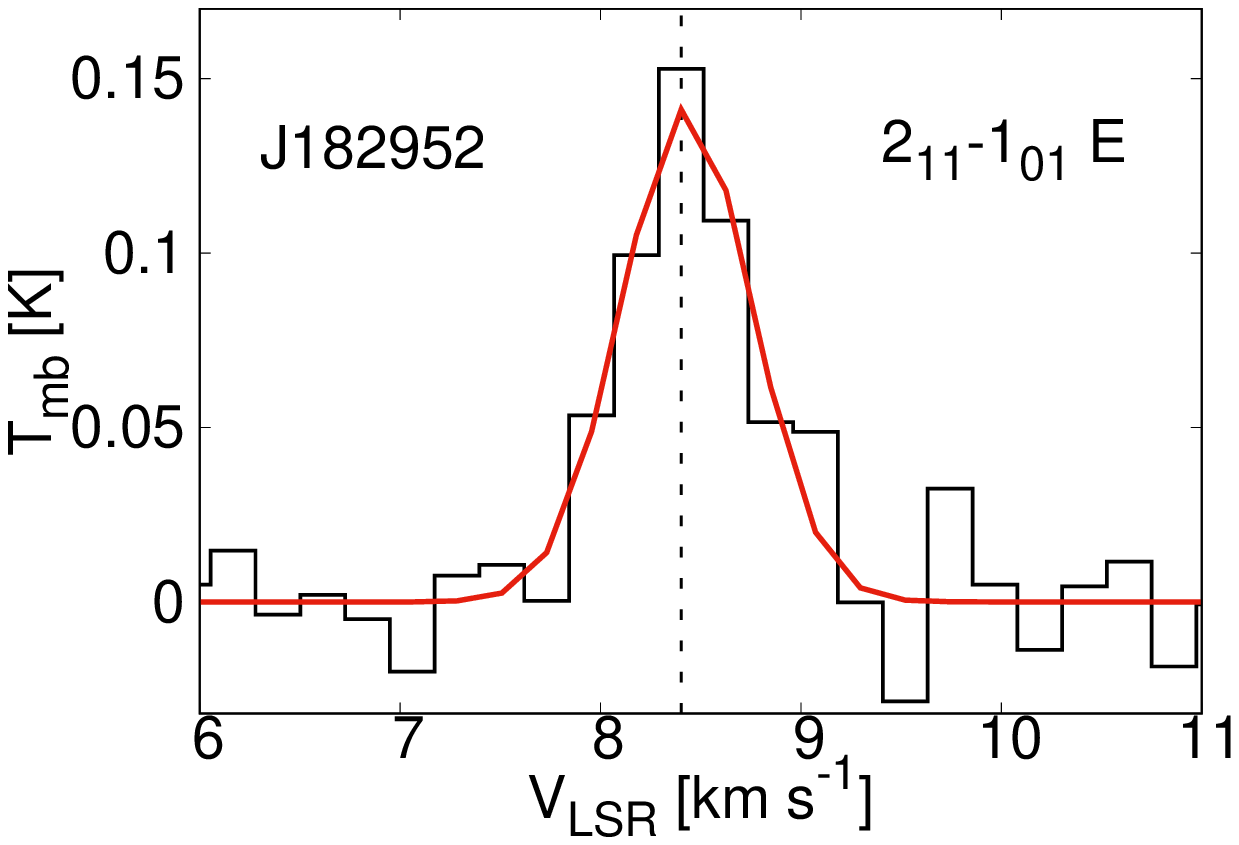}		\\	\vspace{0.3in}
    
    \includegraphics[width=1.5in]{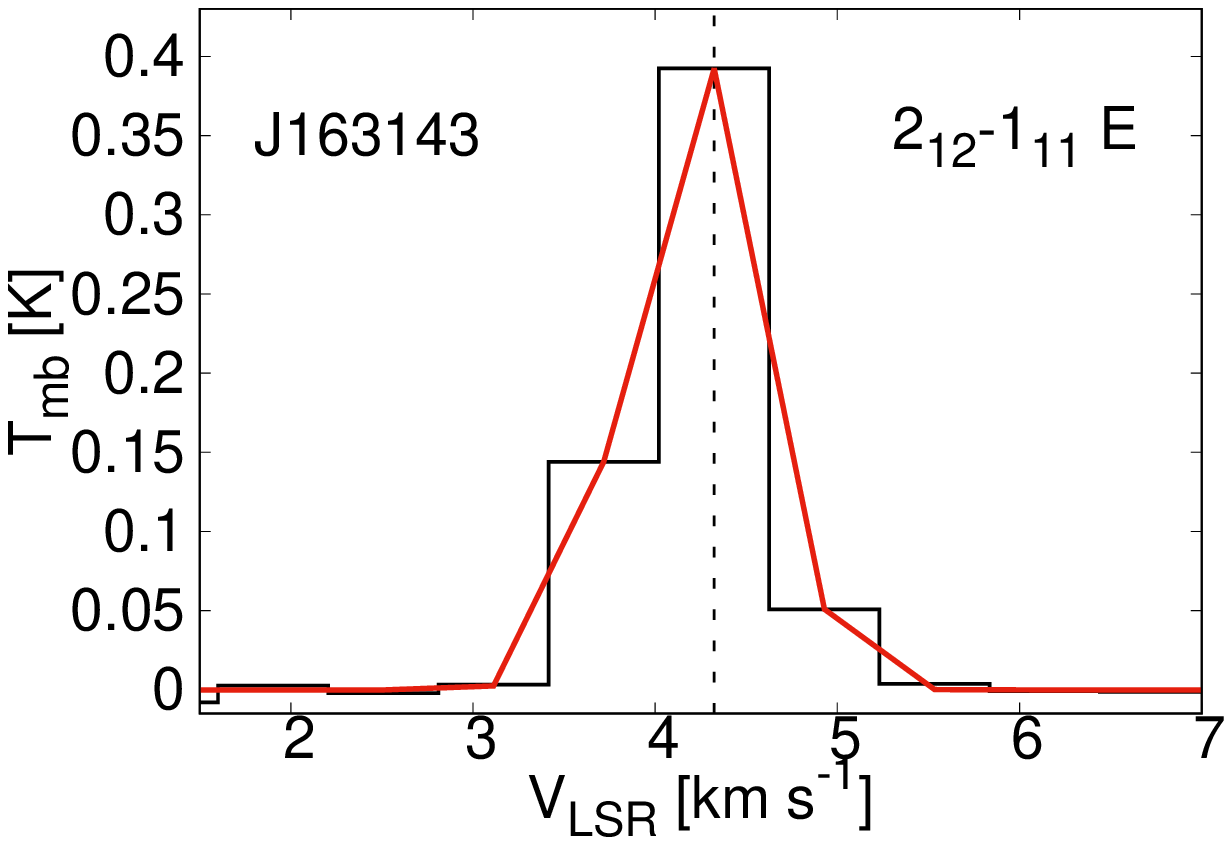}
    \includegraphics[width=1.5in]{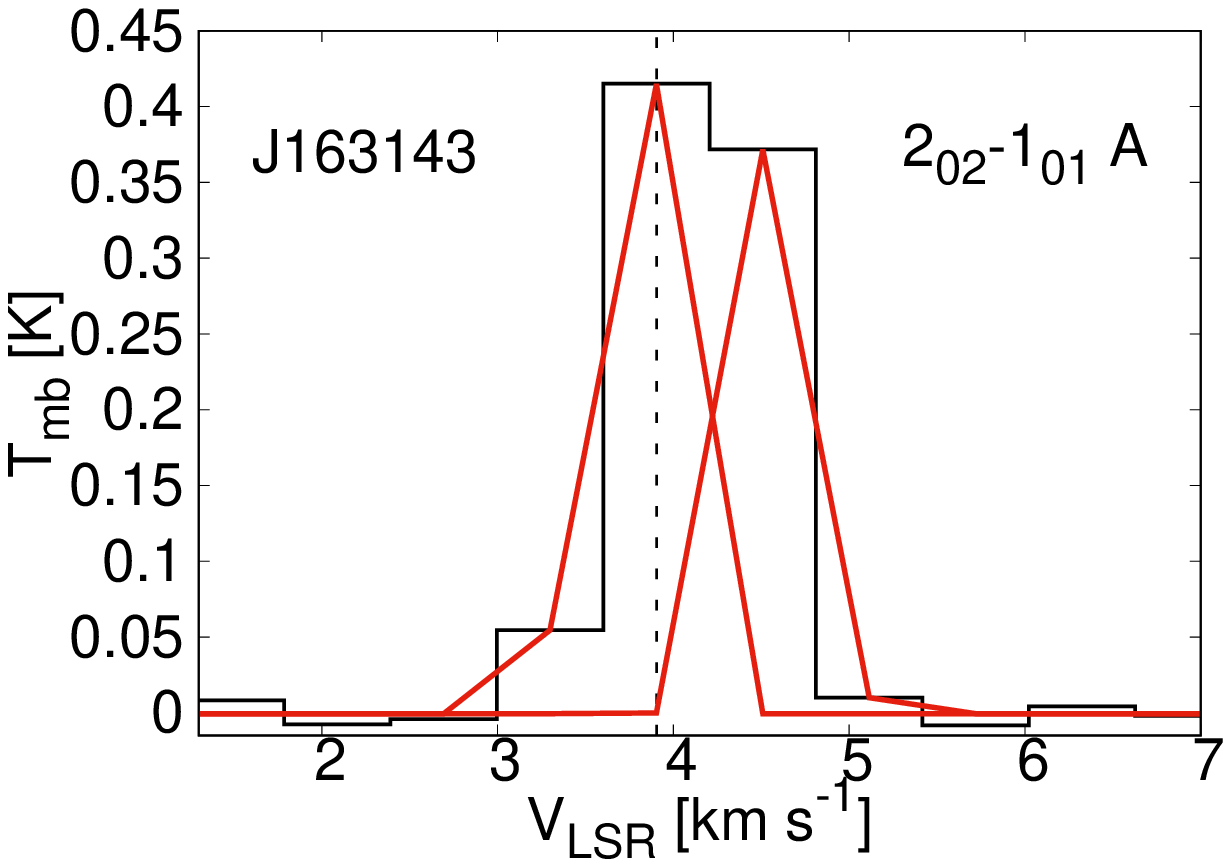}	
    \includegraphics[width=1.5in]{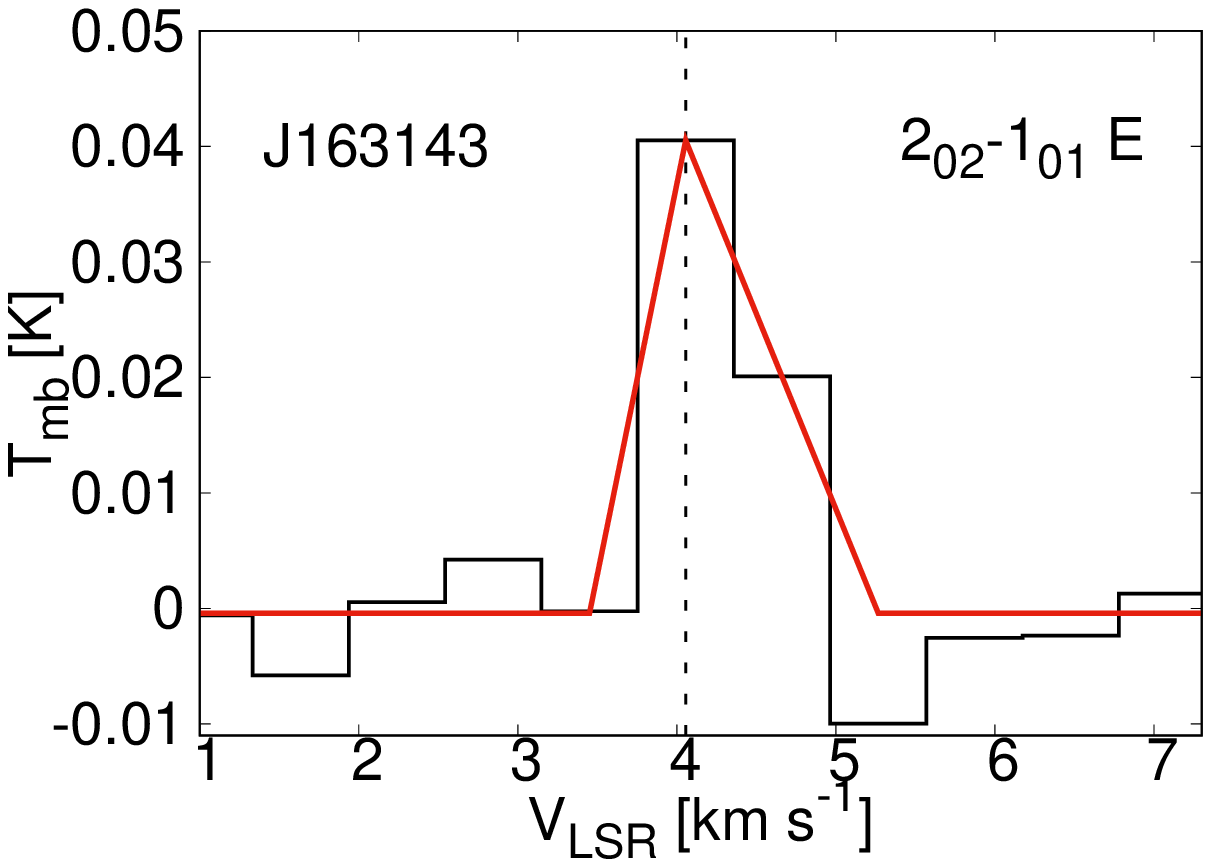}
    \includegraphics[width=1.5in]{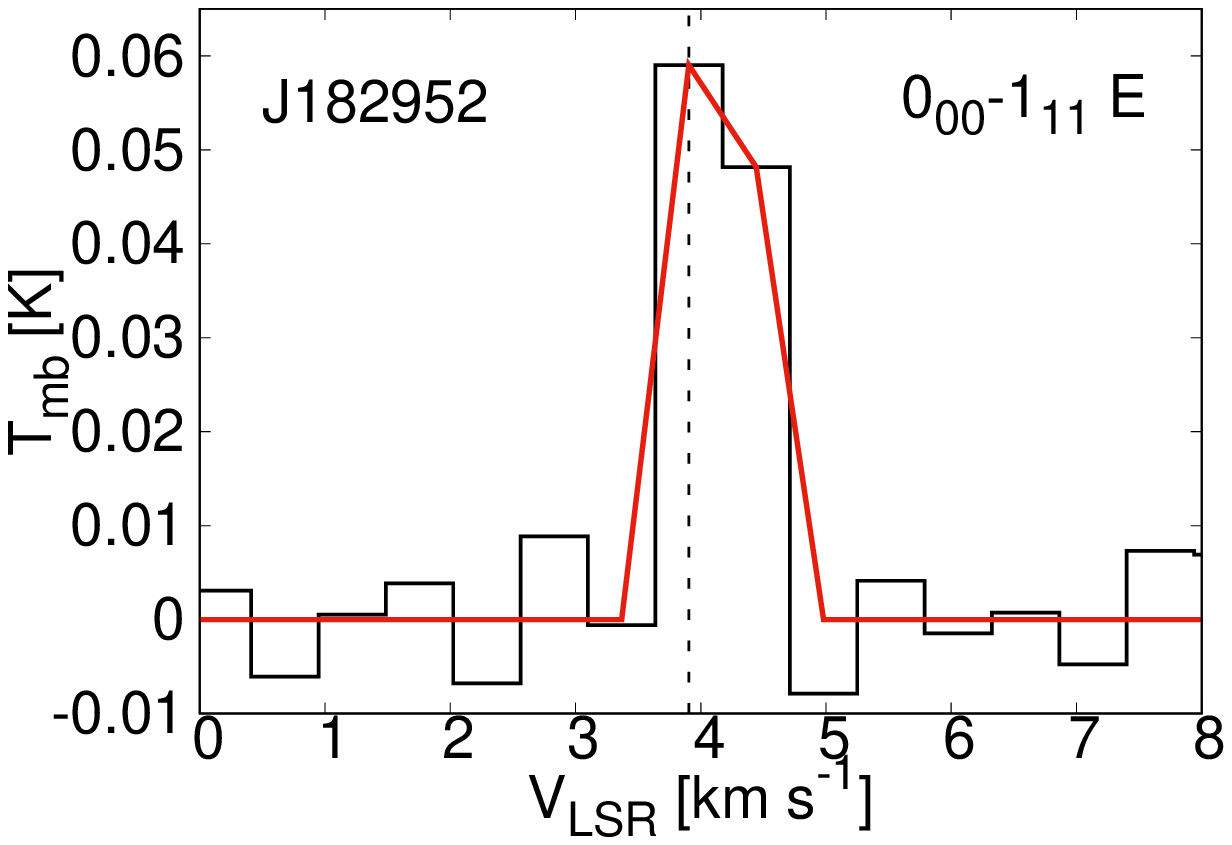}	\\	\vspace{0.3in}
    
    \includegraphics[width=1.5in]{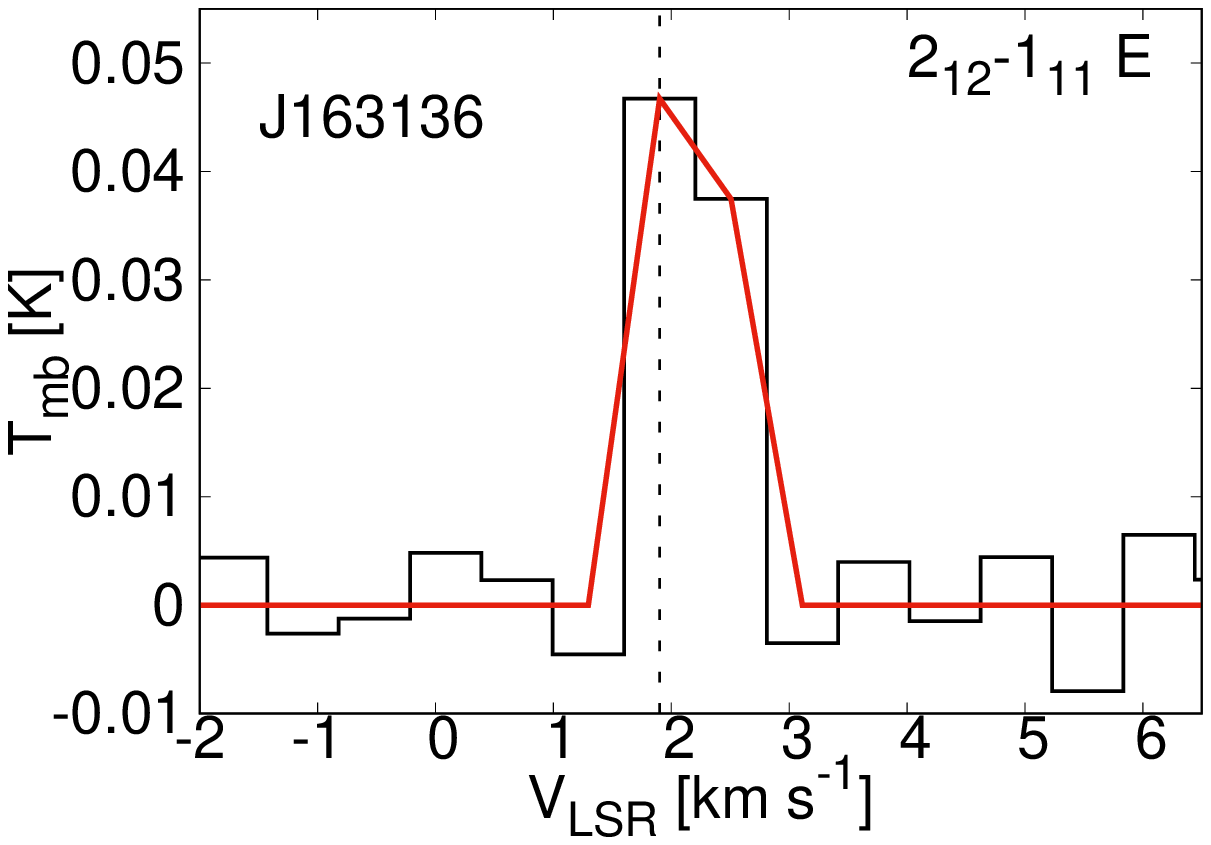}	
    \includegraphics[width=1.5in]{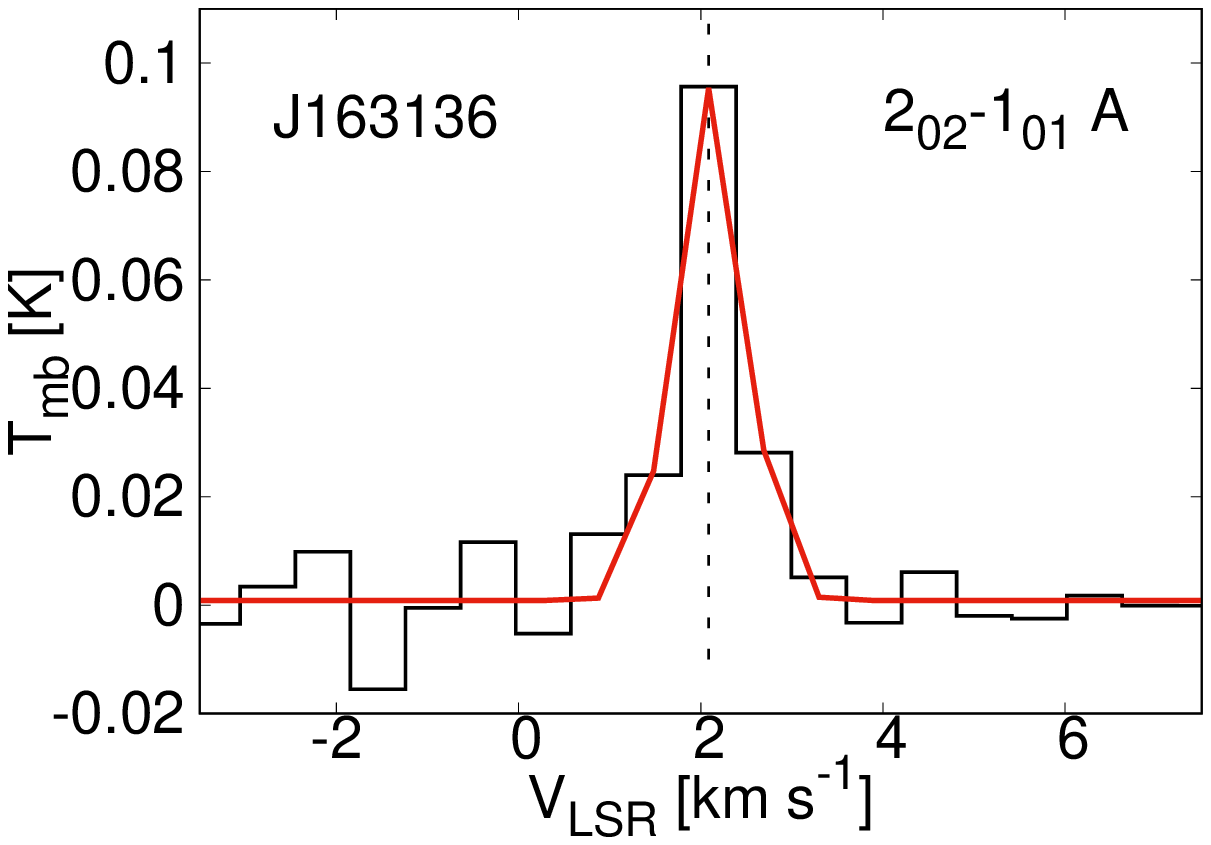}
    \includegraphics[width=1.5in]{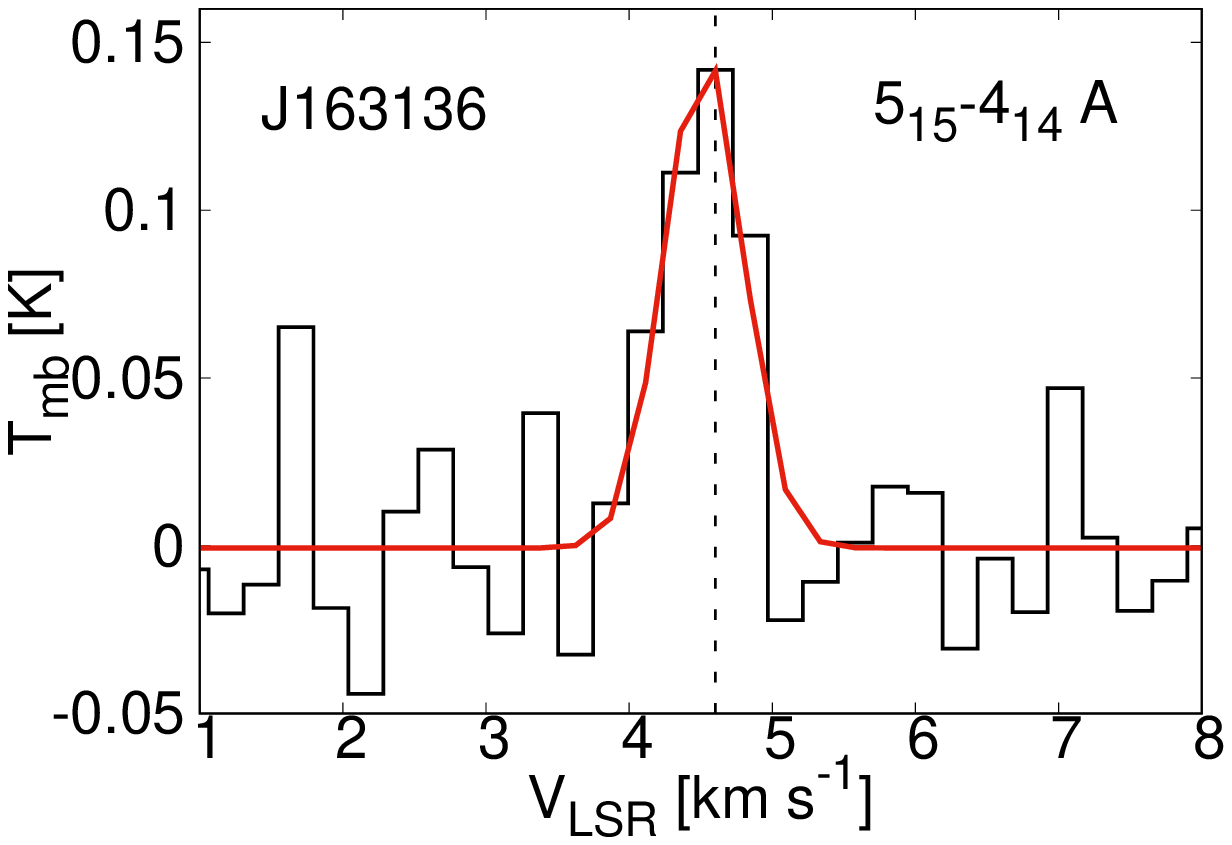}         	\\	\vspace{0.3in}
    
    \includegraphics[width=1.5in]{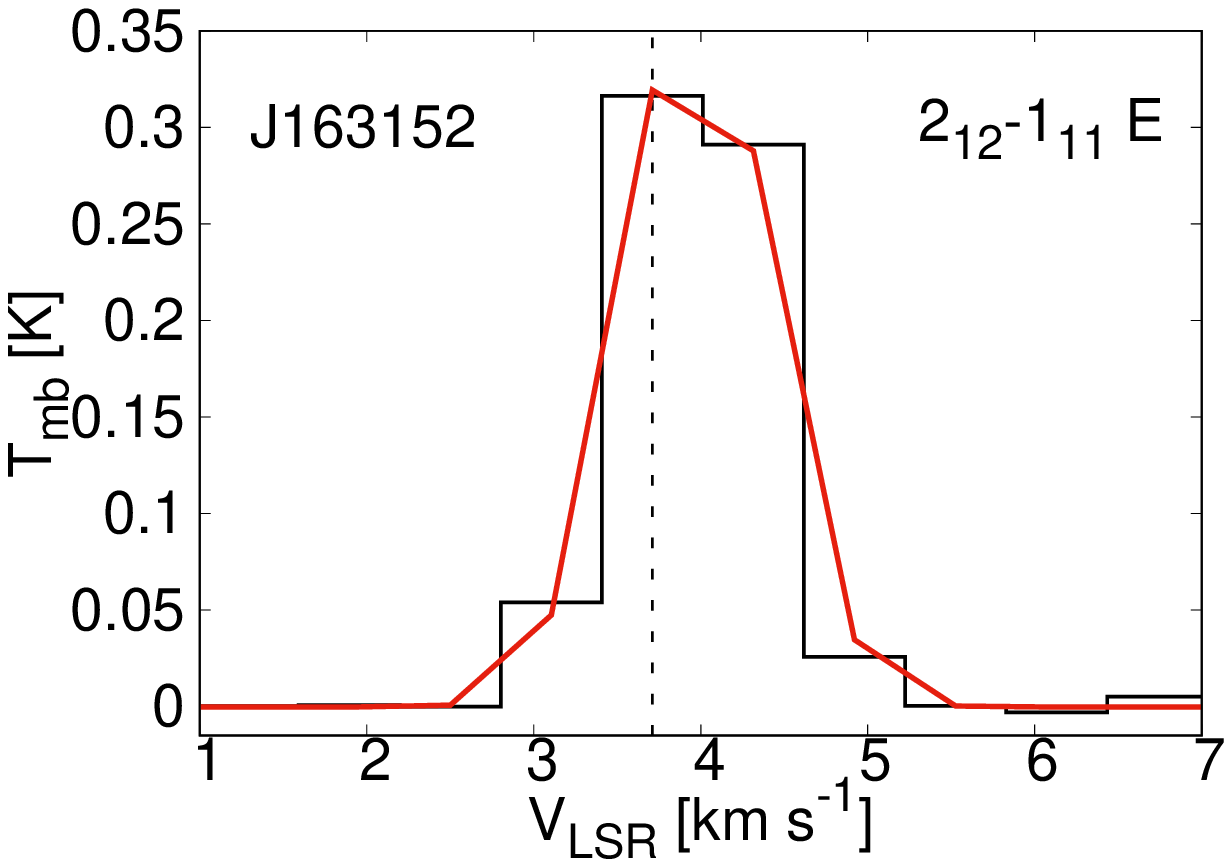}
    \includegraphics[width=1.5in]{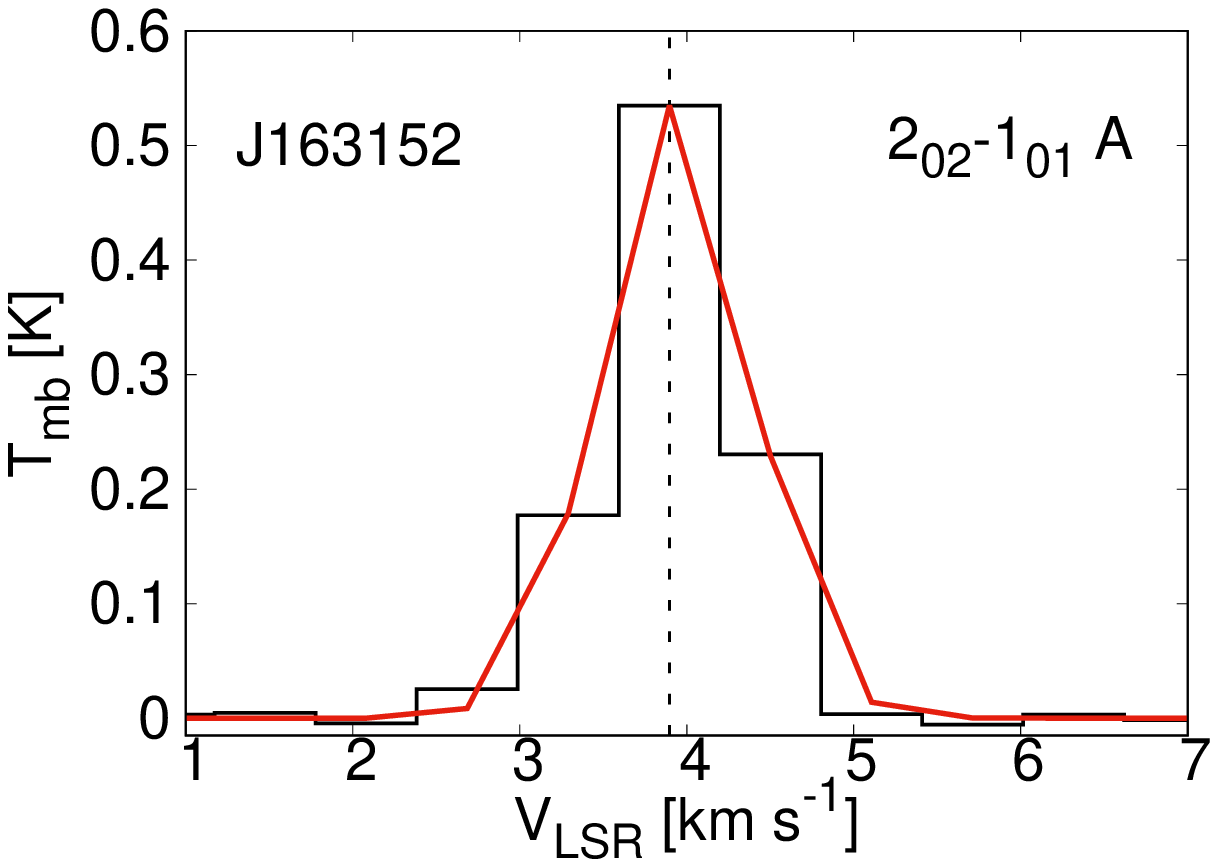}
    \includegraphics[width=1.5in]{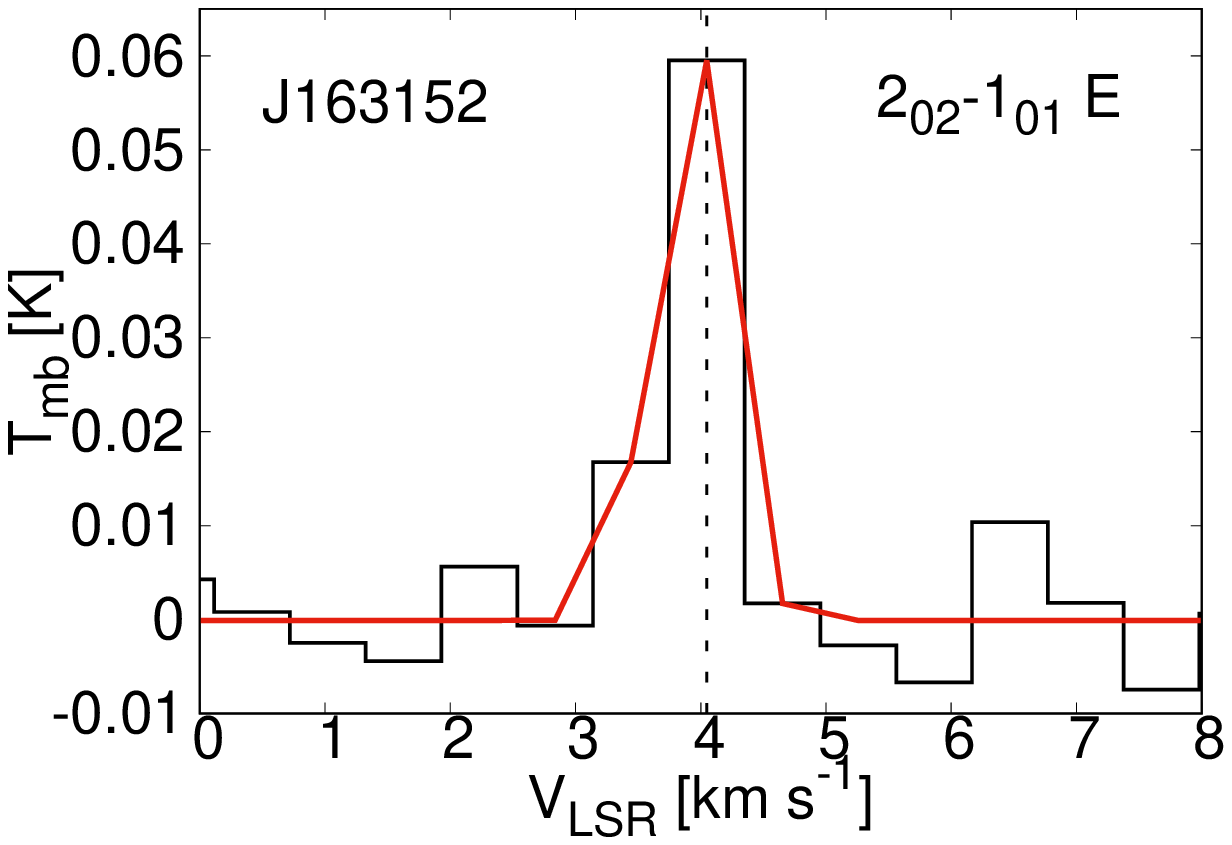}
    \includegraphics[width=1.5in]{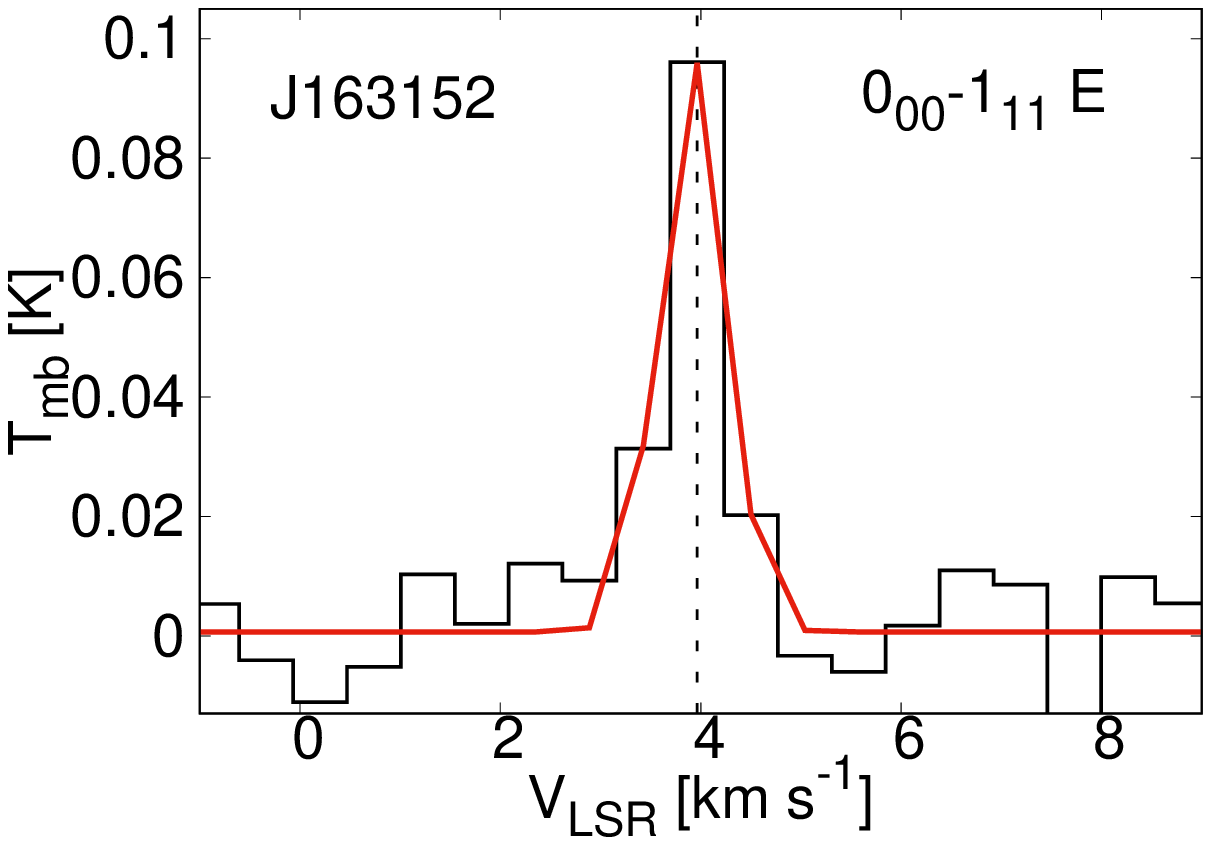}	
    \includegraphics[width=1.5in]{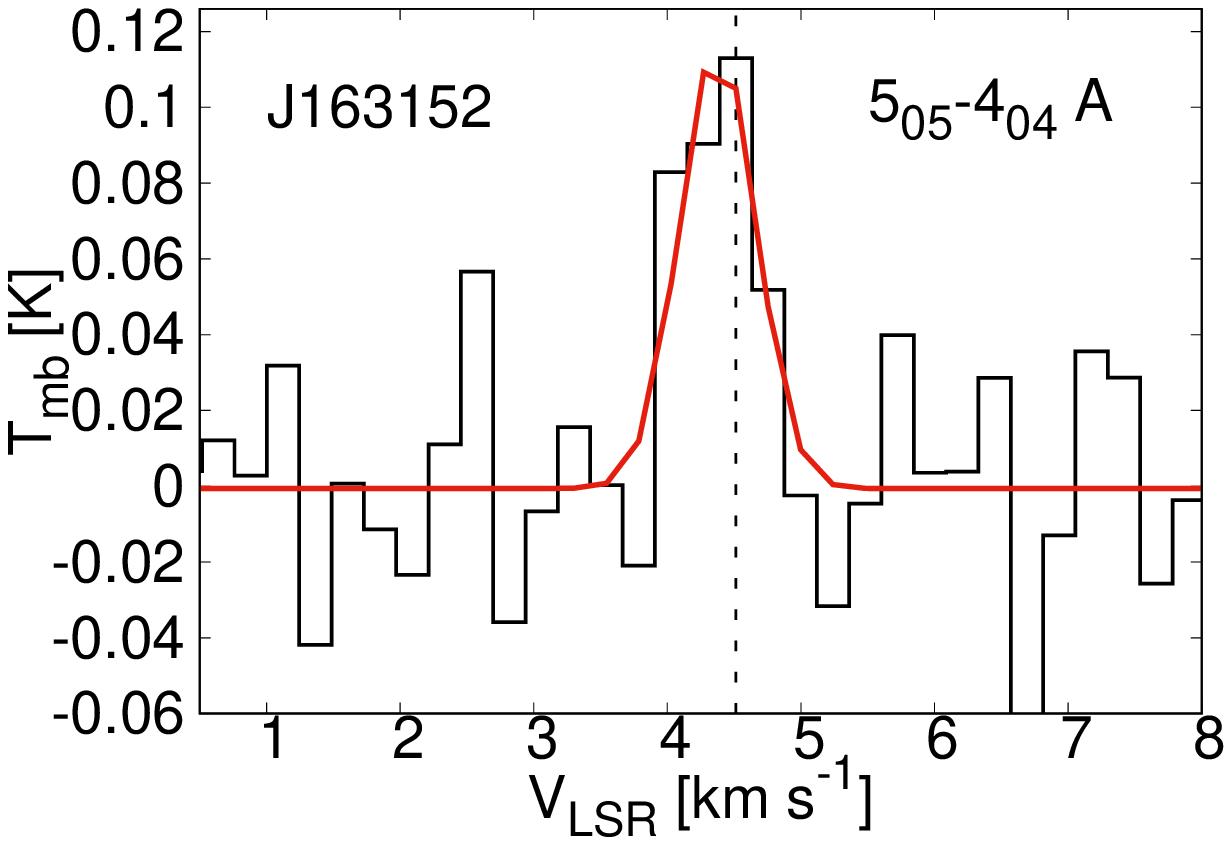}	
    \includegraphics[width=1.5in]{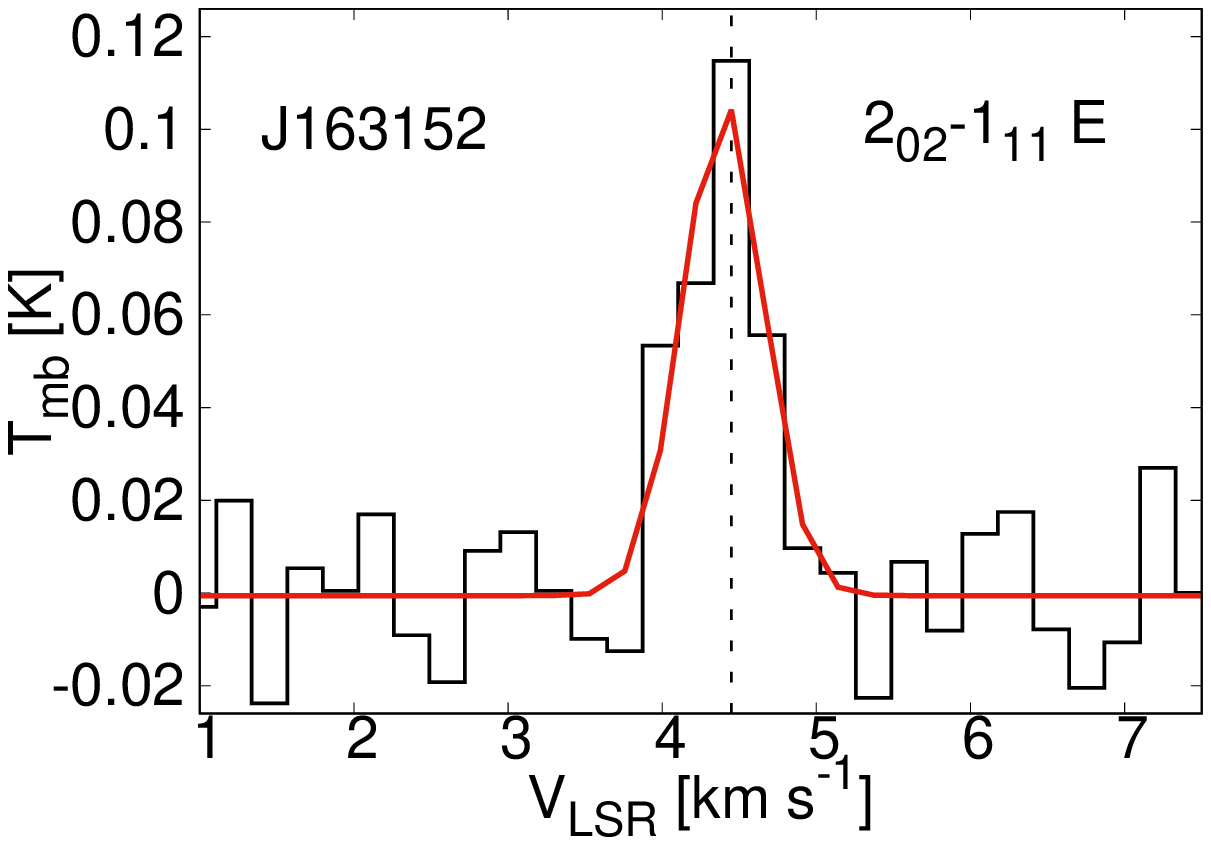}    
    \caption{Continued.}      
  \end{figure*}

\setcounter{figure}{2}    
 \begin{figure*}
  \centering         
    \includegraphics[width=1.5in]{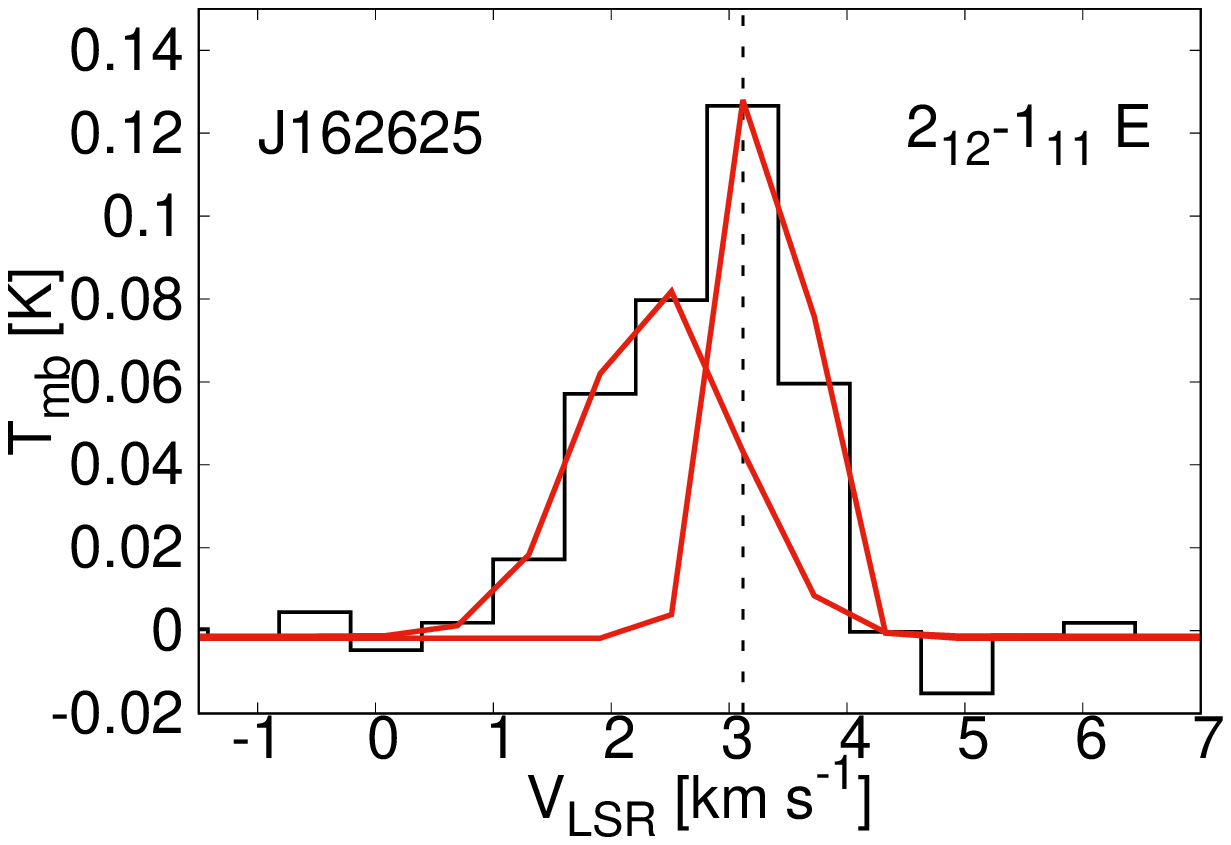}	
    \includegraphics[width=1.5in]{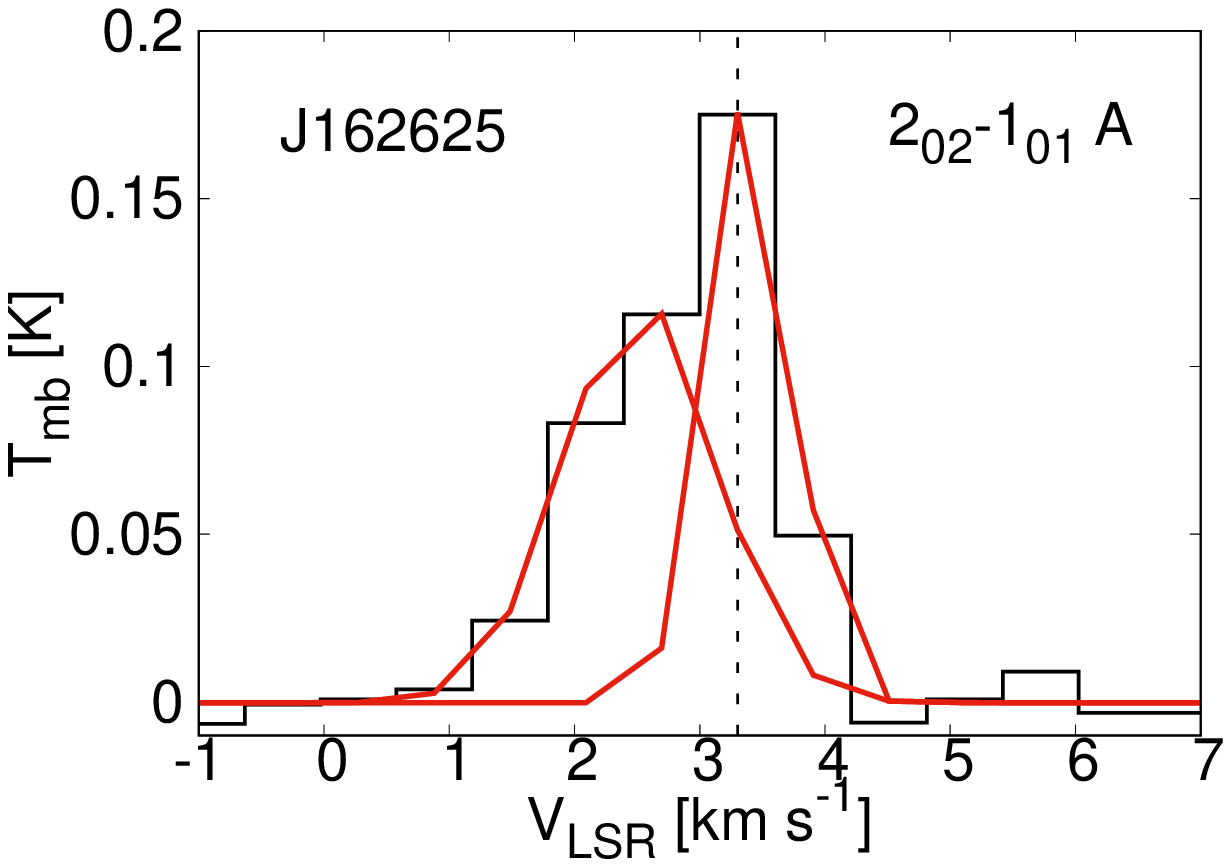}
    \includegraphics[width=1.5in]{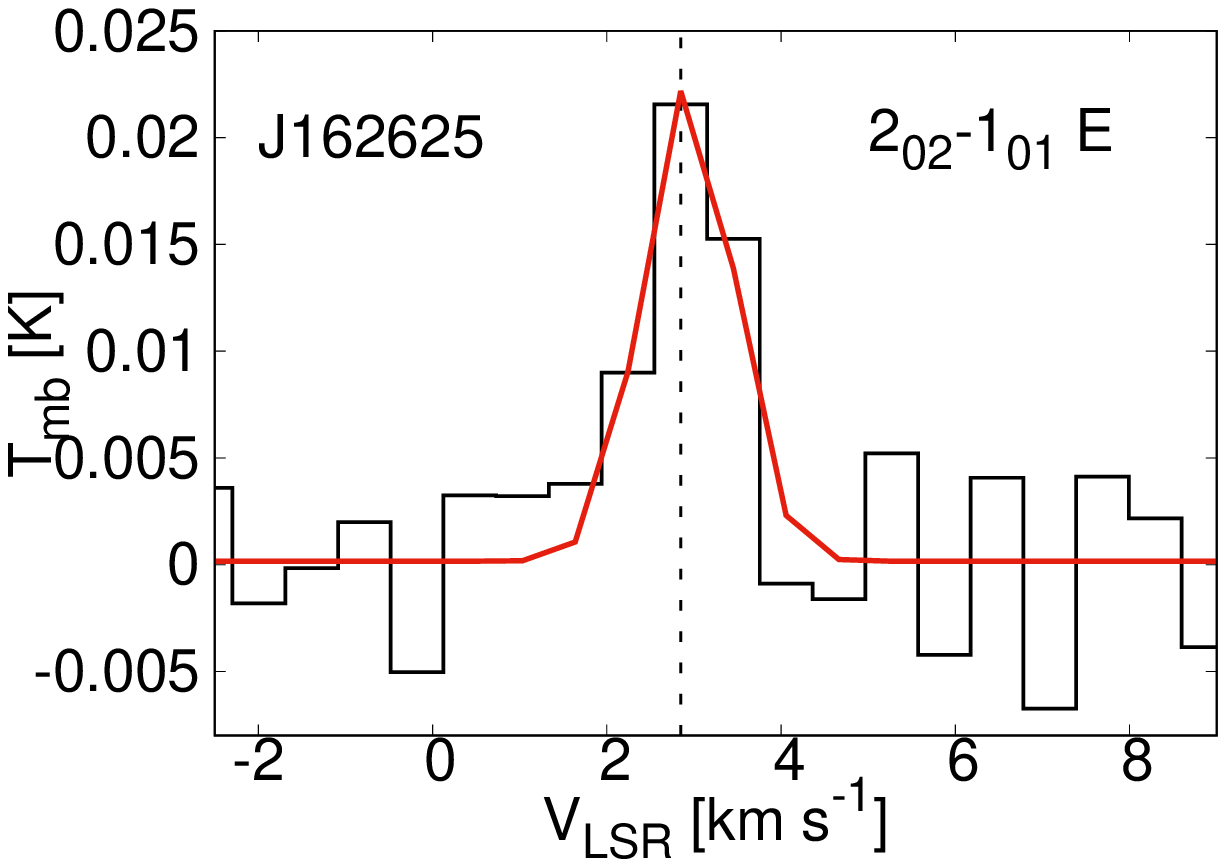}
    \includegraphics[width=1.5in]{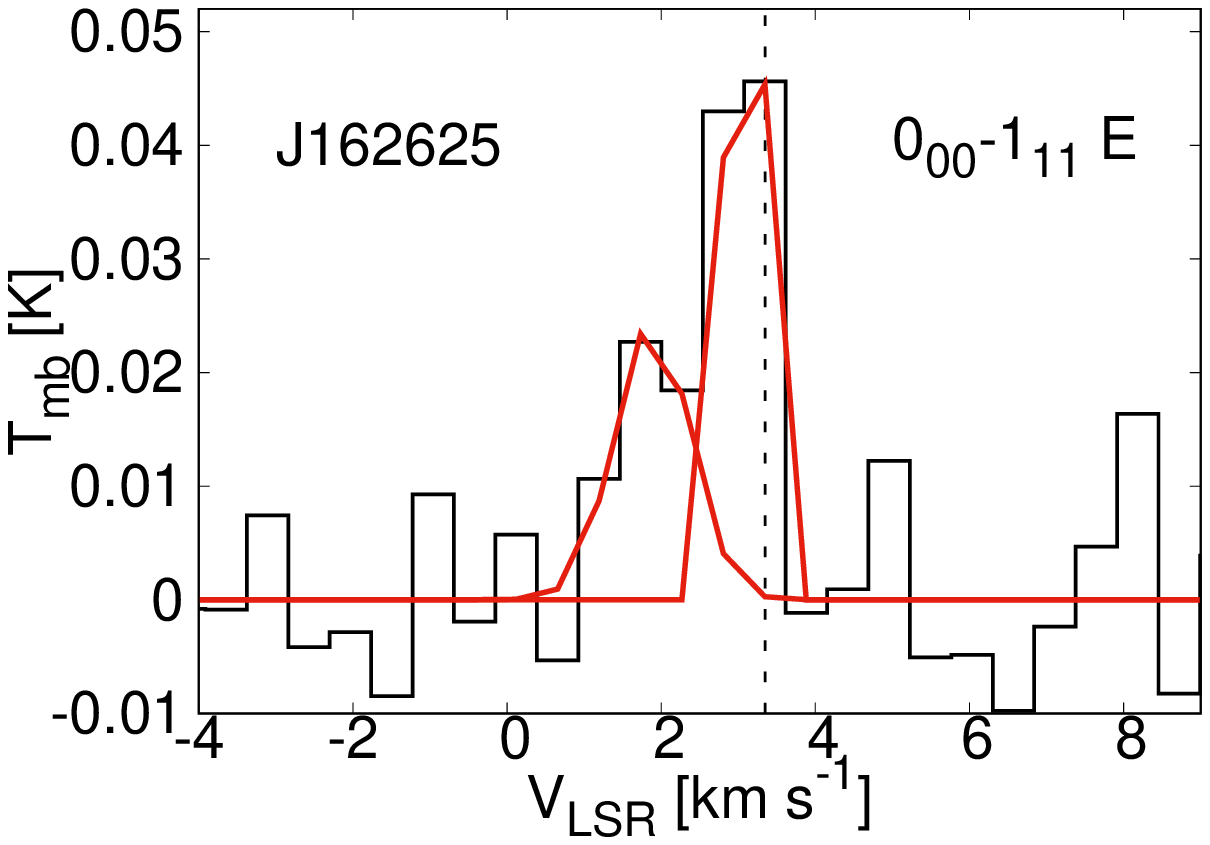}
    \includegraphics[width=1.5in]{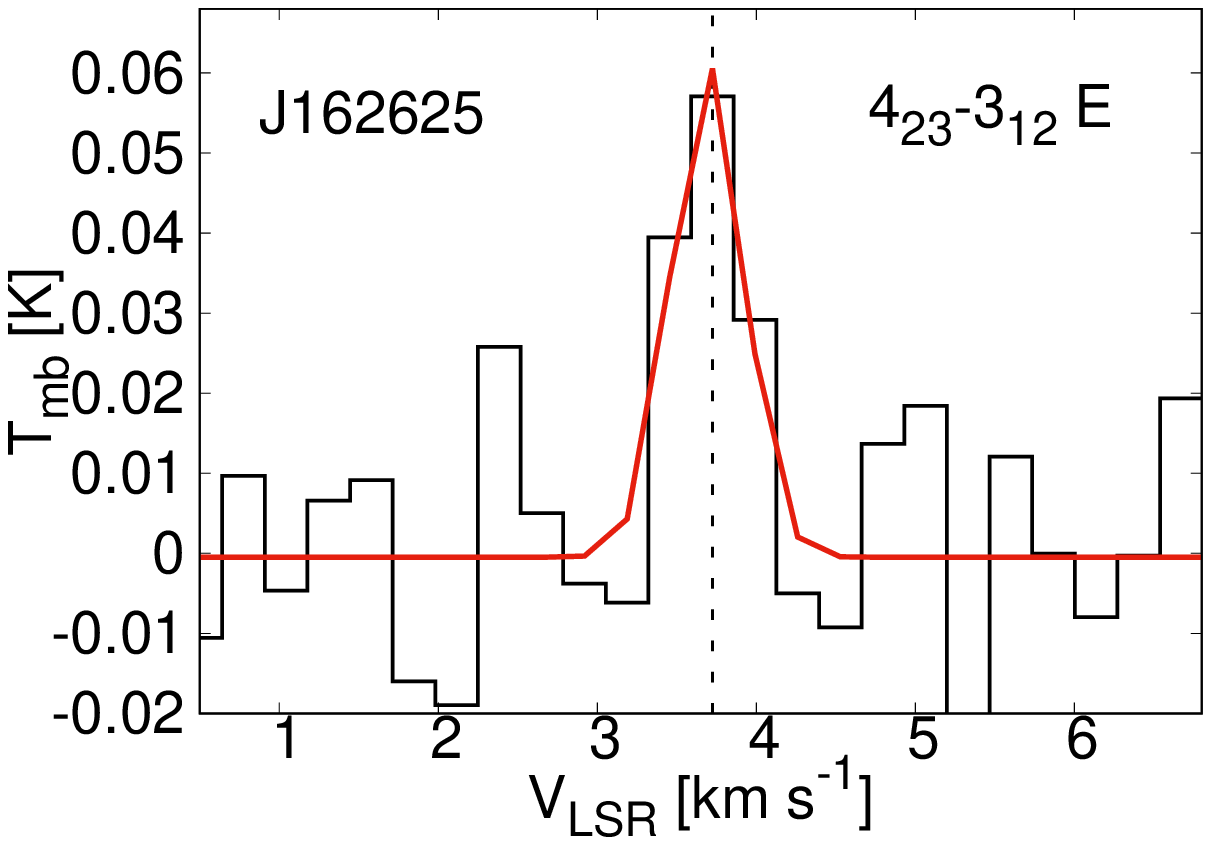}	
    \includegraphics[width=1.5in]{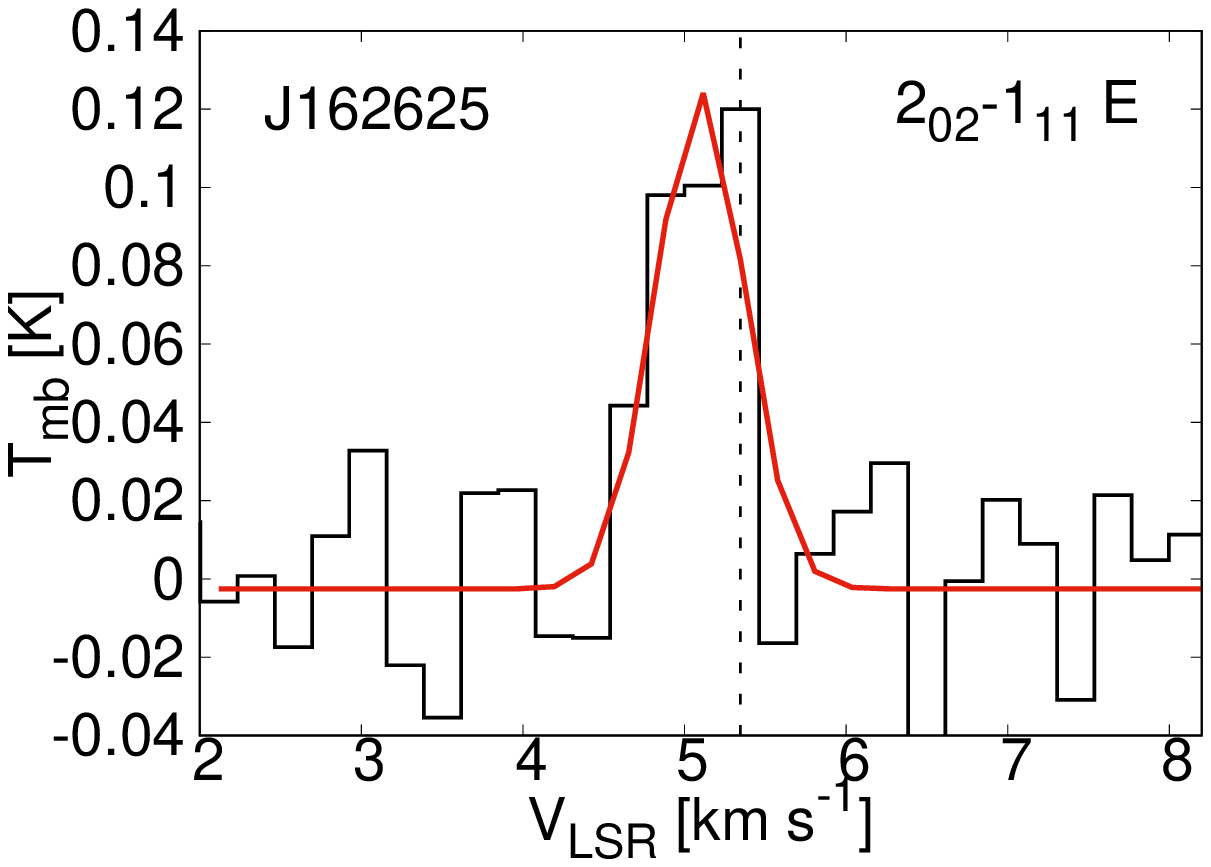}		\\	\vspace{0.2in}

    \includegraphics[width=1.5in]{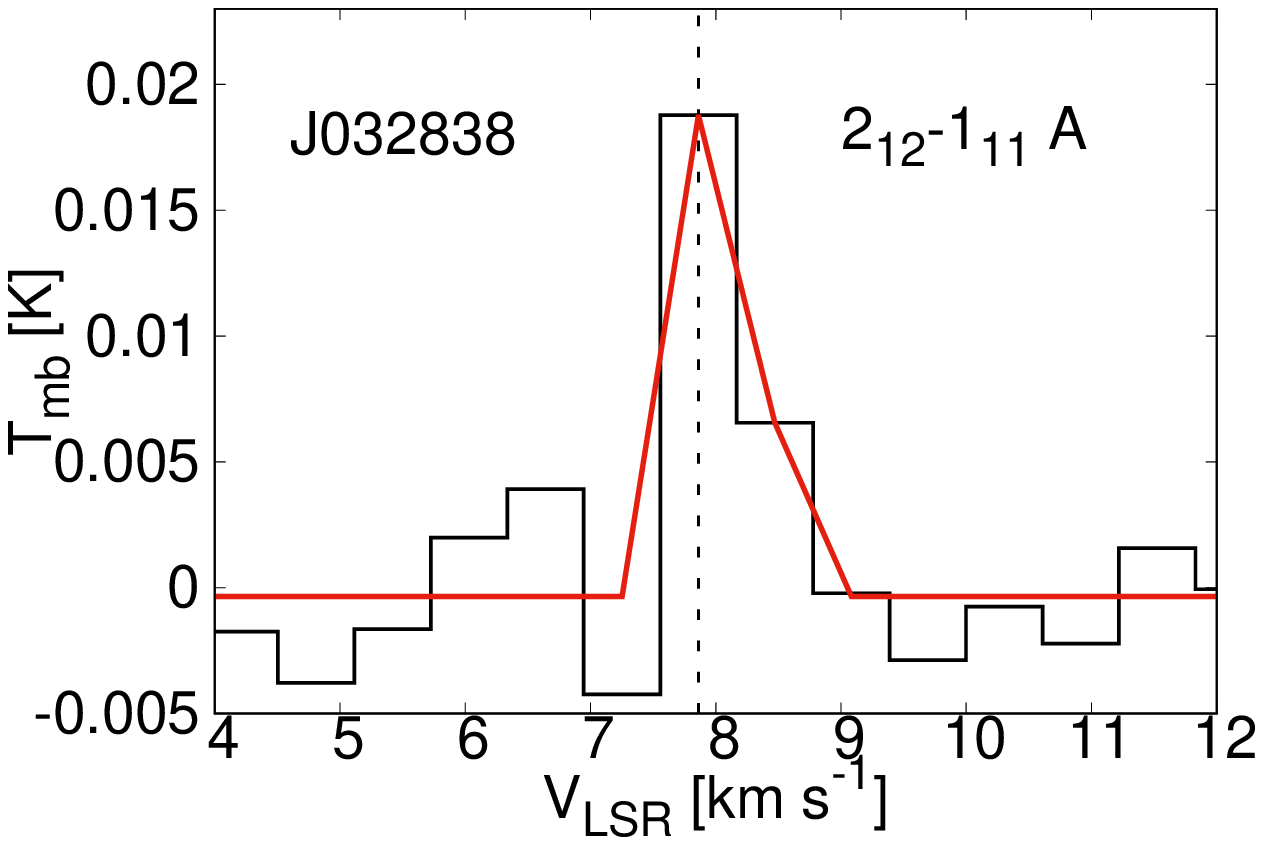}	
    \includegraphics[width=1.5in]{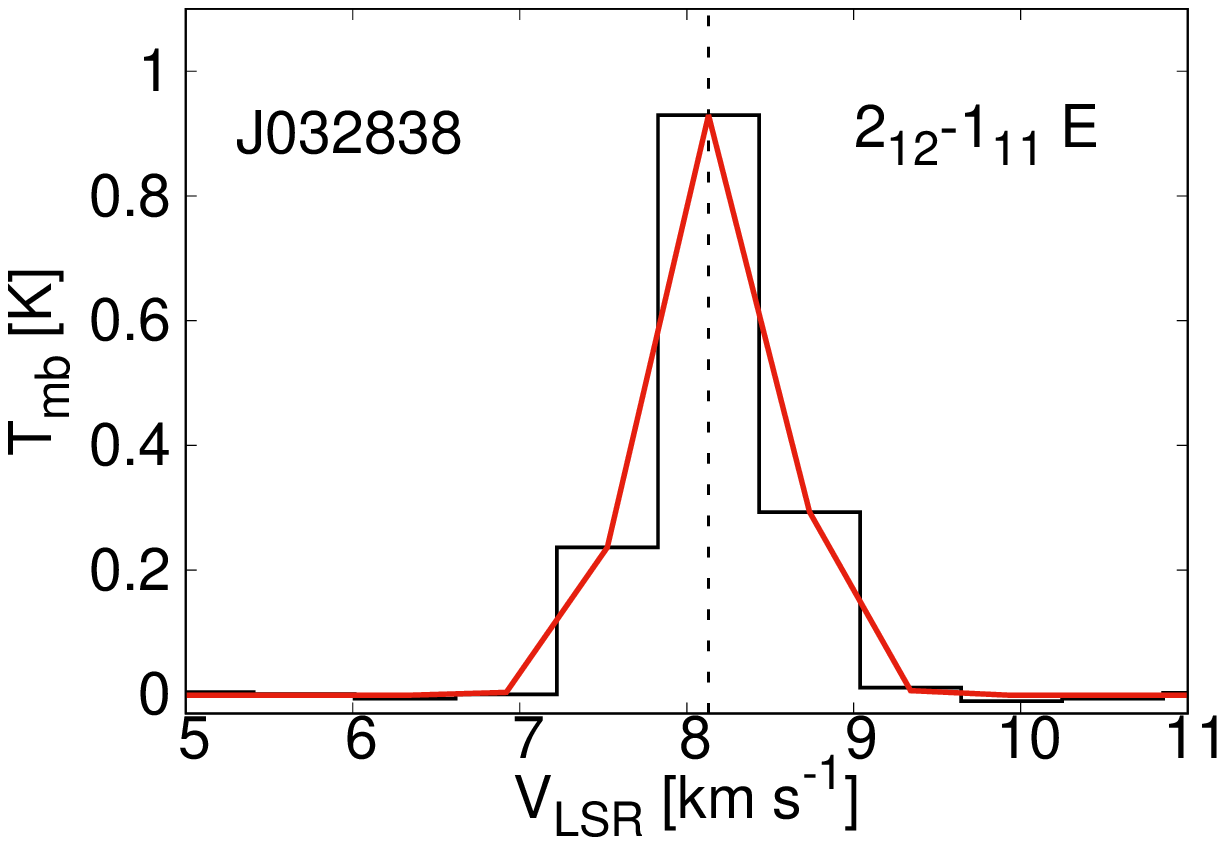}	
    \includegraphics[width=1.5in]{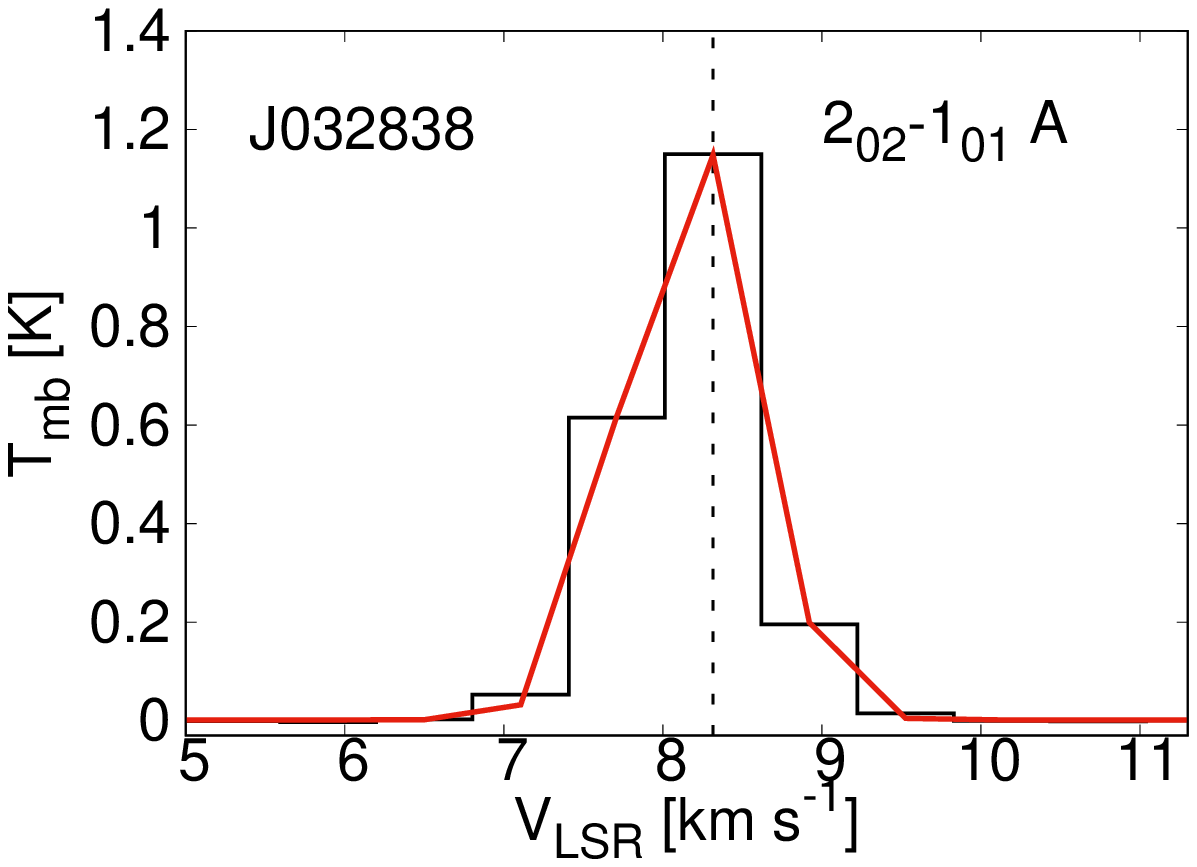}	
    \includegraphics[width=1.5in]{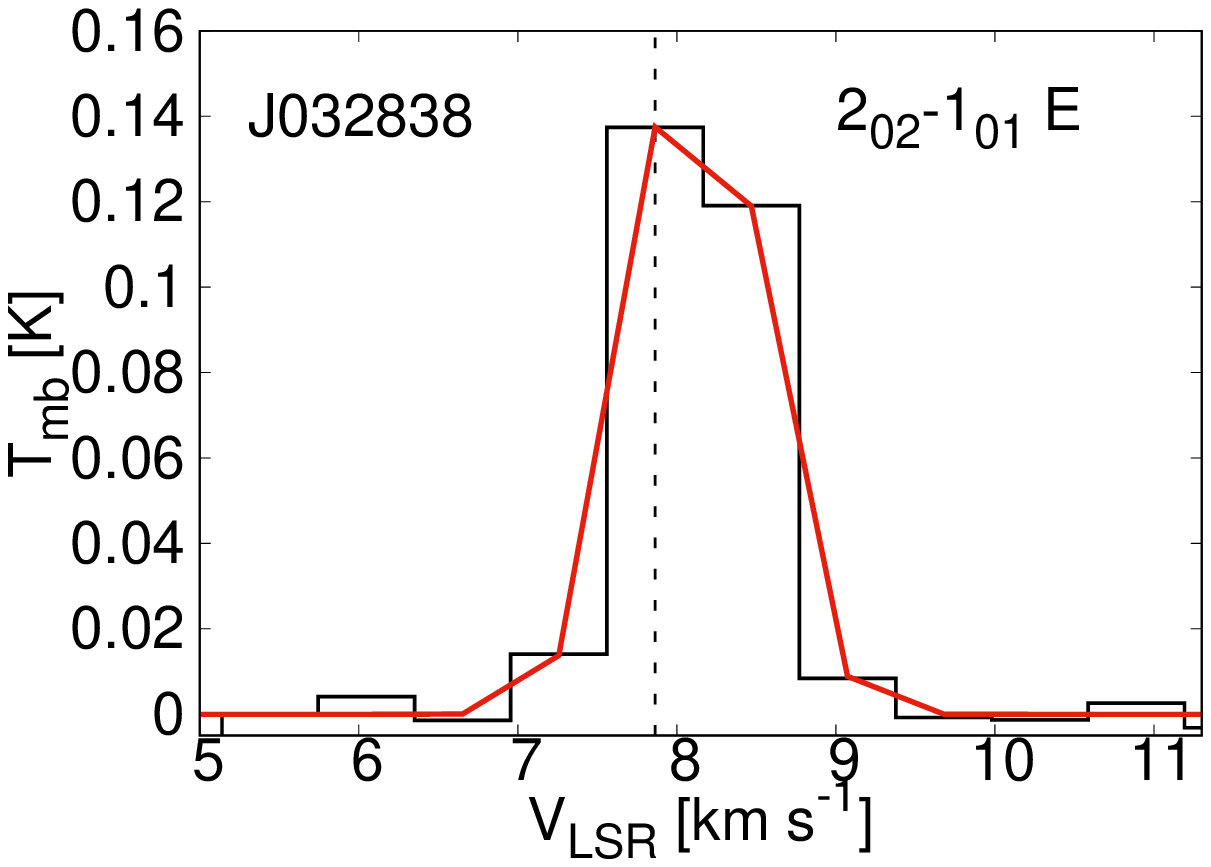}	
    \includegraphics[width=1.5in]{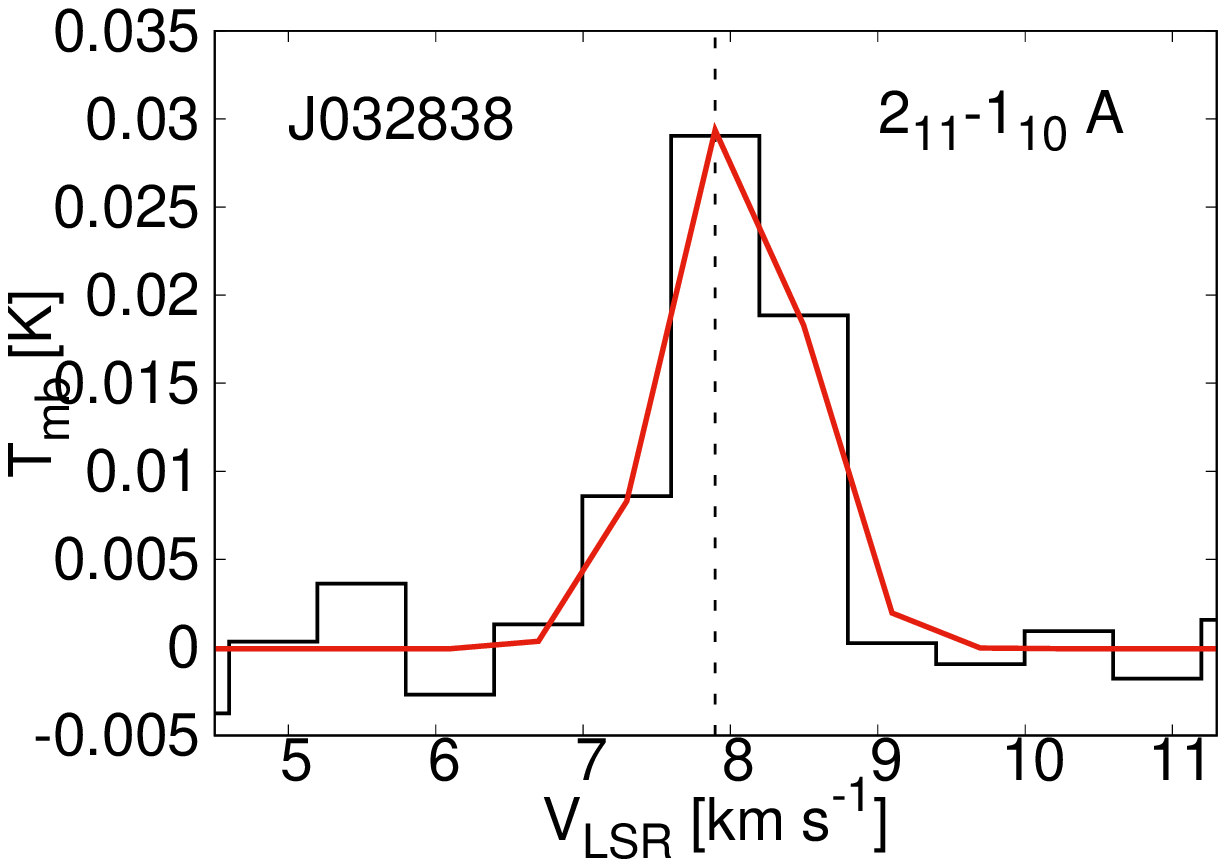}
    \includegraphics[width=1.5in]{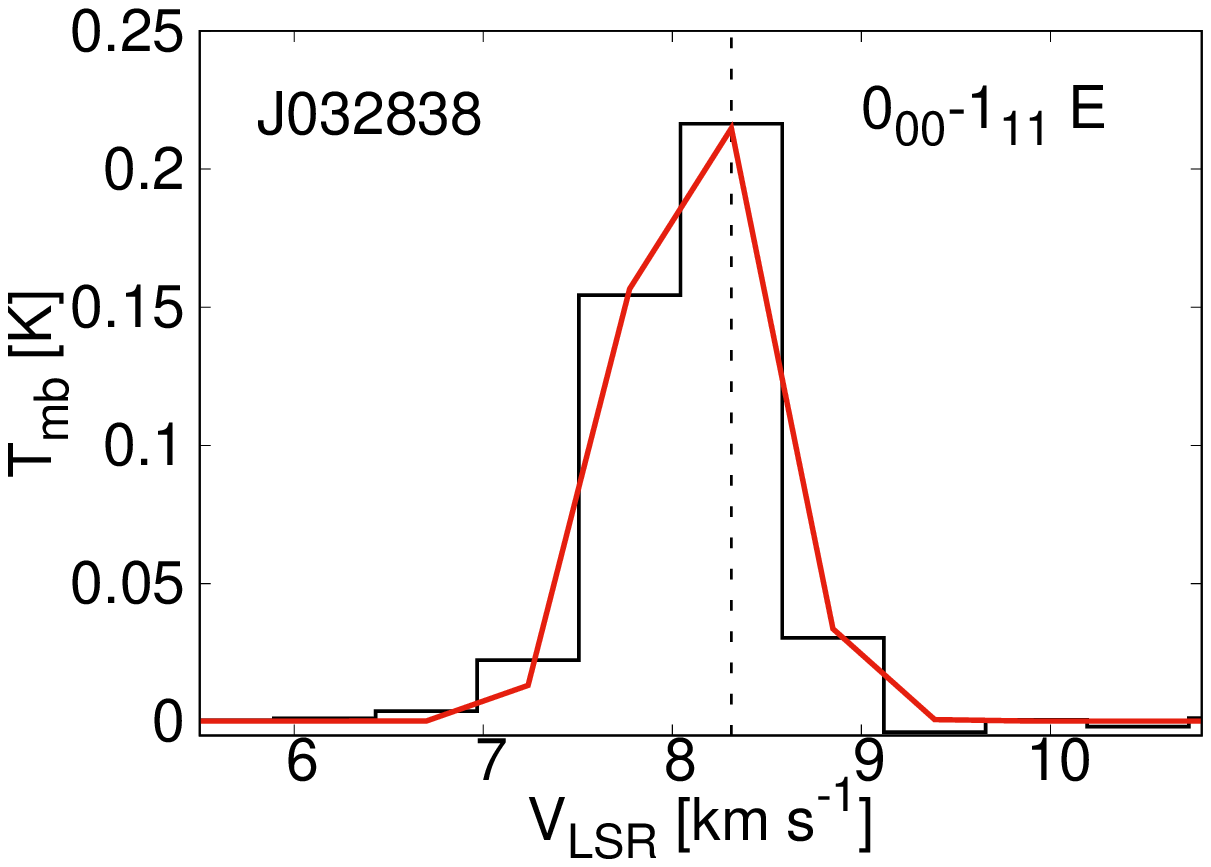}
    \includegraphics[width=1.5in]{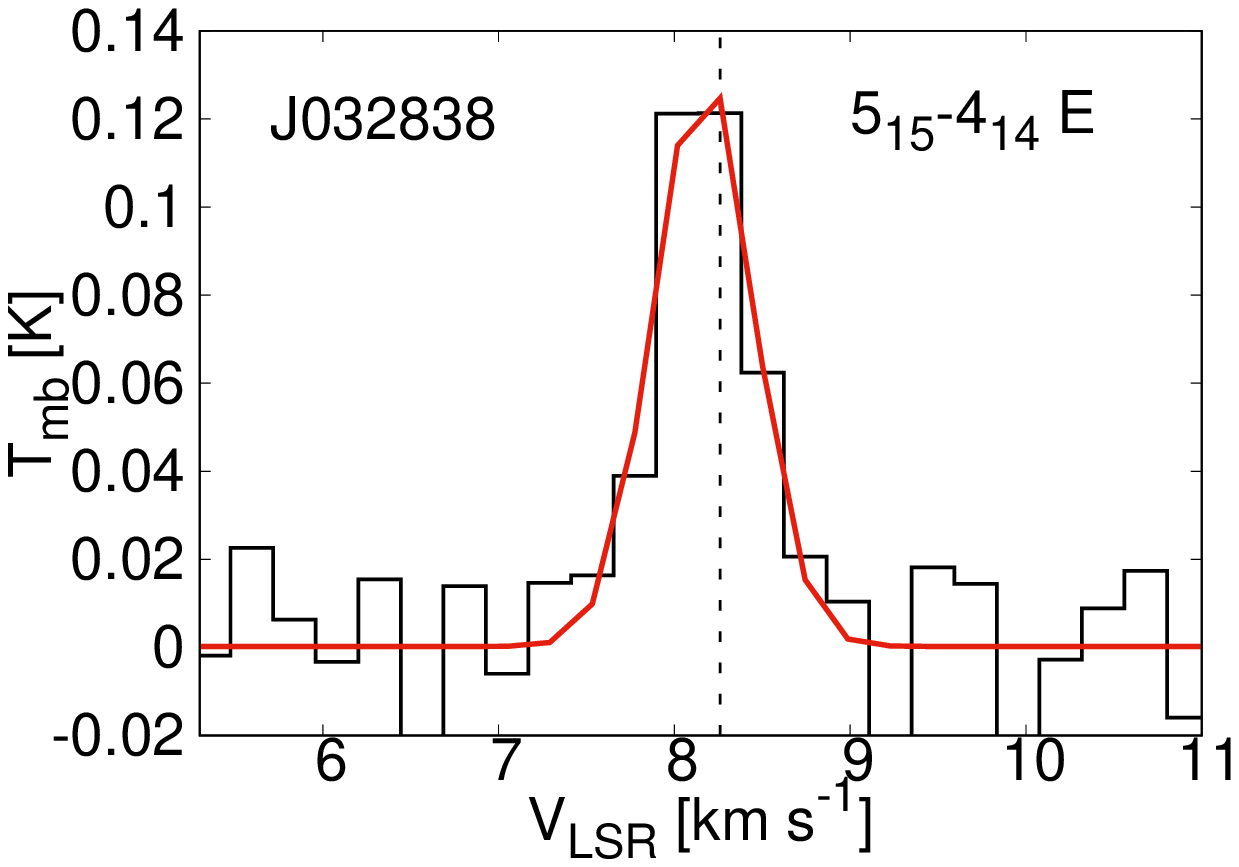}
    \includegraphics[width=1.5in]{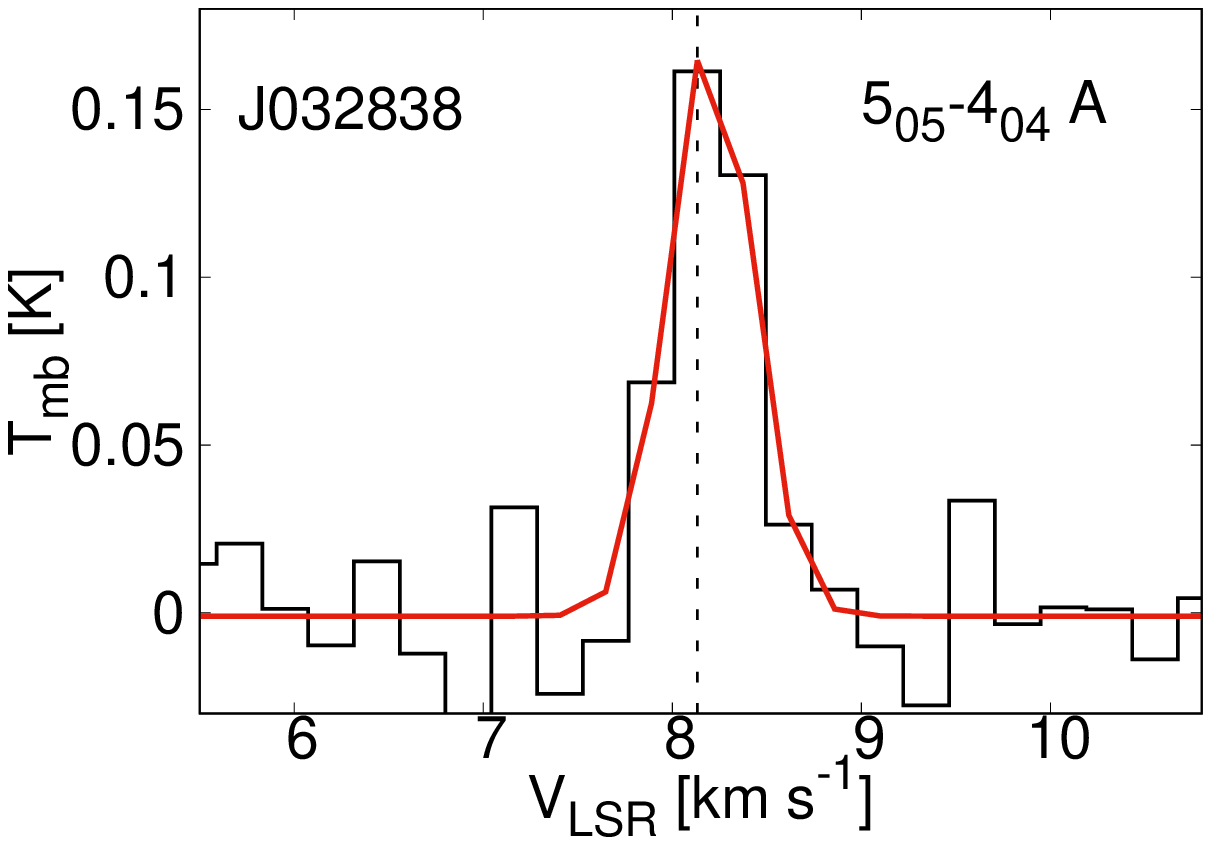}
    \includegraphics[width=1.5in]{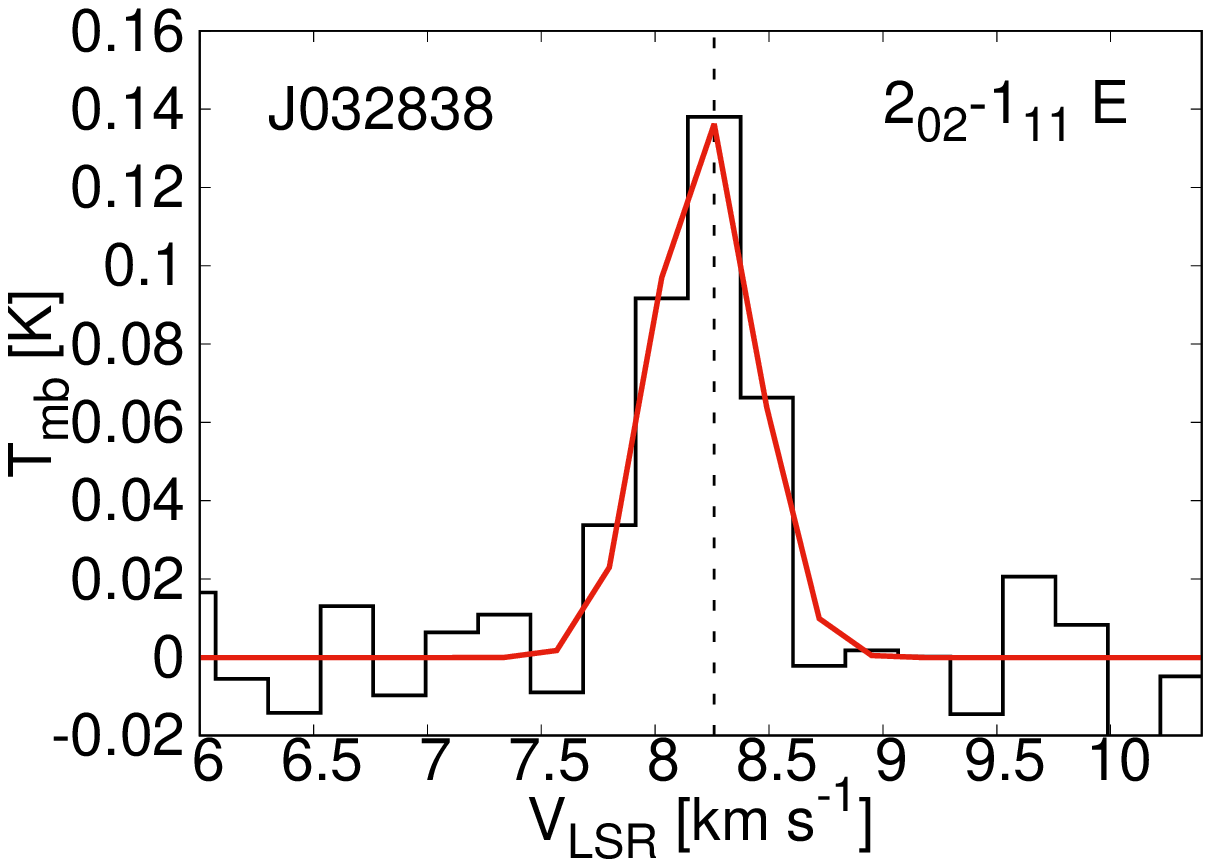}		
    \includegraphics[width=1.5in]{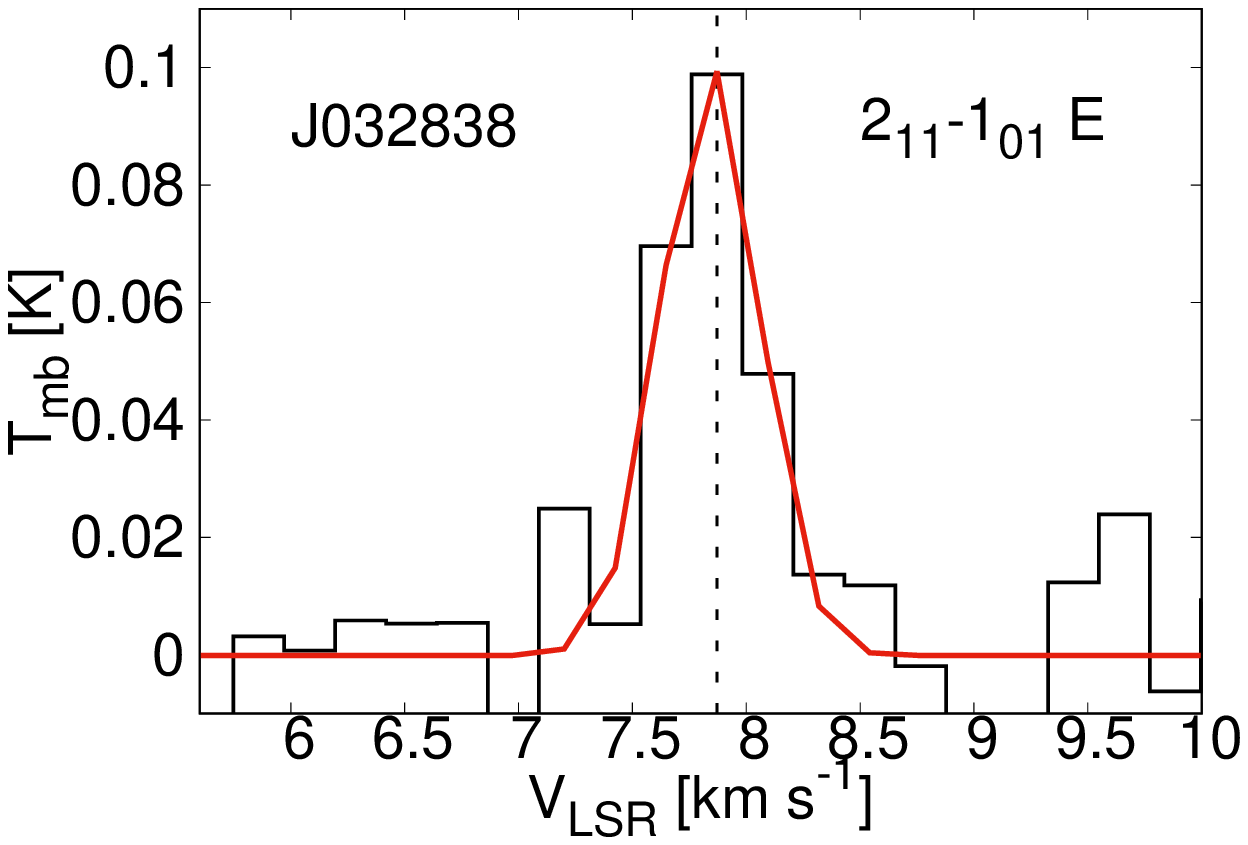}		\\	\vspace{0.2in}
    
    \includegraphics[width=1.5in]{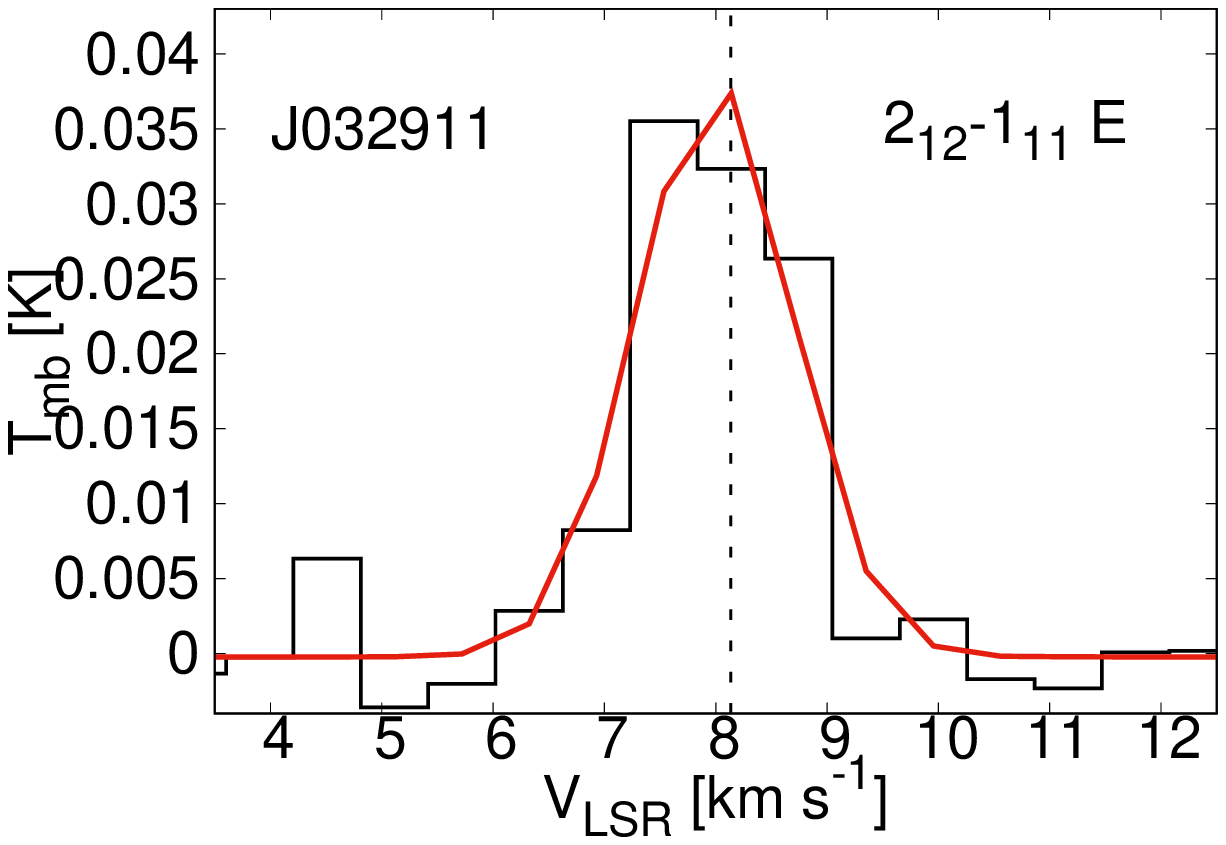}
    \includegraphics[width=1.5in]{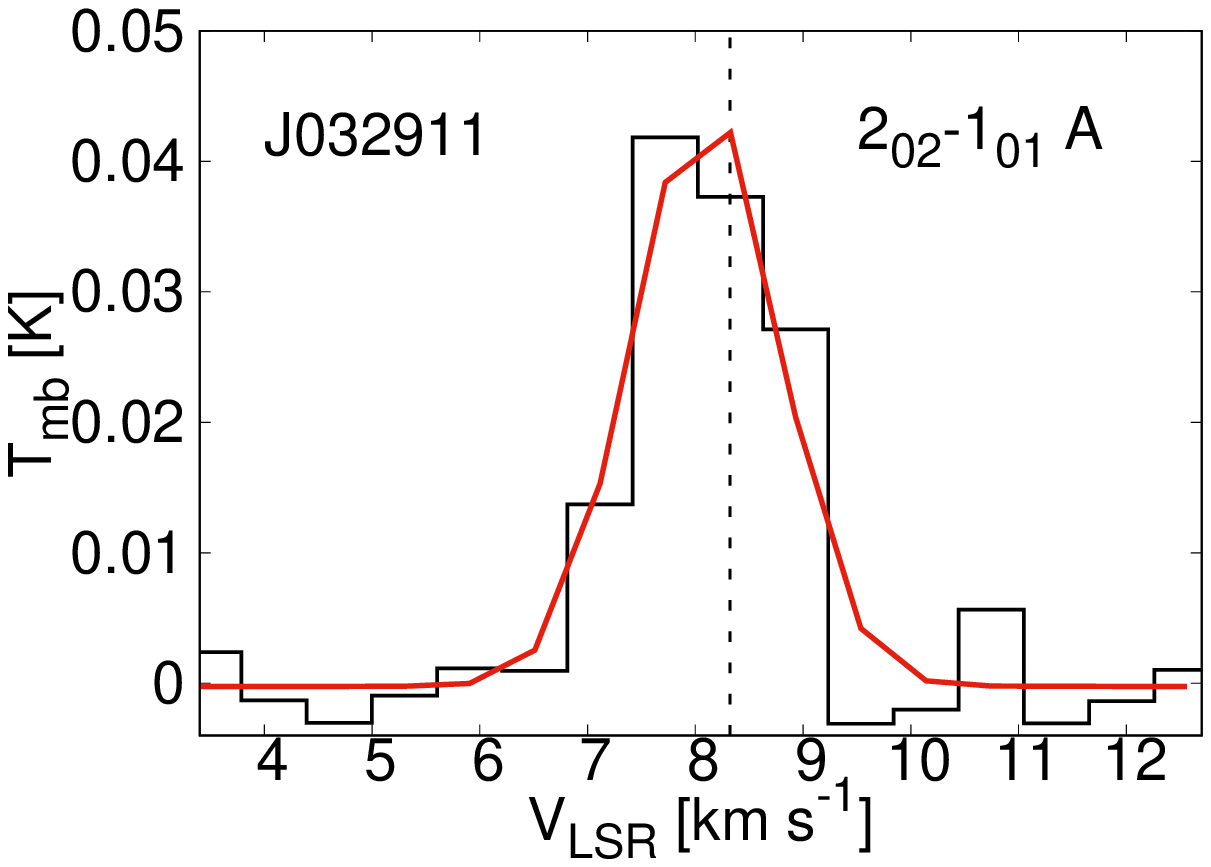}
    \includegraphics[width=1.5in]{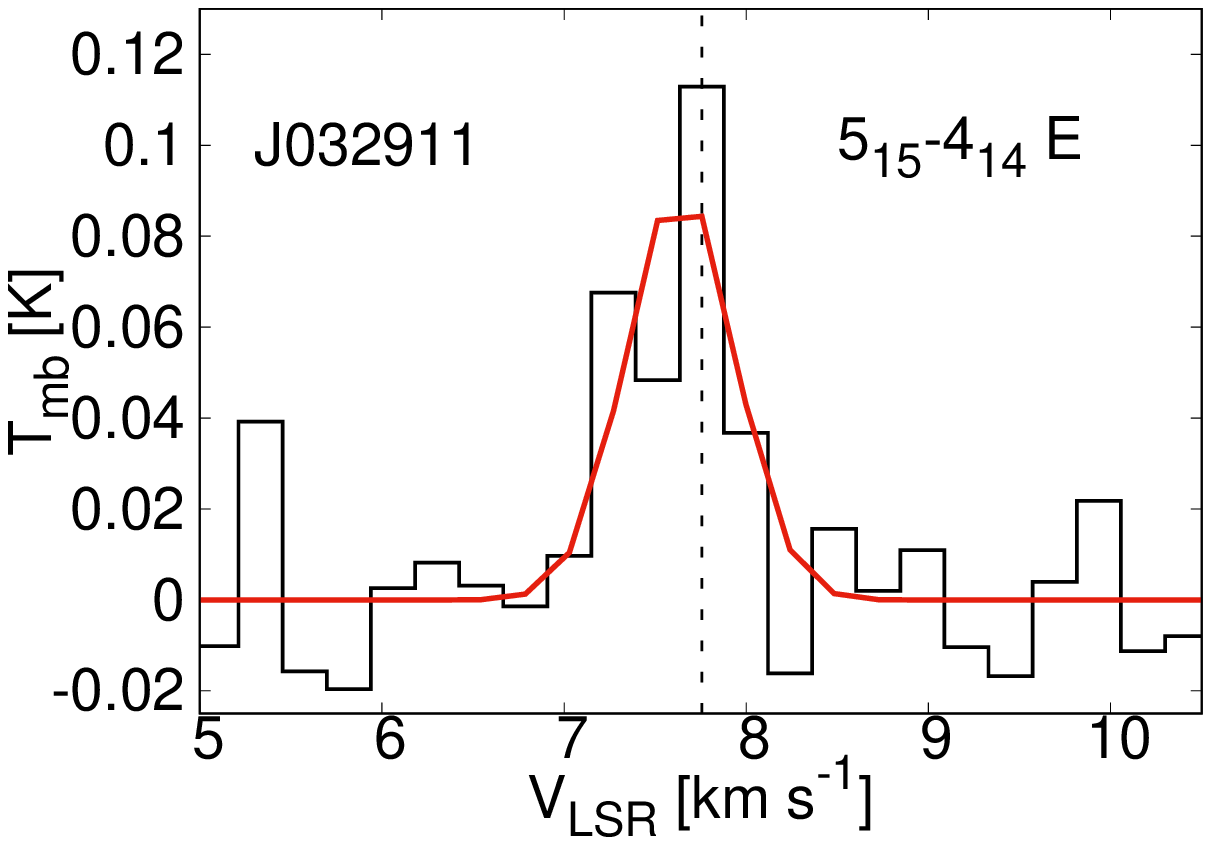}
    \includegraphics[width=1.5in]{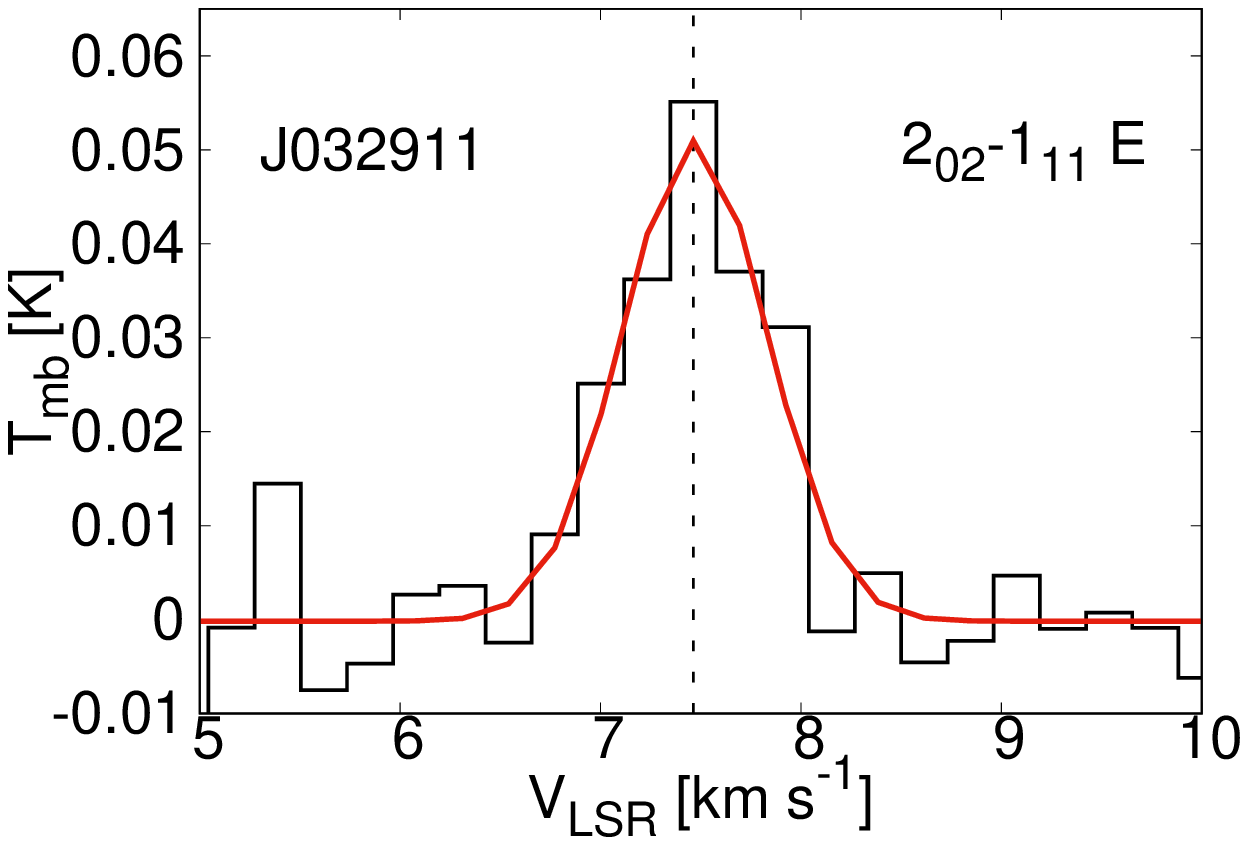}    	\\	\vspace{0.2in}
     
    \includegraphics[width=1.5in]{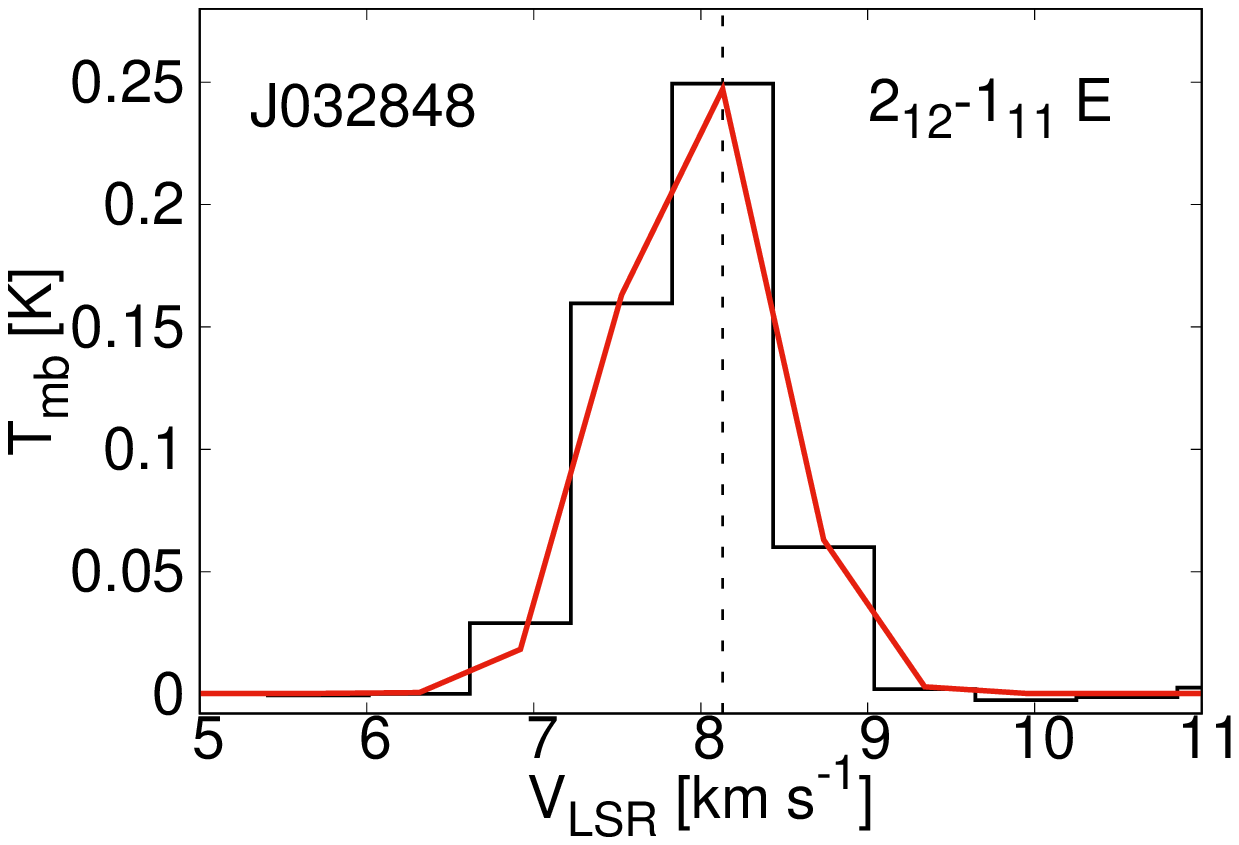}
    \includegraphics[width=1.5in]{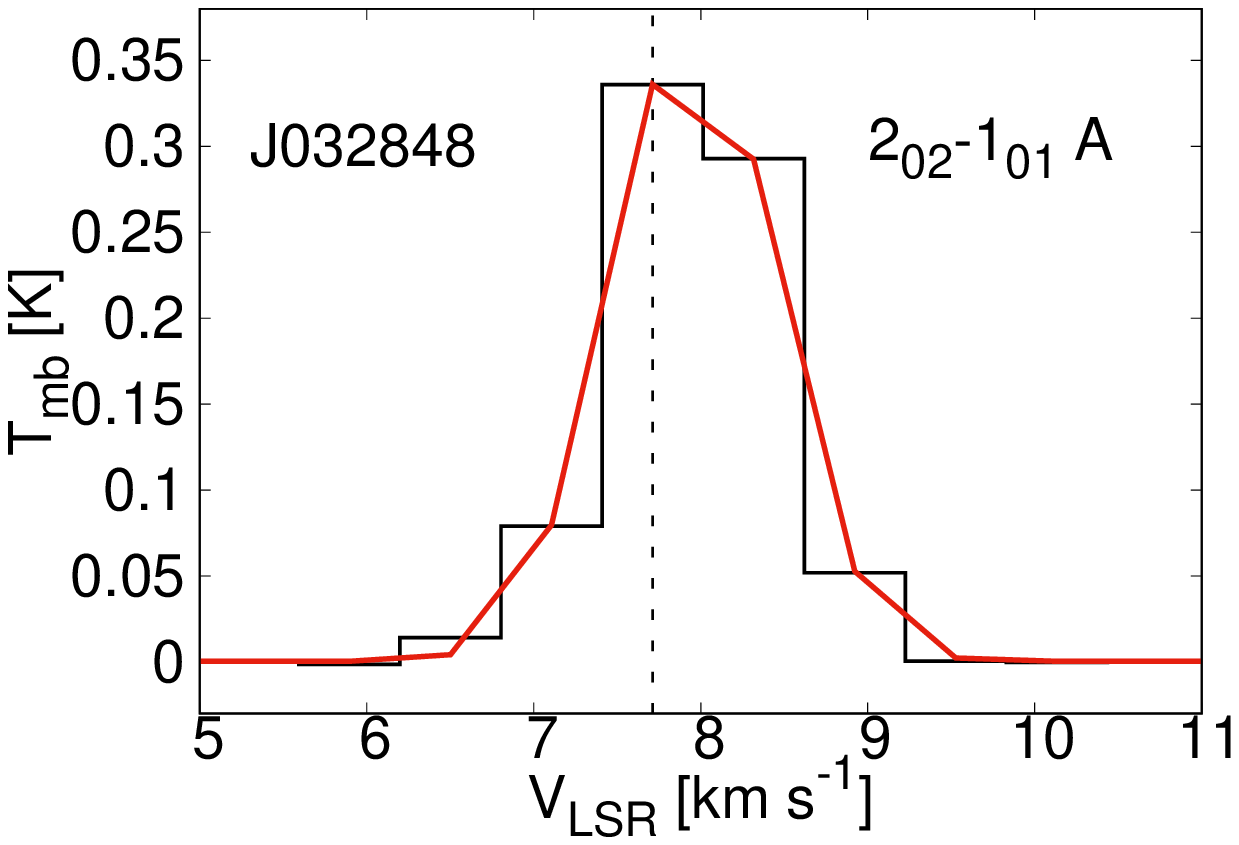}
    \includegraphics[width=1.5in]{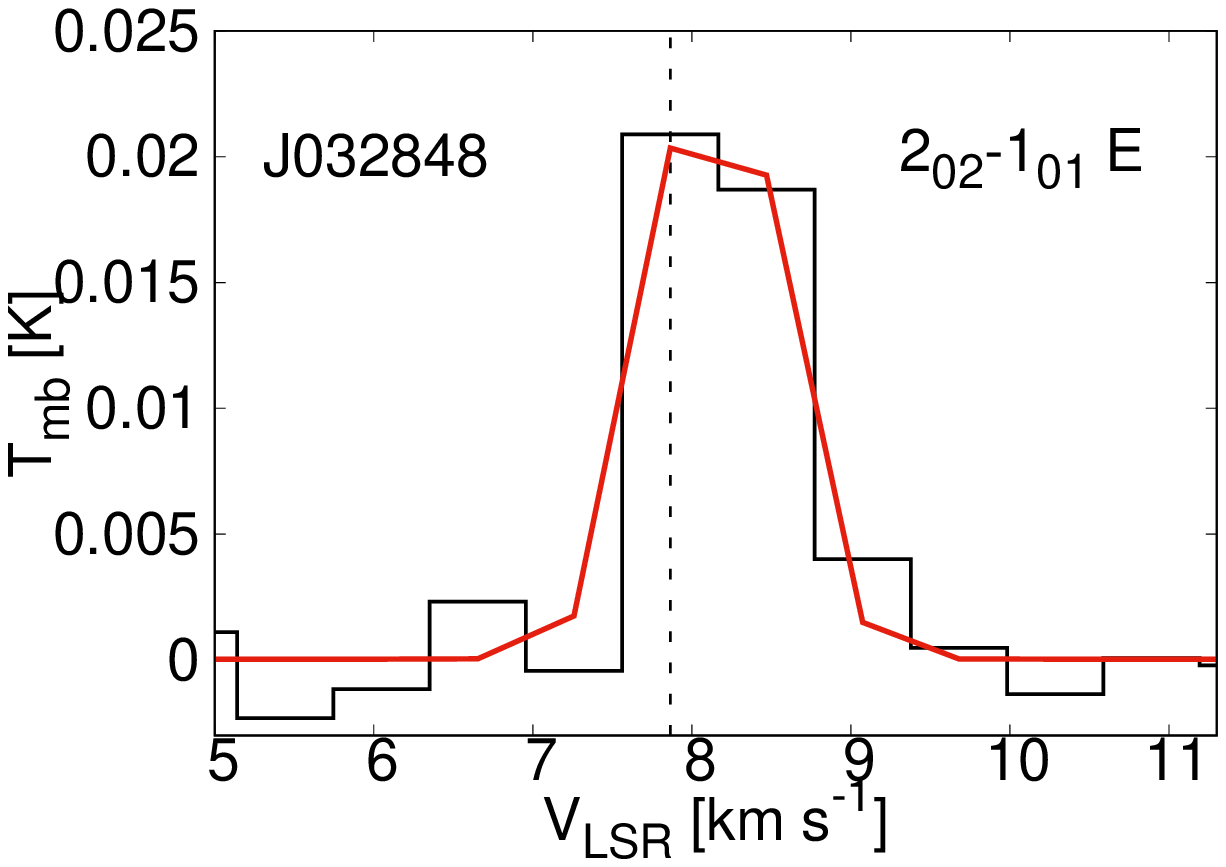}
    \includegraphics[width=1.5in]{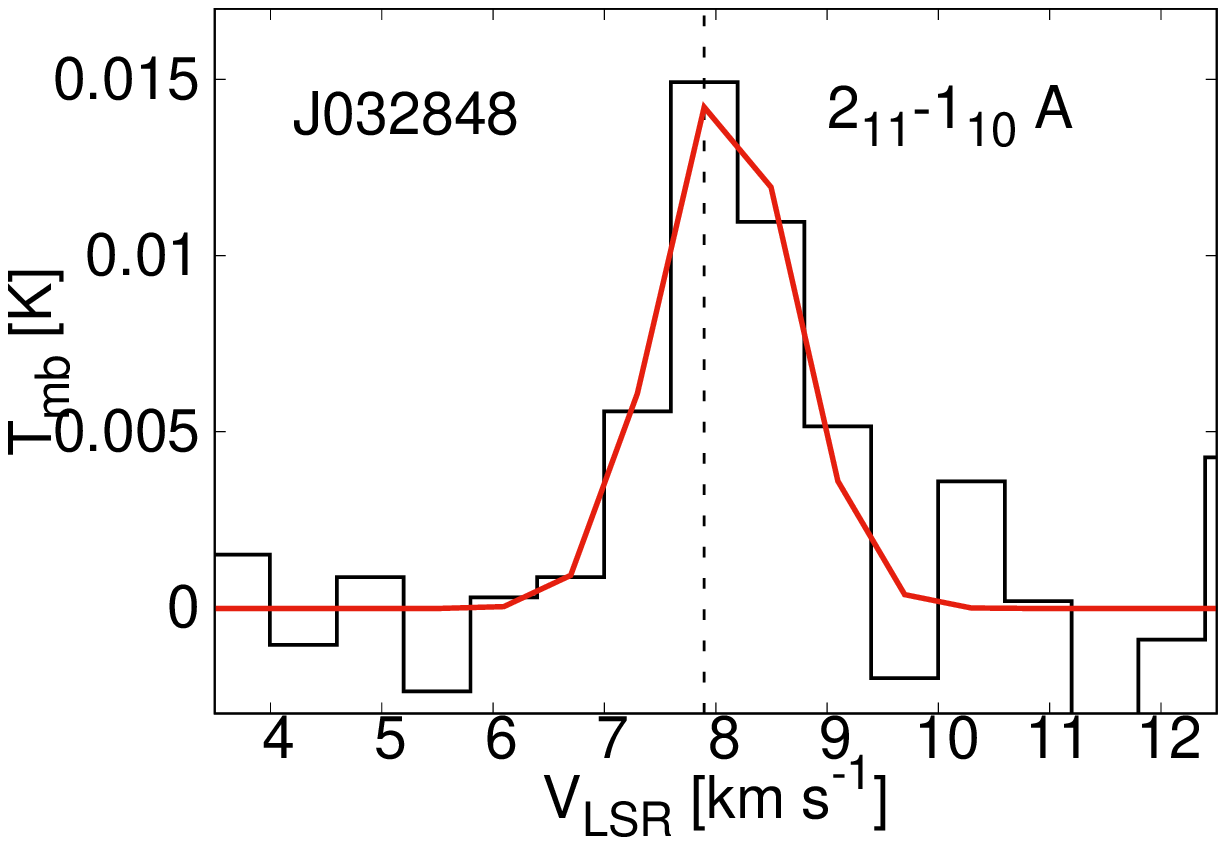}
    \includegraphics[width=1.5in]{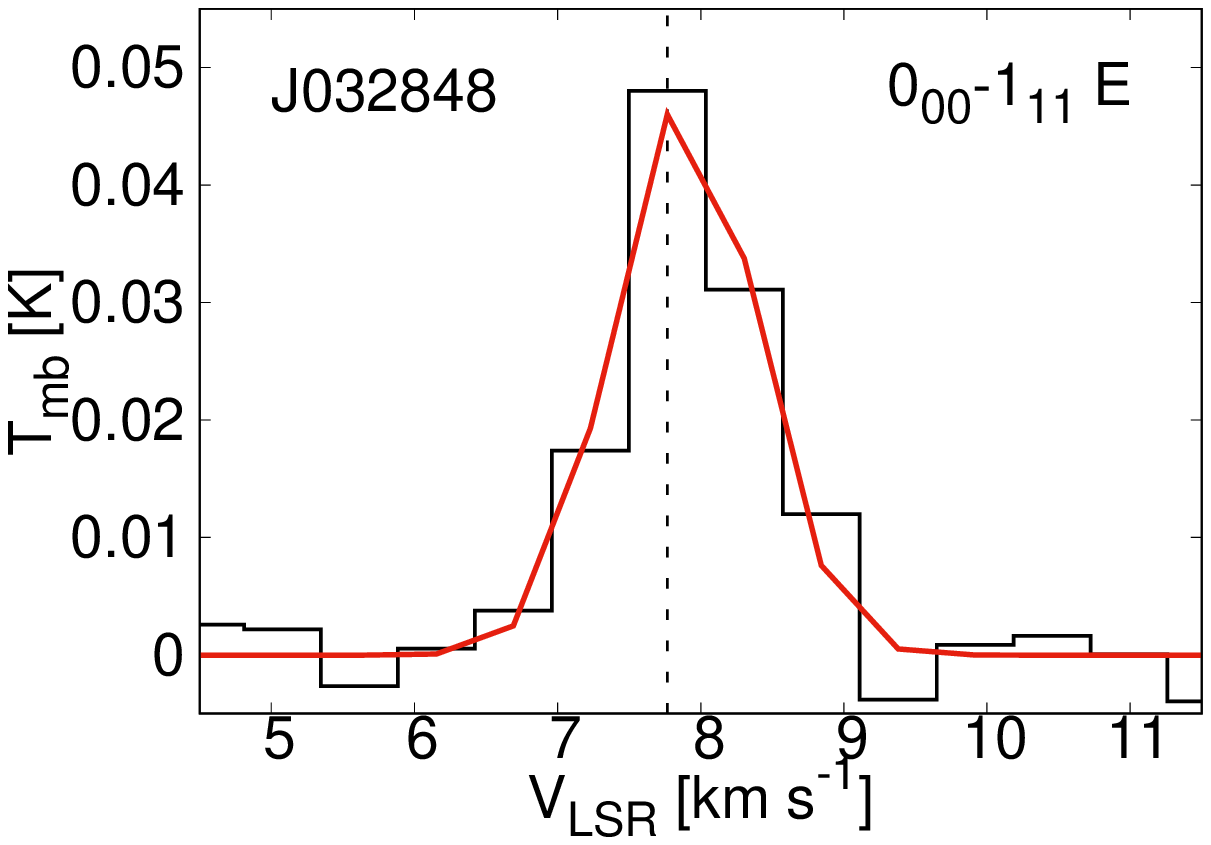}
    \includegraphics[width=1.5in]{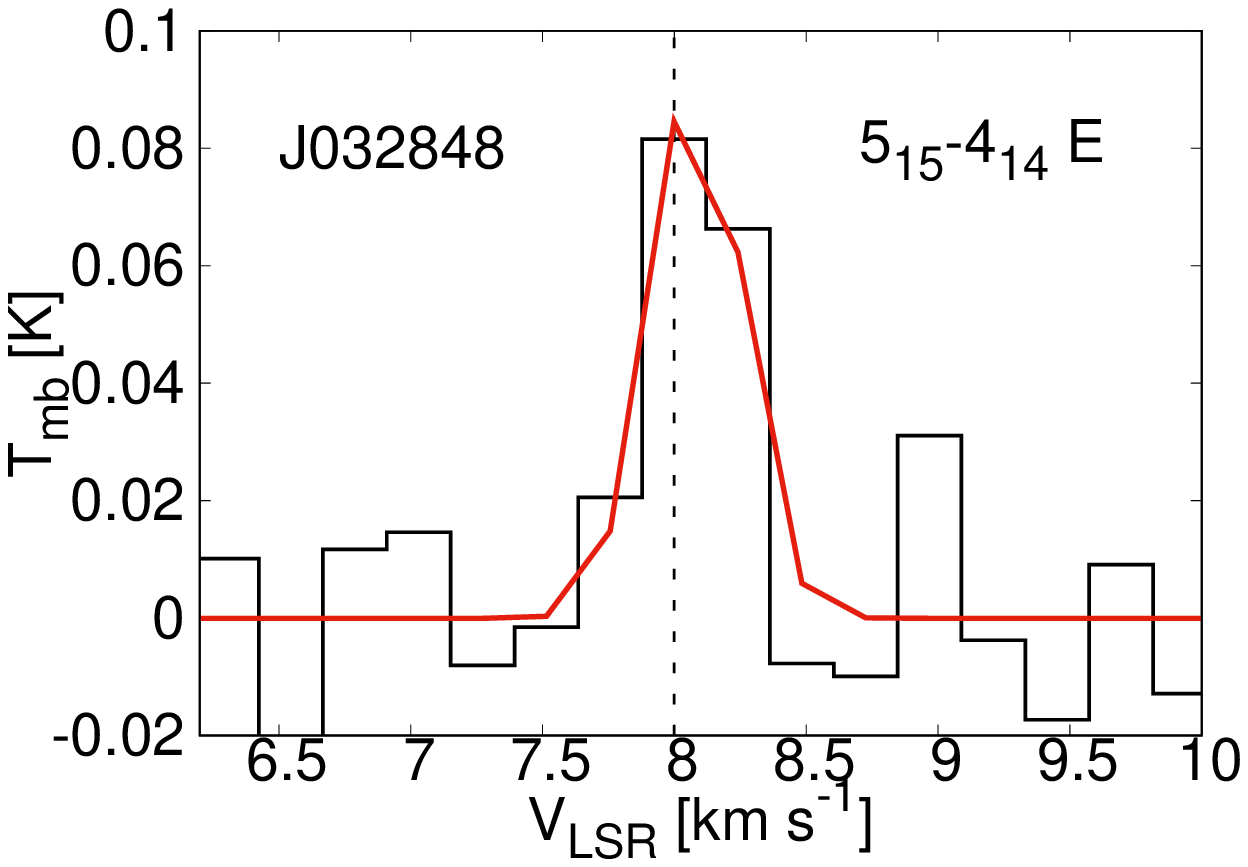}
    \includegraphics[width=1.5in]{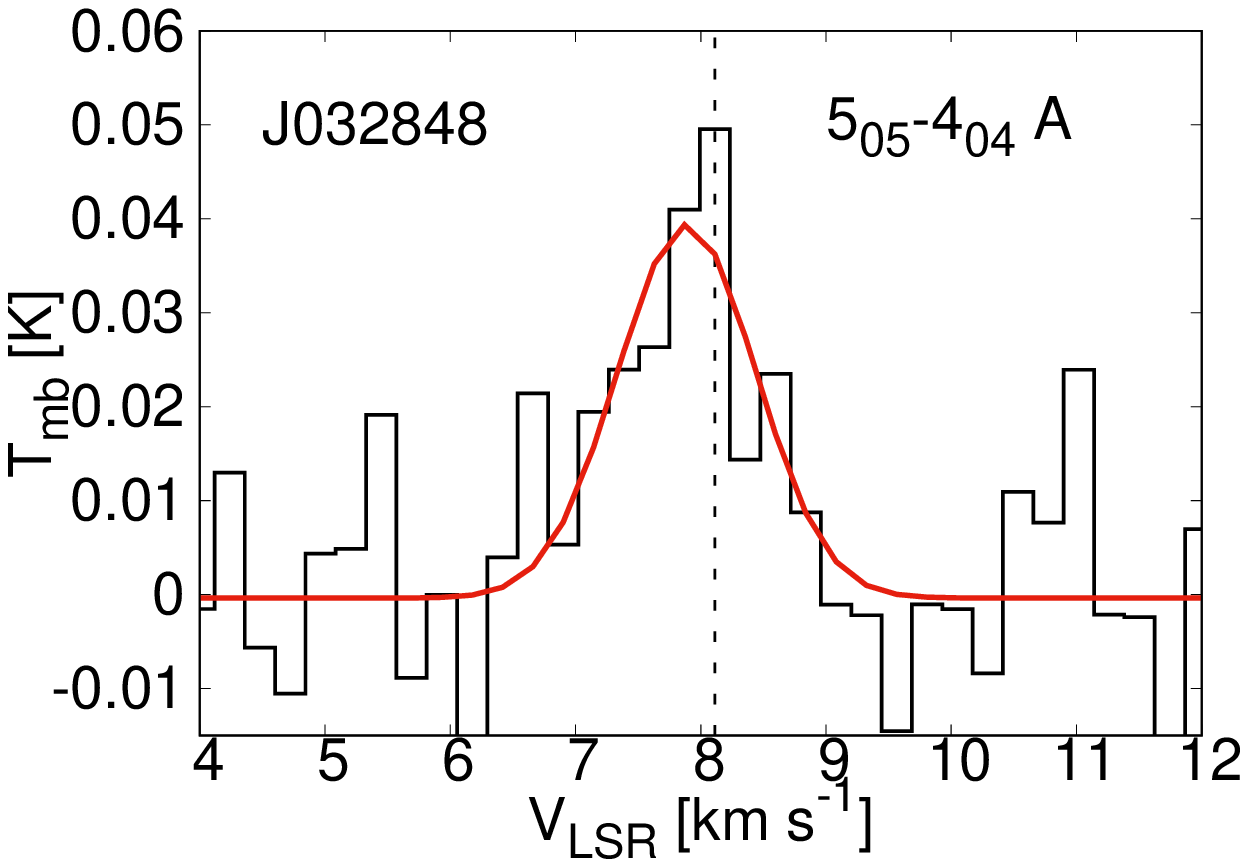}	
    \caption{Continued.}       
  \end{figure*}

\setcounter{figure}{2}    
 \begin{figure*}
  \centering  
    \includegraphics[width=1.5in]{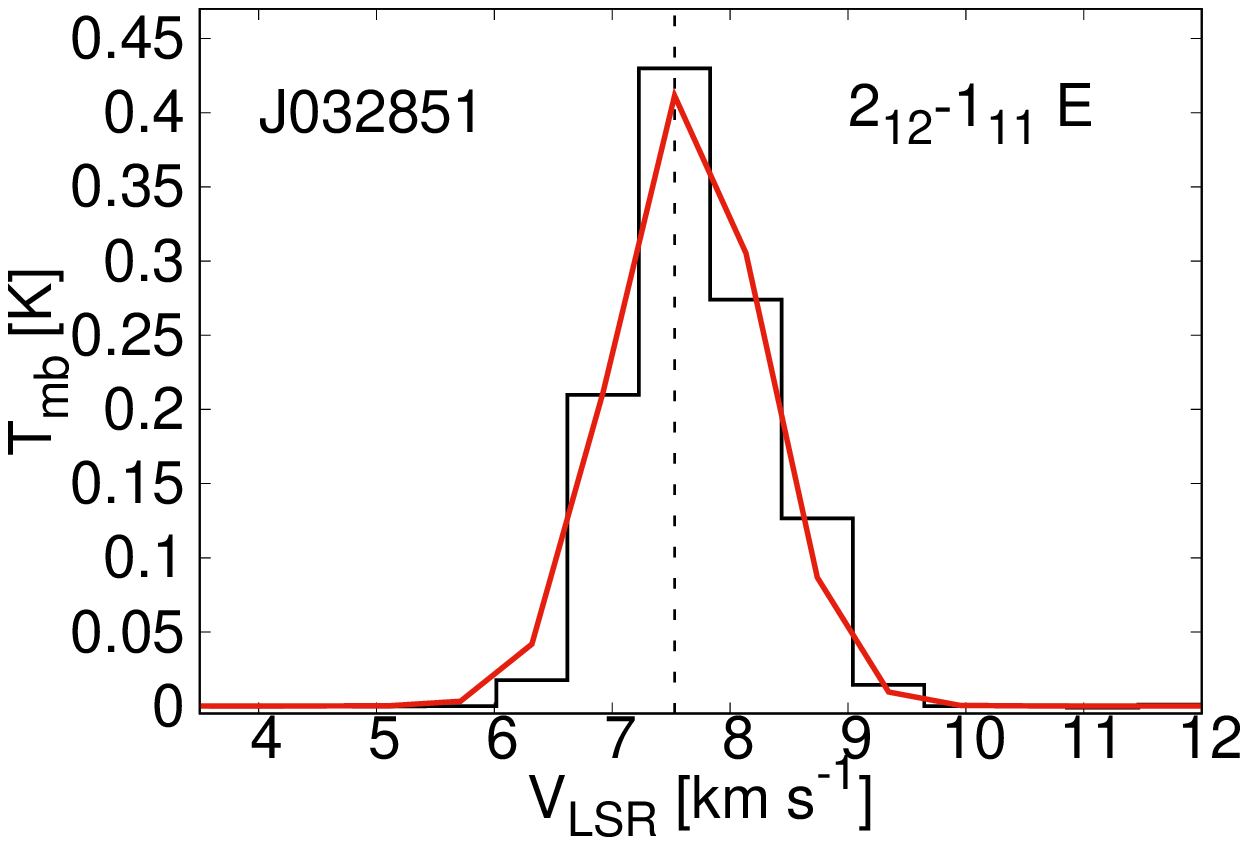}
    \includegraphics[width=1.5in]{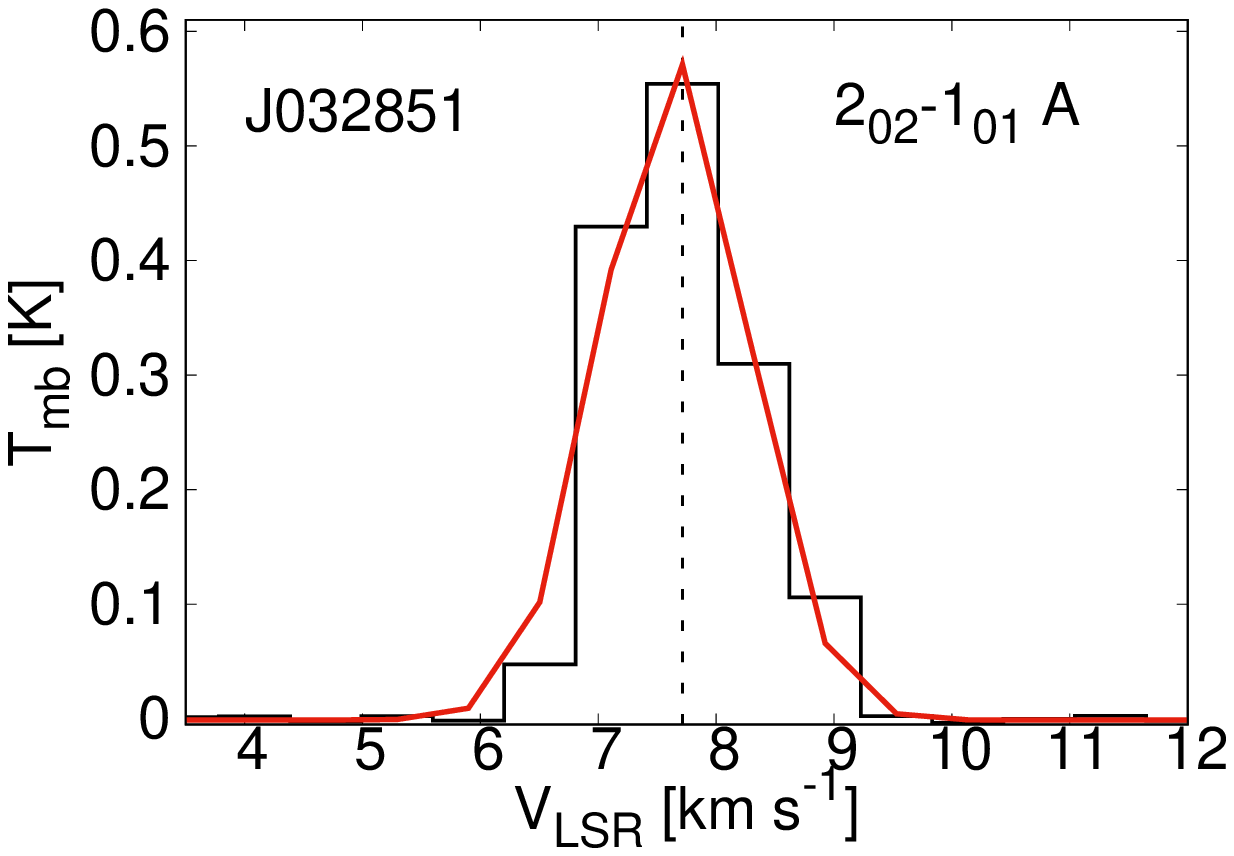}
    \includegraphics[width=1.5in]{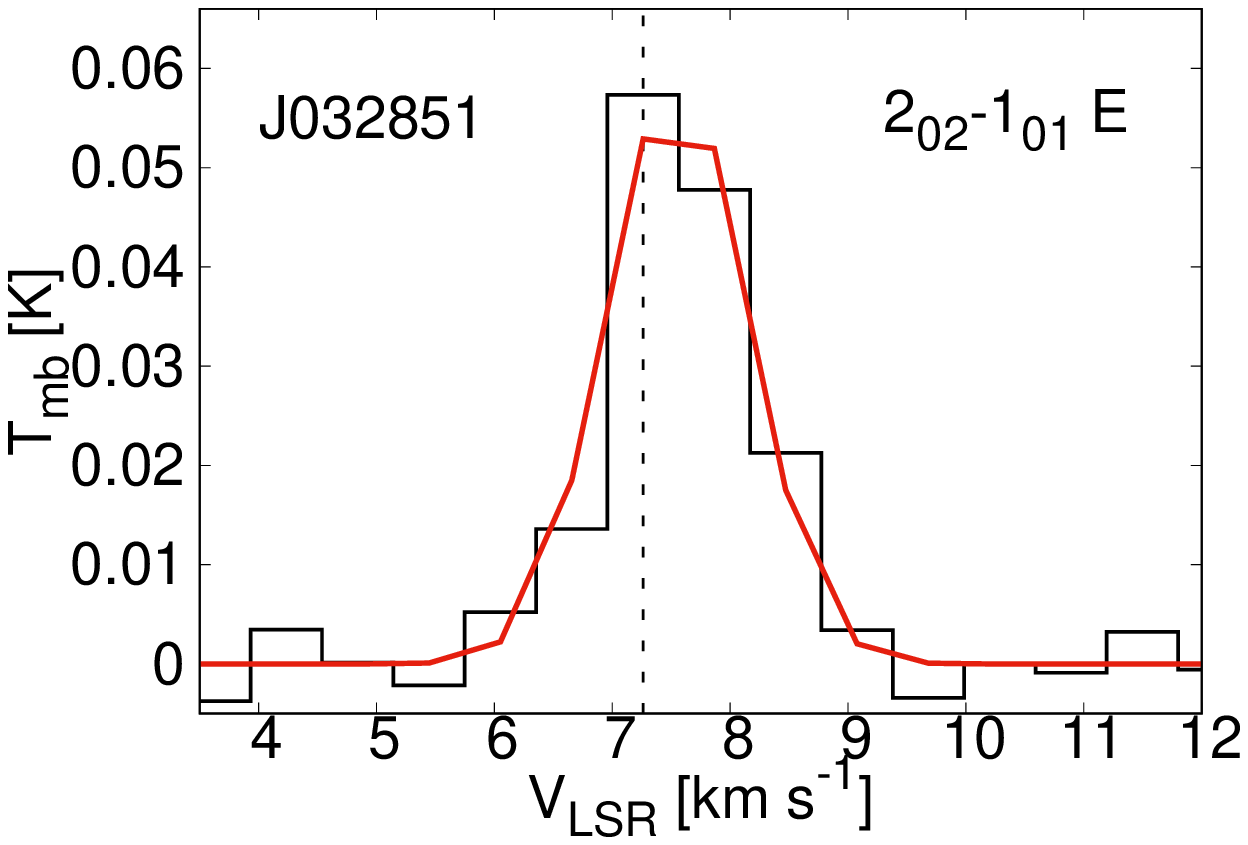}
    \includegraphics[width=1.5in]{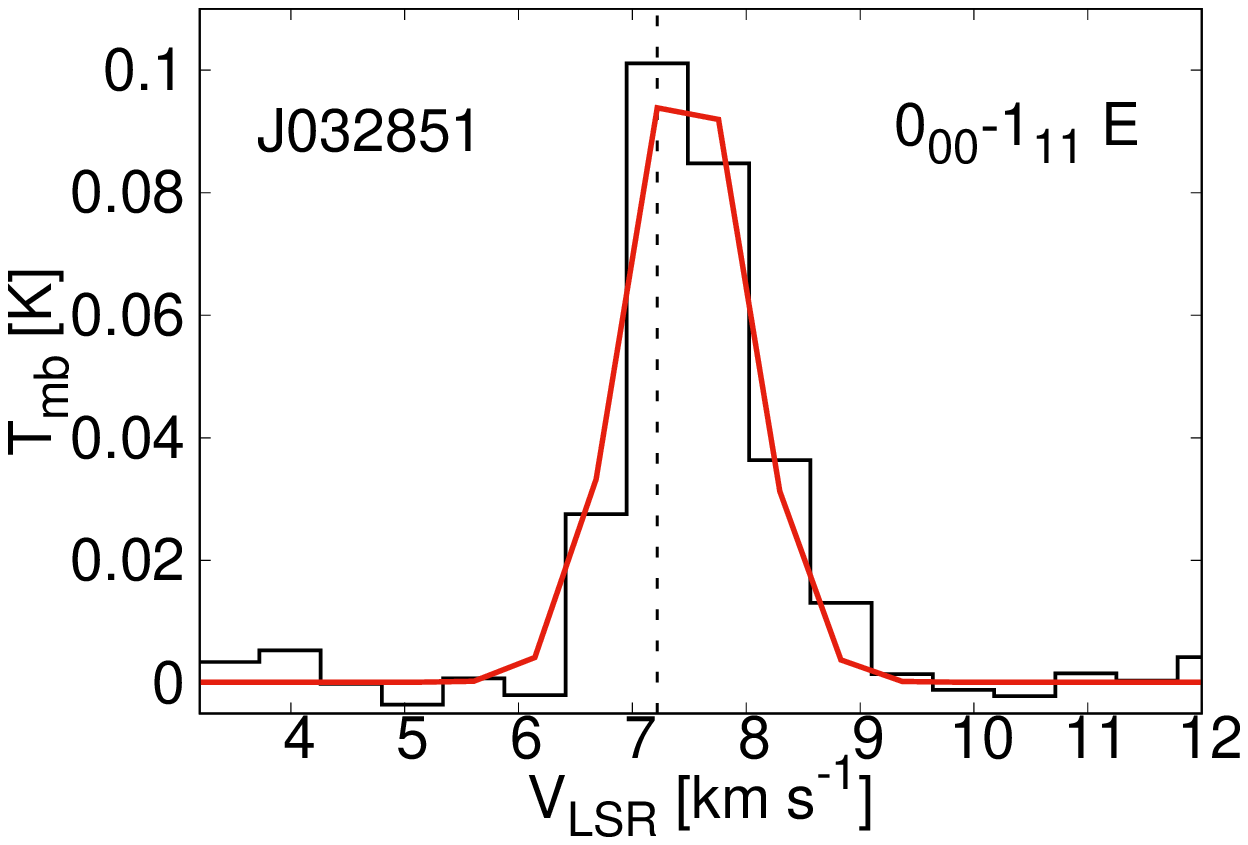}		
    \includegraphics[width=1.5in]{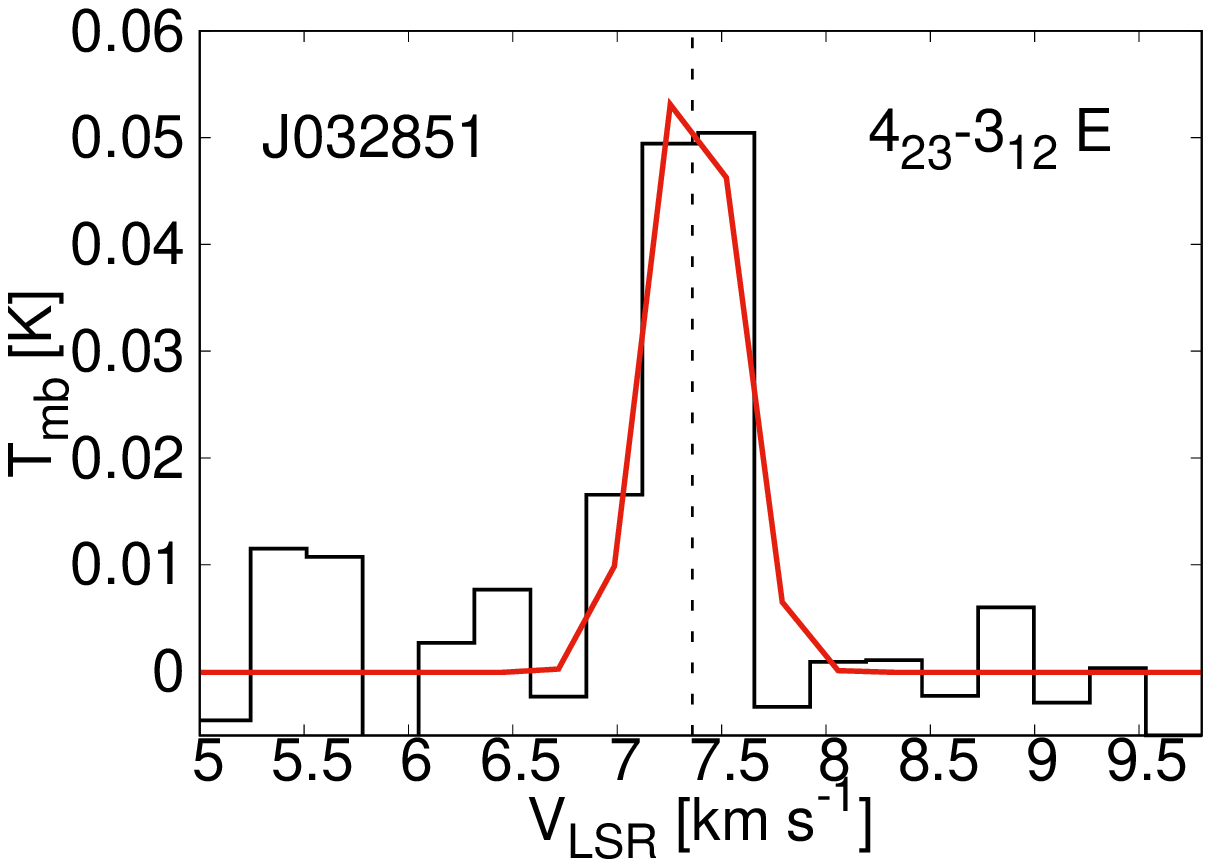}
    \includegraphics[width=1.5in]{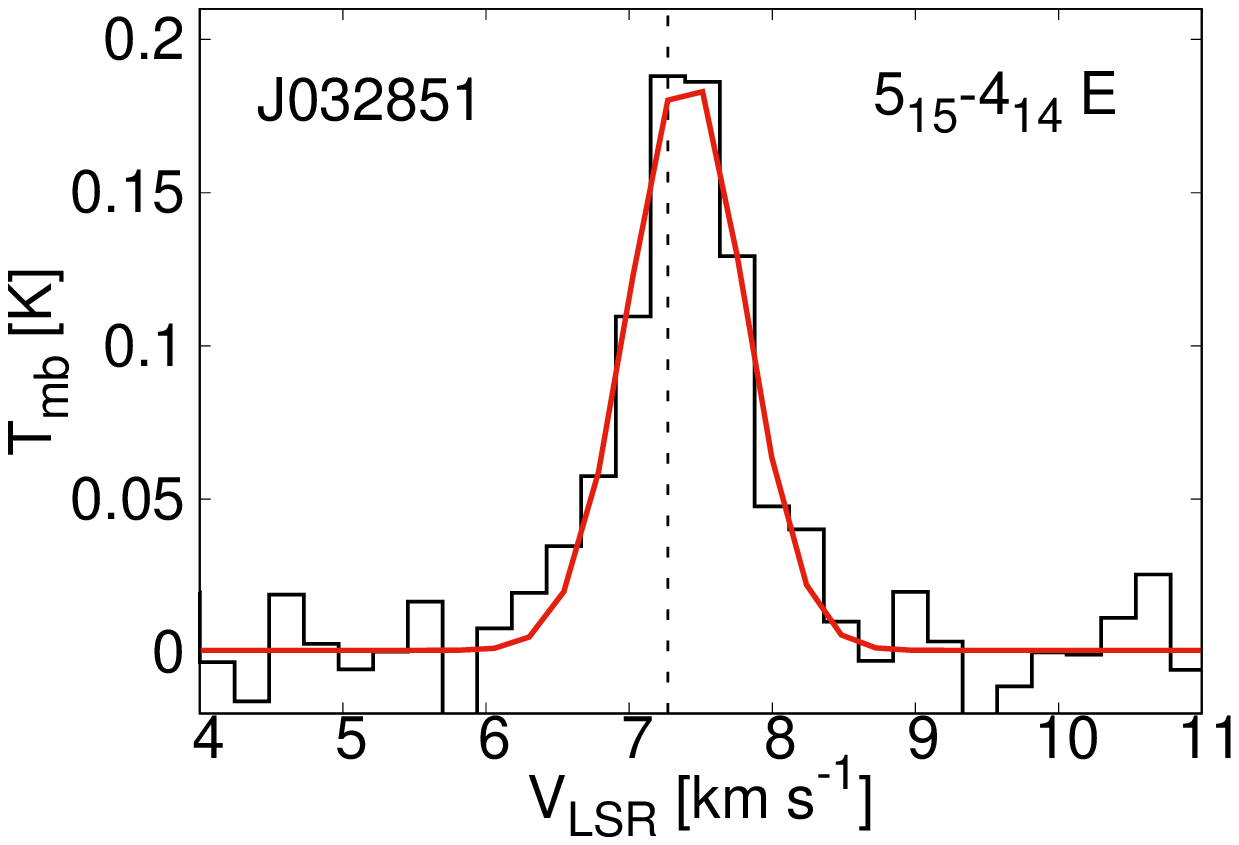}	
    \includegraphics[width=1.5in]{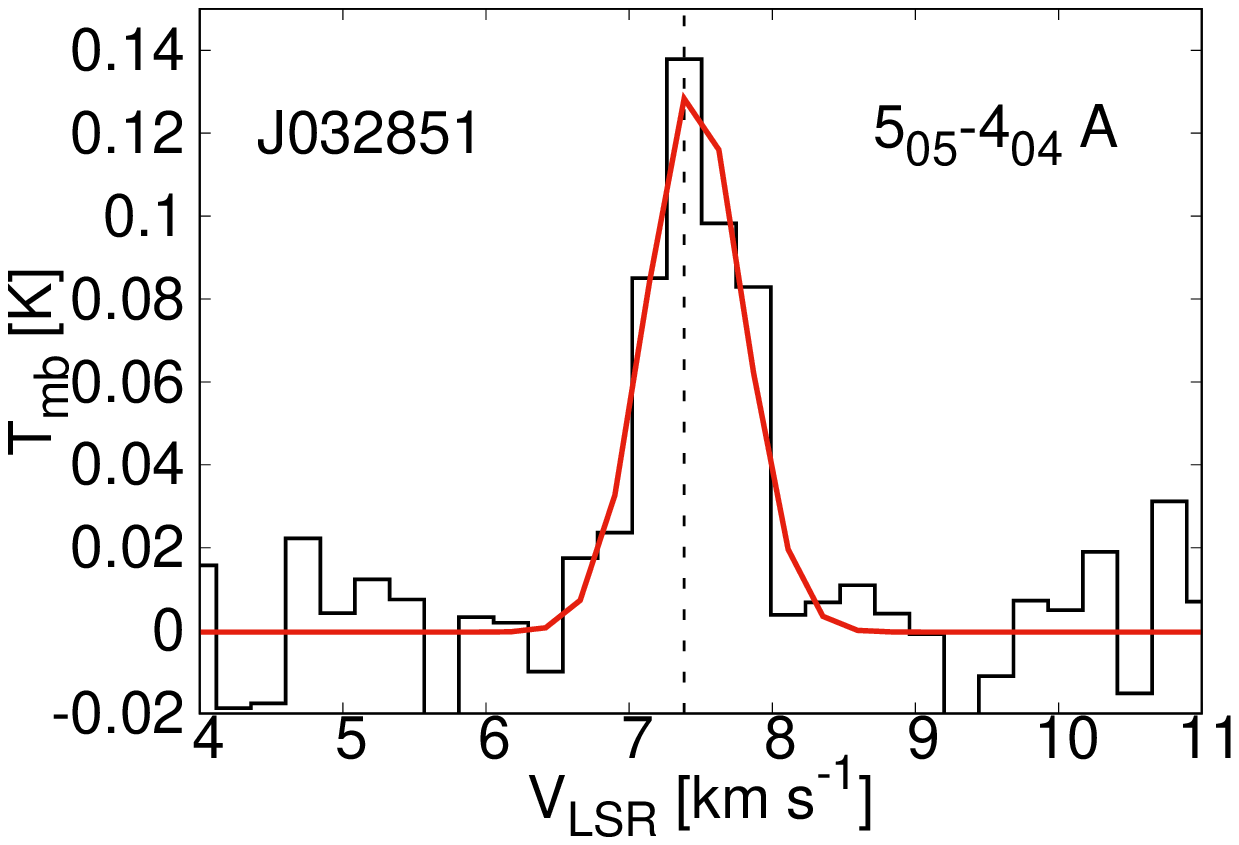}	\\	\vspace{0.3in}

    \includegraphics[width=1.5in]{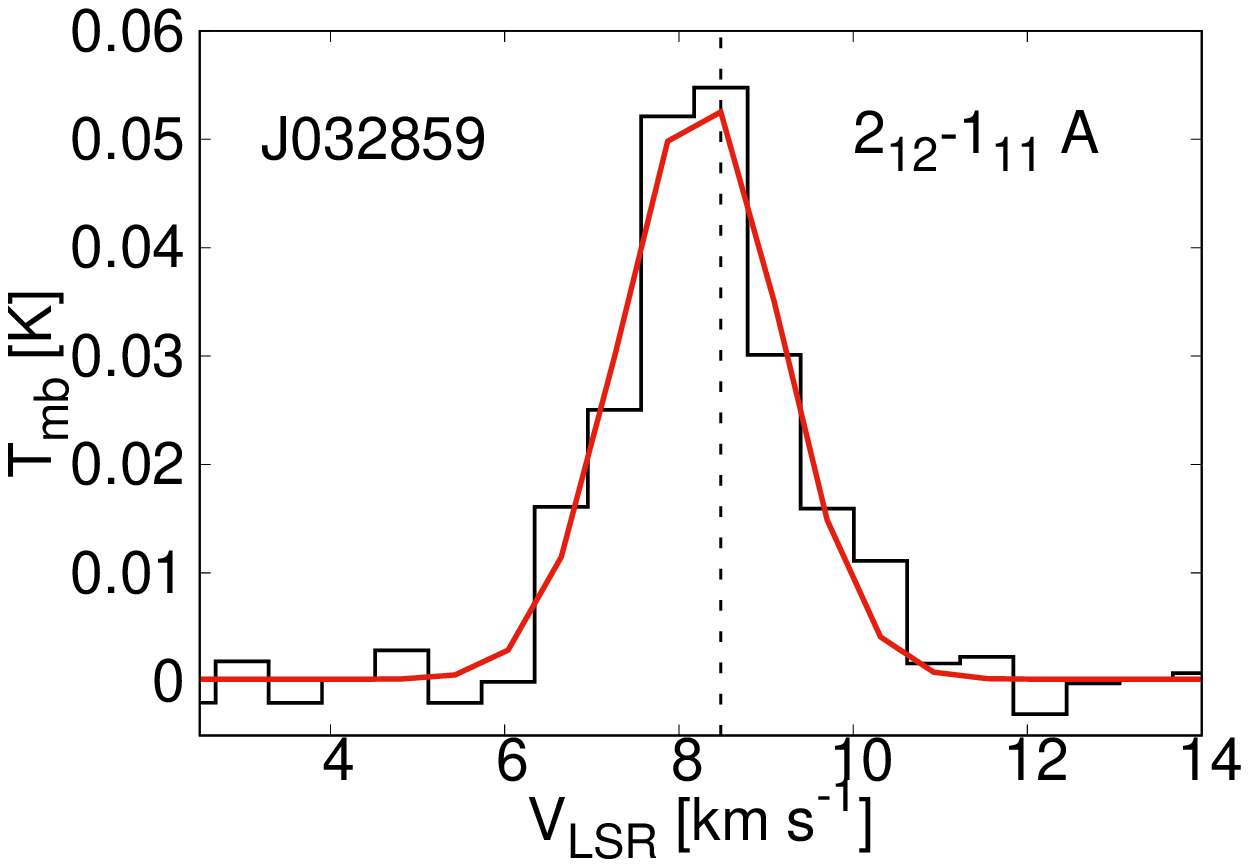}
    \includegraphics[width=1.5in]{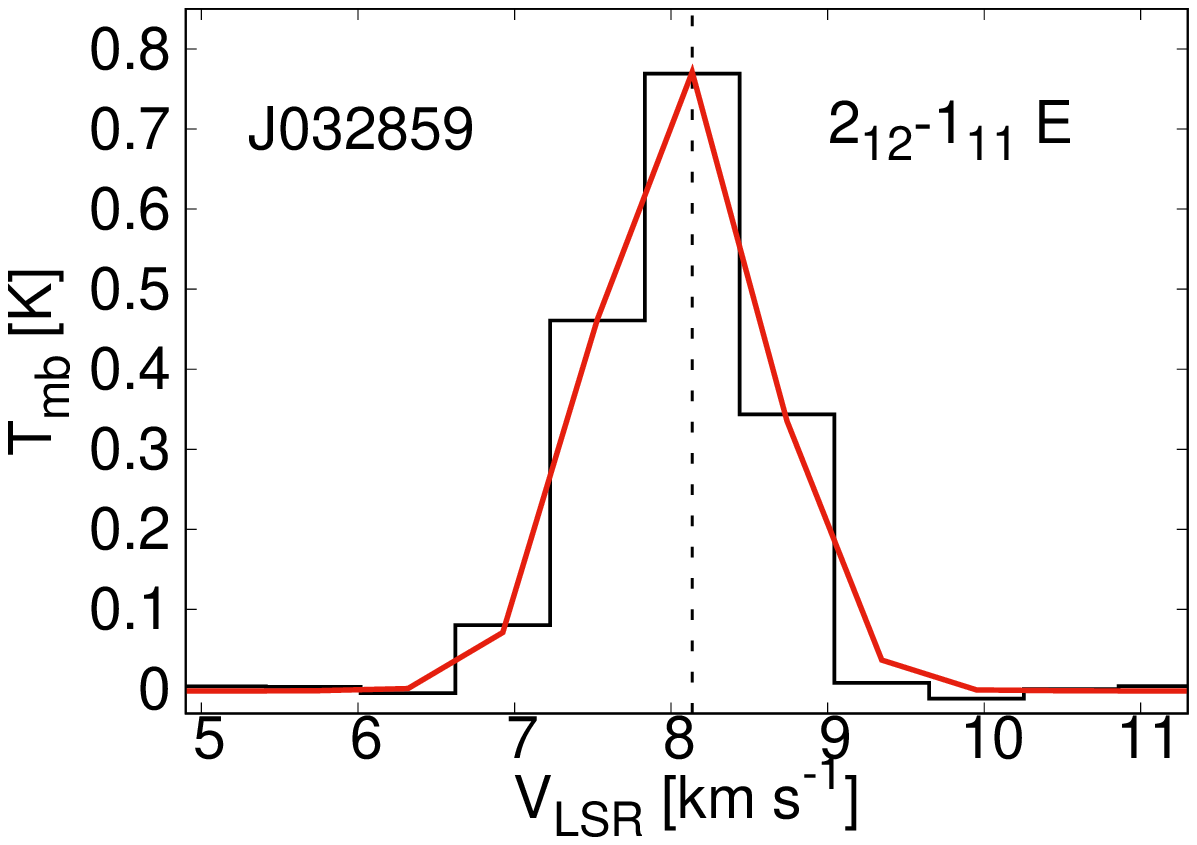}
    \includegraphics[width=1.5in]{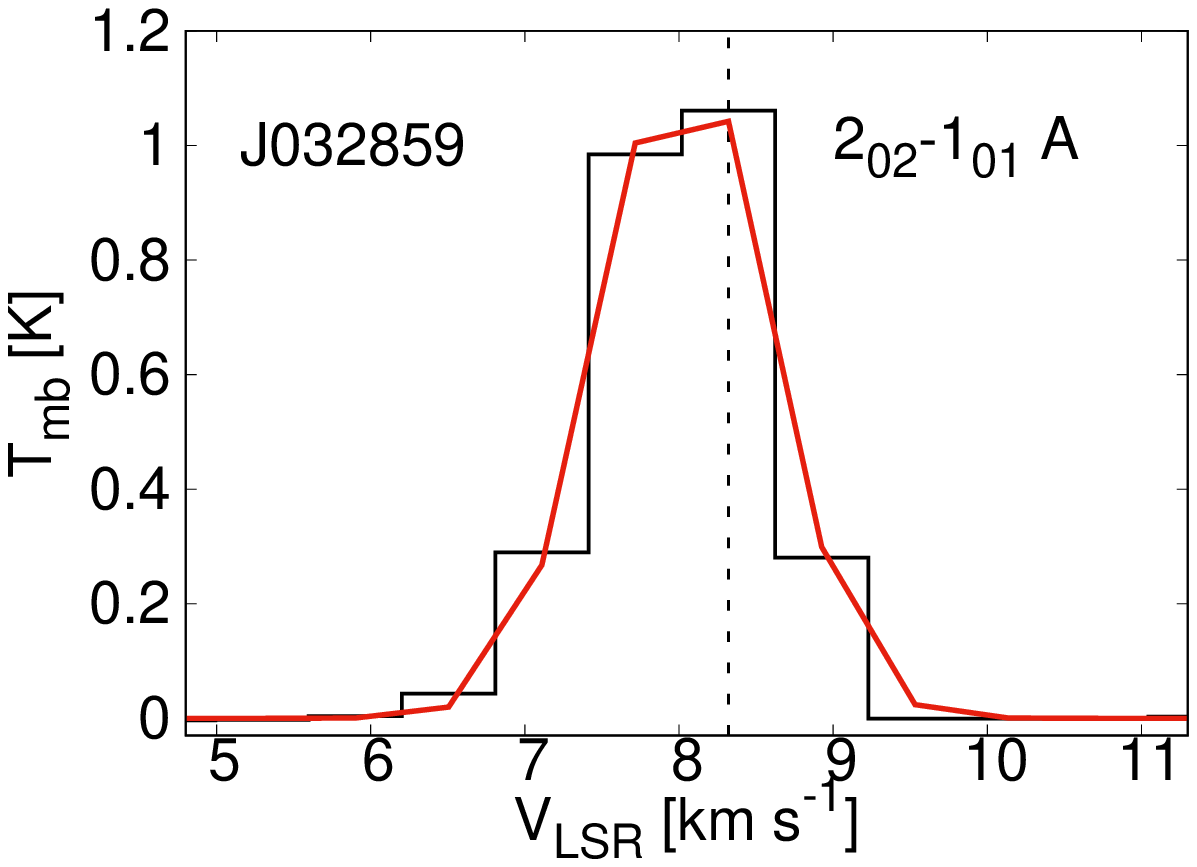}
    \includegraphics[width=1.5in]{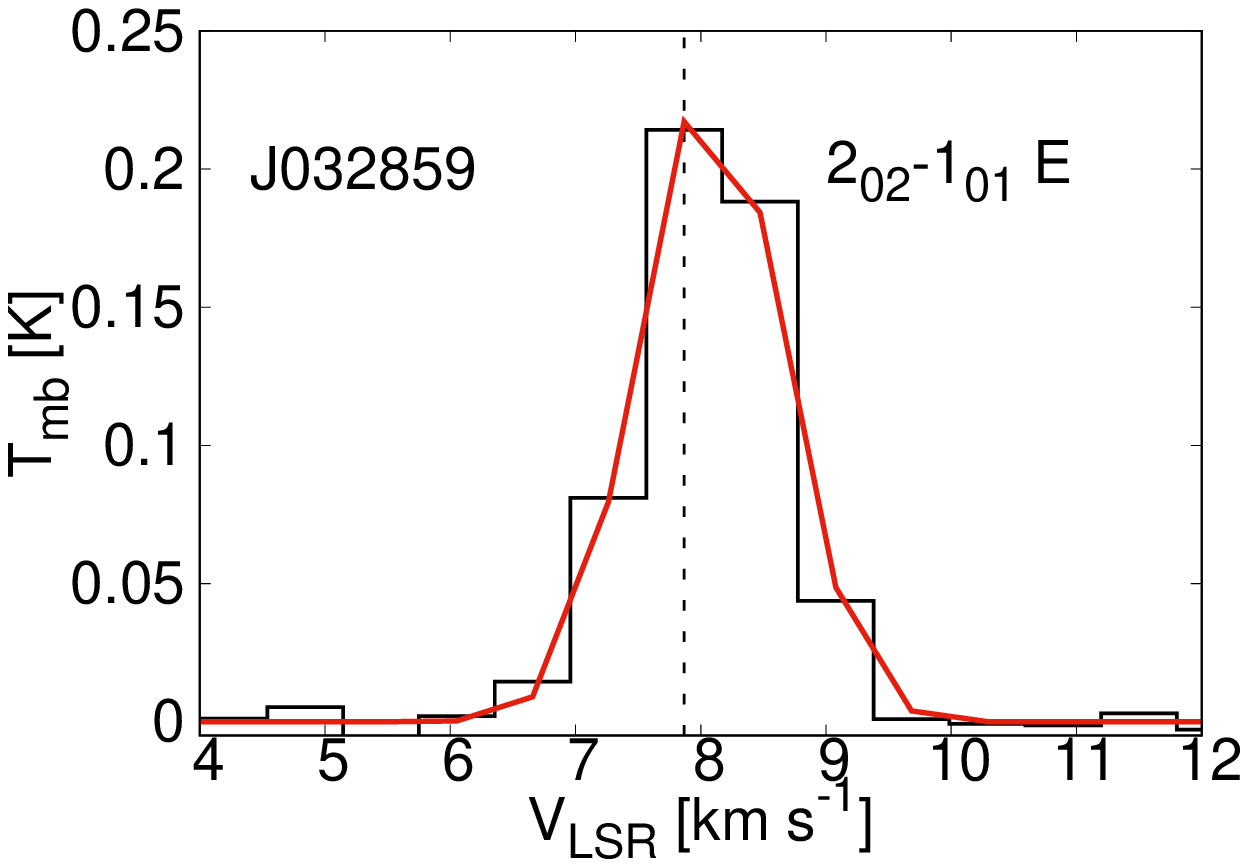}
    \includegraphics[width=1.5in]{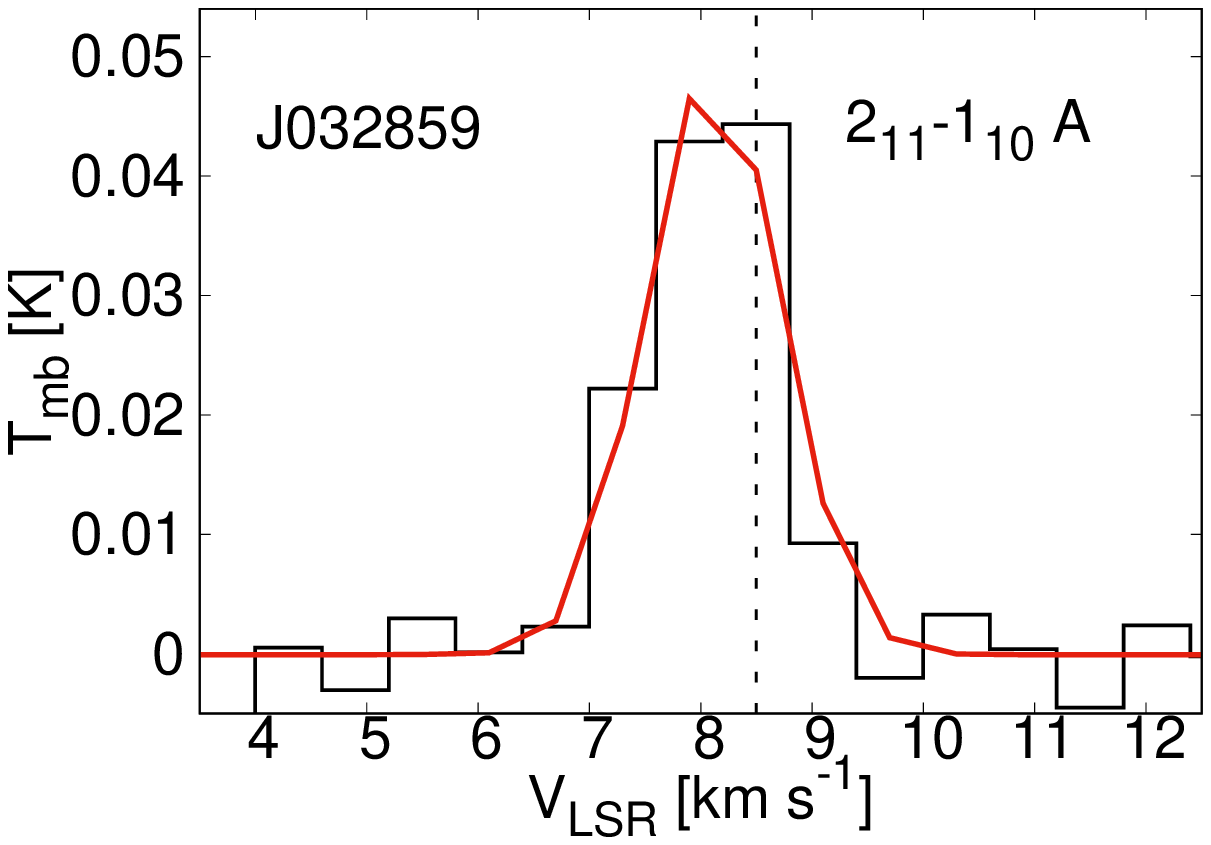}
    \includegraphics[width=1.5in]{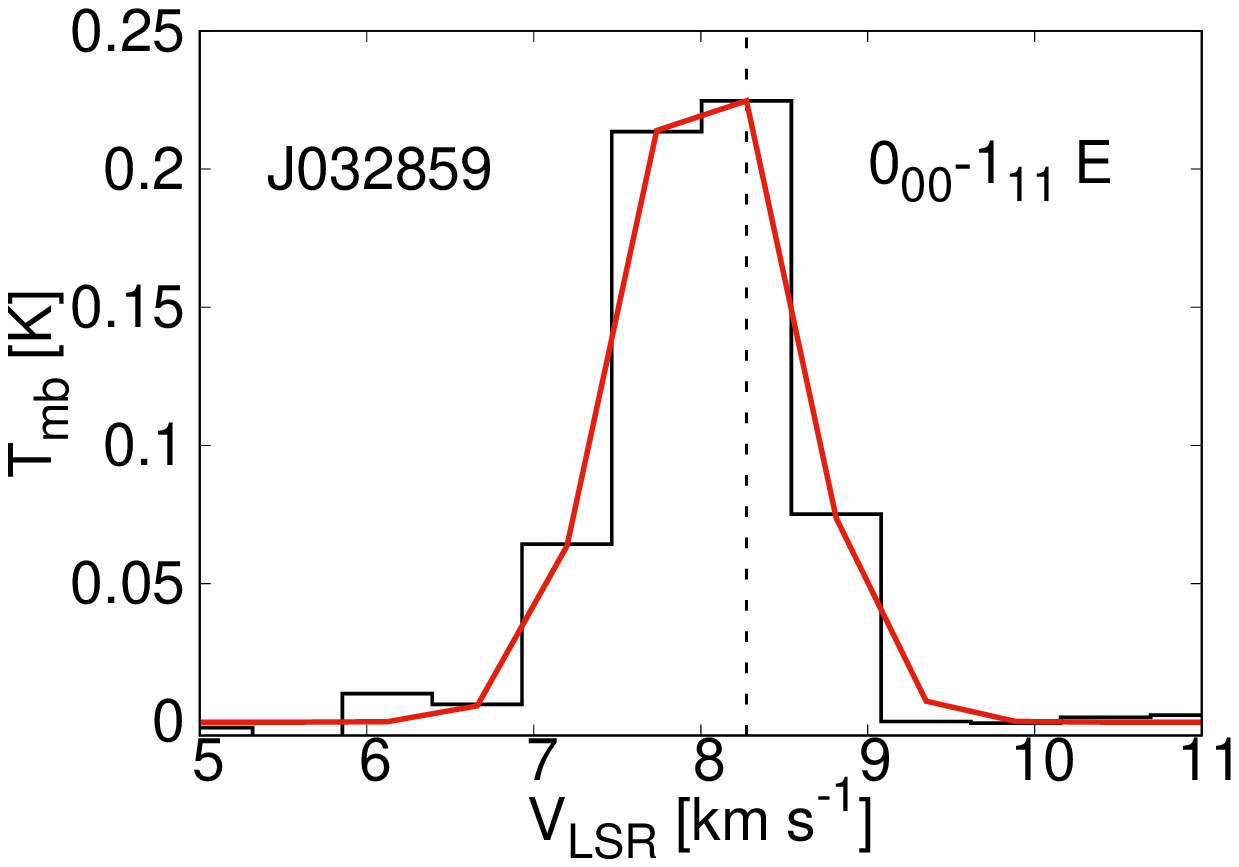}		
    \includegraphics[width=1.5in]{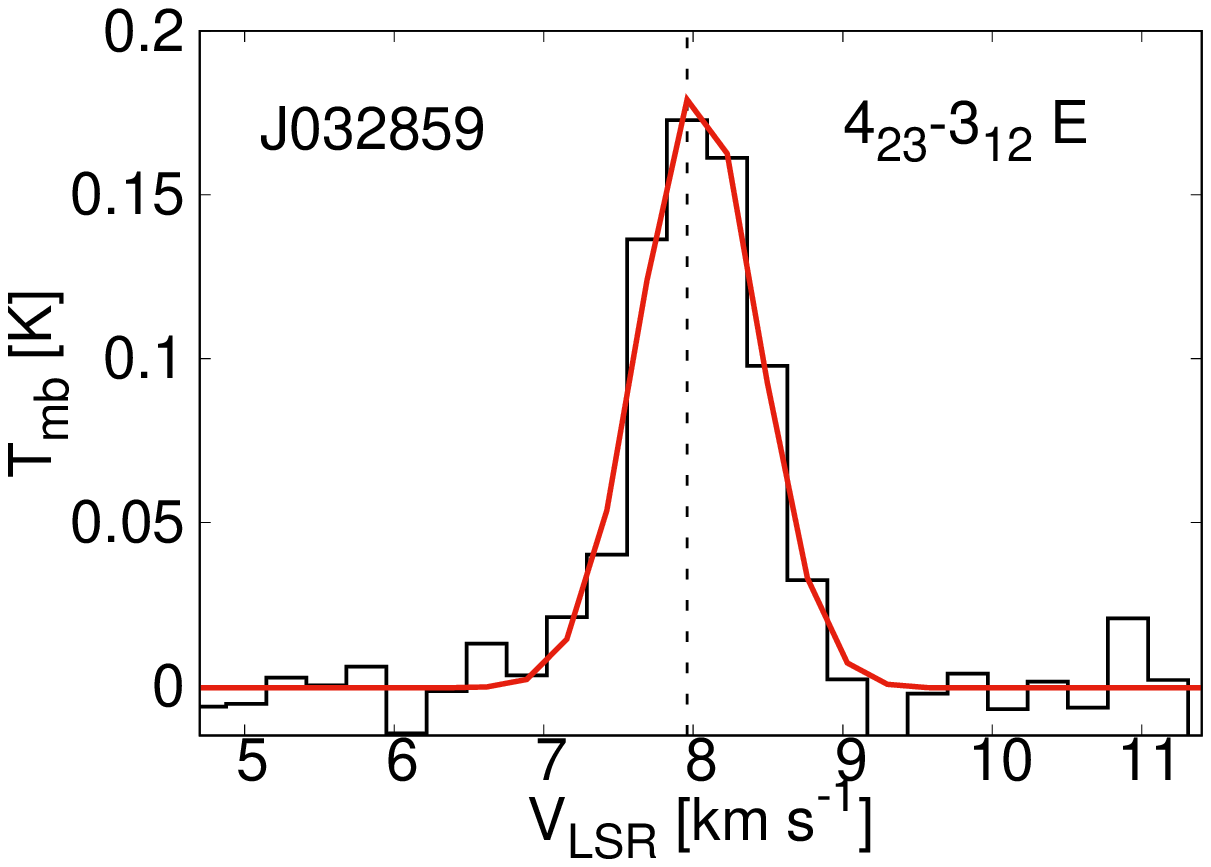}	
    \includegraphics[width=1.5in]{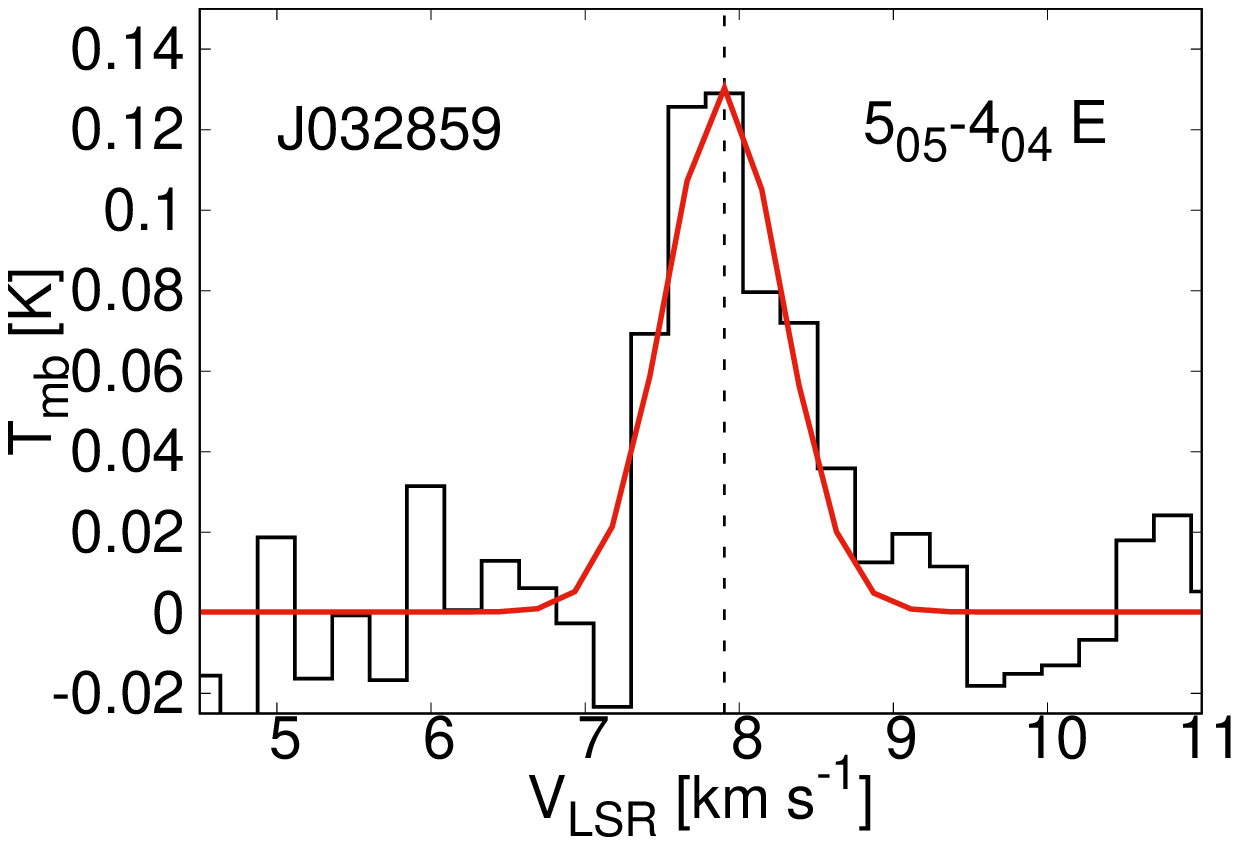}
    \includegraphics[width=1.5in]{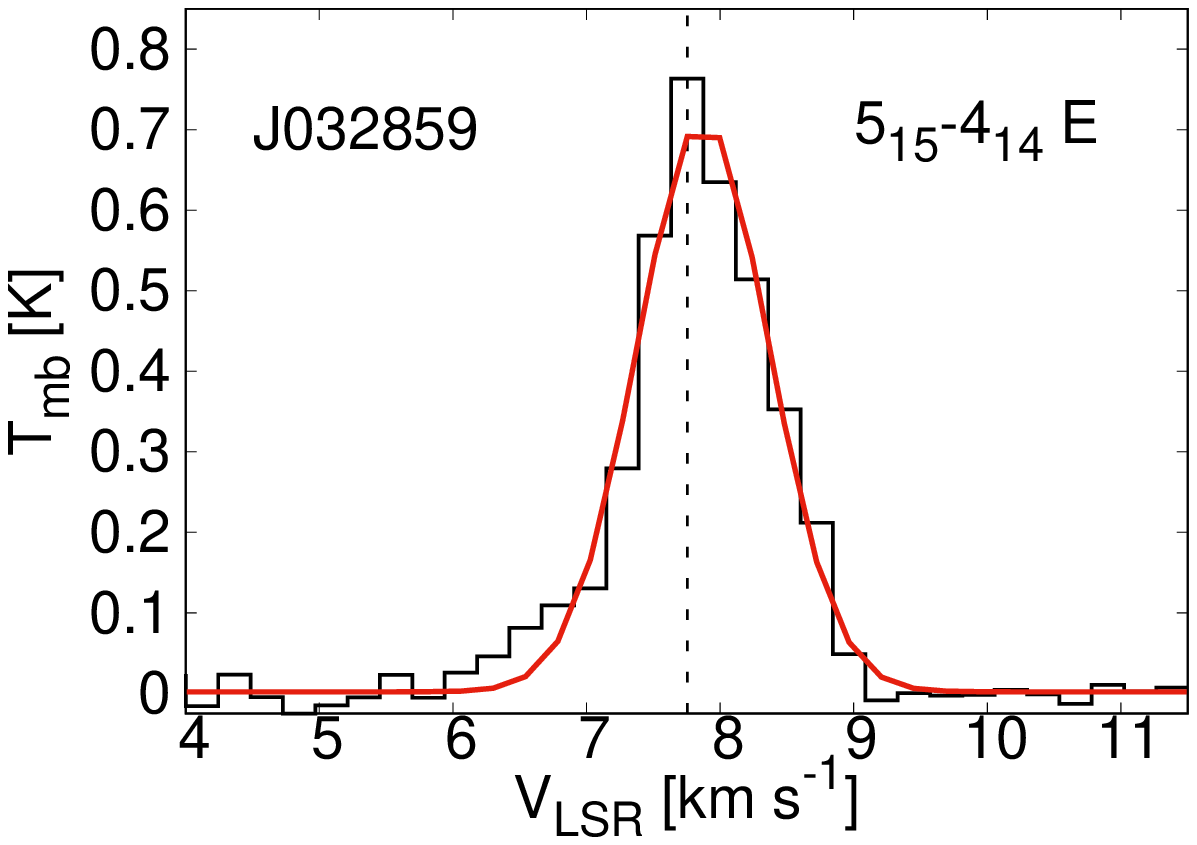}	
    \includegraphics[width=1.5in]{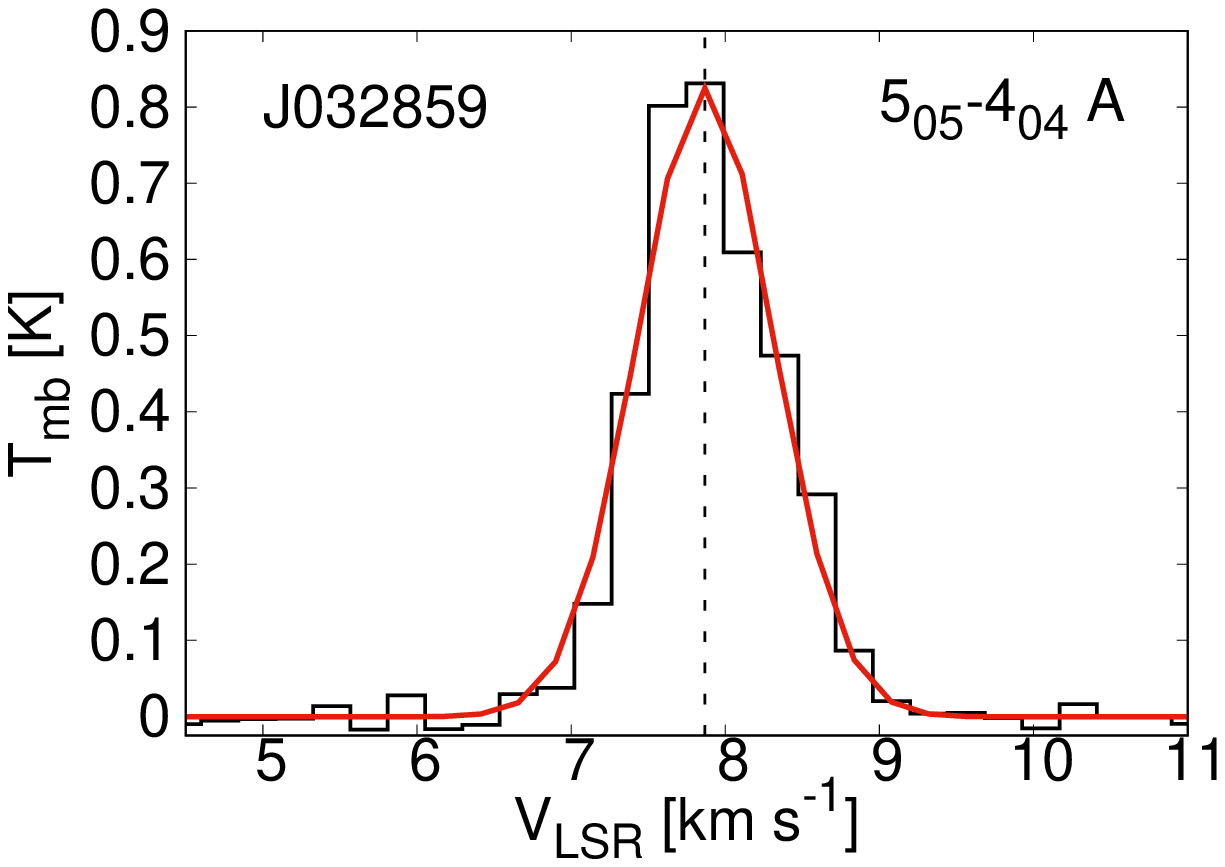}
    \includegraphics[width=1.5in]{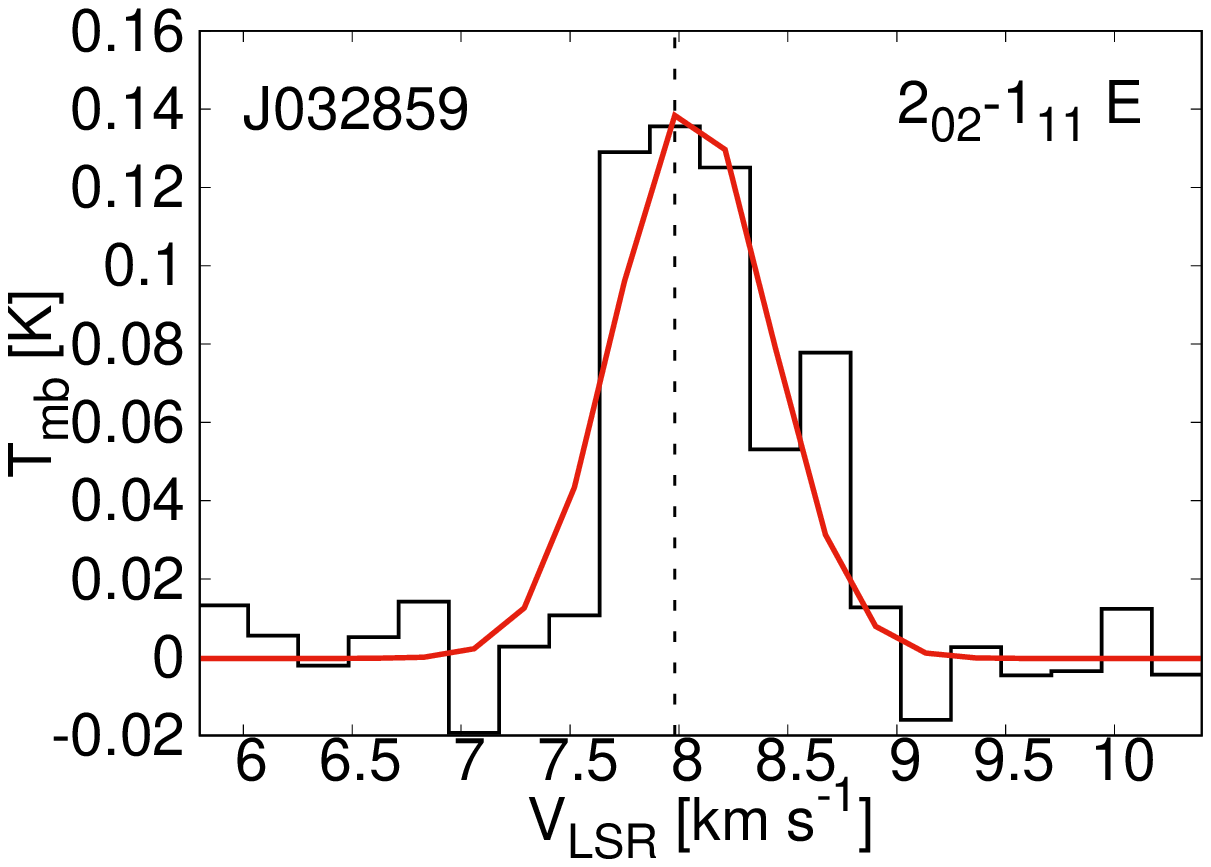}  
    \caption{Continued.}        
  \end{figure*}

\newpage
	
\begin{table}
\centering
\caption{Line Parameters \tnote{a}}
\label{line-pars}
\begin{threeparttable}
\begin{tabular}{llllllll} 
\hline
Transition 		  		& V$_{lsr}$ 	& T$_{mb}$	& $\int{T_{mb} dv}$ 	&  $\Delta$v  		\\
	     				& (km s$^{-1}$)	& (K)			& (K km s$^{-1}$) 	& (km s$^{-1}$)		\\
\hline
\multicolumn{5}{c}{J182844}	\\
\hline
2$_{12}$-1$_{11}$ E		&	7.5		&	0.17		&	0.15		&	0.84		\\
2$_{02}$-1$_{01}$ A 	&	7.7 		&	0.22		&	0.32		&	1.43		\\	
2$_{02}$-1$_{01}$ E 	&	7.2 		&	0.04		&	0.05		&	1.39		\\
0$_{00}$-1$_{11}$ E 	&	7.4 		&	0.04		&	0.01		&	0.16		\\
4$_{23}$-3$_{12}$ E 	&	8.0		&	0.03		&	0.02		&	0.72		\\
3$_{21}$-4$_{14}$ E  	&	8.2		&	0.08		&	0.04		&	0.52		\\
5$_{15}$-4$_{14}$ A 	&	7.4		&	0.06		&	0.07		&	1.12		\\
2$_{02}$-1$_{11}$ E 	&	6.9		&	0.15		&	0.10		&	0.68		\\

\hline 
\multicolumn{5}{c}{J182854}	\\
\hline
2$_{12}$-1$_{11}$ E		&	7.5		&	0.12		&	0.09		& 	0.81		\\
2$_{02}$-1$_{01}$ A 	&	7.7 		&	0.13		&	0.13		& 	0.97		\\
0$_{00}$-1$_{11}$ E 	&	7.4 		&	0.03		&	0.03		& 	1.09		\\

\hline
\multicolumn{5}{c}{J182959}	\\
\hline
2$_{12}$-1$_{11}$ E 	&	8.1 		&	1.12		&	1.45		&	1.28		\\
2$_{02}$-1$_{01}$ A 	&	7.7		&	1.69		&	1.99		&	1.18		\\
2$_{02}$-1$_{01}$ E 	&	7.8		&	0.36		&	0.35		&	0.98		\\
2$_{11}$-1$_{10}$ A 	&	7.9		&	0.07		&	0.06		&	0.92		\\
0$_{00}$-1$_{11}$ E 	&	7.9		&	0.54		&	0.54		&	1.02		\\
4$_{23}$-3$_{12}$ E 	&	8.3 		&	0.09		&	0.06		&	0.67		\\
5$_{15}$-4$_{14}$ A 	&	9.3 		&	0.07		&	0.03		&	0.44		\\
2$_{11}$-1$_{01}$ E 	&	8.4 		&	0.22		&	0.18		&	0.82		\\ 

\hline
\multicolumn{5}{c}{J182856}	\\
\hline
2$_{12}$-1$_{11}$ E 	&	7.5 	 	&	0.78		&	0.97		&	1.24		\\	
2$_{02}$-1$_{01}$ A 	&	7.1 	 	&	0.98		&	1.38		&	1.40		\\	
2$_{02}$-1$_{01}$ E 	&	7.2 	 	&	0.11		&	0.16		&	1.51		\\	
2$_{11}$-1$_{10}$ A 	&	7.3 	 	&	0.02		&	0.02		&	1.02		\\	
0$_{00}$-1$_{11}$ E 	&	7.4 	 	&	0.15		&	0.18		&	1.23		\\	
4$_{23}$-3$_{12}$ E 	&	8.0 	 	&	0.04		&	0.03		&	0.80		\\	
5$_{05}$-4$_{04}$ E 	&	8.0 	 	&	0.05		&	0.03		&	0.42		\\	
5$_{15}$-4$_{14}$ E 	&	7.8 	 	&	0.28		&	0.28		&	0.99		\\	
5$_{05}$-4$_{04}$ A 	&	7.7 	 	&	0.33		&	0.33		&	0.98		\\	
2$_{02}$-1$_{11}$ E 	&	7.7 	 	&	0.17		&	0.20		&	1.14		\\	
2$_{11}$-1$_{01}$ E 	&	7.7 	 	&	0.09		&	0.08		&	0.88		\\	

\hline
\multicolumn{5}{c}{J182952}	\\
\hline
2$_{12}$-1$_{11}$ E 	&	8.1		&	0.99		&	2.10		&	2.08		\\
2$_{02}$-1$_{01}$ A 	&	7.7 		&	1.35		&	2.34		&	1.73		\\
2$_{02}$-1$_{01}$ E 	&	7.8 		&	0.18		&	0.28		&	1.60		\\
2$_{11}$-1$_{10}$ A 	&	7.9 		&	0.04		&	0.11		&	2.80		\\
0$_{00}$-1$_{11}$ E 	&	8.0 		&	0.24		&	0.43		&	1.81		\\
4$_{23}$-3$_{12}$ E 	&	8.0 		&	0.08		&	0.10		&	1.34		\\
5$_{15}$-4$_{14}$ A 	&	8.6 		&	0.28		&	0.28		&	1.01		\\
5$_{15}$-4$_{14}$ E 	&	8.3 		&	0.31		&	0.35		&	1.13		\\
5$_{05}$-4$_{04}$ A 	&	8.4 		&	0.30		&	0.35		&	1.20		\\
2$_{02}$-1$_{11}$ E 	&	8.1 		&	0.12		&	0.08		&	0.68		\\
2$_{11}$-1$_{01}$ E 	&	8.4 		&	0.15		&	0.12		&	0.77		\\
\hline
\multicolumn{5}{c}{J163143}	\\
\hline
2$_{12}$-1$_{11}$ E 	&	4.3 		&	0.40		&	0.36		&	0.91		\\	
2$_{02}$-1$_{01}$ A 	&	3.9 		&	0.41		&	0.51		&	1.23		\\	
2$_{02}$-1$_{01}$ E 	&	4.0 		&	0.04		&	0.03		&	0.80		\\	
0$_{00}$-1$_{11}$ E 	&	3.9 		&	0.06		&	0.06		&	0.98		\\	
\hline
\end{tabular}
\begin{tablenotes}
\item[a] The uncertainty is estimated to be $\sim$10\%-20\% for the peak and integrated intensities and $\Delta${\it v}, and $\sim$0.02-0.04 km s$^{-1}$ for {\it V}$_{lsr}$.
\end{tablenotes}
\end{threeparttable}
\end{table}

\setcounter{table}{0}
\begin{table}
\centering
\begin{threeparttable}
\begin{tabular}{llllllll} 
\hline
Transition 		  		& V$_{lsr}$ 	& T$_{mb}$	& $\int{T_{mb} dv}$ 	&  $\Delta$v  		\\
	     				& (km s$^{-1}$)	& (K)			& (K km s$^{-1}$) 	& (km s$^{-1}$)		\\

\hline
\multicolumn{5}{c}{J163136}	\\
\hline
2$_{12}$-1$_{11}$ E 	& 	1.9 		&	0.05		&	0.05		&	1.08		\\
2$_{02}$-1$_{01}$ A 	& 	2.1 		&	0.10		&	0.10		&	1.05		\\
5$_{15}$-4$_{14}$ A 	& 	4.6 		&	0.14		&	0.09		&	0.66		\\

\hline
\multicolumn{5}{c}{J163152}	\\		
\hline
2$_{12}$-1$_{11}$ E 	&	3.7 		&	0.32		&	0.42		&	1.31		\\
2$_{02}$-1$_{01}$ A 	&	3.9 		&	0.54		&	0.60		&	1.10		\\
2$_{02}$-1$_{01}$ E 	&	4.0 		&	0.06		&	0.05		&	0.77		\\
0$_{00}$-1$_{11}$ E 	&	4.0 		&	0.10		&	0.09		&	0.96		\\
5$_{05}$-4$_{04}$ A 	&	4.5 		&	0.11		&	0.08		&	0.67		\\
2$_{02}$-1$_{11}$ E 	&	4.4 		&	0.11		&	0.06 		&	0.56		\\

\hline
\multicolumn{5}{c}{J162625}	\\		
\hline
2$_{12}$-1$_{11}$ E 	&	 3.1 		&	0.13		&	0.20 		&	1.56		\\
2$_{02}$-1$_{01}$ A 	&	 3.3 		&	0.17		&	0.27 		&	1.54		\\
2$_{02}$-1$_{01}$ E 	&	 2.8 		&	0.02		&	0.03 		&	1.47		\\
0$_{00}$-1$_{11}$ E 	&	 3.3 		&	0.05		&	0.07 		&	1.62		\\
4$_{23}$-3$_{12}$ E 	&	 3.7 		&	0.06		&	0.03 		&	 0.52		\\
2$_{02}$-1$_{11}$ E 	&	 5.3 		&	0.12		&	0.07 		&	 0.62		\\

\hline
\multicolumn{5}{c}{J032838}	\\	
\hline
2$_{12}$-1$_{11}$ A 	&	7.86		&	0.02		&	0.01	 	&	0.66		\\
2$_{12}$-1$_{11}$ E 	&	8.13		& 	0.93		&	0.88		& 	0.95		\\
2$_{02}$-1$_{01}$ A 	&	8.32		& 	1.15		&	1.23		&	1.07		\\
2$_{02}$-1$_{01}$ E 	&	7.86		& 	0.14		&	0.17		& 	1.22		\\
2$_{11}$-1$_{10}$ A 	&	7.89		& 	0.03		&	0.03		& 	1.15		\\
0$_{00}$-1$_{11}$ E 	&	8.31		&	0.22		&	0.23		&	1.05		\\
5$_{15}$-4$_{14}$ E 	&	8.26	 	&	0.12		&	0.09		&	0.78		\\
5$_{05}$-4$_{04}$ A 	&	8.13	 	&	0.16		&	0.08		&	0.52		\\
2$_{02}$-1$_{11}$ E 	&	8.26	 	&	0.14		&	0.08		&	0.55		\\
2$_{11}$-1$_{01}$ E 	&	7.87	 	&	0.10		&	0.05		& 	0.53		\\


\hline
\multicolumn{5}{c}{J032848}	\\	
\hline
2$_{12}$-1$_{11}$ E 	&	8.13		&	0.25		&	0.30	 	&	1.21		\\
2$_{02}$-1$_{01}$ A 	&	7.72	 	&	0.34		&	0.47	 	&	1.44		\\
2$_{02}$-1$_{01}$ E 	&	7.86	 	&	0.02		&	0.03	 	&	1.26		\\
2$_{11}$-1$_{10}$ A 	&	7.89	 	&	0.01		&	0.02	 	&	1.47		\\
0$_{00}$-1$_{11}$ E 	&	7.76	 	&	0.05		&	0.06	 	&	1.22		\\
5$_{15}$-4$_{14}$ E 	&	7.99	 	&	0.08		&	0.03		&	0.41		\\
5$_{05}$-4$_{04}$ A 	&	8.11	 	&	0.05		&	0.05		&	1.08		\\


\hline
\multicolumn{5}{c}{J032851}	\\	
\hline

2$_{12}$-1$_{11}$ E 	&	7.53	 	&	0.43		&	0.65 		&	1.51		\\
2$_{02}$-1$_{01}$ A 	&	7.72	 	&	0.55		&	0.87 		&	1.58		\\
2$_{02}$-1$_{01}$ E 	&	7.26	 	&	0.06		&	0.08	 	&	1.53		\\
0$_{00}$-1$_{11}$ E 	&	7.22	 	&	0.10		&	0.14	 	&	1.38		\\
4$_{23}$-3$_{12}$ E 	&	7.52	 	&	0.05		&	0.03		&	0.61		\\
5$_{15}$-4$_{14}$ E 	&	7.27	 	&	0.18		&	0.20		&	1.06		\\
5$_{05}$-4$_{04}$ A 	&	7.38	 	&	0.14		&	0.10		&	0.76		\\

\hline
\end{tabular}
\end{threeparttable}
\end{table}

\setcounter{table}{0}
\begin{table}
\centering
\begin{threeparttable}
\begin{tabular}{llllllll} 
\hline
Transition 		  		& V$_{lsr}$ 	& T$_{mb}$	& $\int{T_{mb} dv}$ 	&  $\Delta$v  		\\
	     				& (km s$^{-1}$)	& (K)			& (K km s$^{-1}$) 	& (km s$^{-1}$)		\\

\hline
\multicolumn{5}{c}{J032859}	\\	
\hline

2$_{12}$-1$_{11}$ A 	&	8.48	 	&	0.05		&	0.13	 	&	2.30		\\
2$_{12}$-1$_{11}$ E 	&	8.13	 	&	0.77		&	1.00	 	&	1.30		\\
2$_{02}$-1$_{01}$ A 	&	8.32	 	&	1.06		&	1.61	 	&	1.52		\\
2$_{02}$-1$_{01}$ E 	&	7.87	 	&	0.21		&	0.33	 	&	1.53		\\
2$_{11}$-1$_{10}$ A 	&	8.50	 	&	0.04		&	0.07	 	&	1.63		\\
0$_{00}$-1$_{11}$ E 	&	8.27	 	&	0.22		&	0.31	 	&	1.40		\\
4$_{23}$-3$_{12}$ E 	&	7.96	 	&	0.17		&	0.17		&	1.00		\\
5$_{05}$-4$_{04}$ E 	&	7.90		&	0.13		&	0.13		&	0.98		\\
5$_{15}$-4$_{14}$ E 	&	7.75	 	&	0.76		&	0.91	 	&	1.20		\\
5$_{05}$-4$_{04}$ A 	&	7.87	 	&	0.83		&	0.90		&	1.10		\\
2$_{02}$-1$_{11}$ E 	&	7.98	 	&	0.14		&	0.12		&	0.90		\\


\hline
\multicolumn{5}{c}{J032911}	\\	
\hline

2$_{12}$-1$_{11}$ E 	&	7.53	 	&	0.03		&	0.06		&	1.81		\\
2$_{02}$-1$_{01}$ A 	&	7.72	 	&	0.04		&	0.07	 	&	1.70		\\
5$_{15}$-4$_{14}$ E 	&	7.76	 	&	0.11		&	0.06		&	0.60		\\
2$_{02}$-1$_{11}$ E 	&	7.46	 	&	0.05		&	0.04		&	0.81		\\


\hline
\end{tabular}
\end{threeparttable}
\end{table}

\newpage

\section{Critical densities}\label{critical_density}

The critical densities listed in Table~\ref{lines} were computed by considering in the numerator the Einstein coefficient of the transitions and in the denominator the collision rate between the upper and the lower level of the transition. This is an approximation that we tested against a full computation with a radiative transfer code. We can define the critical density as the density of the collision partner (here H$_2$) above which the excitation temperature is equal to the kinetic temperature. We ran a grid of models with the escape probability code RADEX, assuming a kinetic temperature of 10~K, a turbulent width 1 km/s, and a background temperature of 2.73~K. The column density of either methanol-A and methanol-E is 10$^{12}$ cm$^2$ so that the emissions are optical thin. The energy levels, transition frequencies and Einstein A coefficients are taken from CDMS (Endres et al.\ 2016). The collisional rates with H$_2$ for the ground states of A- and E-type methanol were calculated by Rabli \& Flower (2010) for temperatures in the range from 10 to 200 K. Table~\ref{critical_density_a} and \ref{critical_density_e} contains the results of the grid. Compared to the values listed in Table~\ref{lines}, the densities above which $T_{\rm ex}$ $\simeq$ $T_{\rm kin}$ are slightly higher. The values also illustrate the possibility of a level to be overpopulated ($T_{\rm ex}>$10~K).

\begin{table*}
\centering
\caption{Excitation temperatures for the observed methanol-E lines with a column density of 10$^{12}$ cm$^{-2}$, $T_{\rm kin}$=10~K, a turbulent width of 1 km/s, and a background temperature of 2.73~K.}
\begin{tabular}{lrrrrrrrrrr}
\hline
            $n$(H$_2$) &  (2\_-1)-(1\_-1) &  (2\_0)-(1\_0) &  (0\_0)-(1\_-1) &  (4\_2)-(3\_1) &  (3\_-2)-(4\_-1) &  (5\_0)-(4\_0)&  (5\_-1)-(4\_-1) &  (2\_0)-(1\_-1) &  (2\_1)-(1\_0)\\
  $\left[{\rm cm}^{-3} \right]$ &  96.739358 &  96.744545&  108.893945 & 218.440063 & 230.027047 & 241.700159 & 241.767234 & 254.015377 & 261.805675 \\
\hline
  1.0E+02 & 2.7 & 2.7 &  2.7 &  2.8 &  2.7 &  2.9 &   2.9 &  2.7 &  2.7 \\
  1.0E+03 & 2.9 & 2.8 &  2.7 &  3.2 &  2.4 &  3.5 &  3.7 &   2.8 & 2.8 \\
  1.0E+04 & 3.9 & 3.4 &   2.7 &  4.9 & 2.0 & 4.6 &  4.6 & 3.0 &  2.9 \\
  1.0E+05 & 8.0 & 5.8 &   2.7 &   10.2 & 1.7 &  6.1 & 6.4 & 3.8 &  4.4 \\
  1.0E+06 & 9.5 & 9.2 &  3.7 &   12.1 &  2.1 &  8.3 &  9.0 &  5.6 &  7.6 \\
  1.0E+07 & 9.7 & 10.0 & 6.2 &  10.8 &  4.1 &  9.6 &  9.8 &  7.9 &  9.6 \\
  1.0E+08 &  9.9 & 10.0 &  8.6 &  10.2 &  7.8 & 9.9 & 10.0 &   9.4 &  10.0 \\
  1.0E+09 &  10.0 & 10.0 &   9.8 &  10.0 & 9.7 &  10.0 & 10.0 &  9.9 &  10.0 \\
  1.0E+10 &  10.0 & 10.0 &  10.0 &  10.0 &  10.0 &  10.0 & 10.0 &  10.0 &  10.0 \\
\hline
\end{tabular}
\label{critical_density_a}
\end{table*}

\begin{table*}
\centering
\caption{Excitation temperatures for the observed methanol-A lines with a column density of 10$^{12}$ cm$^{-2}$, $T_{\rm kin}$=10~K, a turbulent width of 1 km/s, and a background temperature of 2.73~K.}
\begin{tabular}{lrrrrrr}
\hline
  $n$(H$_2$) &  (2\_1)-(1\_1) &  (2\_0)-(1\_0) &  (2\_-1)-(1\_-1) &  (5\_1)-(4\_1) &  (5\_0)-(4\_0) \\
  $\left[{\rm cm}^{-3} \right]$ &  95.91431&  96.741371 & 97.582798 & 239.746219 & 241.791352\\
\hline
 1.0E+02 &   2.7 &   2.8 &  2.8 &  2.8 &   3.1 \\
  1.0E+03 &   2.9 &  2.9 &  2.9 &  3.3 &  4.0 \\
  1.0E+04 &  3.3 &  4.3 & 4.3 &   4.3 &  5.0 \\
  1.0E+05 &  3.4 &   8.5 &   8.8 &    5.9 &  6.7 \\
  1.0E+06 & 4.3 &   9.6 &   10.4 &   7.6 &   8.9 \\
  1.0E+07 &  7.7 &  9.9 &   9.9 &  9.2 &   9.7 \\
  1.0E+08 &  9.7 &  10.0 &   10.0 &   9.9 &  9.9 \\
  1.0E+09 &  10.0 &   10.0 &   10.0 & 10.0 &   10.0 \\
  1.0E+10 &   10.0 &  10.0 &  10.0 &  10.0 &   10.0 \\
\hline
\end{tabular}
\label{critical_density_e}
\end{table*}

\newpage

\section{RADEX models}
\label{radexgrid}

One of the main assumptions for the LTE rotational diagram analysis and a main unknown in non LTE modelling is the hydrogen gas density of the emitting gas. Here we discuss the different tests and computation made to estimate this gas density. 

We have generated grids of RADEX uniform sphere models and calculated line intensity ratios for values of $n_{\rm H_2}$ between 10$^{3}$ and 10$^{8}$ cm$^{-3}$ and $T_{\rm kin}$ between 10 and 50 K using the python package {\sc spectralradex} (Holdship et al.\ 2021, in prep.). The column density for all the models is set to 5 $\times$ 10$^{13}$ cm$^{-2}$ based on the rotational diagram analysis to keep the emission lines optically thin. In this limit, the intensity of each line is directly proportional to the column density. Therefore, the line ratios are only sensitive to the density of the collision partner H$_2$ and the gas kinetic temperature and not on the assumed column density. The collision partner is H$_2$ with collision rates computed by Rabli \& Flower (2010) for gas between 10 and 200~K. The line width is 1 km/s and the background temperature is 2.73~K.

\begin{figure*}
  \centering 
     \includegraphics[width=3.3in]{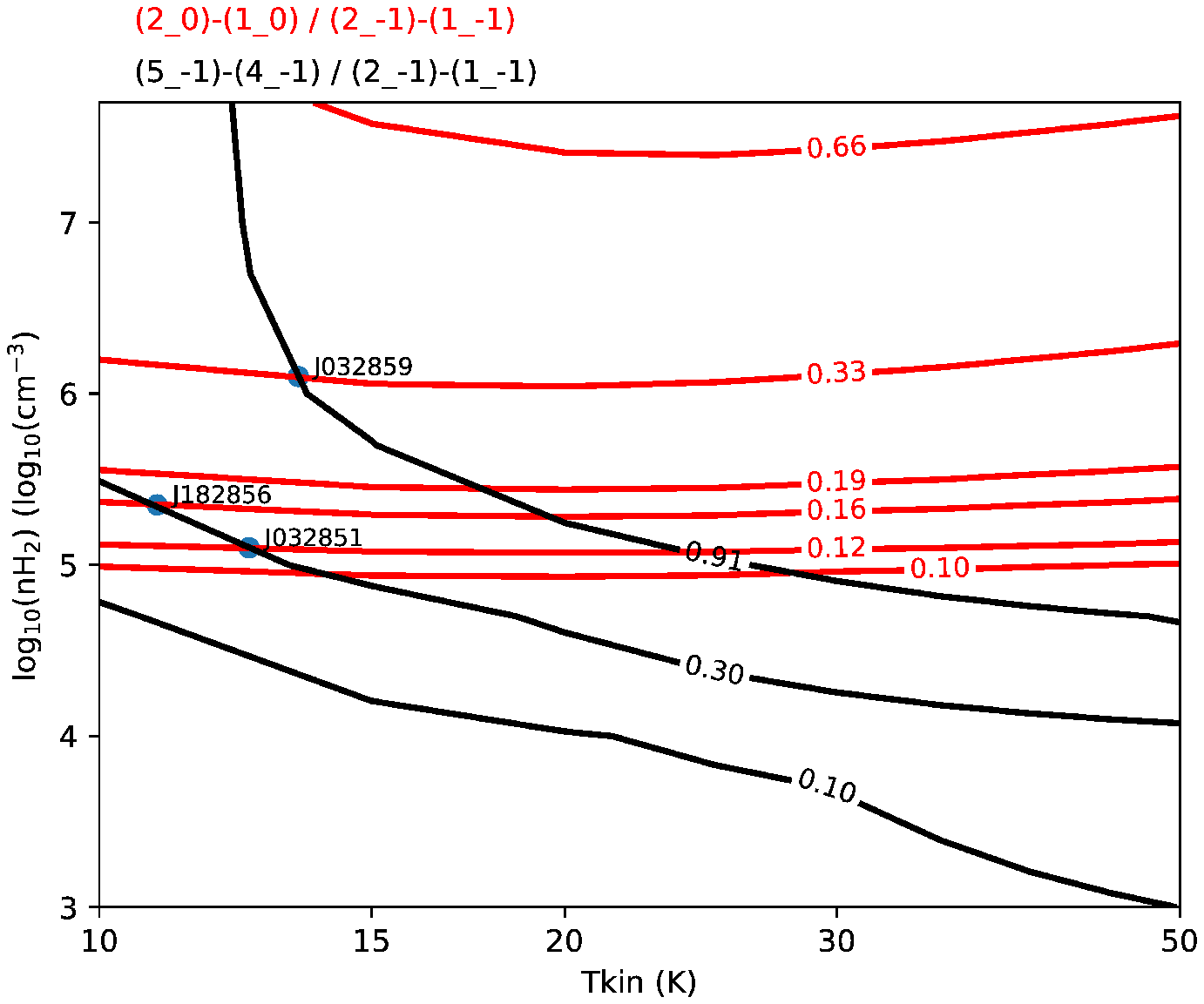}
     \includegraphics[width=3.3in]{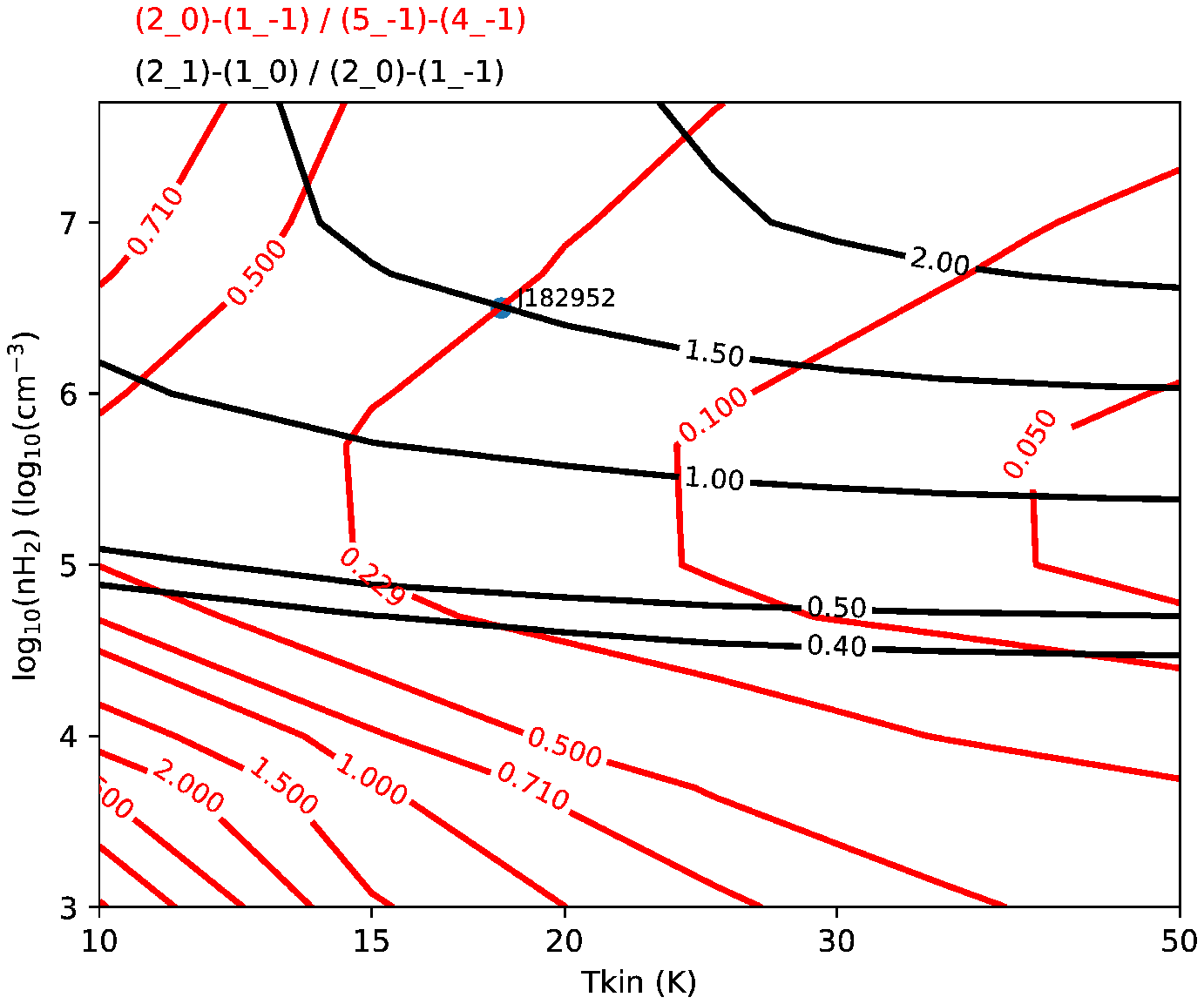}
    \caption{RADEX line ratio diagrams. The left panel includes the ratios 2$_{02}$-1$_{01}$E / 2$_{12}$-1$_{11}$ (indicated by the red lines) and 5$_{15}$-4$_{14}$E / 2$_{12}$-1$_{11}$ (indicated by the black lines) and the right panel includes the ratios 2$_{02}$-1$_{11}$E / 5$_{15}$-4$_{14}$E (red lines) and 2$_{11}$-1$_{01}$E / 2$_{02}$-1$_{11}$E (black lines). Observed line ratios when available are shown by the blue points.}
     \label{radex-diag}     
  \end{figure*}

The diagrams are shown in Fig.~\ref{radex-diag}. We chose methanol-E transitions that are detected in most objects and that span different upper energy levels: 2$_{02}$-1$_{01}$E  2$_{12}$-1$_{11}$, 2$_{02}$-1$_{11}$E,  5$_{15}$-4$_{14}$E, and 2$_{11}$-1$_{01}$E. The first diagram (left panel of Fig.~\ref{radex-diag}) includes the ratios 2$_{02}$-1$_{01}$E / 2$_{12}$-1$_{11}$ and 5$_{15}$-4$_{14}$E / 2$_{12}$-1$_{11}$ and the second diagram (right panel of Fig.~\ref{radex-diag}) includes the ratios 2$_{02}$-1$_{11}$E / 5$_{15}$-4$_{14}$E and 2$_{11}$-1$_{01}$E / 2$_{02}$-1$_{11}$E. 

The first diagram (left panel of Fig.~\ref{radex-diag}) seems to probe low temperature gas for three sources: J032859, J182856, and J032851. The observed line ratios for J032848 and J032838 cannot be put on the diagram suggesting that gas below 10~K is present. Collision data for gas below 10~K is required to test the methanol emission of very cold gas. Although only a handful of sources have sufficient data to be put on the left diagram, the 2$_{02}$-1$_{01}$E / 2$_{12}$-1${_11}$ ratios have values between 0.10 and 0.66 for 12 out of the 14 sources, which indicates that the gas density is between 10$^5$ and 10$^{7}$ cm$^{-3}$. Although high, the derived densities are still below the critical densities derived also from RADEX modelling (Section~\ref{critical_density}.

The second diagram (right panel of Fig.~\ref{radex-diag}) suggests that the gas around J182952 can be as warm as 18~K with a density of $\sim$ 3 $\times$ 10$^{6}$ cm$^{-3}$. Interestingly, J032838 has ratios 2$_{02}$-1$_{11}$E / 5$_{15}$-4$_{14}$E of 0.89 and 2$_{11}$-1$_{01}$E / 2$_{02}$-1$_{11}$E of 0.62, which are inconsistent with the diagram. This suggests that a single density, single temperature RADEX model is insufficient to explain the entire set of lines.

In order to investigate if the T$_{rot}$ (or $T_{ex}$) derived using the rotational diagram is an indicator of the gas kinetic temperature, $T_{kin}$, we have derived column densities of the CH$_{3}$OH 2$_{02}$--1$_{11}$ transition line for J182856 using the non-LTE radiative transfer code RADEX (van der Tak et al. 2007). For a fixed H$_{2}$ number density and line width, we set $T_{kin}$ at 5, 10, 15, 20, 30 K, and varied the column density to match with the observed peak line intensity, $T_{mb}$. This resulted in a range of $T_{ex}\sim$ 4--8 K, and column density in the range of (3--5)$\times$10$^{13}$ cm$^{-2}$. While a two-component analysis is a better fit to the data points for J182856, a single temperature component fit in the rotational diagram results in a $T_{ex}$ of 8.7 K and a column density of 5$\times$10$^{13}$ cm$^{-2}$ (Tables~\ref{Trot};~\ref{Ncol}), which are within the range measured with RADEX. Thus, $T_{kin}$ is close to $T_{ex}$ under the assumption of a single temperature gas component in the rotational diagram. The rotational diagram method could be over-estimating the cold component column density by a factor of $\sim$2 compared to the volume-averaged column density derived using RADEX. A more prominent difference is seen between the $T_{ex}\sim$ 25 K for the warm component determined from the rotational diagram for J182856 and the $T_{ex}\sim$ 4--8 K derived from RADEX. The warm gas component is predicted to have a narrow spatial coverage and would make a much lower fraction by mass compared to the cold component, as discussed in Sect.~\ref{discussion}. Thus it could be diluted in the volume-averaged RADEX calculation while fitting it separately in the rotational diagram could provide a better characterization of the physical conditions in the system (Fig.~\ref{rot-diag}). 

\begin{figure*}
  \centering 
     \includegraphics[width=3in]{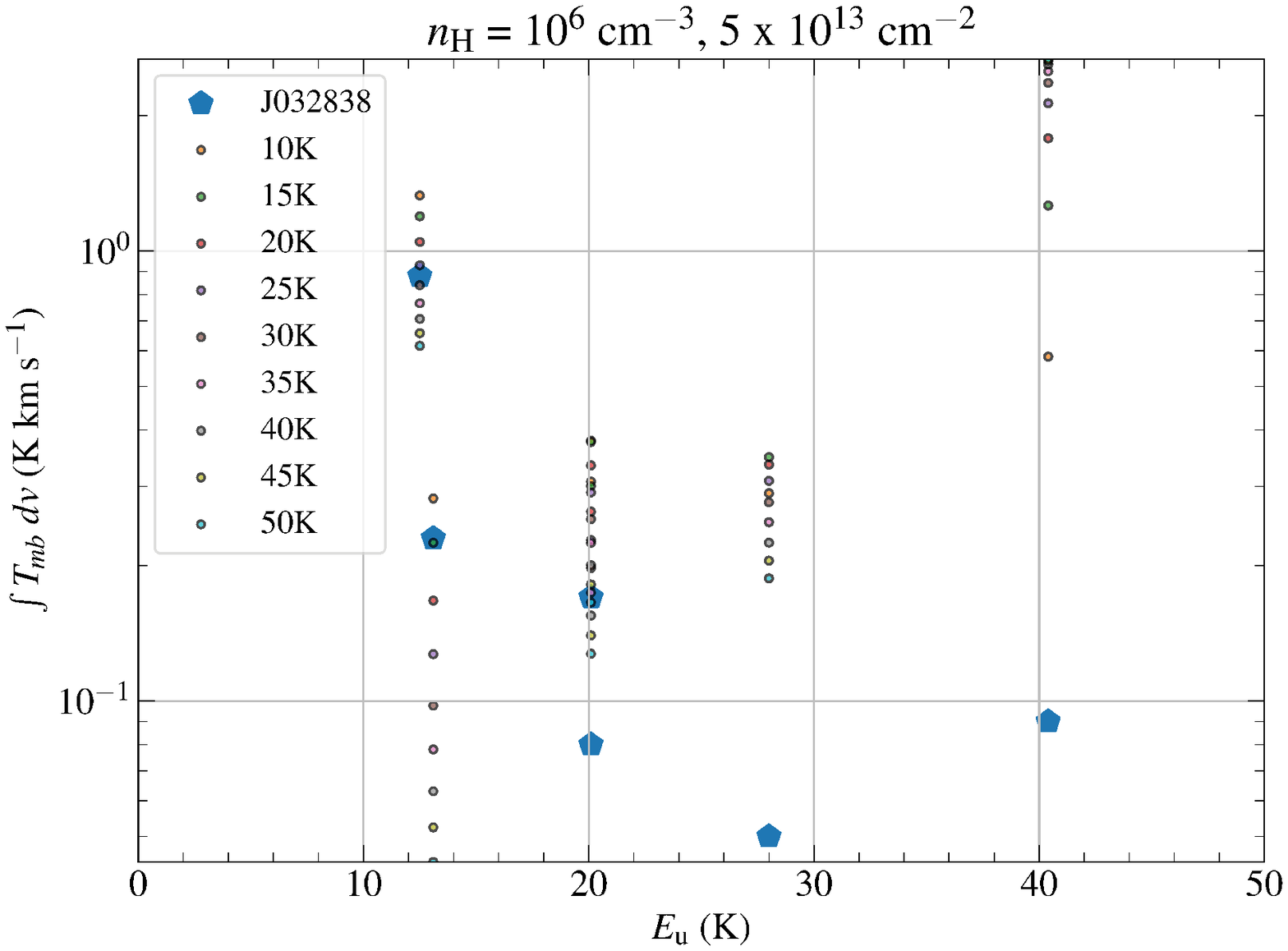}
     \includegraphics[width=3in]{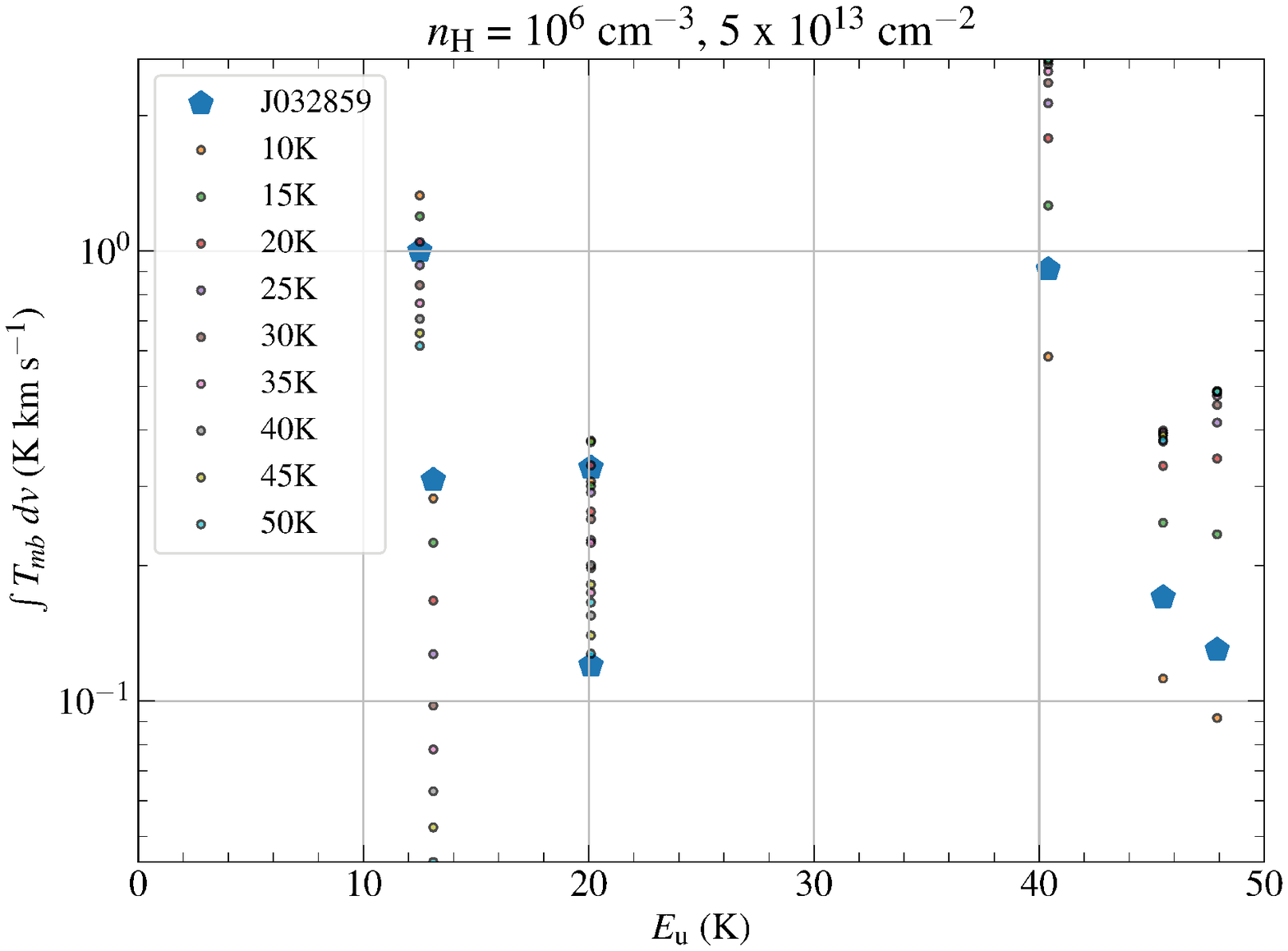}
     \includegraphics[width=3in]{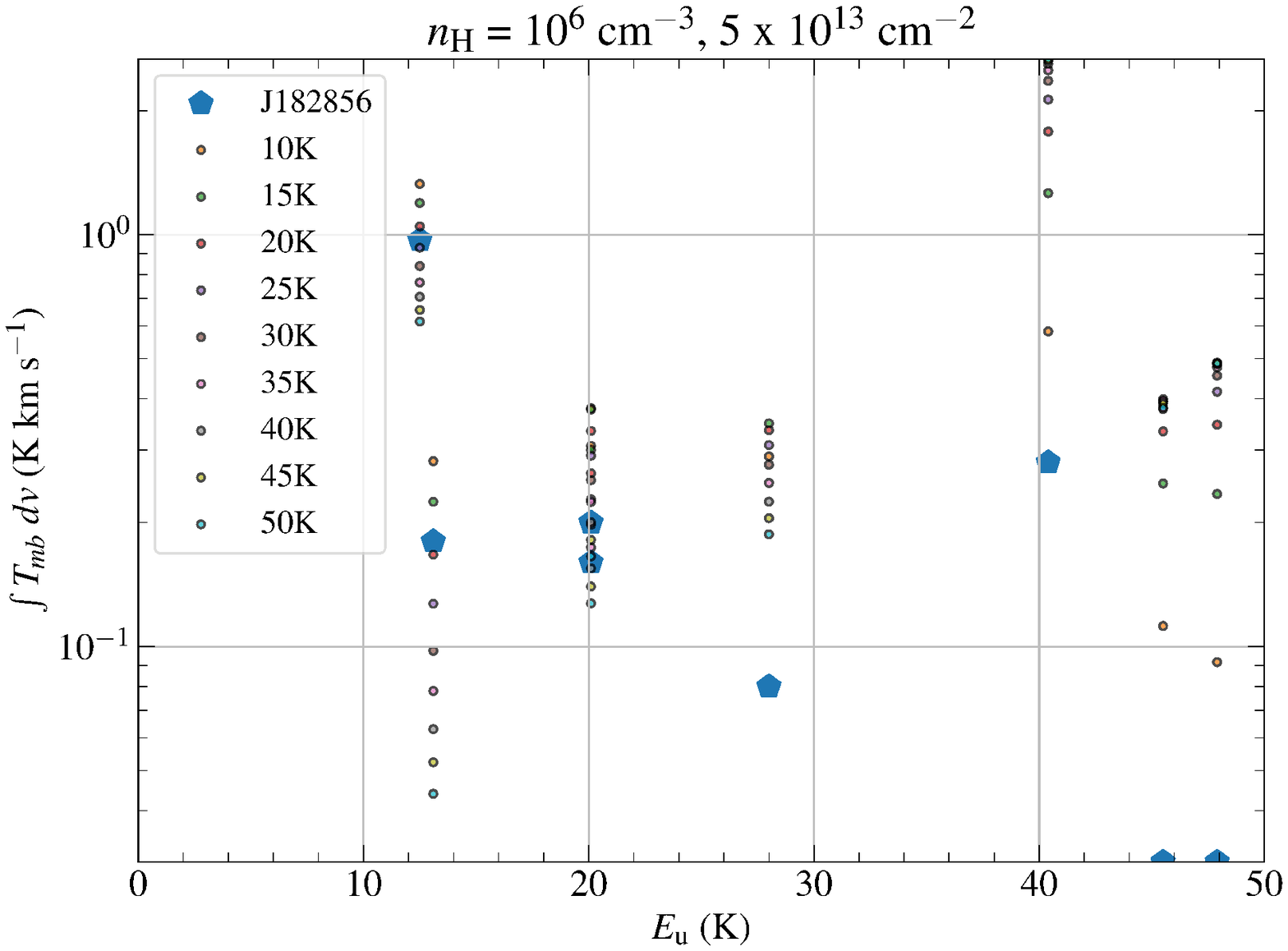}
     \includegraphics[width=3in]{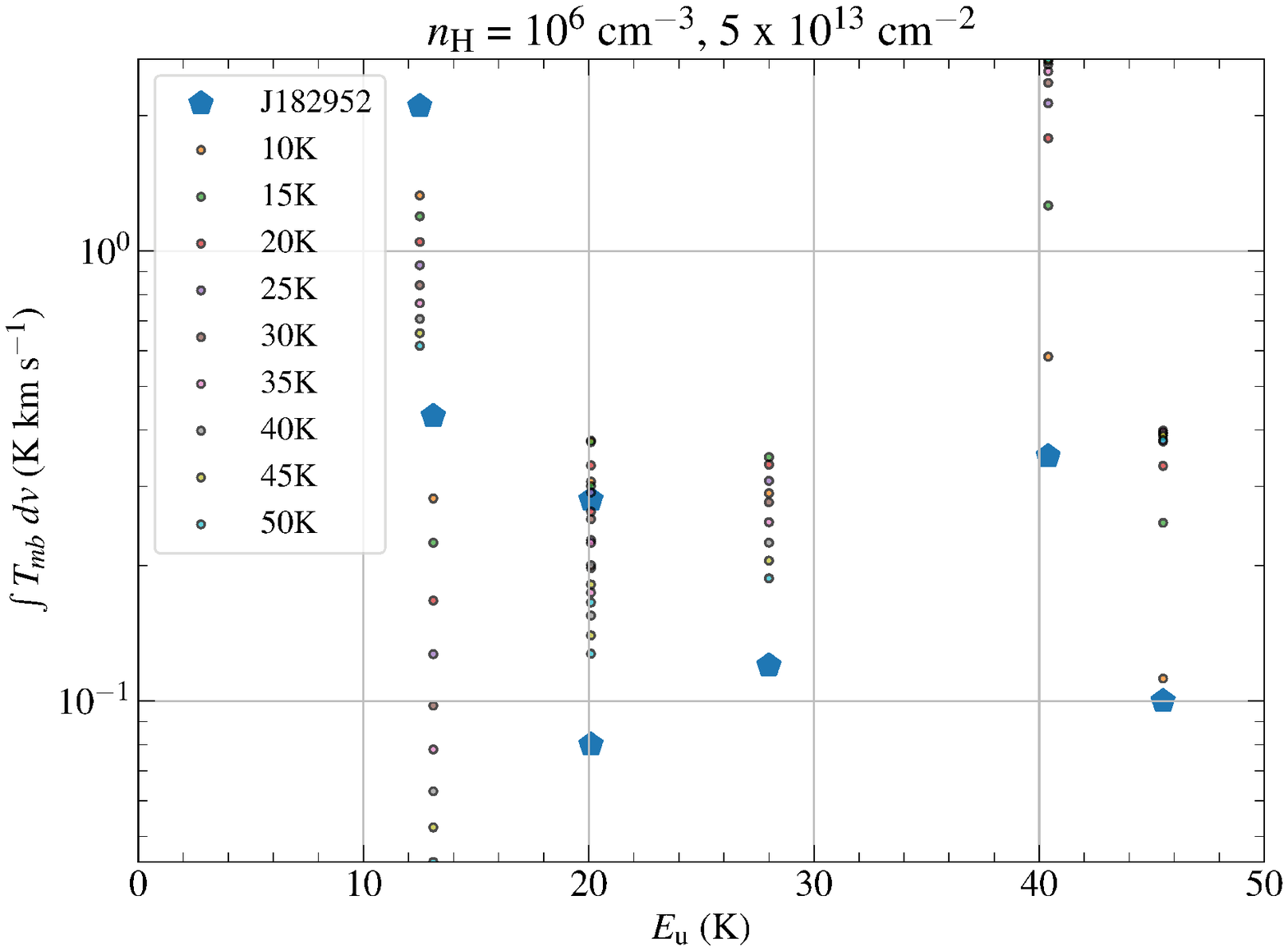}     
   \caption{Comparisons between observed and modelled line intensities using RADEX for different sources. The models (circles) assume a single component with a constant kinetic temperature, a density of 10$^{6}$ cm$^{-3}$, and an average column density for methanol of 5$\times$10$^{13}$ cm$^{-2}$. The temperature is varied between 10 and 50 K in a 5 K increment. The observed values are plotted with a star symbol.}
   \label{radex2}
   \end{figure*}

Figure~\ref{radex2} shows the RADEX model for four sources with sufficient CH$_{3}$OH-E detected transitions. The RADEX analysis uses data in the Leiden-Lambda database, which uses degeneracies without the spin factor. We fixed the density n$_{H}$ at 10$^{6}$ cm$^{-3}$ and the methanol column density at 5$\times$10$^{13}$ cm$^{-2}$. We vary the gas temperature between 10 and 50 K in steps of 5 K. Apart from  J032859, the single density, temperature, column density model cannot fit the data for the targets. All the other sources seem to require more than one component. Interestingly, the RADEX modelling derived parameters for J032859 (T=10-15 K, 10$^{6}$ cm$^{-3}$, N(CH$_{3}$OH) = 5$\times$10$^{13}$ cm$^{-2}$) are consistent with the rotational diagram analysis (Fig.~\ref{rot-diag}; Tables~\ref{Trot};~\ref{Ncol}).

\newpage

\section{Comparison with protostars}
\label{proto-2}

In Fig.~\ref{proto}, we have compared the CH$_{3}$OH column densities and $T_{rot}$ for the proto-BDs with low-mass protostars. The data for the low-mass protostars is also from IRAM 30m observations presented in \"{O}berg et al. (2014) and Graninger et al. (2016). We have tested our rotational diagram methodology by re-analysing the values in Table 1 in the erratum of Graninger et al. (2016) for low-mass protostars that have more than 2 detected transitions. As shown in Table~\ref{gran}, our re-analysis of the Graninger et al. (2016) data is consistent within the numerical uncertainties (for example the partition function value is interpolated) with their calculations. Therefore, we can confidently compare our derived column densities and $T_{rot}$ for the proto-BDs with their measurements for the protostars. 

We note that we have taken the degeneracies and the partition functions from the CDMS database that includes the spin degeneracies in their upper level degeneracy values. Compared to the Leiden-Lambda database, the values we used in rotational diagrams are indeed a factor 4 lower in the CDMS database. Since the partition function values are also taken from the CDMS database, the extra factor cancels-out in the g$_{up}$/partition function ratio. Therefore, the values plotted in the y-axis of the rotational diagrams do not affect our computation.

\begin{table*}
\centering
\caption{Re-analysis of Graninger et al. (2016) data}
\label{gran}
\begin{threeparttable}
\begin{tabular}{llllllll} 
\hline
  & \multicolumn{2}{c}{Our analysis}	 & \multicolumn{2}{c}{Graninger et al. (2016; Erratum Table 1)} 	 	\\
Object & N(CH$_{3}$OH) (10$^{13}$ cm$^{-2}$) & T$_{rot}$ (K) & N(CH$_{3}$OH) (10$^{13}$ cm$^{-2}$) & T$_{rot}$ (K) 		\\

\hline
IRAS 03245+3002	& 	1.5 		&	7.4 	&	1.5	&	7.4	\\
B1-c                          	&      	1.7        	&      7.2   	&	1.7	&	7.2	\\
L1455 IRS3             	&     	1.4         	&      4.4   	&	1.5	&	4.4	\\
IRAS 23238+7401    	&      	2.0         	&      5.3  	&	2.2	&	5.4	\\
L1455 SSM1             &     	1.4        	&     	5.7	&	1.5	&	5.7	\\
L1014 IRS                 &     	0.8         	&      	4.7 	&	0.8	&	4.8	\\
\hline
\end{tabular}
\end{threeparttable}
\end{table*}


\bsp	
\label{lastpage}
\end{document}